\documentclass[fleqn, usenatbib]{mnras}
\pdfoutput=1
\usepackage{pbox}
\usepackage{physics}
\usepackage[english]{babel}
\usepackage[utf8x]{inputenc}
\usepackage[T1]{fontenc}
\usepackage{txfonts}
\usepackage{aecompl}
\usepackage{multirow}
\usepackage{rotating}
\usepackage{float}
\usepackage{amsmath}
\usepackage{amssymb} 
\usepackage{graphicx}
\usepackage[colorinlistoftodos]{todonotes}
\usepackage{pdflscape}
\begin{document}
\label{firstpage}
%\title{Multiwavelength variability and correlation studies of Mrk\,421 during historically low X-ray and $\gamma$-ray activity in 2015--2016}
\title[Mrk\,421 during low activity in 2015--2016]{Multiwavelength variability and correlation studies of Mrk\,421 during historically low X-ray and $\gamma$-ray activity in 2015--2016}
%%%%%%%%%%%%%%%%%%%%%%%%%%%%%%%
%%%% The author list: %%%%%%%%%
%%%%%%%%%%%%%%%%%%%%%%%%%%%%%%%
\author[V.~A.~Acciari~et~al.]{\parbox{\textwidth}{\Large{
MAGIC Collaboration: V.~A.~Acciari$^{1}$,
S.~Ansoldi$^{2,5}$,
L.~A.~Antonelli$^{3}$,
%A.~Arbet Engels$^{4}$,
K.~Asano$^{5}$,
%D.~Baack$^{6}$ 
A.~Babi\'c$^{7}$,
%B.~Banerjee$^{24}$\thanks{\href{mailto:biswajit.banerjee@saha.ac.in}{biswajit.banerjee@saha.ac.in}},
B.~Banerjee$^{24}$\thanks{Corresponding authors: Biswajit~Banerjee, Pratik~Majumdar, David~Paneque,  and Tomislav~Terzi\'c (e-mail:\href{mailto: contact.magic@mpp.mpg.de}{contact.magic@mpp.mpg.de})},
%\thanks{Corresponding authors: Biswajit~Banerjee, Pratik~Madumdar, David~Paneque and Tomislav~Terzi\'c (e-mail:\href{mailto: contact.magic@mpp.mpg.de}{contact.magic@mpp.mpg.de})},
A.~Baquero$^{8}$,
U.~Barres de Almeida$^{9}$,
J.~A.~Barrio$^{8}$,
J.~Becerra Gonz\'alez$^{1}$
% ...
}
\newline
\emph{\normalsize Affiliations are listed at the end of the paper}
}}
%%%%%%%%%%%%%%%%%%%%%%%%%%%%%%%
%%%% The author list: %%%%%%%%%
%%%%%%%%%%%%%%%%%%%%%%%%%%%%%%%
% These dates will be filled out by the publisher
\date{Accepted 2020 November 17. Received 2020 November 6; in original form 2020 June 30}
% Enter the current year, for the copyright statements etc.
\pubyear{2020}
%\linenumbers
% Don't change these lines

\pagerange{\pageref{firstpage}--\pageref{lastpage}}
\maketitle

\clearpage

% Abstract of the paper
\begin{abstract}
We report a characterization of the multi-band flux variability and correlations of the nearby (z=0.031) blazar Markarian\,421 (Mrk\,421) using data from Mets{\"a}hovi, {\it Swift}, {\it Fermi}-LAT, MAGIC, FACT and other collaborations and instruments from November 2014 till June 2016. Mrk\,421 did not show any prominent flaring activity, but exhibited periods of historically low activity above 1\,TeV (F$_{>1\mathrm{TeV}}<$ 1.7$\times$10$^{-12}$\,ph\,cm$^{-2}$\,s$^{-1}$) and in the 2-10\,keV (X-ray) band (F$_{2-10\,\mathrm{keV}}<$3.6$\times$10$^{-11}$\,erg\,cm$^{-2}$\,s$^{-1}$), during which the {\it Swift}-BAT data suggests an additional spectral component beyond the regular synchrotron emission.The highest flux variability occurs in X-rays and very-high-energy (E$>$0.1\,TeV) $\gamma$-rays, which, despite the low activity, show a significant positive correlation with no time lag. The HR$_{keV}$ and HR$_\mathrm{TeV}$ show the {\em harder-when-brighter} trend observed in many blazars, but the trend flattens at the highest fluxes, which suggests a change in the processes dominating the blazar variability.
Enlarging our data set with data from years 2007 to 2014, we measured a positive correlation between the optical and the GeV emission over a range of about 60 days centered at time lag zero, and a positive correlation between the optical/GeV and the radio emission over a range of about 60 days centered at a time lag of  $43^{+9}_{-6}$ days.This observation is consistent with the radio-bright zone being located about 0.2~parsec downstream from the optical/GeV emission regions of the jet. The flux distributions are better described with a LogNormal function in most of the energy bands probed, indicating that the variability in Mrk\,421 is likely produced by a multiplicative process.%\newline
%\textbf{Keywords:}galaxies: active -- BL Lacertae objects: individual: Mrk\,421 -- methods: data analysis -- methods: observational -- radiation mechanisms: non-thermal
\end{abstract}


\section{Introduction}\label{sec:intro}
Markarian 421 (Mrk 421), located
at a redshift z = 0.031
\citep{1975ApJ...198..261U}, is an extensively studied \,TeV source. It was first detected as a TeV emitter by the Whipple telescope in 1992 \citep{1992Natur.358..477P}. 
\textcolor{black}{Mrk\,421 is a BL Lac 
\textcolor{black}{type}
object whose broadband spectrum is characterized by}
a double-peak structure, where the first peak originates from the synchrotron radiation by leptons inside the jet. The origin of the second 
peak is 
believed
to be synchrotron self Compton (SSC) emission \citep{1996ApJ...461..657B,2019ApJ...874...47V}, although hadronic scenarios \citep[e.g.][]{1993A&A...269...67M,2003APh....18..593M} have also been used to explain the high-energy emission of Mrk\,421 \citep[e.g.][]{2011ApJ...736..131A,2016APh....80..115P}. \par

The light curve (LC) of Mrk\,421 is highly 
variable, and it has gone into outburst several times in all bands (radio to TeV) in which 
\textcolor{black}{it}
is observed.  During an outburst, the TeV emission can vary on \textcolor{black}{sub-hour} timescales 
\citep{Gaidos,2020ApJ...890...97A}.
Many attempts have been made to trace the ongoing physical processes inside the jet. The majority of the simultaneous multiwavelength (MWL) observations were performed during flaring activity, 
when the VHE $\gamma$-ray flux of Mrk\,421 exceeded the flux of the Crab Nebula\footnote{The flux of the Crab Nebula, used in this work for reference purposes, is retrieved from \citet{2015JHEAp...5...30A} } (there after 1\,Crab) by 2--3 times,
which is the standard candle for ground-based $\gamma$-ray instruments \citep{1995Malcomb, Whipple19971, Whipple2, 2001W, VERITAS200608, 2015A&A...578A..22A}. Only a handful of attempts have been made to study the broad-band emission of Mrk~421
during non-flaring episodes.
For instance, \citet{2009ApJ...695..596H} report a very detailed study using MWL observations of Mrk\,421 that were not triggered by flaring episodes. But the VHE $\gamma$-ray activity of Mrk\,421 during this observing campaign (mostly in 2006) was twice the typical VHE $\gamma$-ray activity of Mrk\,421, which, according to \citet{14Y}, is half the flux of the Crab \textcolor{black}{N}ebula. Moreover, the data from \citet{2009ApJ...695..596H} actually contained two flaring episodes, 
when the flux from Mrk\,421 was higher than double that of the Crab \textcolor{black}{N}ebula for several days. 
On the other hand,
\citet{2015A&A...576A.126A} performed a study with the data from a MWL campaign in 2009, when Mrk\,421 was at its typical VHE $\gamma$-ray flux
level, and \citet{2016ApJ...819..156B} reported an extensive study with data from 2013 January-March, when Mrk\,421 showed very low-flux at X-ray and VHE.

One of the key aspects that has been investigated 
\textcolor{black}{in}
several past MWL campaigns 
on Mrk\,421 is the correlation between X-rays and VHE $\gamma$-rays. A direct correlation between these two \textcolor{black}{wave-bands} has been reported in several articles \citep[e.g.][]{1995ApJ...449L..99M,1996ApJ...472L...9B,2007ApJ...663..125A,  2008ApJ...677..906F, 2009ApJ...691L..13D, 2011ApJ...736..131A, 2011ApJ...738...25A, 2013PASJ...65..109C, 2015A&A...578A..22A,2016ApJS..222....6B}. However, almost all of these studies 
were carried out during flaring activity.  
\textcolor{black}{There are only two cases which report such a correlation during low activity without flares: \citet{2015A&A...576A.126A} measured the VHE/X-ray correlation with a marginal significance of 3\,$\sigma$, and \citet{2016ApJ...819..156B}, report the VHE/X-ray correlation with high significance despite the low-flux in X-ray and VHE $\gamma$-rays thanks to the very high sensitivity 
\textit{NuSTAR} and stereoscopic data from MAGIC and VERITAS.} 
The emission among the other energy bands appears to be less correlated than that for the X-ray and VHE bands, and \cite{1995ApJ...449L..99M}, \cite{2007ApJ...663..125A}, \cite{2013PASJ...65..109C} and \cite{2016ApJ...819..156B} reported 
no correlation between the optical/UV and X-rays and the optical/UV and TeV bands
during low states of the source. 

Using data taken in 2009, \citet{2015A&A...576A.126A} found a negative correlation between the optical/UV and the X-ray emission. 
The cause of this correlation was the long-term trend in the optical/UV and in X-ray activity; while the former increased during the entire observing campaign, the latter systematically 
decreased. This correlation 
was statistically significant when considering only the 2009 data set but, using data from 2007 to 2015, \citet{2017MNRAS.472.3789C} did not measure any overall correlation between the optical and the X-ray emission.
On the other hand, \citet{2017MNRAS.472.3789C} did find a correlation between the GeV and the optical emission.
This correlation study used the discrete correlation function  \citep[DCF,][]{1988ApJ...333..646E}
and identified a peak 
\textcolor{black}{with a DCF value of}
about 0.4, centered at zero time lag
\textcolor{black}{($\tau$)}
but extending over many tens of days to positive and negative values. However, the statistical significance of this correlation was not reported.
As for the radio bands, the 5\,GHz radio outburst lasting a few days in 2001 February/March, and occurring at approximately the same time as an X-ray and VHE flare, was reported by \cite{2003A&A...410..101K} as evidence of correlation 
without any time lag between the radio and X-ray/VHE emission in Mrk\,421.  But the statistical significance of this positive correlation was not reported. As there were many similar few-day X-ray and VHE flares throughout 2001, but only a single radio flare, the claimed correlation may simply be chance coincidence.  Using the low activity data taken over almost the whole year 2011, \cite{2014A&A...571A..54L} reported a marginally significant 
($\leq 3 \sigma$) correlation between radio very long baseline interferometry (VLBI) and GeV $\gamma$-rays 
for a range of about $\pm$30 days centered at
\textcolor{black}{$\tau$=0.}
\citet{2014MNRAS.445..428M}, however, reported a positive correlation between the GeV and radio emission 
\textcolor{black}{at $\tau \sim$ 40 days.}
However, the correlation reported there 
was only at 2.6\,$\sigma$ significance , and was strongly affected by the large $\gamma$-ray and radio flares from July and September 2012, respectively \citep[][]{2014MNRAS.445..428M}.
\par

Overall, the broadband emission of Mrk\,421 is complex, and a dedicated correlation analysis over many years will be necessary in order to properly characterize it.
It is relevant to evaluate whether the various trends or peculiar behaviours, sometimes reported in the literature with only marginal significance, are repeated over time, and also to distinguish the typical behaviour from the 
%rare/
sporadic events. For the latter, it is important to collect multi-instrument data that are not triggered or motivated by flaring episodes. A better understanding of the low-flux state will not only provide meaningful constraints on the model parameters related to the dynamics of the particles inside the jet, but also will \textcolor{black}{provide a baseline} for explaining the high-state activity of the source. 

\par

The study presented in this paper focuses on the extensive MWL data set collected during the campaigns in the years 2015 and 2016, when Mrk\,421 showed low activity in both X-rays and VHE $\gamma$-rays, and no prominent flaring activity \textcolor{black}{($>$2~Crabs for several days)} was measured.
We characterize the variability using the normalized excess variance of 
the flux \citep{2003MNRAS.345.1271V}
for the X-ray and\,TeV bands split into two hard bands ($2-10$\,keV and $>$1\,TeV) and two soft bands ($0.3-2$\,keV and $0.2 - 1$\,TeV). We use these bands to compute HR$_{keV}$ and HR$_{TeV}$ to 
evaluate the harder-when-brighter
\textcolor{black}{behaviour} of the source. Using this data set, we present a detailed correlation study for different 
combinations
of wave-bands. 
In order to better evaluate the correlations among the energy bands with lower amplitude variability and longer variability timescales, we complemented the 2015--2016 data set with data from previous years (from 2007 to 2014). A fraction of these data had already been published \citep[][]{2012A&A...542A.100A,2015A&A...578A..22A,2016A&A...593A..91A,2016ApJ...819..156B}, 
and the rest were specifically collected and analyzed for the study presented here.

This paper is arranged in the following way: in   \textcolor{black}{Section}  \ref{sec:data_analysis} we describe
the instruments that participated in this campaign, the data analysis methods used for each energy band, and 
\textcolor{black}{and a summary of the}
observed MWL data. In   \textcolor{black}{Section} \ref{sec:Summary1516}, we discuss the main characteristics of the MWL light curves from the 2015--2016 campaign. In   \textcolor{black}{Section} \ref{sec:var} and \ref{sec:corr}, we discuss
the different aspects of 
the MWL variability and correlation 
study that we carried out. In   \textcolor{black}{Section} \ref{sec:typicalstate} we characterize the flux distributions in the different wave-bands, and in   \textcolor{black}{Section} \ref{sec:con} we discuss and summarize the main observational results from our work.

\begin{table*}
\centering
\begin{tabular}{ c c c c c c} 
 \hline \hline
Observation &  \multicolumn{3}{c}{ 3NN }& 4NN      & \multirow{2}{*}{Moon filter}\\
conditions  & Low-Moon & Moderate-Moon & High-Moon   & Low-Moon &                             \\  \hline
Low-zenith $(5^{\circ}-35^{\circ})$ & $\sim$30.0 hrs & $\sim$10.0 hrs & $\sim$6.0 hrs & $\sim$7.0 hrs &    \\ 
Medium-zenith $(35^{\circ}-50^{\circ})$ & $\sim$4.0 hrs & $\sim$1.0 hrs & $\sim$1.0 hrs & $\sim$2.0 hrs & $\sim$3.0 hrs\\  
High-zenith $(50^{\circ}-62^{\circ})$ & 1.0 hrs & &  2.0 hrs    &  \\ \hline
\end{tabular}
\caption{Observation conditions at VHE $\gamma$-rays with the MAGIC telescopes during the 2015--2016 campaign. Apart from the standard data (3NN), a subset was taken without a coincidence trigger and a 4NN single-telescope trigger logic. See Section \ref{sec:MAGICdata} for details.}\label{table:VHE}
\end{table*}

\begin{figure*}
    \includegraphics[height=7.5cm, width=0.9\linewidth]{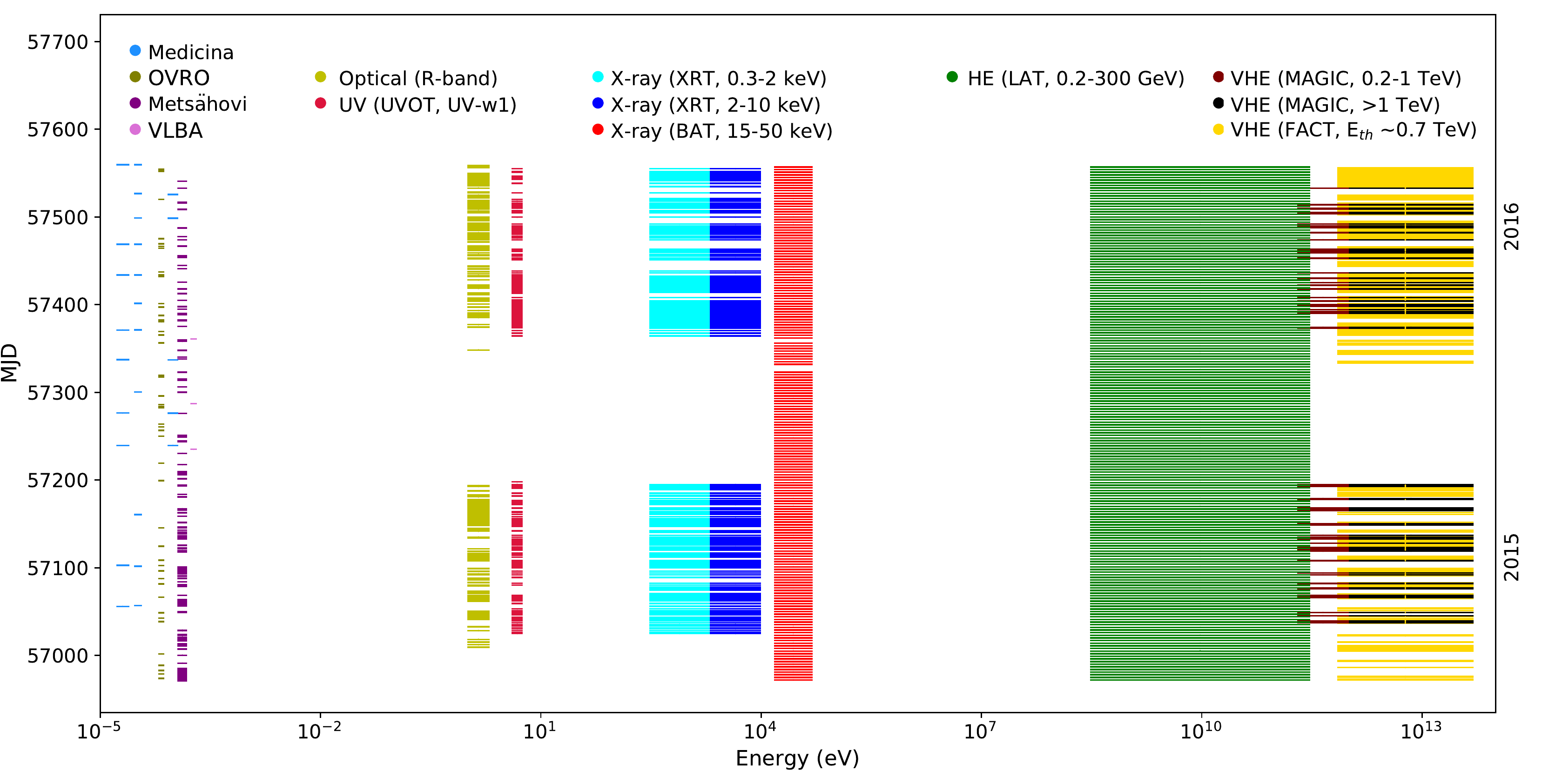}
    \caption{Multi-instrument temporal coverage of Mrk\,421 during the 2015--2016 observation campaign.}
    \label{fig:MYMWLCov}
\end{figure*}

\section{Observations and data analysis}\label{sec:data_analysis}

The temporal and energy coverage provided by the 
  \textcolor{black}{MWL}
observations from the two-year period reported in this paper, i.e., from 2014 November to 2016 June, is depicted in Fig.~\ref{fig:MYMWLCov}. 
We note that there is a period of about 6 months (approximately from 2015 June to 2015 December) when the Sun is too close to Mrk\,421, which prevents observations at optical and VHE $\gamma$-rays (e.g. with MAGIC and FACT), and even observations at soft X-rays with \textit{Swift}. During this half year period,  Mrk\,421
can
only be observed at radio, hard X-rays and HE $\gamma$-rays, as shown in Fig.~\ref{fig:MYMWLCov}.
In the subsections below, we discuss the instrumentation and data analyses used to characterize the emission of Mrk\,421 across the electromagnetic spectrum, from radio to VHE $\gamma$-rays.

\subsection{Radio}
\label{sec:RadioAnalysisDescription}

The study presented here makes use of radio observations from the single-dish radio 
telescopes at the Mets{\"a}hovi Radio Observatory,
which operates at 37\,GHz, at the Owens Valley Radio Observatory (OVRO, at 15\,GHz), and the Medicina radio telescope, which provides multi-frequency data at 5\,GHz, 8\,GHz, and 24\,GHz.  The data from OVRO 
were retrieved directly from the web page of the instrument team\footnote{\url{http://www.astro.caltech.edu/ovroblazars/index.php?page=home}}, while the data from Mets{\"a}hovi and Medicina 
were provided to us directly by
the instrument team. Mrk\,421 is a point source for all of these instruments, and hence the measurements represent an integration of the full source extension, which has a larger size than the emission that dominates the 
\textcolor{black}{highly}
variable X-ray and $\gamma$-ray emission, and possibly also the optical emission.
 Details of the observation and data analysis strategies from OVRO and Medicina are reported in \citet{2011ApJS..194...29R} and \citet{2020MNRAS.492.2807G}, respectively.
As for Mets{\"a}hovi, the detection limit of the telescope at 37\,GHz is 
\textcolor{black}{in}
the order of 0.2\,Jy under optimal conditions. The flux density scale is set by observations of DR~21, and the sources NGC~7027, 3C~274, and 3C~84 are used as secondary calibrators. The error estimate on the Mets{\"a}hovi flux density includes the   contributions from the rms measurement and the uncertainty 
in the absolute calibration. A detailed description of the data reduction and analysis is given in \citet{1998A&AS..132..305T}. In this particular analysis, as is done in most analyses, the measurements that do not survive a quality control (usually due to unfavourable weather) are discarded semi-automatically. In the final data reduction, the measurements are checked manually, which includes ruling out bad weather conditions or other environmental effects such as, e.g., a rare but distinct flux density increase caused by aircraft in the telescope beam. Additionally, the Mets{\"a}hovi team also checked that the general flux levels are consistent for adjacent measurements (i.e. other sources observed before and after the target source).

The study also uses the Very Long Baseline Array (VLBA) total and polarized intensity images of Mrk~421 at 43\,GHz obtained within the VLBA-BU-BLAZAR program of $\sim$monthly monitoring of a sample of $\gamma$-ray blazars\footnote{\url{http://www.bu.edu/blazars/VLBAproject.html}}. 
The source was observed in a short-scan mode along with $\sim$30
other blazars over 24~hrs, with $\sim$45~min on the source. A detailed description of the observations and data reduction can be found in \cite{2017ApJ...846...98J}.
The analysis of  the polarization properties was based on Stokes Q and U parameter images obtained in the same manner as described in \cite{2007AJ....134..799J}.

\subsection{Optical}

In this paper, we use only R-band photometry.
These optical data were obtained with the KVA telescope (at the Roque de los Muchachos), ROVOR, West Mountain Observatory, and the iTelescopes network.  The stars reported in \citet{1998A&AS..130..305V} were used for calibration, and the coefficients given in \citet{2011ApJ...737..103S} were used to correct for the Galactic extinction. The contribution from the host galaxy in the R band, which is about 1/3 of the measured flux, was determined using \citet{2007A&A...475..199N}, and subtracted from the values reported in 
Fig. \ref{fig:MWLC}.
Additionally, a point-wise fluctuation of 2 per cent on the measured flux was added in quadrature to the statistical uncertainties in order to account for potential day-to-day differences in observations with any of the instruments.

\subsection{
Neil Gehrels \textit{Swift} Observatory}
This study uses the following instruments on board the Neil Gehrels \textit{Swift}  
Observatory \citep{2004ApJ...611.1005G}:

\subsubsection{UVOT}

The {\it Swift} UV/Optical Telescope (UVOT; \citealt{2005SSRv..120...95R}) was used to perform observations in the UV range (with the filters W1, M2, and W2).
For all of the observations, 
\textcolor{black}{data were analyzed using}
aperture photometry for all filters using the standard UVOT software distributed within the HEAsoft package (version~6.16), and the calibration files from CALDB version 20130118.  The counts were extracted from an aperture of 5 arcsec radius, and converted to fluxes using the standard zero points from \citet{2011AIPC.1358..373B}. Afterwards, the fluxes were dereddened using $E(B-V)=0.012$ \citep{2011ApJ...737..103S} with $A_{\lambda}/E(B-V)$ ratios calculated using the mean Galactic interstellar extinction curve reported in \citet{1999PASP..111...63F}. Mrk\,421\ is on the ``ghost wings'' \citep{2006PASP..118...37L} 
of the nearby star 51\,UMa in many of the observations, and hence the background 
had to be estimated from two circular apertures of 16 arcsec radius off the source, symmetrically with respect to Mrk\,421,  excluding stray light and shadows from the support structure.

\begin{landscape}
\begin{figure}
\centering
\includegraphics[width=\linewidth, height=16cm]{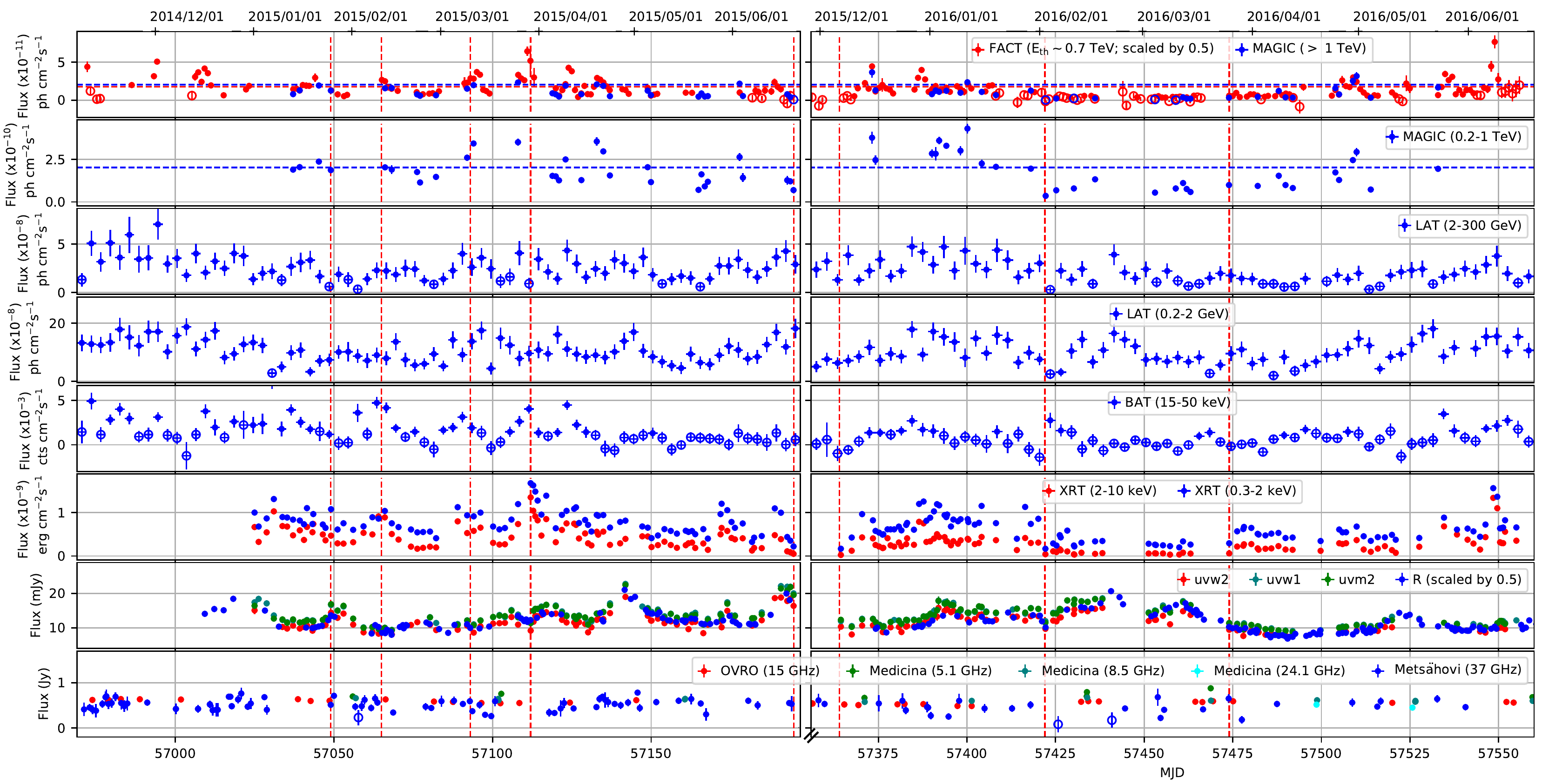}
\caption{From top to bottom: light curves of MAGIC, FACT, {\it Fermi}-LAT, {\it Swift}-BAT, -XRT, and -UVOT, optical R-band telescopes (KVA telescope at the Roque de los Muchachos, ROVOR, Brigham, New Mexico Skies and the  iTelescopes), Mets{\"a}hovi, OVRO, and Medicina from 2014 November to 2016 June. A 3-day temporal bin has been used for the HE $\gamma$-ray fluxes from \textit{Fermi}-LAT and hard X-ray fluxes from \textit{Swift}-BAT. For all of the other energy bands, individual single night observations are depicted. The vertical dashed red lines represent interesting flux variations discussed in   \textcolor{black}{Section}  \ref{sec:Summary1516}. The LCs from FACT and R-band are scaled by a factor of 0.5 for better visibility. In the top panel, horizontal blue and red lines represent \textcolor{black}{the flux of the Crab Nebula} above 1\,TeV and above 0.7\,TeV energy (the latter one scaled by a factor of 0.5), respectively. The horizontal line in the second panel from the top represents the \textcolor{black}{the flux of the Crab Nebula} in the $0.2-1$\,TeV energy band. The open markers in the MAGIC ($>$1\,TeV), FACT, LAT, BAT, and Mets\"{a}hovi light curves are used to display flux measurements with a relative error larger than 0.5.}
\label{fig:MWLC}
\end{figure}
\end{landscape}

\subsubsection{XRT}
The {\it Swift}  X-ray Telescope (XRT; \citealt{2005SSRv..120..165B}) was used to perform observations in the energy range from 0.3\,keV to 10\,keV. All of the {\it Swift}-XRT observations were taken in the Windowed Timing (WT) readout mode. The data were processed using the XRTDAS software package (v.3.2.0), which was developed by the ASI Space Science Data Center (SSDC) and released by HEASARC in the HEASoft package (v.6.19). The event files were calibrated and cleaned with standard filtering criteria with the {\tt xrtpipeline} task using the
calibration files available from the {\it Swift}/XRT CALDB (version 20160609). 
For each observation, the X-ray spectrum was extracted from the summed cleaned event file. Events for the spectral analysis were selected within a circle of 20-pixel ($\simeq$46 arcsec) radius, which encloses about 90 per cent of the \textcolor{black}{point-spread function (PSF),} centered at the source position. The background was extracted from a nearby circular region of 40-pixel radius. The ancillary response files (ARFs) were generated with the {\tt xrtmkarf} task applying corrections for PSF losses and CCD defects using the cumulative exposure map.\par

Before the spectral fitting, the $0.3-10$\,keV source spectra were binned using the {\tt grppha} task to ensure a minimum of 20 counts per bin. The spectra were modeled in XSPEC using power-law and 
log-parabola models that include a photoelectric absorption by a fixed column density estimated to be $N_{\rm H}=1.92\times10^{20}$\,cm$^{-2}$ \citep{2005A&A...440..775K}. The 
log-parabola model typically fits the data better than the power-law model (though statistical improvement is marginal in many cases), and was therefore used to compute the X-ray fluxes in the energy bands 
$0.3-2$\,keV
and $2-10$\,keV, which are reported in Fig.~\ref{fig:MWLC}.

\subsubsection{BAT}
\label{sec:BAT}
A daily average flux in the energy range $15-50$\,keV 
measured by the
\textit{Swift}-BAT 
instrument
was obtained from the BAT website\footnote{\url{http://heasarc.nasa.gov/docs/swift/results/transients/}}. The detailed analysis procedure can be found in \citet{2013ApJS..209...14K}. 
The BAT fluxes related to time intervals of multiple days reported in this 
paper were obtained by performing a standard weighted average of the BAT daily fluxes, which is exactly the same procedure used by the BAT team to obtain the daily fluxes from the orbit-wise fluxes.

\subsection{\textit{Fermi}-LAT}
\label{sec:FermiLAT}
The GeV $\gamma$-ray fluxes related to the 2015--2016 observing campaigns were obtained with the Large Area Telescope (LAT, \citealt{2009ApJ...697.1071A}) onboard the {\it Fermi} \textcolor{black}{Gamma}-ray Space Telescope.
The \textit{Fermi}-LAT data presented 
in this paper were analyzed using the standard \textit{Fermi} analysis software tools 
(version \textit{v11r07p00}), and the \textit{P8R3\_SOURCE\_V2} response function. We used events from $0.2-300$\,GeV selected within a 10$^\circ$ region of interest (ROI) centered 
on Mrk\,421 and having a zenith \textcolor{black}{distance}
below 100$^\circ$ to avoid contamination from the Earth's limb. The diffuse Galactic and \textcolor{black}{isotropic} components were modelled with the files  gll\_iem\_v06.fits and iso\_P8R3\_SOURCE\_V2.txt respectively\footnote{\url{https://fermi.gsfc.nasa.gov/ssc/data/access/lat/BackgroundModels.html}}. All point sources in the third {\it Fermi}-LAT source catalog \citep[3FGL][]{2015ApJS..218...23A} located in the 10$^\circ$ ROI and an additional surrounding 5$^\circ$-wide annulus were included in the model. In the unbinned likelihood fit, the spectral shape parameters were fixed to their 3FGL values, while the normalizations of the eight sources within the ROI identified as variable were allowed to vary, as were the normalisations of the diffuse components and the spectral parameters related of Mrk~421.

Owing to the moderate sensitivity of \textit{Fermi}-LAT to detect
Mrk~421 on daily timescales (especially when the source is not flaring), we performed the
unbinned likelihood analysis on 3-day time intervals
to determine the light curves in the two energy bands $0.2-2$\,GeV and  $2-300$\,GeV reported in  Fig.~\ref{fig:MWLC}.
The flux values were computed using a power-law function with 
\textcolor{black}{the}
index fixed to 1.8, which is the spectral shape that describes Mrk\,421 during the two years considered in this study, as well as the power-law index reported in the 3FGL \textcolor{black}{and 4FGL} \citep[][]{2015ApJS..218...23A,Abdollahi_2020}. \textcolor{black}{The analysis results are not expected to change when using the 4FGL \citep{Abdollahi_2020}
(instead of the 3FGL) for creating the XML file.
This is due to the 3-day time intervals considered here, which are
very short for regular LAT analyses, implying that only bright
sources (i.e. already present in the 3FGL) can significantly contribute to the photon background in the Mrk\,421 RoI.} We repeated the same procedure fixing the photon indices to 1.5 and 2.0, and found no significant change in the flux values, indicating that the results are not sensitive to the selected photon index used in the differential energy analysis. For the multi-year (2007-2016) correlation study reported in Section \ref{sec:corr}, where the GeV flux is compared to the radio and optical fluxes, we applied the same analysis described above, but this time for all events above 0.3\,GeV in the time interval MJD~54683--57561.

\subsection{MAGIC}\label{sec:MAGICdata}
The MAGIC telescope system \citep{2016APh....72...76A} consists of two
\textcolor{black}{Cherenkov}
telescopes of 17\,m diameter situated 
on
the Canary island of La Palma (28.7$^{\circ}$ N, 17.9$^{\circ}$ W)
\textcolor{black}{at 2200 m a.s.l.}
The MAGIC telescopes are sensitive to $\gamma$-rays of 
energies from
50\,GeV to 50\,TeV using the standard trigger when observing at low zenith \textcolor{black}{distances}
under
dark conditions. \par
Here, we report on the Mrk\,421 data 
gathered by the MAGIC telescope
during the 2015--2016  (MJD\,57037--57535) MWL campaign. The observations with the MAGIC telescope system were performed under varying observational conditions which are shown in Table \ref{table:VHE}. 
During this MWL campaign, Mrk\,421 was observed in the zenith \textcolor{black}{distance} range 
from 5$^\circ$ to 62$^\circ$. The data were separated in the following sub-samples: a) Low zenith \textcolor{black}{distance} range (5$^\circ$ to 35$^\circ$), b) Medium zenith \textcolor{black}{distance} range (35$^\circ$ to 50$^\circ$), and c) High zenith 
\textcolor{black}{distance} range
(50$^\circ$ to 62$^\circ$). Depending on the influence of the 
\textcolor{black}{night sky} background light, the data were separated in the following sub-samples: i) dark condition, ii) low-moon condition and iii) high-moon condition, as defined in \citet{Ahnen:2017uny}. 
For analysing data in different background light conditions, the prescriptions from \citet{Ahnen:2017uny} 
were followed. \par

Most of the data in this campaign were taken  in stereoscopic mode with the standard trigger settings, including 
a coincidence trigger between telescopes and a 3NN \textcolor{black}{single-telescope} trigger logic (event registered when three next-neighbor pixels are triggered;  \citealt{2016APh....72...76A}). A minor subset was taken in the so-called mono mode (without coincidence trigger) and a 4NN \textcolor{black}{single-telescope} trigger logic. The data taken with the latter settings 
\textcolor{black}{were}
analysed following the standard analysis procedure with a fixed size cut of 150 photo-electrons (phe) instead of 50\,phe (used in standard data analysis). This size cut has been optimised by crosschecking the spectrum of the Crab \textcolor{black}{N}ebula observed in the same mode.\par
Since the analysis energy threshold increases with the background light and 
\textcolor{black}{larger}
zenith \textcolor{black}{distance} observations, we set a \textcolor{black}{uniform} \textcolor{black}{minimum energy} of 200\,GeV for the entire data sample. The data (in all observation conditions) were analysed using the MAGIC Analysis and Reconstruction Software \citep[MARS;][]{mars2013}.

\subsection{FACT}
\label{FACTDescription}
The First G-APD Cherenkov Telescope (FACT) is an imaging atmospheric Cherenkov telescope with a mirror area of 9.5 m$^2$. It is located next to the two MAGIC telescopes
at the Observatorio del Roque de los Muchachos 
\citep{2013JInst...8P6008A}. Operational since 2011 October, FACT observes $\gamma$-rays in an energy range from a few hundreds of GeV up to about 10\,TeV. The observations are performed in a fully remote and automatic way allowing for long-term monitoring of bright TeV sources at low cost.\par
Owing to a camera using silicon-based photosensors (SiPM, aka Geiger-mode Avalanche Photo Diodes or G-APDs) and a feedback system, to keep the gain of the photosensors stable,
\textcolor{black}{FACT achieves a good and stable performance}
\citep{2014JInst...9P0012B}.
The possibility of performing observations during bright ambient light along with \textcolor{black}{almost robotic} operation allows for a high instrument duty cycle, minimizing the observational gaps in the light curves \citep{2017ICRC...35..609D}.
Complemented by an unbiased observing strategy, this renders FACT an ideal instrument for long-term monitoring.\par
Between 2014 November 10 and 2016 June 17 (MJD 56972 to 57556), FACT collected 884.6 hours of 
data
on
Mrk 421. 
The data were analysed using the Modular Analysis and Reconstruction
Software 
\citep[MARS;][]{2010apsp.conf..681B} with the analysis as described in \citet{2019ICRC...36..630B}.\par
To select data with good observing conditions, a data quality selection cut based on the cosmic-ray rate was applied \citep{2017ICRC...35..779H}. For this, the artificial trigger rate R750 was calculated, \textcolor{black}{adopting} a threshold of 750 DAC-counts, since above this value no effect from the ambient light is found.
The dependence of R750 on the zenith distance was determined as described in \citet{2017ICRC...35..612M} and \citet{2019APh...111...72B} providing a corrected rate R750$_{\mathrm{cor}}$. To account for cosmic ray rate variations due to seasonal atmoshpheric changes, a reference value R750$_{\mathrm{ref}}$ was determined for each moon period. Data with good quality were selected using a cut of 0.93 $<$ R750$_{\mathrm{cor}}$/R750$_{\mathrm{ref}}$ $<$ 1.3. 
Furthermore, 
nights
\textcolor{black}{with less than 20 minutes of good-quality data} 
were rejected.
\par
This results in a total data sample of 637.3 hours of Mrk\,421 from 239 nights after data quality selection. We further discard nights where Mrk\,421 was not significantly (2\,$\sigma$) detected, resulting in 513.6 hours of data from 180 nights.\par
Based on the $\gamma$-ray rate measured from the Crab \textcolor{black}{N}ebula, the dependence of the $\gamma$-ray rate 
on zenith distance and trigger threshold 
was determined and 
the data were corrected accordingly.
For the conversion to flux, the energy threshold was determined using simulated data. The light curve as measured by FACT is shown in the top panel of Fig.~\ref{fig:MWLC}.

\begin{figure*}
\includegraphics[height=10cm, width=\linewidth]{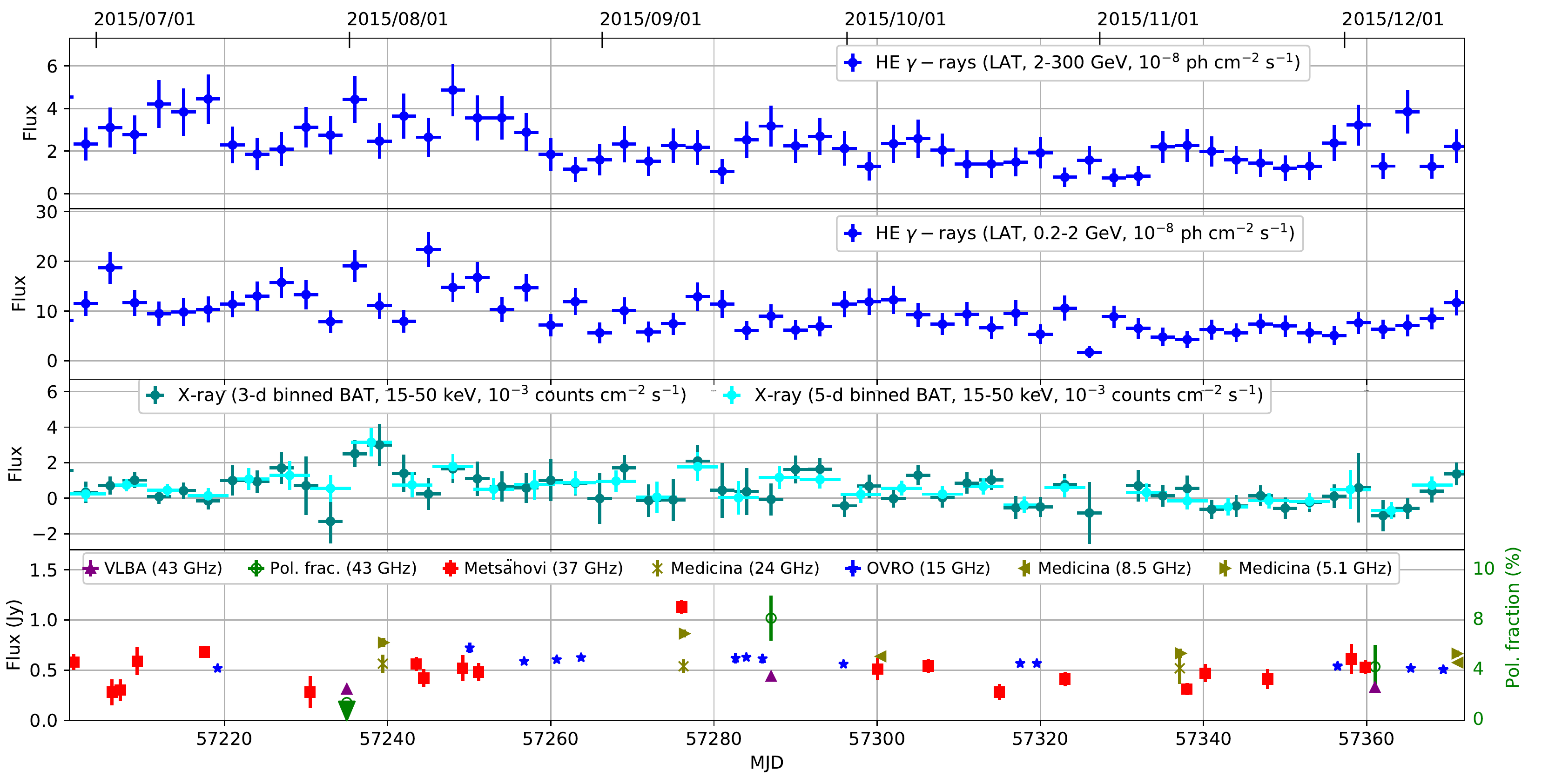}
\caption{
  \textcolor{black}{MWL} light curves around 2015 September, when Mets\"{a}hovi measured a large flux density increase at 37\,GHz. 
(From the top to the bottom panels)
the {\it Fermi}-LAT $\gamma$-ray flux in two energy bands, the {\it Swift}-BAT X-ray flux, the 
single-dish Mets\"{a}hovi 37\,GHz, OVRO 15\,GHz and the Medicina 5\,GHz, 8\,GHz and 24\,GHz flux densities, and the interferometric VLBI core fluxes at 43\,GHz for
the
three measurements performed on August 1, September 22, and December 5.
The \textcolor{black}{linear} polarization fraction for three VLBA measurements is also reported in the bottom panel \textcolor{black}{with an upper limit on August 1, marked by an inverted triangle in green}. See \textcolor{black}{Section}  \ref{sub:MetFlare} for details.}\label{fig:MeFlare}
\end{figure*}

\begin{figure}
\includegraphics[height=16cm, width=\linewidth]{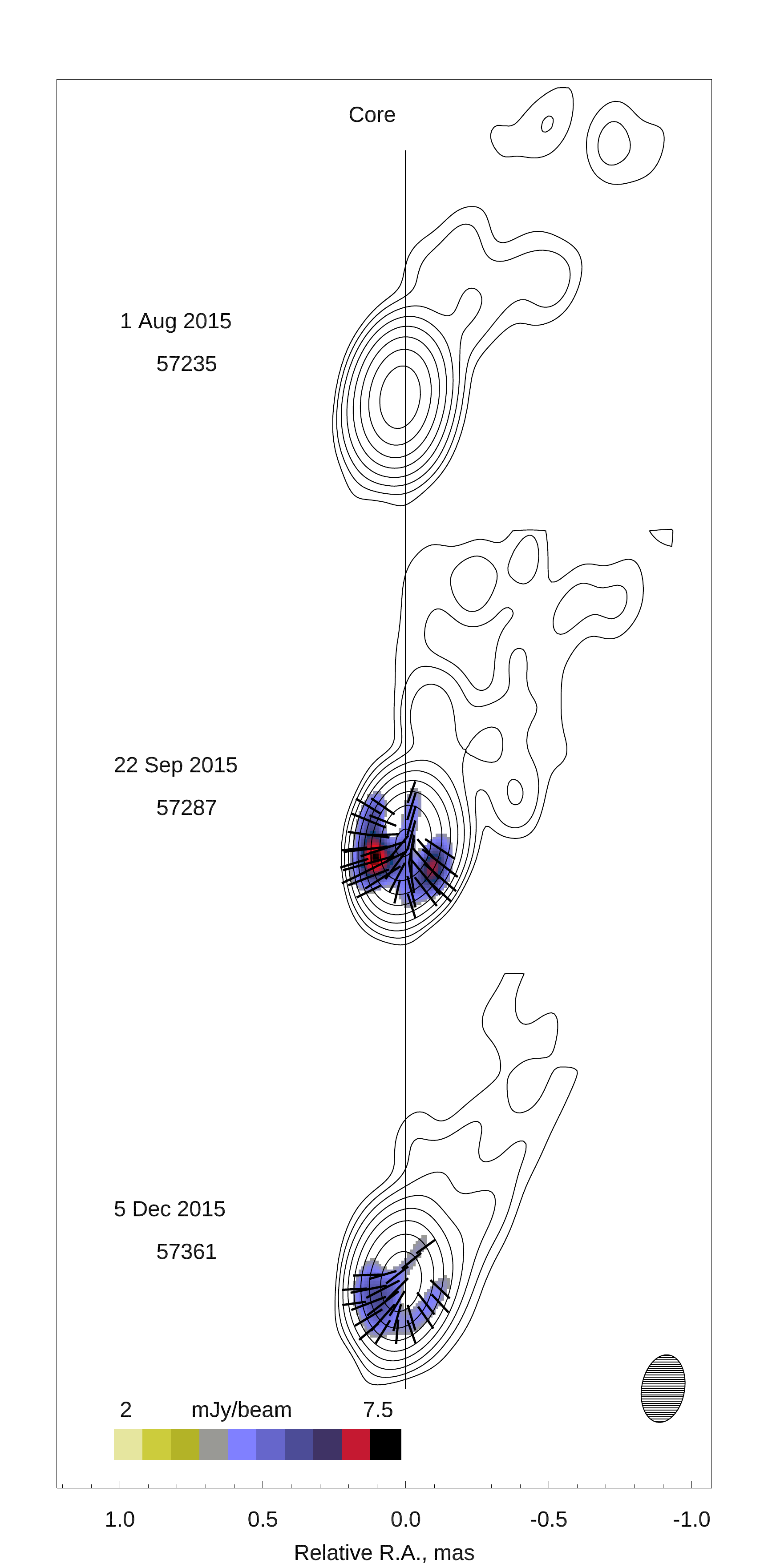}
\caption{
A sequence of total (contours) and polarized (color scale) intensity
images of Mrk\,421 at 43\,GHz obtained with 
the
VLBA. The image is convolved with a FWHM of 0.24$\times$0.15~mas$^2$ along PA=-10$^\circ$. The global total intensity peak is 329\,mJy/beam, and the contours 0.35, 0.70, 1.4, etc. up to 89.6 per cent of the global intensity peak. The color scale is the polarized intensity, and the black line segments within each image show the direction of the
polarization electric vector, with the length of the segment being
proportional to the polarized intensity values. The black vertical
line indicates the position of the core. See   \textcolor{black}{Section}  \ref{sub:MetFlare} for details.}\label{fig:MeFlare2}
\end{figure}

\section{Overall  MWL activity}\label{sec:Summary1516}
During the observation periods \textcolor{black}{November 2014 to June 2015 (MJD $57037-57195$) and December 2015 to June 2016 (MJD $57364-57525$)}, Mrk\,421 showed mostly low activity 
in the X-ray and VHE $\gamma$-ray bands. Fig.~\ref{fig:MWLC} shows the MWL LCs from radio to TeV energies observed within this period. In these two MWL campaigns, no large VHE flares (VHE flux $>4$\,Crabs) or extended VHE flaring activities (VHE flux $>2$\,Crabs for several consecutive days)
were seen. A slower flux variation in the optical and UV 
\textcolor{black}{emissions}
along with \textcolor{black}{stable radio emission} have also been seen. In this section, we first report
on interesting features of the fluxes measured in different \textcolor{black}{wave-bands} during the 2015--2016 campaign, and then discuss a peculiar radio flare.

\subsection{Identification of notable characteristics}

The multi-instrument LC from Fig.~\ref{fig:MWLC} shows several unusual characteristics, which are 
indicated with red vertical line
\textcolor{black}{and are discussed in the paragraphs below.}

{\bf Intra-night variability on 2015 January 27 \& March 12 (MJD~57049~\&~57093):} %\newline
The VHE $\gamma$-ray data set  was checked for intra-night variability (INV).  \textcolor{black}{From the 61 observations with MAGIC and 180 observations with FACT reported here,} INV was observed in only two nights,  2015 January 27 (MJD~57049), found in the MAGIC data, and 2015 March 12 (MJD~57093), found in the FACT data. The LCs and details of the INV study are reported
{in the supplementary online material (Appendix~\ref{sec:INV}).}
In the first case, the VHE flux  from Mrk\,421 dropped from \textcolor{black}{$\sim$1.3\,Crab down to $\sim$0.8\,Crab, while in the second one, where the statistical uncertainties are larger, it decreased from $\sim$2 Crab down to $\sim$1\,Crab. } As depicted in Fig.~\ref{fig:MWLC}, both nights show enhanced X-ray  \textcolor{black}{flux}, but no particularly high  \textcolor{black}{flux} in the GeV, optical or radio bands.

{\bf Spectral hard state on 2015 February 12 (MJD 57065)}: %\newline
\textcolor{black}{This is the only night in the 2015-2016 campaign in which the $2-10$\,keV flux was higher than the $0.3-2$\,keV flux. The respective flux values are} \textcolor{black}{F}$_{2-10\,\mathrm{keV}}$=$(9.12\pm0.12)\times 10^{-10}$\,erg\,cm$^{-2}$\,s$^{-1}$ and \textcolor{black}{F}$_{0.3-2\,\mathrm{keV}}$=$(8.61\pm0.05)\times10^{-10}$\,erg\,cm$^{-2}$\,s$^{-1}$. 
This state is associated with a high hard X-ray flux observed with \textit{Swift}-BAT and a low state in optical R- and UV-bands. 

{\bf Highest X-ray flux during 2015--2016 on 2015 March 31 (MJD 57112):} %\newline 
\textcolor{black}{On this day, the highest flux in the X-ray band during this 2015-2016 campaign was observed. The corresponding fluxes are F$_{0.3-2\,\mathrm{keV}}$=$(1.68\pm0.06)\times 10^{-9}$\,erg\,cm$^{-2}$\,s$^{-1}$ and   F$_{2-10\,\mathrm{keV}}$=$(1.35\pm0.01)\times10^{-9}$\,erg\,cm$^{-2}$\,s$^{-1}$. This means that the flux increased by a factor of about five (two) compared to the average X-ray flux in the 2-10\,keV (0.3-2\,keV) energy band during the 2015-2016 campaign.}
The contemporaneous VHE $\gamma$-ray data from FACT showed a high flux state. \par

{\bf Low X-ray flux on 2015 June 22 \& 2015 December 8 (MJD 57195 \& 57364):} %\newline
The lowest flux in the 2015--2016 campaign in the X-ray band was observed on 2015 December 8 (MJD~57364), with the integrated flux in the $0.3-2$\,keV and $2-10$\,keV  bands being $(1.67 \pm 0.03) \times 10^{-10}$\,erg\,cm$^{-2}$\,s$^{-1}$ and  $(2.41 \pm 0.15) \times 10^{-11}$\,erg\,cm$^{-2}$\,s$^{-1}$, respectively. This is the lowest flux ever reported in  the $2-10$\,keV band.  Previously, to the best of our knowledge, the \textcolor{black}{lowest} flux in the $2-10$\,keV band was $(3.5 \pm 0.2) \times 10^{-11}$\,erg\,cm$^{-2}$\,s$^{-1}$, observed on 2013 January 20 (6$^{th}$ orbit) and reported in \citet{2016ApJ...819..156B}.

On \textcolor{black}{2015 June 22} (MJD~57195), the source showed similar low-flux levels in the $2-10$\,keV and $0.2-1$\,TeV bands to MJD 57364, with measured fluxes of $(4.95 \pm 0.23) \times 10^{-11}$\,erg\,cm$^{-2}$\,s$^{-1}$ and $(0.7 \pm 0.1) \times 10^{-10}$\,ph\,cm$^{-2}$\,s$^{-1}$, respectively.

%%%%%%%%%%%%%
{\bf Low flux states during 2016 February 4--March 27 (MJD 57422--57474):}
On MJD\,57422, the source  evolved into a state where the flux remained very low  in the X-ray and VHE $\gamma$-ray bands, as measured with \textit{Swift}-XRT and MAGIC. MAGIC observed the lowest flux state in the $0.2-1$\,TeV energy band with a flux value of ($3.56\pm 0.91)\times10^{-11}$\,ph\,cm$^{-2}$\,s$^{-1}$. However, there are a few days (e.g., MJD 57422--57429) with high flux at hard X-ray ($15-50$\,keV), as measured with the \textit{Swift}-BAT instrument. This will be further discussed in Section 4.3.
%%%%%%%%%%%%%

\subsection{Peculiar radio flaring activity in 2015 September}
\label{sub:MetFlare}

On 2015 September 11 (MJD 57276), the 13.7-meter diameter Mets{\"a}hovi radio telescope measured a 37\,GHz flux from Mrk\,421 of $1.13\pm0.07$\,Jy, one of the highest fluxes ever observed at this wavelength and about twice that of any other observation from this campaign, as shown in Fig.~\ref{fig:MeFlare}. 
Only during the flaring episode  from September 2012 was a similar high flux state observed in the 15\,GHz radio bands, along with a flare in the HE $\gamma$-rays and optical R-band about 40 days before the radio flare \citep{2015MNRAS.448.3121H}.
\par
There were several 37\,GHz measurement attempts of Mrk\,421 in late August, late September, and early October, but all of them had to be discarded due to bad weather conditions (for details, see \ref{sec:RadioAnalysisDescription}), leaving only the September~11 data point, and making it stand out as the only indication of a high state in that time period. However, 
a flux increase is also suggested by the OVRO 15\,GHz data, in which the flux density level is slightly elevated in late August and September. There are no simultaneous data at 15\,GHz and 37\,GHz.
There are, however, data at 5\,GHz and 24\,GHz from the Medicina radio telescope on the same date. The 5\,GHz flux density is higher than the average value at this frequency, while the 24\,GHz does not show any evidence of significant variability. 

Within the regular monitoring program 
of the Boston University group, the VLBA performed three observations around the 2015 September 11 radio flare, namely on August 1 (MJD 57235), September 22 (MJD 57287), and December 5 (MJD 57361).
The core VLBA fluxes and \textcolor{black}{linear} polarization fraction are displayed in Fig.~\ref{fig:MeFlare}, while the images yielded by these observations are reported in Fig.~\ref{fig:MeFlare2}. Within the statistical uncertainties of the VLBA measurements, one does not see any change in the core VLBA radio flux (even though one observation happened only 11 days after the Mets{\"a}hovi flare), yet there is a clear change in the polarization fraction, from less than 2 per cent for the observation from August 1, to about 8 per cent for the observation from September 22. Additionally, in the image related to the observation from September 22, there is a radial polarization pattern across the 
\textcolor{black}{Southern}
half of the core region. This suggests that the magnetic field $B$ is roughly circular and centered on the brightness peak of the core, as one might expect from a helical field when one views it down the axis. This polarization pattern remained through 2016 March. 
The $\gamma$-ray light curve from  the {\it Fermi}-LAT and X-ray light curve from the {\it Swift}-BAT do not show any obvious flux enhancement during the time of the radio flaring activity, although they show some activity (both BAT and LAT) about 40 days before the radio flare. 
During this time, there 
\textcolor{black}{were}
no optical or VHE observations because of the Sun.

The low polarization fraction at 43\,GHz on 2015 August 1 implies that the magnetic field was very highly disordered in the core at this epoch. A radial polarization pattern, as measured at 43\,GHz on 2015 September 22, can result from turbulent plasma flowing across a conical standing shock, as found in the simulations of 
%Cawthorne et al. (2013)
\citet{2013ApJ...772...14C}
and \citet{2016Galax...4...37M}.
%Marscher (2016).
However, in such a scenario the linear polarization pattern is always present, since it is created by the partial ordering of the magnetic field by the shock front. Periods of polarization $<2\%$ across the entire core should not be observed.

An alternative picture ascribes the radial polarization pattern to the circular appearance of a magnetic field with a helical or toroidal geometry that is viewed within $\sim0.2/\Gamma$ radians of the axis of the jet 
\citep{2002ApJ...577...85M},
where $\Gamma$ is the bulk Lorentz factor of the emitting plasma. In this case, the ratio of the observed polarization measured in the image, $\sim8\%$, to the value for a uniform magnetic field direction, $\sim75\%$\footnote{
The linear polarization of synchrotron radiation is proportional to the value for a uniform magnetic field, $(1+\alpha)/(5/3+\alpha)$, where $\alpha$ is the spectral index \citep[see classical book by][]{1970ranp.book.....P}.
For typical spectral indices from 0.5 to1.5, the uniform-field linear polarization fraction ranges from 68\% to 79\%.}, implies that the helical field is superposed on a highly disordered field component that is $\sim 10$ times stronger. If the helical field becomes disrupted by a current-driven kink instability, particle acceleration could cause a flare 
\citep{2017Galax...5...64N, 2017ApJ...835..125Z, 2018PhRvL.121x5101A}.
The polarization pattern then becomes complex, with the possibility that the polarization becomes very low at some point 
\citep{2020arXiv200307765D}.
Such a flare would be expected to start at X-ray and VHE $\gamma$-ray energies upstream of the core, then propagate downstream so that it appears later at radio frequencies 
\citep{2017Galax...5...64N}.
The disruption of the helical field by the instability could lead to the disordered component inferred from the VLBA images.

\section{Variability study}
\label{sec:var}
Mrk\,421 is known to exhibit significant flux variations from radio to VHE $\gamma$-rays. 
In this work, we quantify different aspects of variability by computing 
the fractional variability (F$_\mathrm{var}$) and 
the hardness ratio (HR). 
\subsection{Fractional variability}
\label{sec:Fvar}
We use fractional variability (F$_\mathrm{var}$) as a tool to characterize 
the
variability of the source in different wave-bands. It is defined as the normalized excess variance of 
the
flux \citep{2003MNRAS.345.1271V}:
\begin{equation}
F_\mathrm{var}= \sqrt[]{\dfrac{S^{2} -<\sigma_\mathrm{err}^{2}>}{<F_{\gamma}>^{2}}},
\end{equation}\label{eq:fvar}
where \newline $S$ is the standard deviation of $N$ flux measurements,
$<\sigma_\mathrm{err}^{2}>$ is the mean squared error,
$<F_{\gamma}>$ is the average photon flux.
The uncertainty in the fractional variability (F$_\mathrm{var}$) has been estimated using the formalism described in \citet{2008MNRAS.389.1427P}:
\begin{equation}
\Delta F_\mathrm{var} = \sqrt[]{F_\mathrm{var}^2 + err(\sigma_\mathrm{NXS}^2)}-F_\mathrm{var},
\end{equation}
where
\begin{equation}
err\left(\sigma_\mathrm{NXS}^2\right)=\sqrt{ \left( \sqrt{\frac{2}{N}} \cdot \frac{\langle \sigma_\mathrm{err}^{2}\rangle}{\langle F_{\gamma} \rangle^{2}} \right)^{2} + \left(\sqrt{\frac{\langle\sigma_\mathrm{err}^{2}\rangle}{N}} \cdot \frac{2 F_\mathrm{var}}{\langle F_{\gamma}\rangle} \right)^{2}}.
\end{equation}

\begin{figure}
\centering
\includegraphics[height=7cm, width=\linewidth]{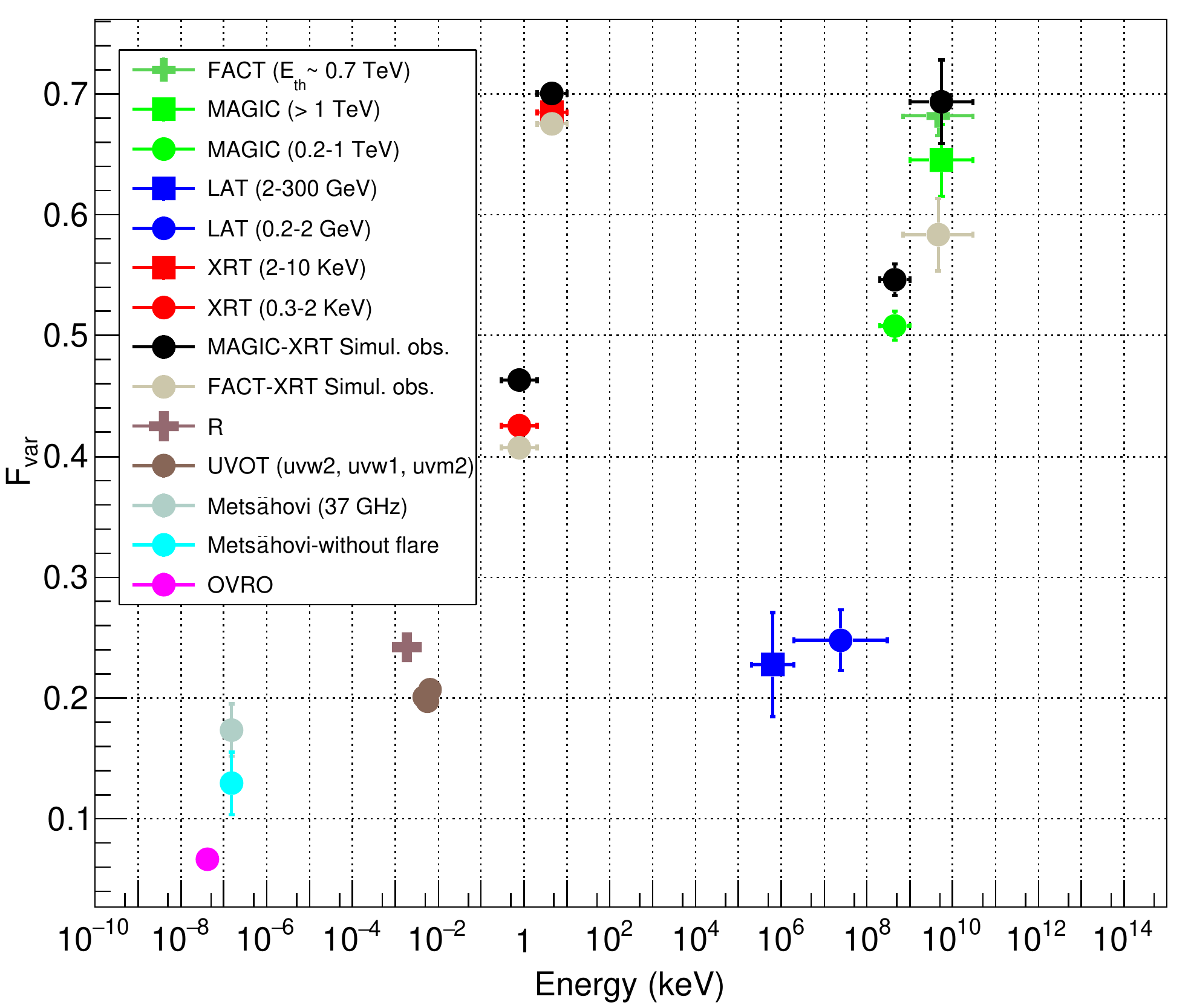}
\caption{Fractional variability as a function of 
energy for the MWL LCs presented in Fig.~\ref{fig:MWLC}.
The horizontal error bars represent the energy bin and the vertical error bars denote the 1\,$\sigma$ uncertainties on the calculated fractional variability 
(not visible for some of the data sets). For the X-ray and VHE data, we show the results derived with all data from 2015--2016 campaigns, and also the results obtained with simultaneous X-ray ({\it Swift}) and VHE data (MAGIC or FACT). See   \textcolor{black}{Section}  \ref{sec:Fvar} for details.}\label{fig:Fvar}
\end{figure}

The F$_\mathrm{var}$ computed 
from the multi-band light curves of Fig.~\ref{fig:MWLC} are shown in Fig.~\ref{fig:Fvar}. 
In order to ensure the use of reliable flux measurements, we only consider fluxes with relative errors (flux-error/flux) smaller than 0.5, i.e. Signal-to-Noise-Ratio (SNR) larger than 2. This is done to avoid dealing with systematic uncertainties that could arise in very-low-significance measurements, when we mostly deal with background that may not be well modelled. This cut discards only a small fraction of the full data set (see open markers in Fig.~\ref{fig:MWLC}). \textcolor{black}{The only instrument that is substantially affected is \textit{Swift}-BAT, whose data are not used for the variability studies reported here.}

The highest variability was measured with FACT and MAGIC at energies above 1\,TeV, with F$_\mathrm{var}$ close to 0.7. The MAGIC data in the energy range $0.2 - 1$\,TeV show variability at the level of 0.5. \textcolor{black}{These values} are about a factor of two higher than that reported during the 2009 campaign \citep{2015A&A...576A.126A}. \par

\begin{figure}
\centering
   \includegraphics[width=\linewidth,height=7cm]{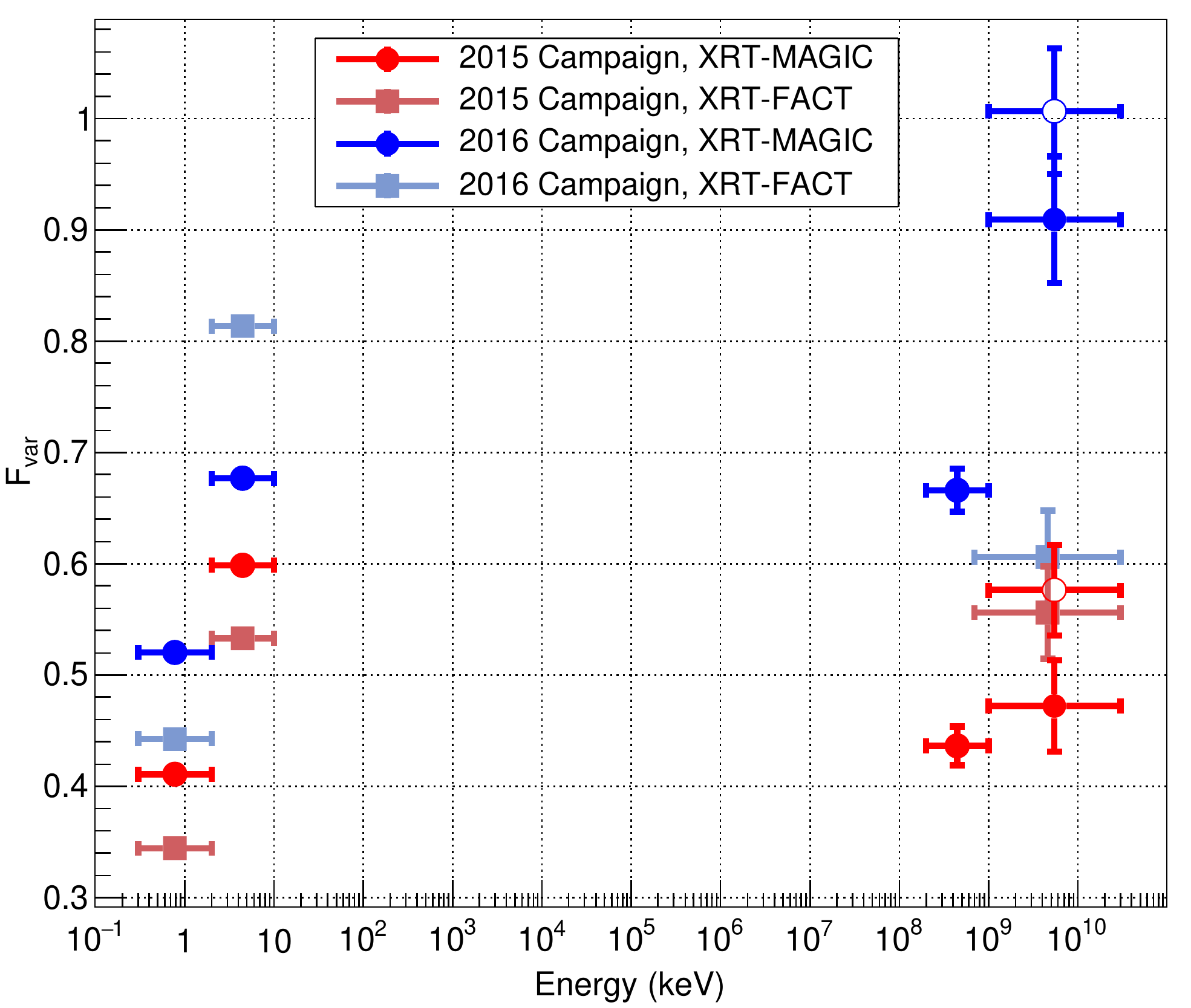}
   \caption{Fractional variability for X-rays and VHE $\gamma$-rays for two subsets of data, the {\em 2015 campaign} (MJD range 56970--57200) and the {\em 2016 campaign} (MJD range 57350--57560). The open markers display the F$_\mathrm{var}$ above 1\,TeV for the two subsets when \textcolor{black}{adding} four flux measurements with SNR below 2 (see text for details).
   }
   \label{fig:201516a2keV}
   \end{figure}

In order to quantify the  variability for different levels of emission, we further divide the X-ray and VHE $\gamma$-ray  data (the two energy bands with the highest variability) into two data subsets, the {\em 2015 campaign} (MJD range 56970--57200) and the {\em 2016 campaign} (MJD range 57350--57560). For this study, we only use simultaneous X-ray and VHE $\gamma$-ray observations. Most of the MAGIC and \textit{Swift}-XRT observations occurred within 2 hours, but owing to the lack of intra-night variability for most of the nights, for this study we consider simultaneous observations those taken within the same night (within 0.3 day). 
This results in 21
pairs of XRT/MAGIC observations for the 2015 campaign subset and 24
for the  2016 campaign subset. The average X-ray flux in the \mbox{$2-10$\,keV} energy range for the first data set is \mbox{$4.1 \times$\,10$^{-10}$\,erg\,cm$^{-2}$\,s$^{-1}$,} while it is \mbox{$2.1 \times$\,10$^{-10}$\,erg\,cm$^{-2}$\,s$^{-1}$,} for the second one, while the 2-year average flux is \mbox{$3.1\times$\,10$^{-10}$\,erg\,cm$^{-2}$\,s$^{-1}$}.
Therefore, the 
2016 
data tell us about the
activity of Mrk\,421 during the lowest fluxes, while the 2015 campaign tells us about predominantly higher fluxes within the 2-year data set considered here. For each X-ray/VHE pair, we have four flux measurements, two at X-rays ($0.3-2$\,keV and $2-10$\,keV) and two at VHE (0.2-1\,TeV and $>$1\,TeV). The F$_\mathrm{var}$ for these two subsets is reported in  Fig.~\ref{fig:201516a2keV}.
All flux measurements have a SNR$>$2.0, apart from four VHE flux
measurements above 1\,TeV: MJD~57195, MJD~57422, MJD~57430 and
MJD~57453. These four flux values were excluded from the calculation of F$_\mathrm{var}$ above 1\,TeV. The first day belongs to the 
2015 campaign
subset, while
the other three belong to the 
2016 campaign
subset. All of them are related to time intervals with very low
X-ray and VHE $\gamma$-ray  \textcolor{black}{flux} (see Fig.~\ref{fig:MWLC}).
Because of the low number of XRT/MAGIC pairs, for completeness,  \textcolor{black}{Fig.~\ref{fig:201516a2keV} also reports the F$_\mathrm{var}$ 
when the four  excluded measurements above 1\,TeV with SNR$<$2 are included in the calculations. Because of the addition of 1+3 flux
points with very low-flux, the F$_\mathrm{var}$ increases slightly, compared
to the F$_\mathrm{var}$ computed using only the measurements with SNR$>$2. } We repeated the same exercise using {\it Swift}-XRT and FACT observations taken within 0.3 days, which yielded 37 and 34 XRT/FACT pairs of observations (with flux measurements with SNR$>$2) for
the 2015 and 2016 campaigns, respectively. The calculated F$_\mathrm{var}$ values for these data subsets are also shown in Fig.~\ref{fig:201516a2keV}. 
The F$_\mathrm{var}$ calculated with the simultaneous XRT/FACT data is, in general, somewhat lower than that
calculated with the XRT/MAGIC simultaneous data. The reason behind this lower variability is the requirement for SNR$>2$ in the VHE flux measurements by FACT, which removes simultaneous XRT/FACT pairs with X-ray fluxes that are well below the average flux for each of the two campaigns (see Fig.~\ref{fig:MWLC}), and hence decreases the overall F$_\mathrm{var}$. 
On the other hand, the F$_\mathrm{var}$ for the simultaneous XRT/FACT in the $2-10$\,keV band is higher than that computed with the simultaneous XRT/MAGIC data in the same energy band. This is due to the XRT/FACT data covering time intervals in 2015 December and 2016 June, which are not covered by the XRT/MAGIC data, where the $2-10$\,keV flux in the X-rays was several times higher (up to factor of $\sim$5) than the average 2-10 keV
flux in the 2016 campaign. Two conclusions can be derived from this exercise with this data set. First, the F$_\mathrm{var}$ is higher during the 2016 campaign (lower X-ray and VHE fluxes) than during the  2015 campaign. Second, for the 
2015 campaign, the variability is similar in\,keV and in\,TeV energies, while for the  2016 campaign, the variability in\,TeV is somewhat higher than in\,keV energies.

\subsection{Hardness ratio}
\label{sec:HR}

\begin{figure*}
    \includegraphics[height=7cm, width=0.95\linewidth]{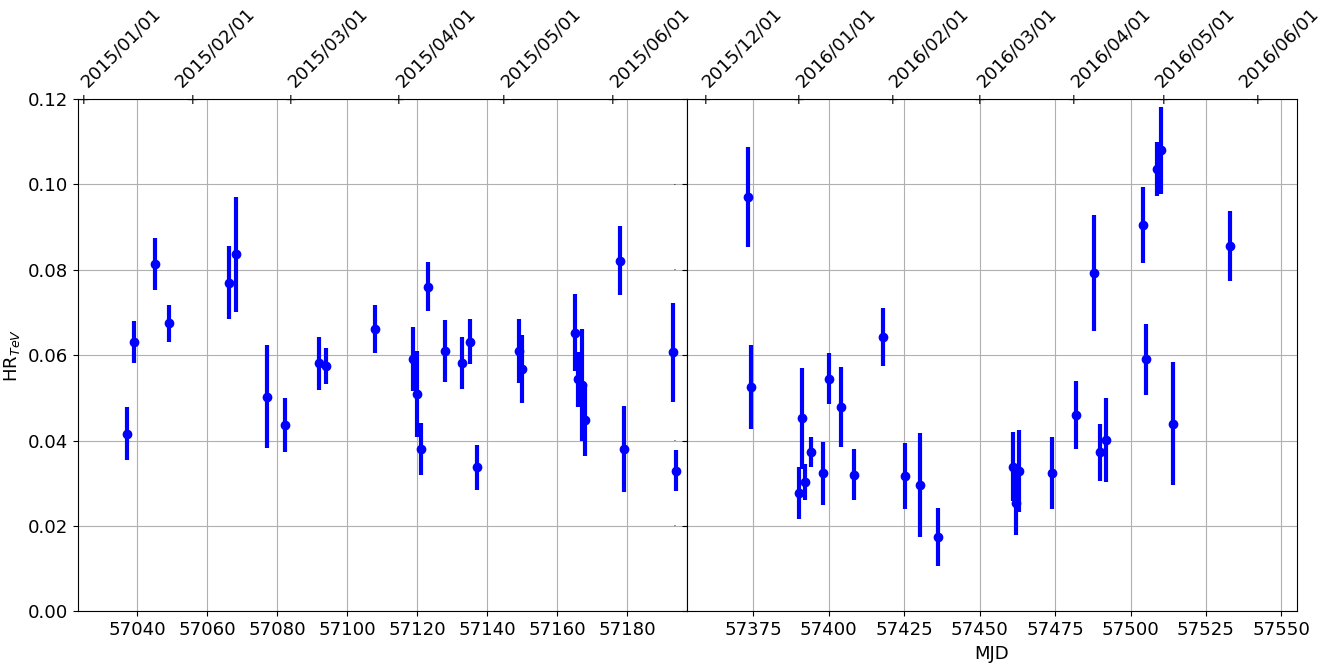}
    \includegraphics[height=6cm, width=0.48\linewidth]{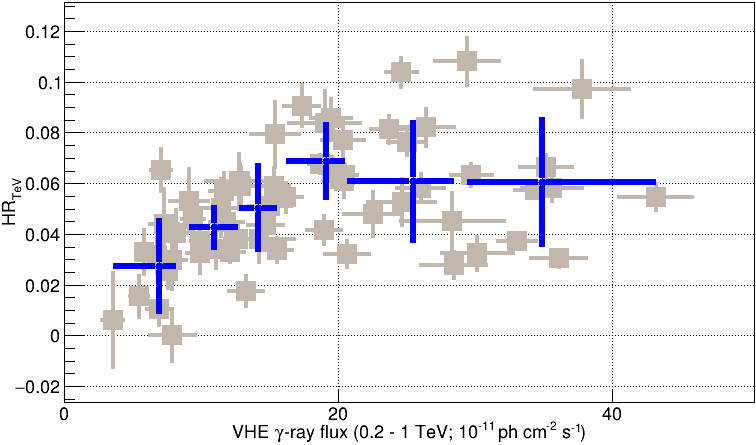}
    \includegraphics[height=6cm, width=0.48\linewidth]{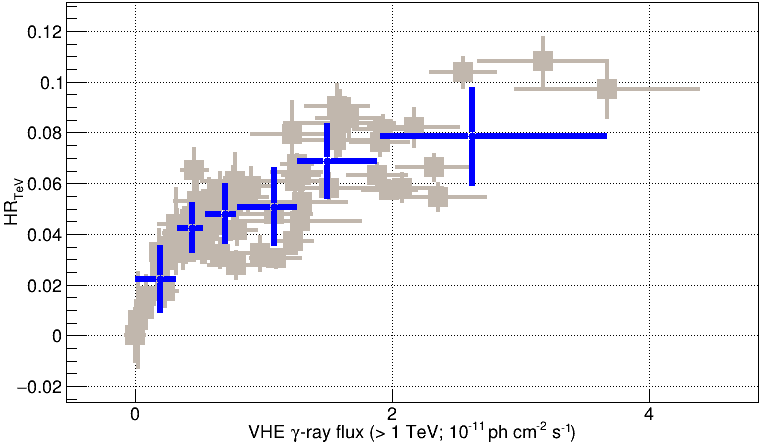}
    \caption{HR as a function of time (top panel) and flux (bottom panels) during 2015--2016 for two TeV energy bands, namely $0.2-1$\,TeV and above 1\,TeV. 
    The blue markers 
    report the average and standard deviation of the   \textcolor{black}{HR$_{\mathrm{TeV}}$} data binned with 10 entries. See   \textcolor{black}{Section}  \ref{sec:HR} for details.}\label{fig:HRXT}
\end{figure*}

\begin{figure*}
\centering
\includegraphics[height=8cm, width=1.0\linewidth]{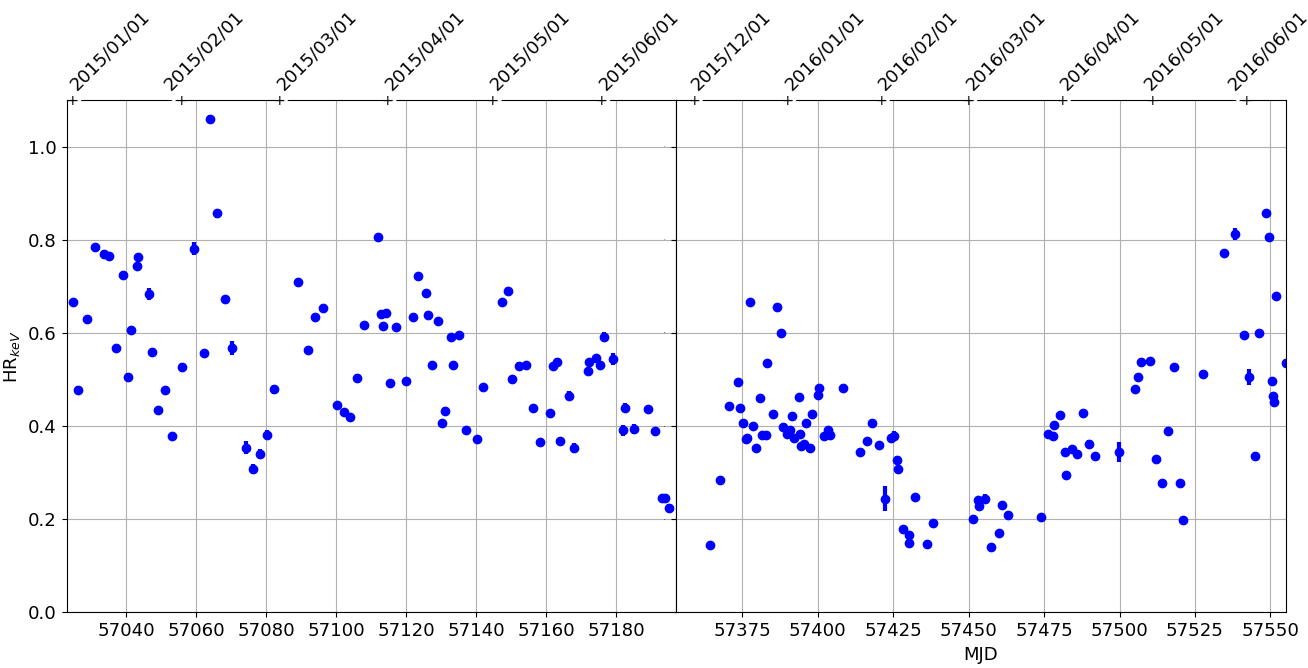}
\includegraphics[height=6cm, width=0.48\linewidth]{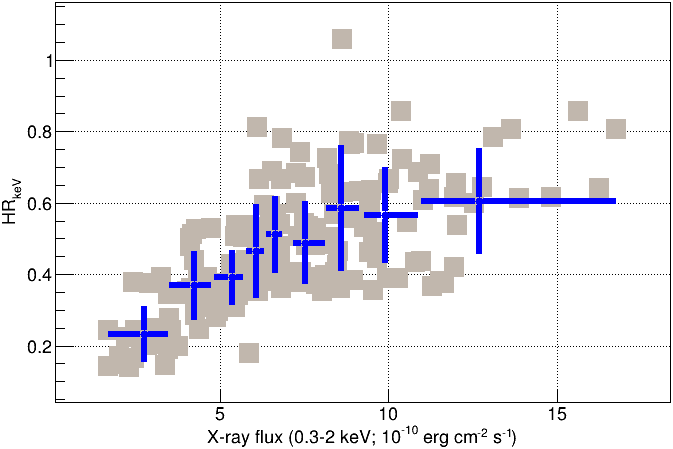}
\includegraphics[height=6cm, width=0.48\linewidth]{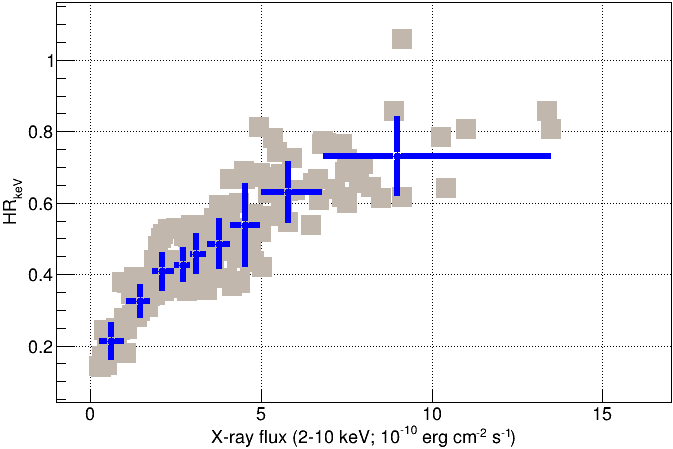}
\caption{HR as a function of time (top panel) and flux (bottom panels) in the X-rays in two energy bands, namely $0.3-2$\,keV and $2-10$\,keV observed with \textit{Swift}-XRT. 
The blue markers depict the 
the average and standard deviation of the   \textcolor{black}{HR$_{\mathrm{keV}}$} data binned with 20 entries. See   \textcolor{black}{Section}  \ref{sec:HR} for details.}\label{fig:HRXlong1}
\end{figure*}

In X-rays and VHE $\gamma$-rays, we define 
the hardness ratio as the ratio of the integral flux in the \textcolor{black}{high-energy (hard)} band 
to the integral flux in the \textcolor{black}{high-energy (soft)} band: 

$\mathrm{HR}_\mathrm{TeV} = \dfrac{\mathrm{F}_{> 1\,\mathrm{TeV}}}{\mathrm{F}_{0.2 - 1\,\mathrm{TeV}}}$; \quad 
$\mathrm{HR}_\mathrm{keV} = \dfrac{\mathrm{F}_{2 - 10\,\mathrm{keV}}}{\mathrm{F}_{0.3 - 2\,\mathrm{keV}}}$,\newline
where 
%\newline
$\mathrm{F}_\mathrm{E}$ is the integrated flux in the energy band E.
%\newline

The upper panel of Fig.~\ref{fig:HRXT} shows the variation of HR$_{\mathrm{TeV}}$ calculated from the 2015--2016 data. During the low-flux state \textcolor{black}{(MJD\,57422 to 57474)}, the   \textcolor{black}{HR$_{\mathrm{TeV}}$} is 
$\leq$0.03. The bottom two panels 
show the variation of \textcolor{black}{HR$_{\mathrm{TeV}}$} with the integral flux in two energy bands namely $0.2-1$\,TeV and above 1\,TeV observed with MAGIC. 
Additionally,  the bottom panel of Fig.~\ref{fig:HRXT} also depicts the average and the standard deviation of data subsets of 10 observations\footnote{The exact number of measurements for grouping the data is not relevant. For the MAGIC data we used 10 measurements, which provides sufficient event statistics, and allows one to visualize different segments of the HR vs. Flux relation.}, binned according to their flux. This is done for a better visualization of the overall trend in the   \textcolor{black}{HR$_{\mathrm{TeV}}$}-flux plot, as well as the dispersion of the data points.
In both plots, one can see a bending in the $HR$ vs. flux trend. This distortion is particularly important for the \textcolor{black}{HR$_{\mathrm{TeV}}$} vs. soft-band VHE flux 
(left panel), where one can see a flattening in the $HR$ beyond 20$\times$10$^{-11}$\,ph\,cm$^{-2}$\,s$^{-1}$.
\par

Figure \ref{fig:HRXlong1} shows the variation of   \textcolor{black}{HR$_{\mathrm{keV}}$} with time and flux.
The   \textcolor{black}{HR$_{\mathrm{keV}}$} 
\textcolor{black}{ranges from} 0.15 to 1.05 ($\Delta$ $HR$ = 0.9). The   \textcolor{black}{HR$_{\mathrm{keV}}$} observed on 2016 March 10 
(MJD 57457)
is the lowest reported   \textcolor{black}{HR$_{\mathrm{keV}}$} so far, which is 0.14$\pm$0.01.
The low-flux state mentioned in Fig.~\ref{fig:MWLC} from 
\textcolor{black}{MJD\,57422 to 57474}
can be identified in Fig.~\ref{fig:HRXlong1} with a sustained   \textcolor{black}{HR$_{\mathrm{keV}}$}$<0.3$, smaller than the lowest   \textcolor{black}{HR$_{\mathrm{keV}}$} previously reported (HR=0.47, \citealt{2017ApJ...848..103K}) where the source was claimed to be in a historical low-flux state observed by {\it NuSTAR} \citep{2016ApJ...819..156B}. 
\textcolor{black}{The lower panels of}
Fig.~\ref{fig:HRXlong1} show the variation of the \textcolor{black}{HR$_{\mathrm{keV}}$ 
with F$_\mathrm{0.3-2\,keV}$  and F$_\mathrm {2-10\,keV}$.}
\textcolor{black}{The hardest} X-ray state can be identified on MJD \textcolor{black}{57065} with \textcolor{black}{HR$_{\mathrm{keV}}$}=1.05, which is the only occasion of $HR_{keV} > 1$, and consistent with the X-ray spectrum peaking around 10\,keV, previously reported in \citet{2017ApJ...848..103K}. \textcolor{black}{Apart from the high flux observed at hard X-rays by \textit{Swift}-BAT, no exceptionally high flux  is observed in any of the other 
energy bands.} As in the bottom panel of Fig.~\ref{fig:HRXT}, we also depict here the average
and the standard deviation of the data binned in 20 observations\footnote{Owing to the larger number of XRT observations, in comparison with that of MAGIC observations, we decided to bin the XRT data in groups of 20, instead of the 10 used for the MAGIC data.} according to their flux, which also show \textcolor{black}{the} flattening in the $HR$ vs. flux relation.

Overall, the $HR$ vs. flux plots in Fig.~\ref{fig:HRXT} and Fig.~\ref{fig:HRXlong1} show a clear hardening-when-brightening trend in both the X-ray and VHE $\gamma$-ray energy ranges. However, for the highest activities, one can observe that the spectral hardening trend flattens, which is more evident when reporting the HR as a function of the flux in the 
\textcolor{black}{lower}
band from each of the two energy ranges, namely \mbox{$0.3-2$\,keV} and \mbox{$0.2-1$\,TeV.} 
\citet{2016ApJ...819..156B} had already reported a saturation in the X-ray spectral shape variations of Mrk\,421 for very-low and very-high  \textcolor{black}{flux}. \textcolor{black}{The saturation at high fluxes} appears to be consistent with what is reported here, i.e., a flattening in the X-ray spectral shape starting for $2-10$\,keV fluxes above 8$\times$10$^{-10}$\,erg\,cm$^{-2}$\,s$^{-1}$. On the other hand, the flattening in the $HR$ vs. flux relation at VHE $\gamma$-rays has not been reported previously.

\subsection{Appearance of a new component at hard X-ray energies}\label{sec:BATex}
\begin{figure*}
    \centering
    \includegraphics[width=0.8\linewidth,height=10cm]{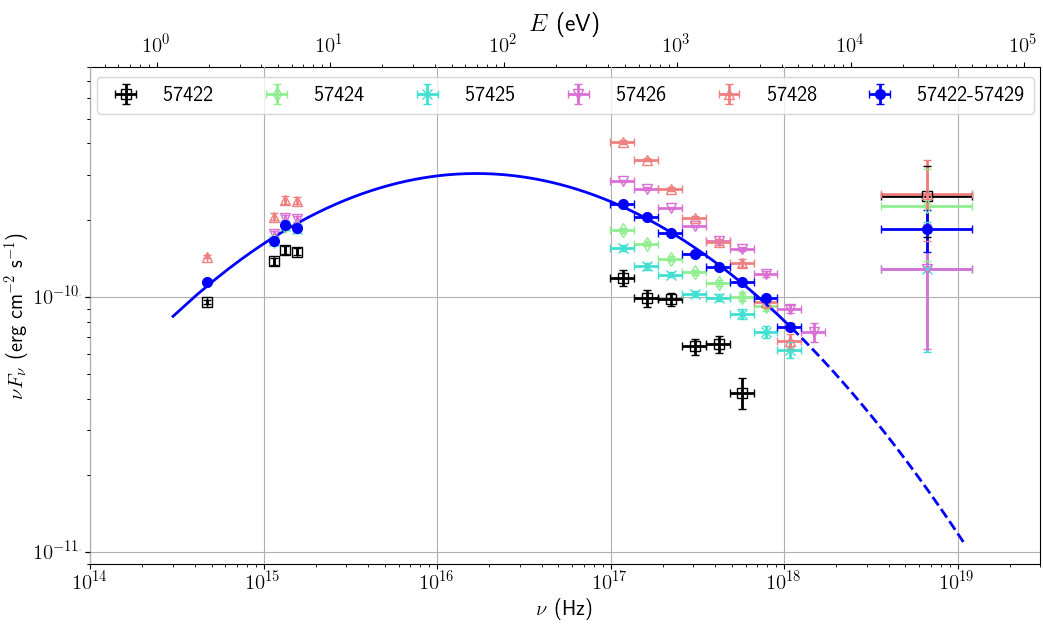}
    \caption{Characterization of the low-energy SED bump of Mrk\,421 during the time interval  \mbox{MJD~57422--57429} (2016 February 4--11). The five available daily observations (from optical to hard X-rays) during this 7-day time interval are reported with open markers, while the 5-day weighted-averaged fluxes are reported with blue-filled markers. The blue solid line depicts the resulting fit with a log-parabola function in the energy range from 1\,eV to 10\,keV, and the dashed line shows the extrapolation of this log-parabola function to the hard X-ray energy range. See   Section \ref{sec:BATex} for details.}
    \label{fig:BATexcess}
\end{figure*}

In this section we report a characterization of the shape of the low-energy SED bump (presumably the synchrotron bump) for the time interval \mbox{MJD~57422--57429} (2016 February 4--11), which is a time interval with a very low X-ray  \textcolor{black}{flux} and a very low HR (see Fig.~\ref{fig:MWLC} and Fig.~\ref{fig:HRXlong1}). Fig.~\ref{fig:BATexcess} shows 
the fluxes in the optical (R-band), UV (W1,M2,W2), soft X-rays ($0.3-10$\,keV) and hard X-rays ($15-50$\,keV) for five days (out of 7-day interval) and the related 5-day combined fluxes obtained with a standard weighted average procedure. The daily fluxes were obtained as described in   \textcolor{black}{Section}  \ref{sec:data_analysis}. The weighted-averaged BAT fluxes (daily and combined) are converted into energy fluxes (in units of erg\,cm$^{-2}$\,s$^{-1}$) using the prescription in \citet{2013ApJS..209...14K}. The hard X-ray BAT fluxes (both the daily fluxes and the 5-day combined flux) appear to be inconsistent with the simple extrapolation from the soft X-ray XRT fluxes. In order to evaluate this, we fit the 5-day combined optical to soft X-ray spectra (solid blue markers in Fig.~\ref{fig:BATexcess}) with a log-parabola function  
\textcolor{black}{F($\nu$)=N$_0$($\nu$/$\nu_0$)$^{-\alpha-\beta log_{10}(\nu/\nu_0)}$,}
where
\textcolor{black}{$\nu_0$}
has been fixed to 3.0$\times$10$^{16}$\,Hz and $N_0$, $\alpha$,
and $\beta$ are the free parameters of the fit. 
Because of the very small uncertainties in the 5-day weighted average of the flux values (typically in the order of $\sim$1\%), a regular fit to the data would be affected by the small spectral distortions (wiggles) caused by small systematics in merging data sets from different instruments and with somewhat different spectral shapes. 
We find that we can smooth out these small spectral distortions by adding a relative flux error of 3\% in quadrature to the actual flux error resulting from the weighted average procedure. The resulting spectral fit, performed in the $\nu F_\nu$ vs. $\nu$ representation, yields a $\chi^2$ of 11.6 for 9 degrees of freedom, with the following parameter values:  $N_0$, $\alpha$, and $\beta$ as (2.91$\pm$0.07)$\times$10$^{-10}$erg\,cm$^{-2}$\,s$^{-1}$, (9.11$\pm$0.39)$\times$10$^{-2}$, and (1.77$\pm$0.06)$\times$10$^{-1}$ respectively. Therefore,
the log-parabola
function provides a good representation of the synchrotron emission averaged over 5-day, from eV to 10\,keV energies. 
The weighted average of the 1-day BAT fluxes over these 5 days with XRT/UVOT observations is (1.84$\pm$0.34) $\times$ 10$^{-10}$\,erg\,cm$^{-2}$\,s$^{-1}$ \footnote{This number is derived from the 5-day weighted average of the BAT count rate, (3.21$\pm$0.59)$\times$ 10$^{-3}$ cts\,cm$^{-2}$\,s$^{-1}$, and the counts-to-energy conversion stated in \citet{2013ApJS..209...14K}.}.
As shown in Fig.~\ref{fig:BATexcess}, the extrapolation of this log-parabola function to the $15-50$\,keV band goes well below the BAT \mbox{5-day} weighted-averaged flux point (5 times the error bar). \textcolor{black}{If instead of using the prescription of \citet{2013ApJS..209...14K} to convert the BAT count rate to energy flux, which employs the spectral shape of the Crab Nebula in the energy range 15-50 keV (i.e., a power-law shape with index 2.15), we employ the spectral shape given by the above-mentioned log-parabola function (which in the 15-50 keV band could be approximated with power-law function with index $\sim$2.5 ), the BAT energy flux would be only 10\% lower than the one reported above (and displayed in Fig.~\ref{fig:BATexcess}), and hence it would not change the overall picture in any significant way.} This observation suggests the presence of an additional component, 
beyond that of the synchrotron emission of the main emitting region. See Section \ref{sec:con} for further discussion about it.

\section{Correlation study}
\label{sec:corr}

\begin{figure*}
\centering
\includegraphics[width=0.49\linewidth,height=5.5cm]{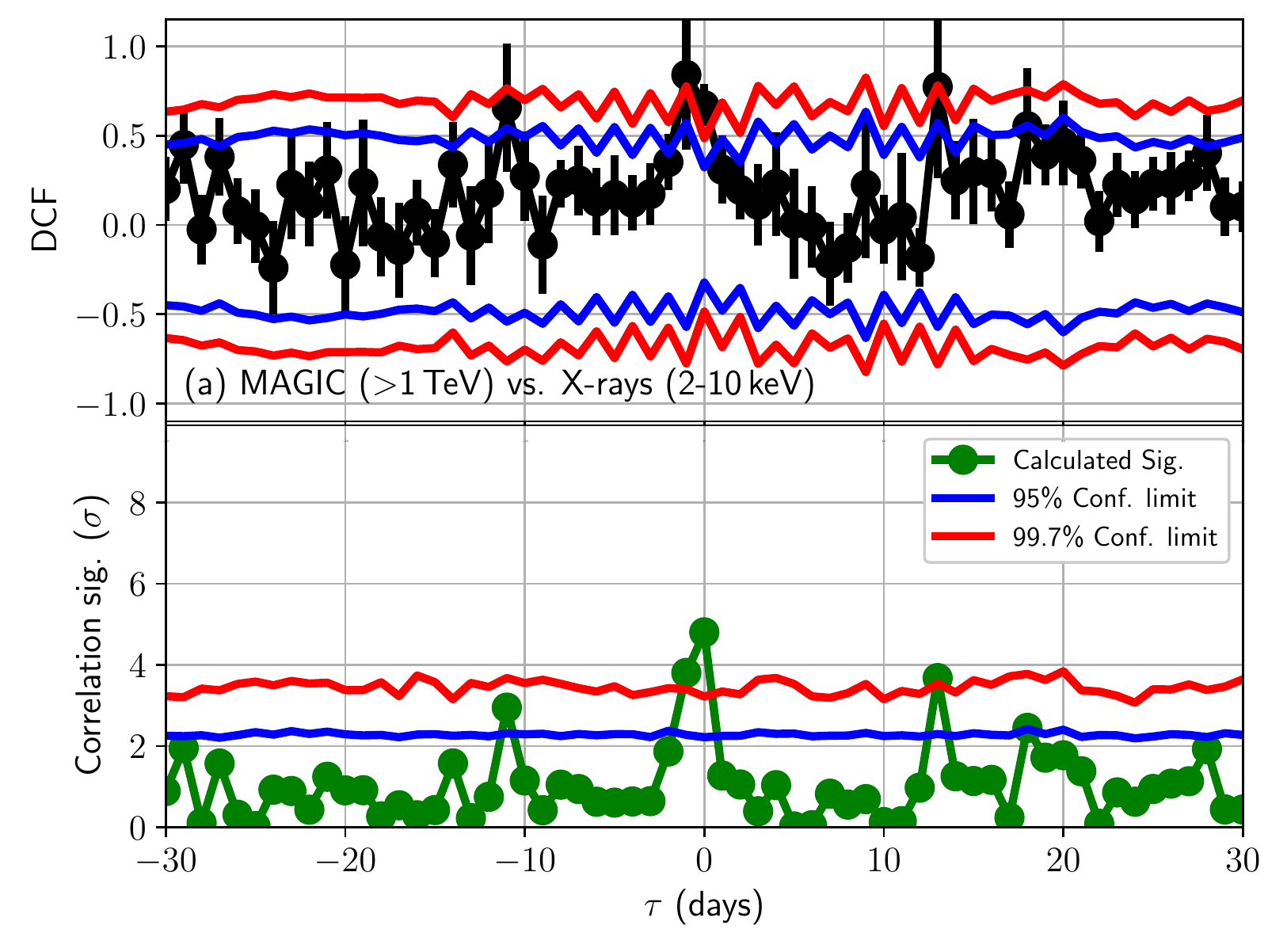}
\includegraphics[width=0.49\linewidth,height=5.5cm]{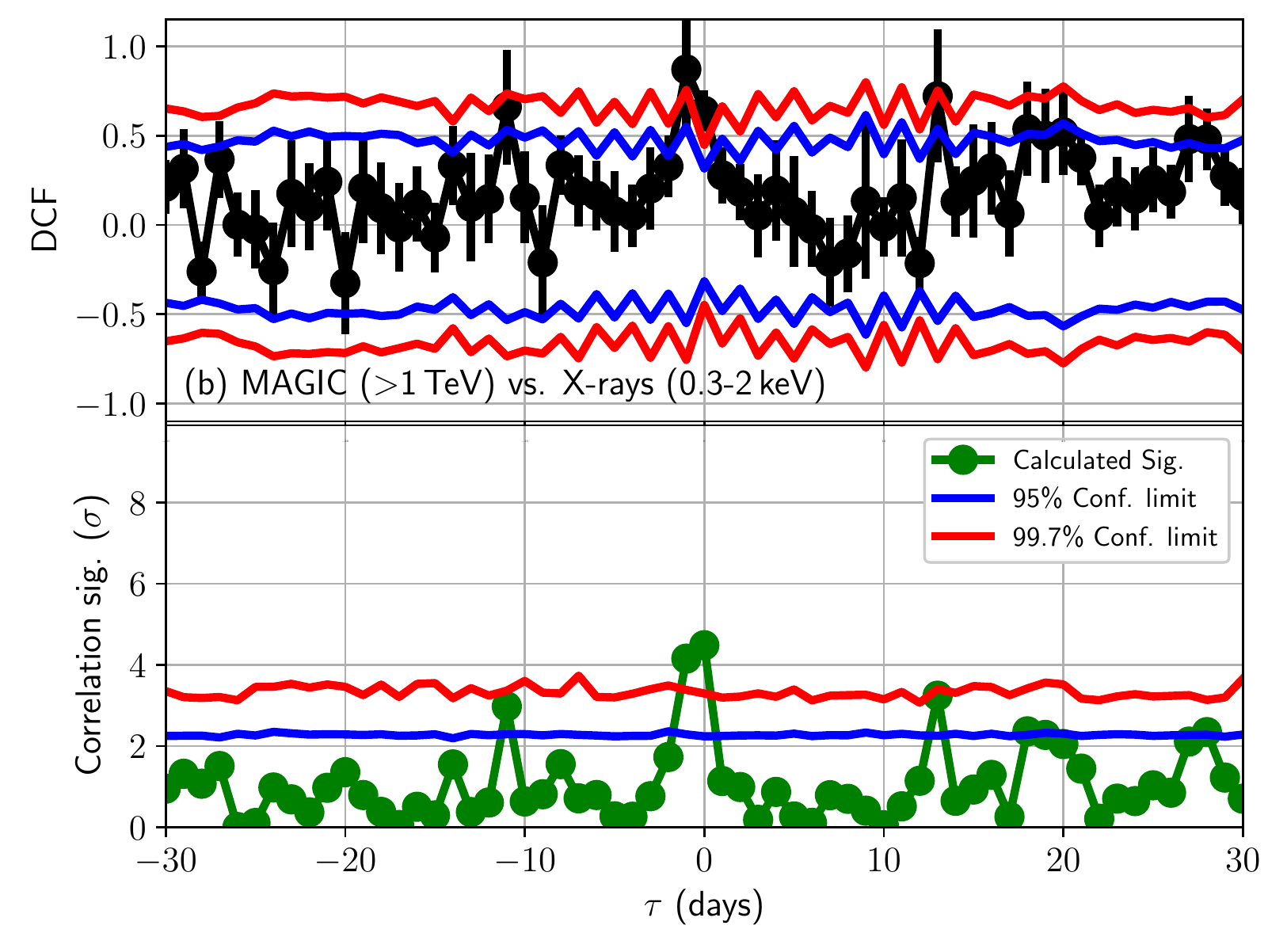}
\includegraphics[width=0.49\linewidth,height=5.5cm]{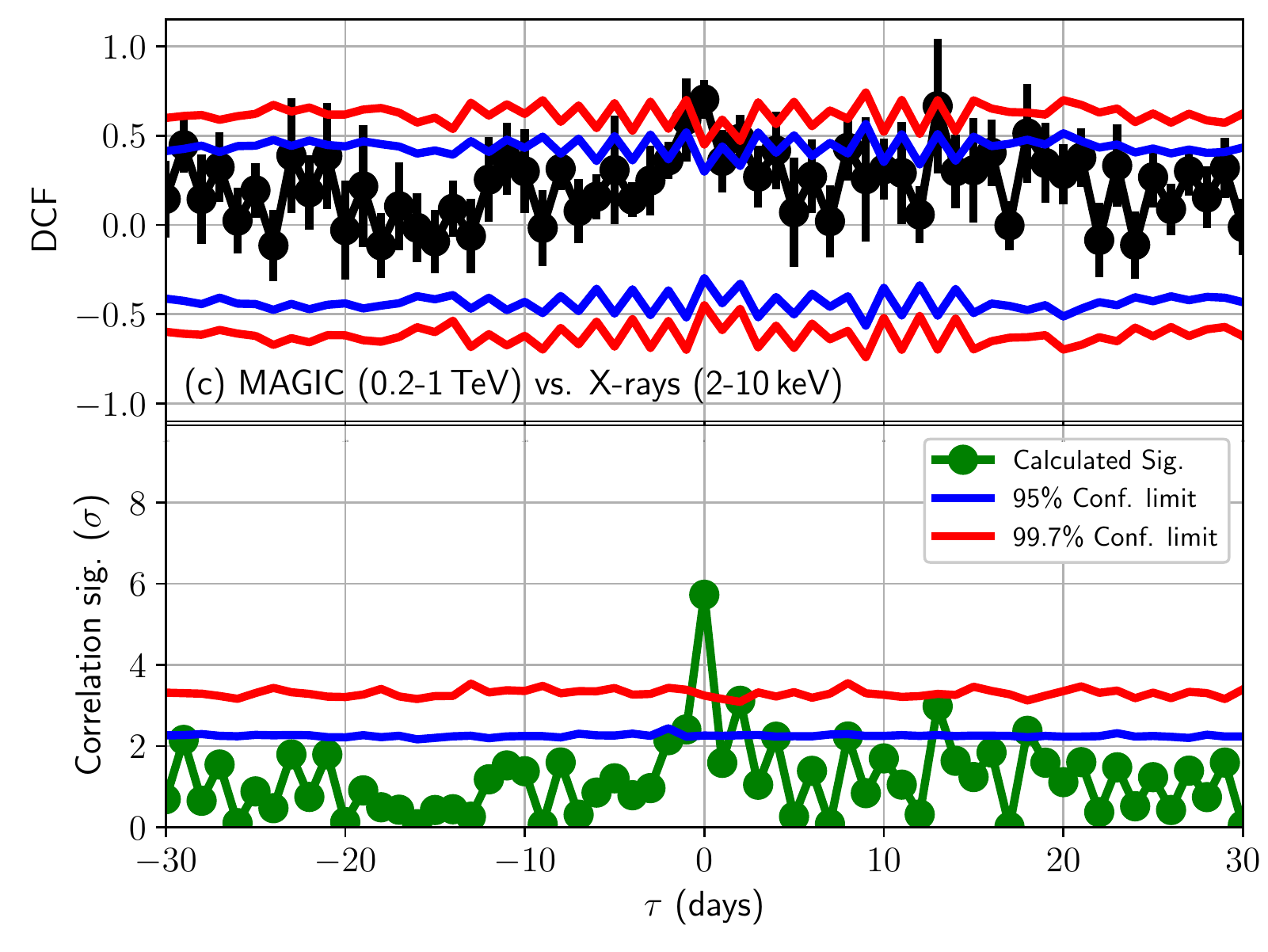}
\includegraphics[width=0.49\linewidth,height=5.5cm]{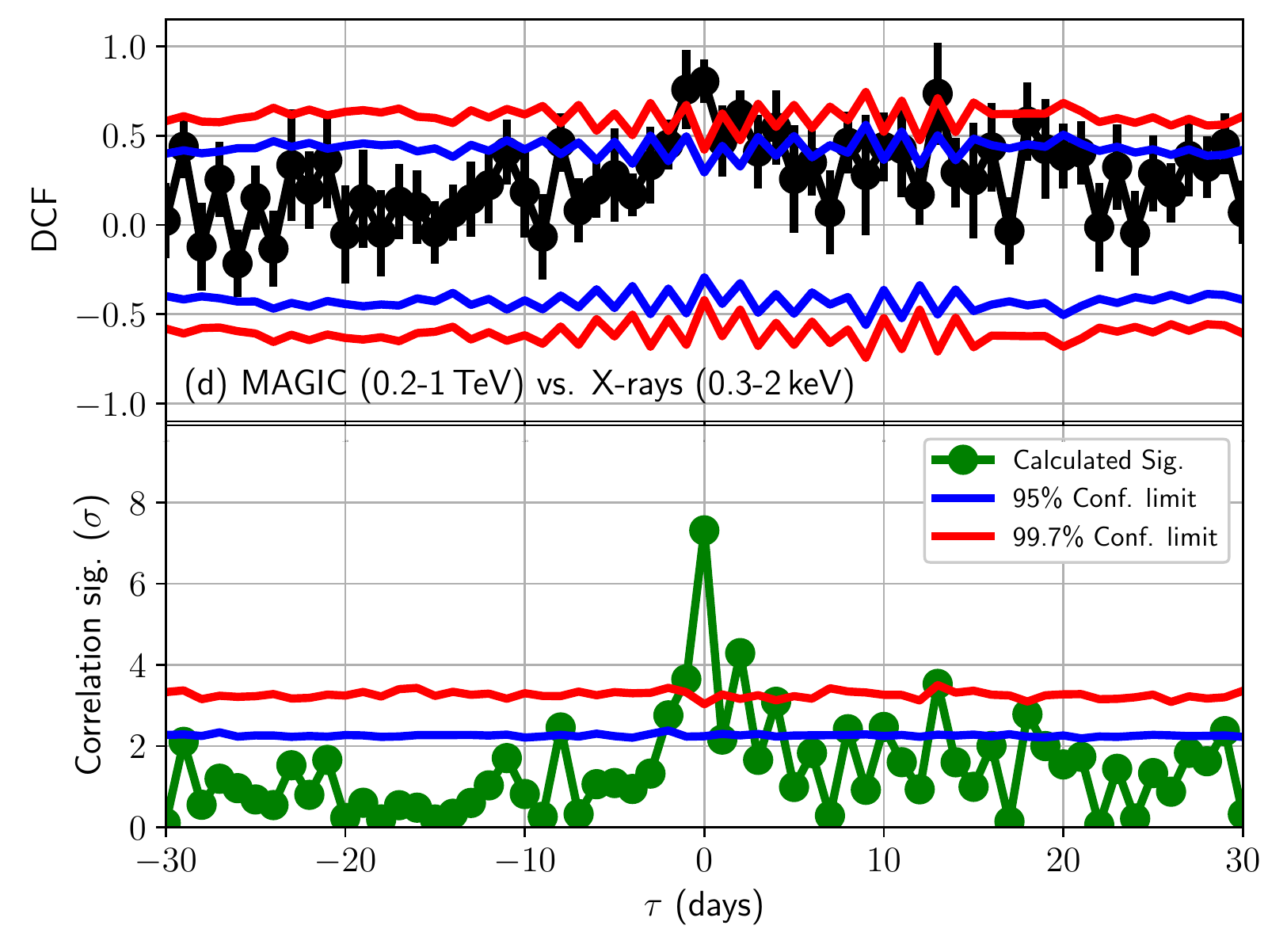}
\includegraphics[width=0.49\linewidth,height=5.5cm]{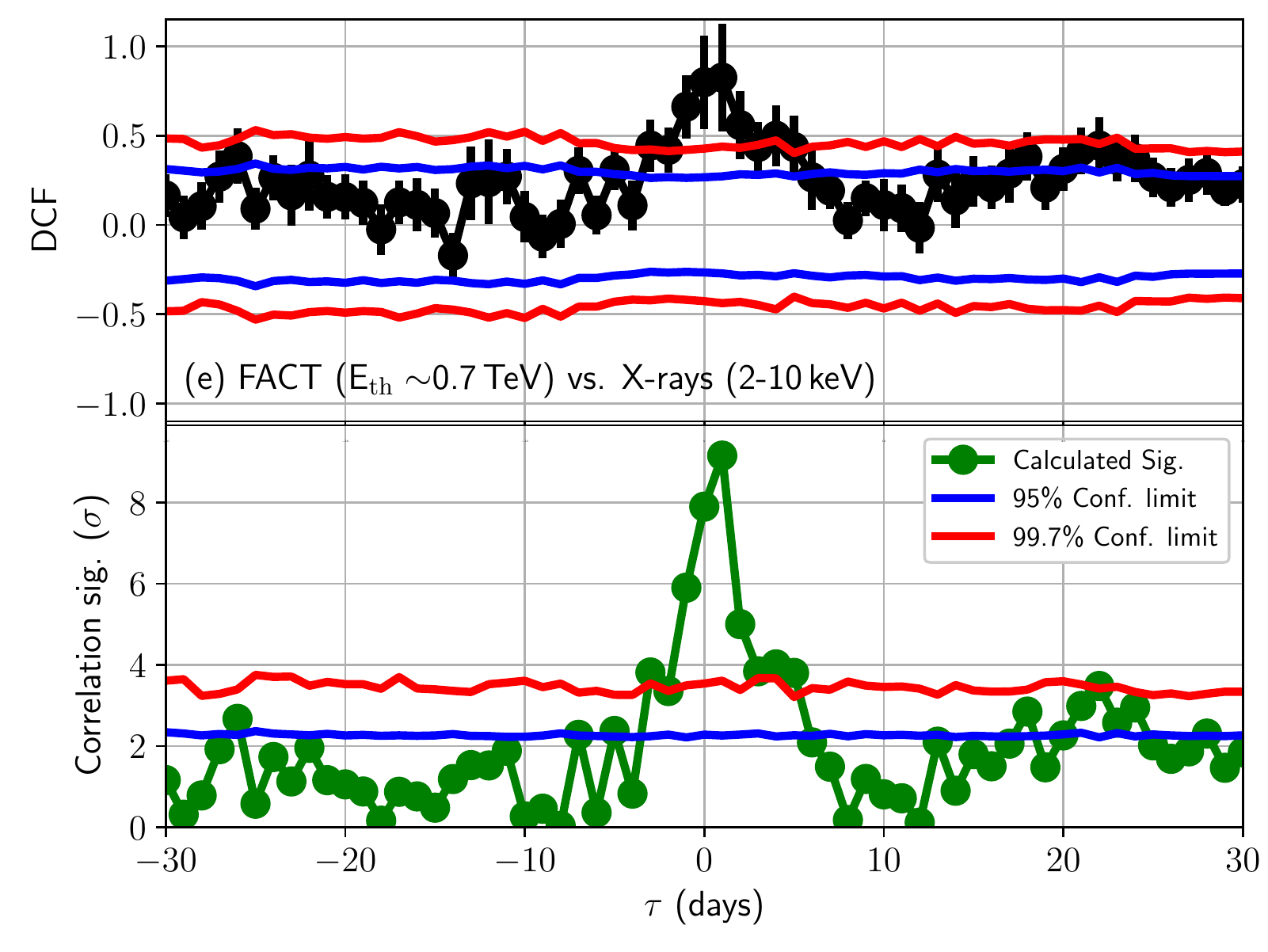}
\includegraphics[width=0.49\linewidth,height=5.5cm]{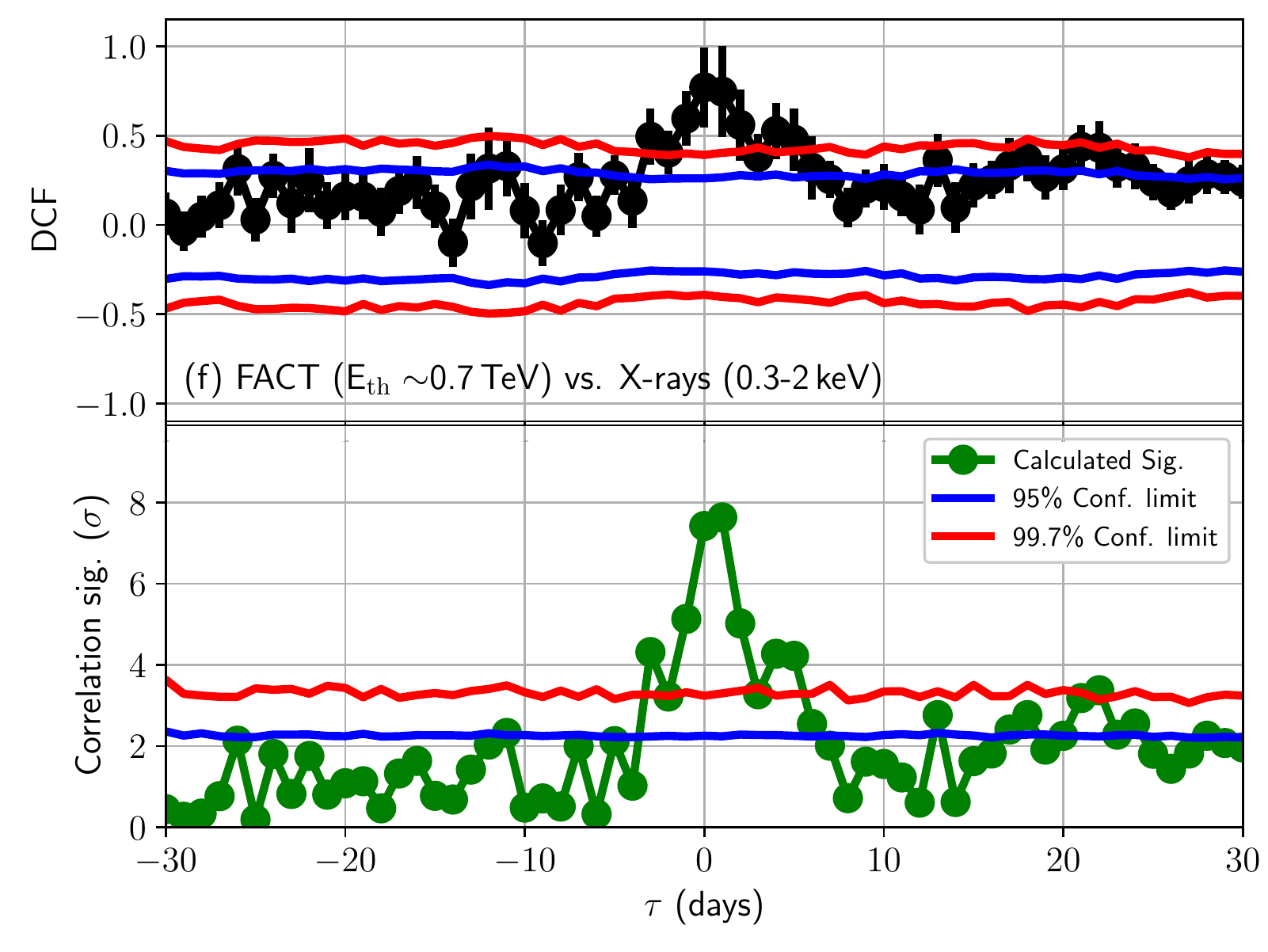}
\caption{Correlation between \textcolor{black}{VHE $\gamma$-rays and X-rays} during 2015--2016 from Mrk\,421 using
the
DCF and the Pearson correlation functions. The top and bottom blocks of each panel show the 
DCF and related errors, and the significance of the Pearson correlation, respectively. 
A positive time lag indicates a lag in the emission of the second (lower) energy band with respect to the first (higher) energy band.
The blue- and red-lines indicate the 95 and 99.7 per cent confidence intervals estimated from the 
Monte Carlo
simulations described in Section   \textcolor{black}{Section}  \ref{sec:corr}.}
\label{fig:DCF_main}
\end{figure*}

\begin{figure*}
\centering
\includegraphics[width=0.49\linewidth,height=5.5cm]{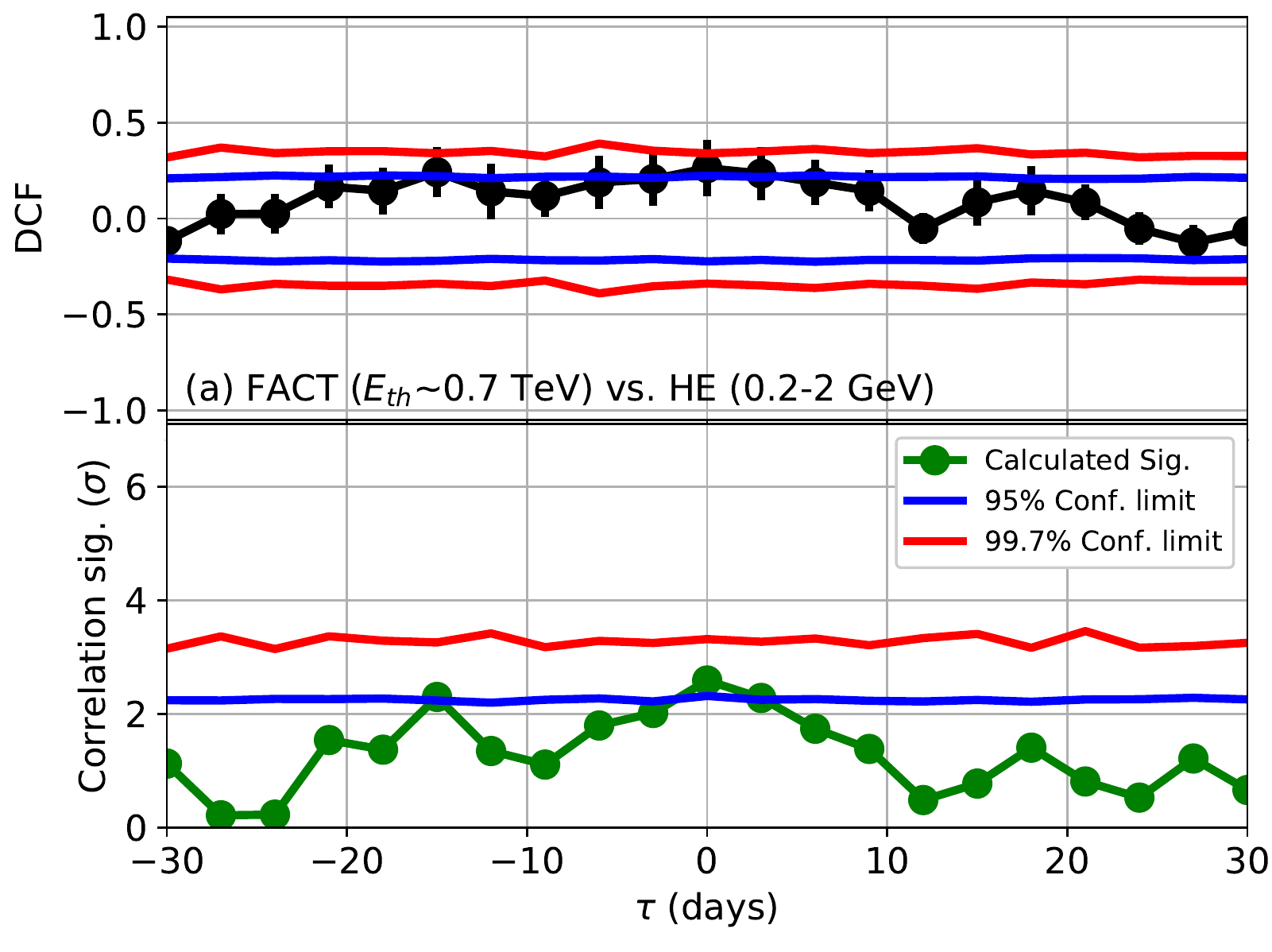}
\includegraphics[width=0.49\linewidth,height=5.5cm]{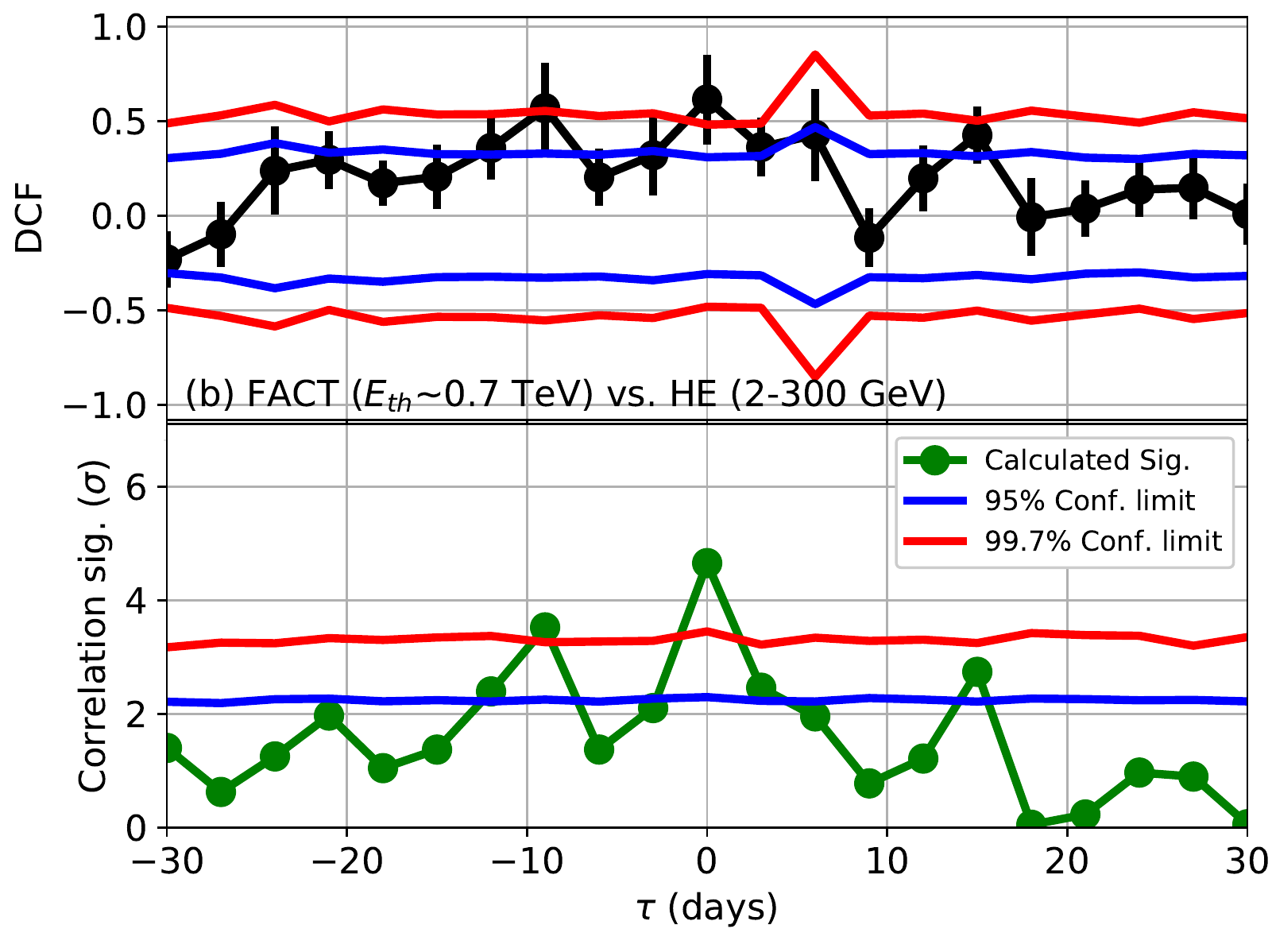}
\caption{Correlation between VHE $\gamma$-rays and HE $\gamma$-rays {using fluxes for 3-day time intervals from 2012 December to 2016 June.} See caption of Fig.~\ref{fig:DCF_main} for further explanations about the panel contents. }
\label{fig:DCF_main1_1}
\end{figure*}

In this section, we discuss
the potential
correlations between the different LCs presented in Fig.~\ref{fig:MWLC}. The correlation between two energy bands (two LCs) is quantified using two methods: the Pearson correlation coefficient with its related 1\,$\sigma$ error and correlation significance  \citep[calculated from][]{2002nrca.book.....P}, and the discrete correlation function \citep[DCF,][]{1988ApJ...333..646E}. The Pearson correlation is widely used in the community, but the DCF has the advantage over the Pearson correlation that it also uses the uncertainties in the individual flux measurements, which also contribute to the dispersion of the flux values, and hence affect the actual correlation between the two LCs. The DCF and Pearson correlation between two energy bands is computed with one LC and with a second shifted in time by zero or more {\em time lags}.
\textcolor{black}{We only consider the time lags where we have more than 10 simultaneous observations.}
As in   \textcolor{black}{Section}  \ref{sec:Fvar}, we only consider fluxes with SNR$>$2 (i.e. filled markers in Fig.~\ref{fig:MWLC}) for the characterization of the correlations. This ensures the usage of reliable flux measurements, and minimizes unwanted effects related to non-accounted (systematic) errors.

The calculated significance of the Pearson correlation and the uncertainties of the DCF do not necessarily relate to the actual significance of the correlation, because the correlation can be affected by the way the emission in the two bands has been sampled. A LC may have many data points in some time interval with some specific features (either real or due to fluctuations), and this may artificially boost the significance of the correlation. In order to better assess the reliability of the significance of the correlated \textcolor{black}{behaviour} computed with the measured LCs, we performed the same calculations using Monte Carlo simulated LCs. Each simulated LC is produced from the actual measured LC by randomly shuffling the temporal information of the flux data points, which ensures the resemblance to the actual measured LC in terms of flux values and flux uncertainties. For each correlation we want to study, we generate 10000 Monte Carlo simulated LCs, compute the DCF and Pearson correlations, and derive the 95 per cent \textcolor{black}{(2\,$\sigma$)} and 99.7 per cent  \textcolor{black}{(3\,$\sigma$)} confidence intervals by searching for the correlation values within which 9500 and 9970 cases are confined, respectively. The simulated LCs are not correlated, by construction, and hence the DCF and Pearson correlation values that lie outside the 3\,$\sigma$ contours can be considered as statistically significant (i.e. not produced by random fluctuations).

Despite the low  \textcolor{black}{flux} in the X-ray and VHE $\gamma$-ray bands in the 2015--2016 campaign, the related flux measurement uncertainties are relatively small, and the variability amplitudes in these bands are large, which allows relatively good accuracy in quantifying the correlation.  These correlations are computed using simultaneous observations (performed within 0.3 days)\footnote{In a few cases, there were more than one {\it Swift}-XRT short observations within the 0.3 days of the MAGIC or FACT observation. In these situations, we selected the X-ray observation that is closest in time to the VHE observation.}, and can be quantified on time lags of 1 day. We note that, as shown in the VHE and X-ray LCs from Fig~\ref{fig:MWLC}, there are substantial flux variations on timescales of 1--2 days, and hence it is important to be able to perform the correlation study for time lags of 1 day so that the study takes into account these relatively fast flux variations. However, when quantifying the correlation between the VHE emission measured with MAGIC and FACT and the HE emission measured with \textit{Fermi}-LAT, the study is limited by the 3-day time bins from the \textit{Fermi}-LAT light curves. The LAT analysis could be performed using time intervals of 1 day (instead of 3 days), but the limited sensitivity of LAT to measure Mrk\,421 during non-flaring activity would lead to large flux uncertainties, as well as many time intervals without significant measurements 
(we used SNR$>$2 for this study), which would affect the correlation study.

The radio, optical and the GeV emission of Mrk\,421 show a substantially lower amplitude variability (see Fig.~\ref{fig:Fvar}) and longer timescales for the flux variations (see Fig.~\ref{fig:MWLC}), in comparison to the keV and TeV bands. Because of that, the 2015--2016 data set is not large enough to evaluate reliably the possible correlations among these energy bands. In order to better quantify the correlations among these bands, we complemented the 2015–2016 data set with data from previous years (from 2007 to 2014). Some of these data have already been reported in previous papers \citep[][]{2012A&A...542A.100A,2015A&A...578A..22A,2016A&A...593A..91A,2016ApJ...819..156B}, while other data were specifically analyzed (or collected) for this study. A description of these complementary data sets is provided 
{in the supplementary online material (see Fig. \ref{fig:LongtermMWLC} in Appendix~\ref{sec:LongtermLC}).}
Differently to what occurs for the X-ray and VHE fluxes, the lower variability and longer variability timescales in the radio/optical/GeV emissions allow us to use the observations that are not strictly simultaneous, but only contemporaneous within a few days. For this study, we quantified the observations in temporal bins of 15 days, as done in \citet{2017MNRAS.472.3789C}. The study is performed in the same fashion as for the simultaneous X-ray/VHE fluxes, but with time-bins of 15 days instead of 1 day.

The following subsections report the results obtained from this correlation study, and in Section~\ref{sec:con}, we provide some discussion and interpretation of these results.

\begin{table*}
\centering
\begin{tabular}{c c c c c c } 
\hline \hline
\multirow{2}{*}{Light curve 1} & \multirow{2}{*}{Light curve 2}  &  \multirow{2}{*}{DCF}  &  \multirow{2}{*}{\pbox{2cm}{Pearson \\ Corr. Coeff. ($\sigma$)}} & \multicolumn{2}{c}{Normalized slope of fit}\\ \cline{5-6}
& & & & unbinned & binned \\
\hline
\vspace{0.1cm}
MAGIC; $0.2-1$\,TeV & XRT; $0.3-2$\,keV & 0.80$\pm$0.12 & $0.81\substack{+0.05\\-0.06}$ (7.3)&0.86$\pm$0.02 &  0.96$\pm$0.30\\ 
\vspace{0.1cm}
MAGIC; $0.2-1$\,TeV & XRT; $2-10$\,keV & 0.70$\pm$0.1 &$0.71\substack{+0.07\\-0.08}$ (5.7)& 0.56$\pm$0.02 &  0.60$\pm$0.21\\ 
\vspace{0.1cm}
MAGIC; $>1.0$\,TeV & XRT; $0.3-2$\,keV & 0.64$\pm$0.12 & $0.62\substack{+0.1\\-0.11}$ (4.5)& 0.96$\pm$0.05 &  1.15$\pm$0.38\\
\vspace{0.1cm}
MAGIC; $>1.0$\,TeV & XRT; $2-10$\,keV & 0.67$\pm$0.12 &$0.65\substack{+0.08\\-0.10}$ (4.8)&  0.73$\pm$0.04 &  0.82$\pm$0.25\\ 
\vspace{0.1cm}
FACT; E$_\mathrm{th}\sim0.7$\,TeV & XRT; $0.3-2$\,keV & 0.76$\pm$0.22 & $0.72\substack{+0.05\\-0.06}$ (7.4)& 1.00$\pm$0.05  &  1.20$\pm$0.53\\ 
\vspace{0.1cm}
FACT; E$_\mathrm{th}\sim0.7$\,TeV & XRT; $2-10$\,keV & 0.80$\pm$0.26 & $0.74\substack{+0.05\\-0.06}$ (7.9)& 0.72$\pm$0.04  &  0.80$\pm$0.30\\ 
\hline
\end{tabular}
\caption{Correlation results 
for the X-rays and VHE $\gamma$-rays during 2015--2016 campaign.
This table reports the correlation results for 
\textcolor{black}{$\tau$=0}
(simultaneous emission).
The discrete correlation function (DCF) and the corresponding errors are calculated following \citet{1988ApJ...333..646E}. The 1\,$\sigma$ Pearson correlation errors are calculated following \citet{2002nrca.book.....P}. The slopes of fit for the unbinned (grey markers) and binned data (blue markers), presented in Fig. \ref{fig:FF}, are normalized with the average flux of the bands under consideration. See Section~\ref{sec:vhe-vs-xrays} for details.}\label{table:DCF}
\end{table*}

\begin{table*}
\centering
\begin{tabular}{c c c c c c } 
\hline \hline
\multirow{2}{*}{Light curve 1} & \multirow{2}{*}{Light curve 2}  &  \multirow{2}{*}{DCF}  &  \multirow{2}{*}{Pearson Corr. Coeff. ($\sigma$)} & \multicolumn{2}{c}{Normalized slope of fit}\\ \cline{5-6}
& & & & unbinned & binned \\
\hline
MAGIC; $0.2-1$\,TeV & LAT; $0.2-2$\,GeV & {0.57$\pm$0.21} & ${0.37}\substack{{+0.11}\\{-0.12}}$ {(2.9)}& {3.28$\pm$0.74} & {0.67$\pm$0.59} \\
\vspace{0.1cm}
MAGIC; $0.2-1$\,TeV & LAT; $2-300$\,GeV &  {0.86$\pm$0.35}& ${0.41}\substack{{+0.13}\\{-0.15}}$ {(2.5)} & {2.84$\pm$1.03} & {0.39$\pm$0.56} \\
\vspace{0.1cm}
MAGIC; $>1$\,TeV & LAT; $0.2-2$\,GeV & {0.42$\pm$0.24} & ${0.26}\substack{{+0.14}\\{-0.15}}$ {(1.7)}& {5.30$\pm$1.71} & {0.41$\pm$1.09}\\
\vspace{0.1cm}
MAGIC; $>1$\,TeV & LAT; $2-300$\,GeV &  {-0.03$\pm$0.34} & ${-0.01}\substack{{+0.18}\\{-0.18}}$ {(0.1)} & 
{--}
& {--} \\
\vspace{0.1cm}
FACT; E$_\mathrm{th}\sim0.7$\,TeV & LAT; $0.2-2$\,GeV & {0.48$\pm$0.17} & ${0.32}\substack{{+0.09}\\-{0.10}}$ {(3.0)}& {4.98$\pm$1.04} &  {0.64$\pm$0.55}\\
\vspace{0.1cm}
FACT; E$_\mathrm{th}\sim0.7$\,TeV & LAT; $2-300$\,GeV &  {0.88$\pm$0.35} &${0.53}\substack{{+0.08}\\{-0.09}}$ {(4.9)} & {3.29$\pm$0.75} & {0.71$\pm$0.52} \\
\vspace{0.1cm}
{FACT; E$_\mathrm{th}\sim0.7$\,TeV; 2013--2016} & 
{LAT; $0.2-2$\,GeV; 2013--2016} & {0.26$\pm$0.15} & ${0.22}\substack{{+0.08}\\{-0.08}}$ {(2.6)}& {5.67$\pm$0.92} &  {0.84$\pm$0.40}\\
\vspace{0.1cm}
{FACT; E$_\mathrm{th}\sim0.7$\,TeV; 2013--2016} & 
{LAT; $2-300$\,GeV; 2013--2016} &  {0.61$\pm$0.24} &${0.41}\substack{{+0.07}\\{-0.08}}$ {(4.7)} & {3.69$\pm$0.62} & {0.65$\pm$0.48} \\
\hline
\end{tabular}
\caption{{Correlation results for HE and VHE $\gamma$-rays for 
$\tau$=0 (simultaneous emission) for 2015--2016 campaign, except for the last two rows, where the correlation results are computed using the data from 2012 December to 2016 June (2013--2016). See Section \ref{sec:vhe-vs-he} for details.}}\label{table:DCF_he_vhe}
\end{table*}

\subsection{VHE \texorpdfstring{$\gamma$}--rays and X-rays}
\label{sec:vhe-vs-xrays}

The quantification of the correlations between the VHE $\gamma$-rays and X-rays for a range of $\pm$30\,days, examined in steps of 1 day, is reported in the panels (a)--(f)
of
Fig.~\ref{fig:DCF_main}. All
of
the panels report the DCF vs. the time lag and the significance of the Pearson correlation vs. the time lag. The panels (a)--(d) show the correlation for the two energy bands ($0.2-1$\,TeV and $>$1\,TeV) measured with MAGIC and the two energy bands ($0.3-2$\,keV and $2-10$\,keV) observed with \textit{Swift}-XRT, and the panels (e) and (f) show the correlations obtained using the VHE flux with E$_{th}$ $\sim$0.7\,TeV measured with FACT, and the two energy bands from the \textit{Swift}-XRT.

All the panels (all the energy bands probed) show a positive correlation above 3\,$\sigma$ for 
\textcolor{black}{$\tau$=0,}
which drops quickly for negative and positive lags.  While the shape of the DCF peak is similar for all the bands, the peak in the significance of the Pearson correlation is narrower when using MAGIC than when using FACT. This is produced by the rapid drop in the number of available flux-flux pairs when examining time lags different from zero (simultaneous observations), which critically affects the significance with which a correlation is measured. In the case of MAGIC ,
the number of flux-flux pairs for
$\tau$=0 is  
\textcolor{black}{45},
while the number drops to 
\textcolor{black}{14}
for
\textcolor{black}{\mbox{$\tau$= $-1$ day}}
(X-ray LC shifted 1 day earlier) and 
\textcolor{black}{20}
for \textcolor{black}{$\tau$= $+1$ day}
(X-ray LC shifted 1 day later). On the other hand, when using FACT,  the number of flux-flux pairs for 
\textcolor{black}{$\tau$= 0} is 
\textcolor{black}{71},
and the number is 
\textcolor{black}{71 (72)}
for \mbox{$\tau$= -1} (+1) day, which ensures the same resolution to evaluate the correlation for these different time lags. Table~\ref{table:DCF} reports the DCF and the Pearson correlation, with their related 1\,$\sigma$ uncertainties, and the significance of the Pearson correlation for 
\textcolor{black}{\mbox{$\tau$= 0}}
(simultaneous observations). This table also reports the normalized slopes that relates the VHE $\gamma$-ray and the X-ray fluxes in the various energy bands (see Fig.~\ref{fig:FF}
{in the supplementary online material  Appendix~\ref{sec:FluxFlux}).}
\subsection{VHE $\gamma$-rays and HE $\gamma$-rays }
\label{sec:vhe-vs-he}

In this study, the daily LCs from MAGIC and FACT were 
{prepared}
to match the three-day cadence of the HE $\gamma$-ray LC from \textit{Fermi}-LAT.
{The DCF and Pearson correlation values for the various combinations of
bands from MAGIC, FACT and \textit{Fermi}-LAT are reported in
Table~\ref{table:DCF_he_vhe} for $\tau$= 0. We do not find any significant correlation
between the MAGIC and the LAT energy bands. In this case, the ability to see
correlation is limited by the statistical uncertainties in the LAT
fluxes (for 3-day time intervals) and by the low number of VHE-HE pairs with fluxes that have a SNR$>$2, which are 37 and 33 
when comparing the MAGIC bands $0.2-1$\,TeV and above 1\,TeV with the LAT flux above 2\,GeV, respectively.}

{Despite the larger flux uncertainties from FACT in comparison with those from MAGIC, the number of FACT-LAT data pairs (with SNR$>$2) is about twice as large as MAGIC-LAT: 85 and 71 for the LAT bands $0.2-2$\,GeV and $2-300$\,GeV {respectively}. This is due to the larger sampling and larger temporal coverage from FACT with respect to that from MAGIC.  This includes the additional temporal coverage provided by FACT in 2014 November-December and 2016 June, when Mrk\,421 showed 
{an}
enhanced VHE flux, which appears to have a counterpart in the GeV range (see Fig. \ref{fig:MWLC}). Because of the low
fractional variability in the GeV range, 
the additional temporal coverage provided by FACT {proved beneficial} for accumulating valid information for the understanding of this correlated behaviour.
We find that the Pearson correlation between the FACT VHE flux
(E$_\mathrm{th}\sim0.7$\,TeV) and the LAT HE flux above 2\,GeV is about 0.5 with a significance of almost 5$\sigma$ (with a DCF=$0.88\pm0.35$). The correlation, however, is not significant when using the LAT band $0.2-2$\,GeV, {which} yields only a Pearson correlation {coefficient} of 0.3 with a significance of 3$\sigma$ (with a DCF=$0.48\pm0.17$). }
{In order to better evaluate the correlation between the VHE FACT fluxes and LAT, we decided to complement the FACT data set with the fluxes obtained during the previous years, altogether enlarging the data set to cover the period from 2012 December to 2016 June 
(see {the supplementary online material} in 
Appendix~\ref{sec:LongtermLC}).
The results obtained for this data set of relatively continuous
coverage during 3.5 years (apart from bad weather and periods of no visibility due to the
Sun) are reported in the last
two rows of Table~\ref{table:DCF_he_vhe}. In this case, the number of VHE-HE data
pairs (with SNR$>$2) is 140 and 118 for the LAT bands $0.2-2$\,GeV and
$2-300$\,GeV, respectively. The results are similar to those obtained for the time period from 2014 November to 2016 June. The correlation is not significant for the band $0.2-2$\,GeV, {which yields} a Pearson correlation {coefficient} value of 0.2 with a significance of 2.6$\sigma$ (DCF=$0.26\pm0.15$), while it is marginally significant for the fluxes above 2\,GeV, {which} a Pearson correlation {coefficient} of 0.4 with a significance of 4.7$\sigma$ (DCF=$0.61\pm0.24$). We also studied the magnitude of the correlation for different time lags, for a range of $\pm$30\,days in three-day steps, including a toy MC to evaluate the 2$\sigma$ and 3$\sigma$ confidence intervals. The results are shown in Fig.~\ref{fig:DCF_main1_1}, 
{leading}
to the conclusion that the correlation is only (marginally) significant for the fluxes above 2\,GeV and for $\tau$= 0. The flux-flux correlation plots for the FACT VHE fluxes (E$_\mathrm{th}\sim0.7$\,TeV) and the two \textit{Fermi}-LAT energy bands are shown in 
the supplementary online material (Fig.~\ref{fig:FF-VHE-HE} in Appendix~\ref{sec:FluxFlux}).}

A similar correlation had been previously reported in \citet{2016ApJS..222....6B} 
for VHE $\gamma$-rays measured with the ARGO-YBJ at TeV energies and the HE $\gamma$-rays measured with \textit{Fermi}-LAT above 0.3\,GeV. They quantified the correlation with the DCF analysis, obtaining a correlation for 
\textcolor{black}{$\tau$= 0}
with DCF=$0.61\pm0.22$. The main differences with respect to the
result presented here are the somewhat different energy bands
involved, and the very different temporal scales used for these two
correlation studies. While \citet{2016ApJS..222....6B} used data from
mid 2008 to 2013 in \textcolor{black}{30-day bins}, {we performed the
study with data {from} the end of 2012 to mid 2016 } in time bins of 3 days. Additionally, in
this paper, we also quantify the correlation using the Pearson
correlation function and Monte Carlo simulations to better evaluate
the reliability  of the significance of the correlation. 

\begin{figure*}
\centering
\includegraphics[width=0.51\linewidth,height=5.5cm]{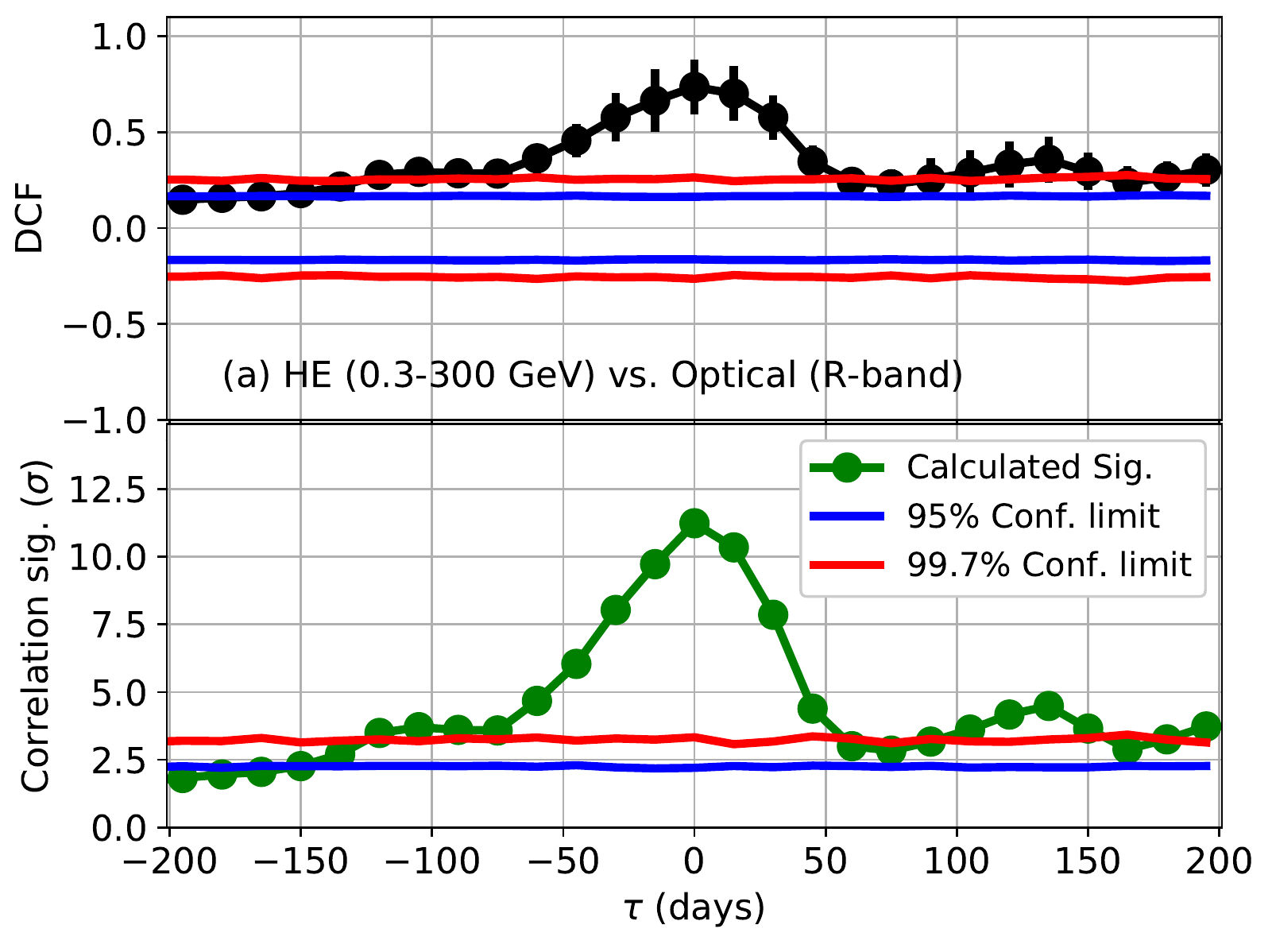}
\includegraphics[width=0.49\linewidth,height=5.5cm]{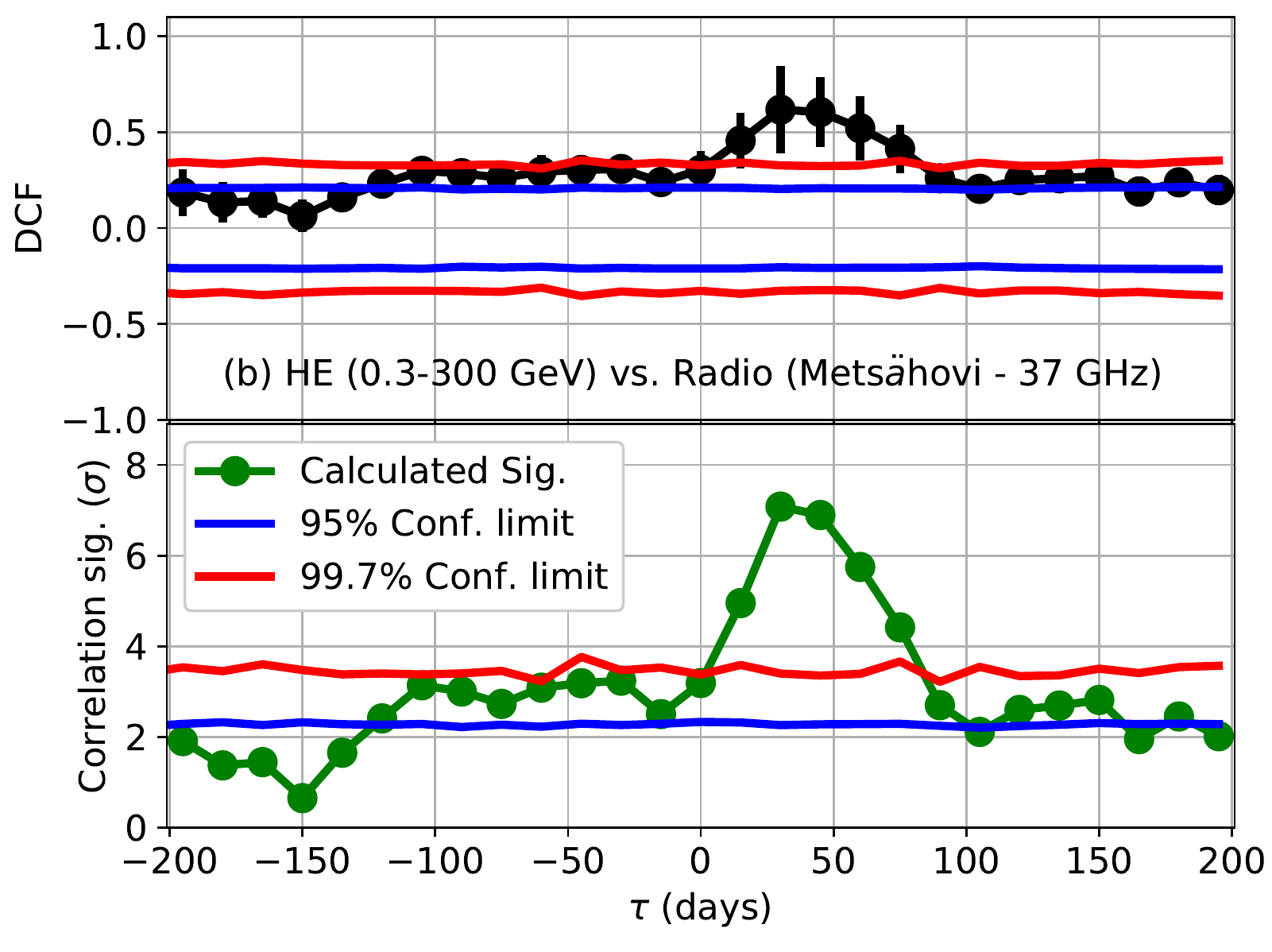}
\includegraphics[width=0.49\linewidth,height=5.5cm]{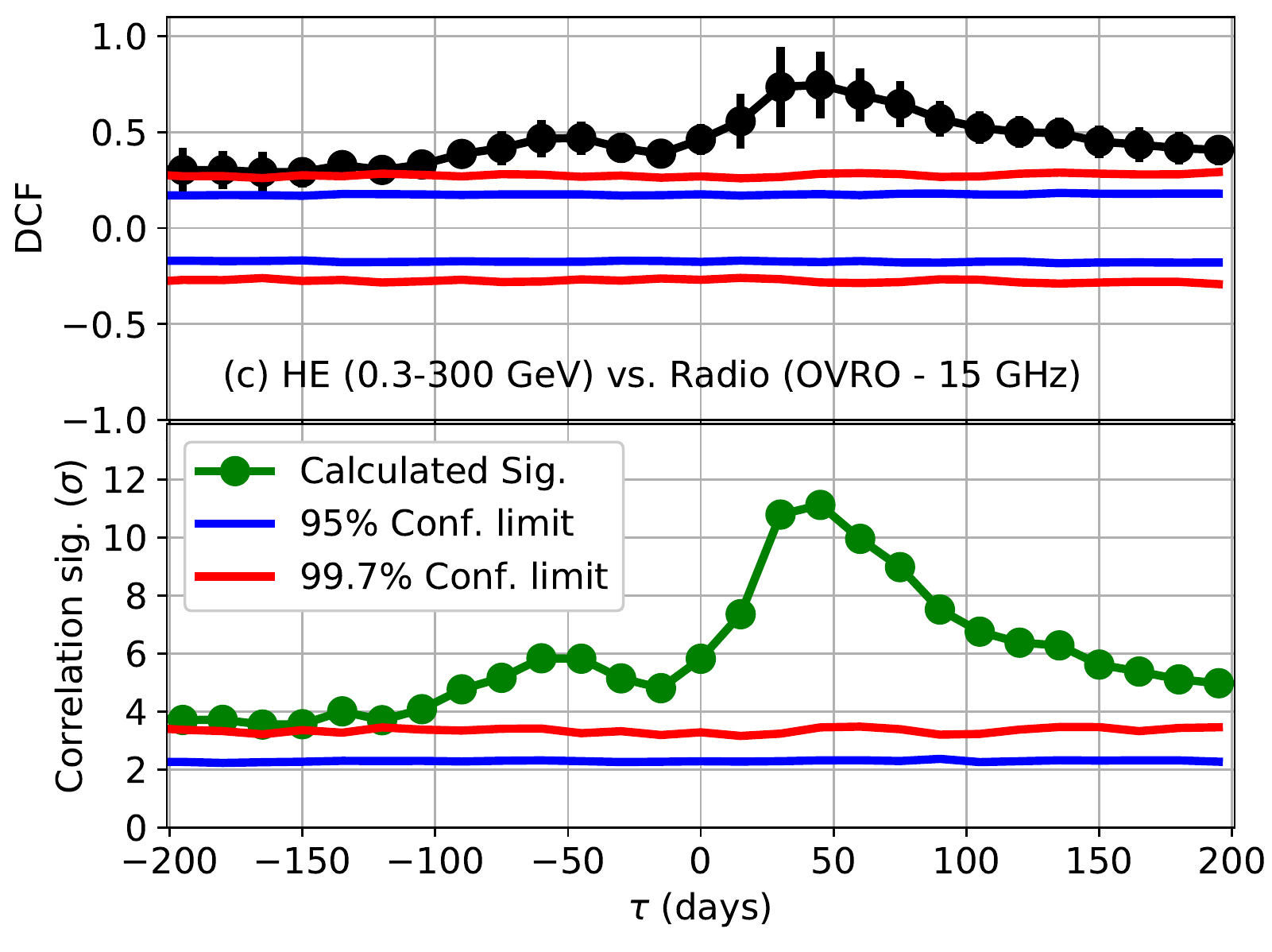}
\includegraphics[width=0.49\linewidth,height=5.5cm]{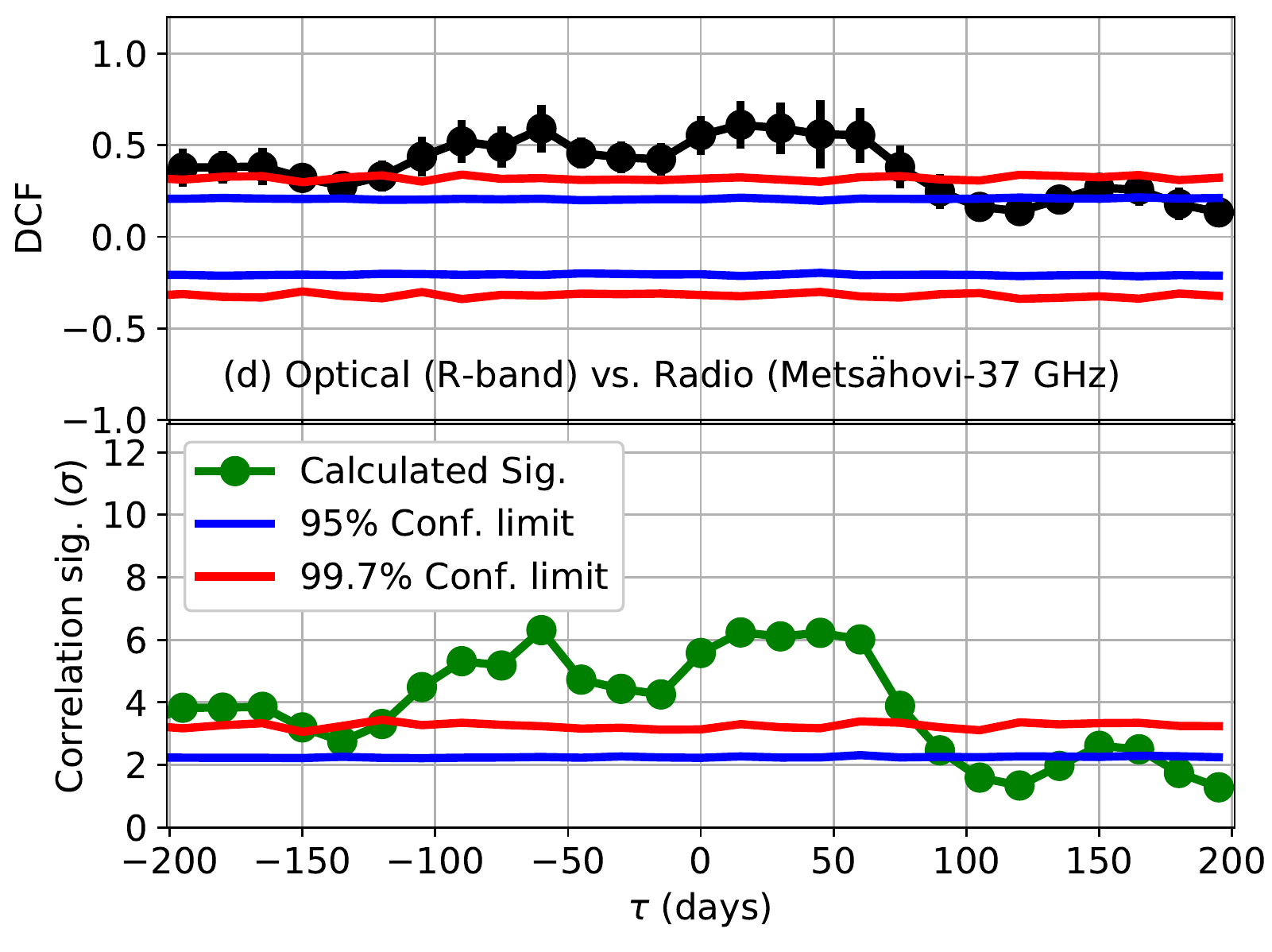}
\includegraphics[width=0.49\linewidth,height=5.5cm]{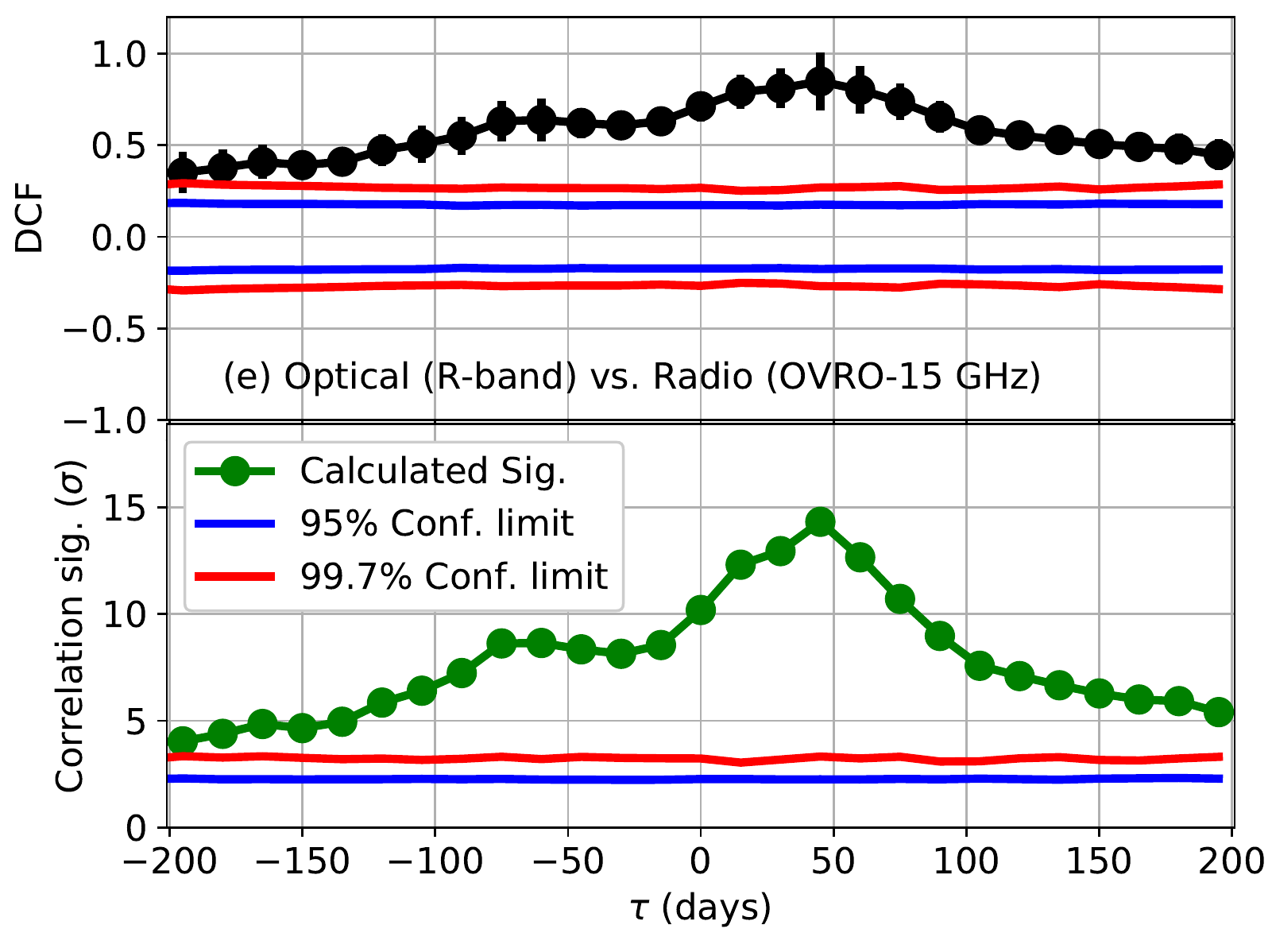}
\caption{Correlation between the HE $\gamma$-rays, optical (R-band) and two radio bands using fluxes for 
15-day time intervals 
from 2007 to 2016. See caption of Fig.~\ref{fig:DCF_main} for further explanations about the panel contents.}
\label{fig:DCF_main1}
\end{figure*}

\subsection{HE $\gamma$-rays and optical band}
\label{sec:he-vs-optical}

The panel (a) of Fig.~\ref{fig:DCF_main1} shows the quantification of the correlation between the HE fluxes in the $0.3-300$\,GeV energy band measured with \textit{Fermi}-LAT and the optical fluxes in the R-band, as measured by a large number of instruments \textcolor{black}{over a time range spanning from 2007 to 2016 }(see {the supplementary online material in}
Appendix~\ref{sec:LongtermLC}). The correlation is computed for a time lag range of $\pm$200\,days in steps of 15\,days, with the HE and R-band fluxes computed in 15-day temporal bins. The plot shows a correlation peak of about 60 days FWHM, and centered at 
\textcolor{black}{$\tau$= 0.}
As reported in Table~\ref{table:DCF1}, the Pearson correlation coefficient is $0.72\pm0.04$, with a correlation significance of about 11\,$\sigma$, and the DCF is $0.74\pm0.17$. 
Because of the 15-day fluxes and 15-day time steps, the resolution with which we can estimate the time lag with the highest correlation is somewhat limited. Following the prescription from \citet{1998PASP..110..660P}, we estimate the time lag with the highest correlation is $3^{+5} _{-9}$ days (see 
{the supplementary online material in }
Appendix~\ref{sec:DCFLag} for details), which is perfectly consistent with no time lag, suggesting that the emission in these two energy bands is simultaneous.
Panel (a) of Fig.~\ref{fig:FF1} shows that 
the relation between the GeV and R-band fluxes can be approximated by a linear function with a normalized slope of 0.6--0.7 (see Table~\ref{table:DCF1}).\par
A positive correlation between the multi-year \textit{Fermi}-LAT $\gamma$-ray flux and the optical R-band flux had been first reported in Fig.~25 of \citet{2017MNRAS.472.3789C}. The DCF from that study, also performed in steps of 15\,days, shows a broad peak of many tens of days around 
\textcolor{black}{$\tau$= 0,}
with the highest DCF value being around 0.4, for the multi-year data set.  However, the significance of the correlation was not quantified in \citet{2017MNRAS.472.3789C}. In this 
paper, we show that a DCF of 0.4 is not necessarily related to a significant ($>$3\,$\sigma$) correlation. We also show that the \textit{Fermi}-LAT $\gamma$-ray flux and optical R-band emissions are positively correlated with a DCF of about 0.8, and with a very high significance ($>$12\,$\sigma$), hence confirming and further strengthening the claims made in \citet{2017MNRAS.472.3789C}.

\begin{table*}
\centering
\begin{tabular}{c c c c c c c c }
\hline \hline
\multirow{2}{*}{Light curve 1} & \multirow{2}{*}{Light curve 2} & \multirow{2}{*}{Time-shift [days]} &  \multirow{2}{*}{DCF}  &  \multirow{2}{*}{Pearson Corr. Coeff. ($\sigma$)} & \multicolumn{2}{c}{Normalized slope of fit}\\ \cline{6-7}
 & & & & & unbinned & binned \\
\hline
\vspace{0.1cm}
HE $\gamma$-ray (LAT; $0.3-300$\,GeV) & Optical (R-band) & 0 & 0.74$\pm$0.14 & $0.72\substack{+0.04\\-0.04}$ (11.2)& 0.66 $\pm$ 0.03 & 0.63 $\pm$ 0.21 \\
\vspace{0.1cm}
HE $\gamma$-ray (LAT; $0.3-300$\,GeV) & Radio (Mets{\"a}hovi; 37\,GHz) & 45 & 0.60$\pm$0.18 &$0.53\substack{+0.06\\-0.06}$ (6.9)& 2.63$\pm$0.17 & 0.79$\pm$0.33 \\ 
\vspace{0.1cm}
HE $\gamma$-ray (LAT; $0.3-300$\,GeV) & Radio (OVRO; 15\,GHz)& 45 & 0.75$\pm$0.17 &$0.72\substack{+0.04\\-0.04}$ (11.1)& 1.53$\pm$0.06 & 1.32$\pm$0.41 \\
\vspace{0.1cm}
Optical (R-band) & Radio (Mets{\"a}hovi; 37\,GHz) & 45 & 0.56$\pm$0.18 & $0.50\substack{+0.06\\-0.07}$ (6.2) &  2.93 $\pm$ 0.17 & 0.83 $\pm$ 0.33\\ 
\vspace{0.1cm}
Optical (R-band) & Radio (OVRO; 15\,GHz) & 45 & 0.85$\pm$0.16 & $0.84\substack{+0.02\\-0.03}$ (14.3)& 2.82 $\pm$ 0.02 & 2.0 $\pm$ 0.35\\
\hline
\end{tabular}
\caption{
Correlation results between the low-variability radiation components of Mrk\,421. The long-term (2007--2016) data 
have been used for the correlation results of the radio, optical, and 
HE $\gamma$-rays.
The column time-shift reports the temporal shift applied to the second light curve with respect to the first one. This time shift corresponds to the time lag with the highest DCF in Fig.~\ref{fig:DCF_main1}. The various columns report the same quantities as in Table~\ref{table:DCF}.
The slopes of fit for the unbinned (grey markers) and binned data (blue markers), presented in Fig.~\ref{fig:FF1}, are normalized with the average flux in the corresponding bands. See   \textcolor{black}{Section}  \ref{sec:corr} for details.}
\label{table:DCF1}
\end{table*}

\subsection{HE $\gamma$-rays and radio band}
\label{sec:he-vs-radio}
Panels (b) and (c) of Fig.~\ref{fig:DCF_main1} show the correlation between the HE $\gamma$-rays in the $0.3-300$\,GeV energy band, measured with \textit{Fermi}-LAT, and the 37\,GHz and 15\,GHz radio flux densities, as measured with Mets{\"a}hovi and OVRO {over a time range spanning from 2007 to 2016 }(see 
{the supplementary online material, Fig. \ref{fig:LongtermMWLC} in }
Appendix~\ref{sec:LongtermLC}). In both cases, one finds a positive correlation characterized by a wide peak, of about 60 days, centered at 
\textcolor{black}{$\tau\sim45$ days.}

{The supplementary online material}
(Appendix~\ref{sec:DCFLag}) reports an estimation of the time lag between these energy bands, obtained with the prescriptions from \citet{1998PASP..110..660P}. We estimate that the time lag between the HE $\gamma$-rays and the 37\,GHz radio flux is 
$41\substack{+10\\-11}$ days, while for the 15\,GHz radio flux it is $47^{+5} _{-9}$ days. 
The panels (b) and (c) of Fig.~\ref{fig:FF1} show that, for a time shift of 45 days, the relation between the GeV and the radio fluxes can be approximated by a linear function. 
As reported in Table~\ref{table:DCF1}, for a time shift of 45\,days, the Pearson correlation coefficient is about 0.5--0.7, with a correlation significance of 7\,$\sigma$ for Mets{\"a}hovi and 11\,$\sigma$ for OVRO, and the DCF is $0.6\pm0.2$ and $0.7\pm0.2$, respectively for Mets{\"a}hovi and OVRO.  Therefore, the correlation between these bands is robustly measured.

The radio emission of blazars has been found to be correlated to the $\gamma$-ray emission using EGRET data \citep[e.g.][]{2001ApJS..134..181J, 2003ApJ...590...95L}  and \textit{Fermi}-LAT data \citep[e.g.][]{2011A&A...532A.146L, 2011ApJ...741...30A}, very often with the radio emission delayed with respect to the $\gamma$-ray emission by tens and hundreds of days \citep[e.g.][]{2015MNRAS.452.1280R}. As for the specific case of Mrk\,421,  \citet{2014MNRAS.445..428M} had first reported a positive correlation between $\gamma$-rays from \textit{Fermi}-LAT and radio from OVRO for a time lag that, 
using the recipe from \citet{1998PASP..110..660P}, was estimated to be 40$\pm$9 days.  However, the correlation reported in that paper was only at the level of 2.6\,$\sigma$ ($p$-value of 0.0104), quantified with a dedicated MC simulation, and strongly affected by the large $\gamma$-ray and radio flares 
from July and September 2012, respectively \citep[][]{2014MNRAS.445..428M}.
\citet{2015MNRAS.448.3121H}, which considered also data from another (smaller) radio flare in 2013, reported a positive correlation for a range of 
\textcolor{black}{$\tau$}
of about 40--70\,days, but did not assign any significance to this measurement. 
In the study reported
upon here our dedicated MC simulations show that the significance of the correlation between \textit{Fermi}-LAT and OVRO is well above 
the 3$\sigma$ contour, and, when using the prescription from \citet{2002nrca.book.....P} to quantify it, we obtained  11\,$\sigma$. Moreover, because of {a data set twice as large as the data used in \citet{2015MNRAS.448.3121H},}
it is not dominated by the large $\gamma$-ray and radio flares in 2012.
In order to better understand this correlation, we removed this large $\gamma$-ray and radio flare 
from 2012 by generously excluding the time interval MJD~56138--56273 from both the $\gamma$-ray and radio LCs,  and repeated the test. We obtained a positive correlation with a significance of 9\,$\sigma$, with a peak that extends over a range of about 60 days, centered at
\textcolor{black}{$\tau\sim45$ days.}
Therefore, we confirm and further strengthen the correlation reported in \citet{2014MNRAS.445..428M}, 
stating with reliability that this is an intrinsic characteristic in the multi-year emission of Mrk\,421, and not a particularity of a rare flaring activity.

\subsection{Optical band and radio band}
\label{sec:optical-vs-radio}

Panels (d) and (e) of Fig.~\ref{fig:DCF_main1} show the correlation between the flux in the optical R-band from GASP-WEBT and the 37\,GHz and 15\,GHz radio flux densities measured with Mets{\"a}hovi and OVRO, respectively. In the case of OVRO, one finds that the highest correlation occurs for 
\textcolor{black}{$\tau\sim45$ days,}
and it is characterized by a wide peak that resembles the one obtained for the GeV vs. 15\,GHz band, as depicted in the panel (c) of Fig.~\ref{fig:DCF_main1}. In the case of Mets{\"a}hovi, the DCF shows much wider structure, without any clear peak, but with high DCF values also around 
\textcolor{black}{$\tau\sim45$ days.}
As done above, we followed the prescriptions 
of
\citet{1998PASP..110..660P} to estimate the time lag between these bands (see 
{the supplementary online material in}
Appendix~\ref{sec:DCFLag}). We obtained 
\textcolor{black}{$\tau$=}
$33^{+19} _{-11}$ days for the R-band and the 37\,GHz radio flux, and 
\textcolor{black}{$\tau$=}
$39^{+6} _{-2}$ 
\textcolor{black}{days}
for the R-band and the 15\,GHz radio flux.  
The panels (d) and (e) of Fig.~\ref{fig:FF1} show that, for a time shift of 45 days, the relation between the R-band and the radio fluxes can be approximated by a linear function. 
As reported in Table~\ref{table:DCF1}, for a time shift of 45\,days, the Pearson correlation coefficient is 0.5 and 0.8, with a correlation significance of 6\,$\sigma$ and 14\,$\sigma$ for Mets{\"a}hovi  and for OVRO, respectively. The DCF is about 0.6 and 0.9 for them, hence indicating a very clear and significant correlated behaviour for these two bands.

\begin{figure*}
    \centering
    \includegraphics[width=\linewidth, height=12cm]{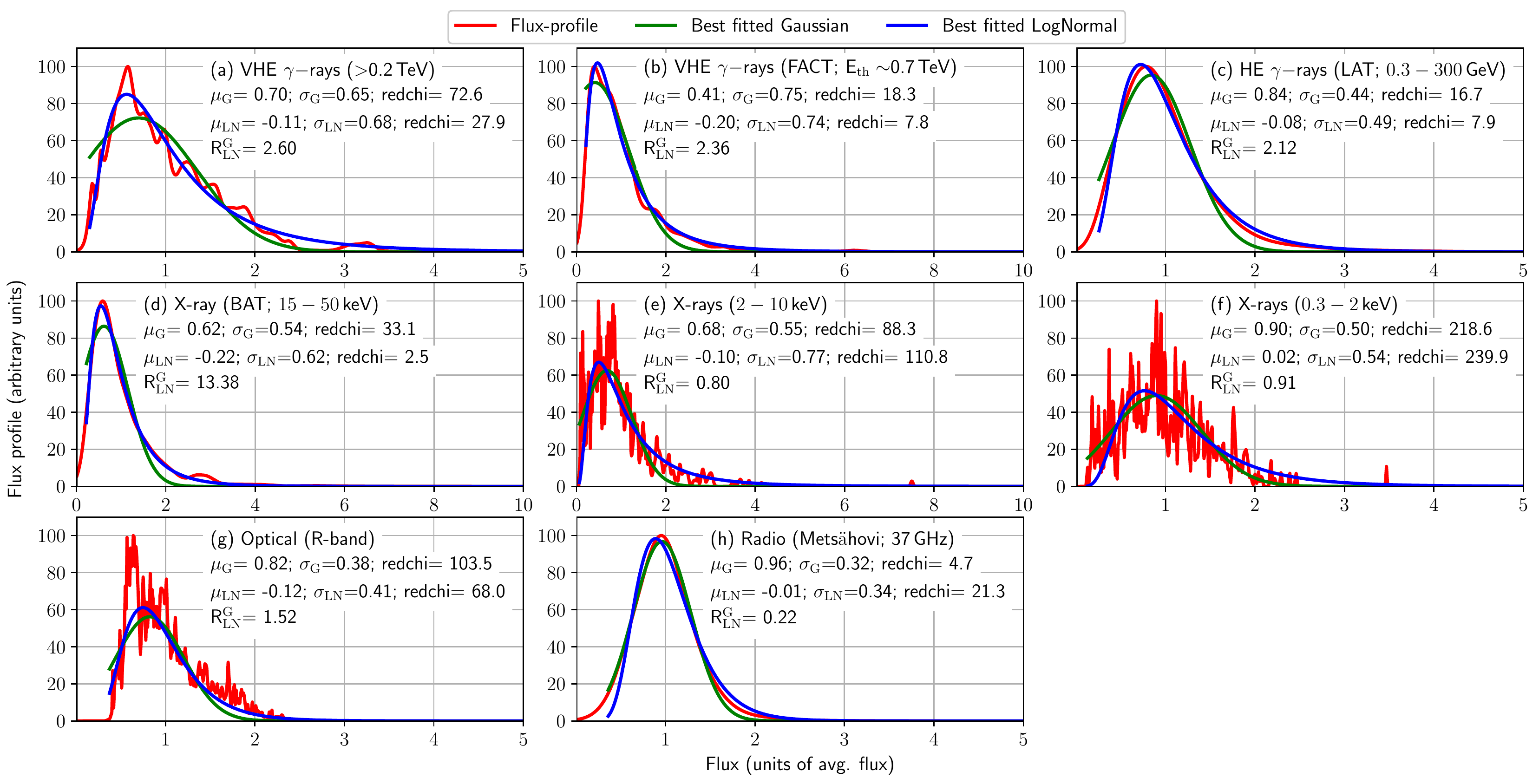}
    \caption{
      Flux distributions of Mrk\,421 in the 2007--2016 period in different energy bands, except for FACT, where only data from {2013}--2016 period were used (see
      {the supplementary online material in}
      Appendix~\ref{sec:LongtermLC}), {where we used} the flux profile method (see {the supplementary online material in} Appendix \ref{sec:A1}).}%
    \label{fig:SS1}%
\end{figure*}

\begin{table*}
\centering
\begin{tabular}{ c  c c c c c c c c}  \hline
\hline
\multirow{2}{*}{Energy-bands} & \multicolumn{2}{c}{G}&   \multicolumn{2}{c}{LN} & \multirow{2}{*}{MPS (Avg. flux)} &  \multicolumn{2}{c}{
redchi} & \multirow{2}{*}{R$^{G}_\mathrm{LN}$ (p)}\\ \cline{2-5} \cline{7-8}
   & $\mu_g$ & $\sigma_g$ &$\mu_\mathrm{LN}$ & $\sigma_\mathrm{LN}$ &   & G & LN \\ \hline
VHE $\gamma$-rays (MAGIC; $> 0.2$\,TeV)      & 0.70 & 0.65 & -0.11 & 0.68 &  0.56 (2.09$\times$10$^{-10}$\,ph\,cm$^{-2}$s$^{-1}$) & 72.6 & 28.0 & 2.6 ({ 4.4$\times$10$^{-2}$}) \\
VHE $\gamma$-rays (FACT; E$_\mathrm{th}\sim0.7$\,TeV) & {0.41} & {0.75} & {-0.20} & {0.74} &  {0.47} {(2.73$\times$10$^{-11}$\,ph\,cm$^{-2}$s$^{-1}$)}& {18.3} & {7.8} & {2.4 (2.7$\times$10$^{-1}$)} \\
HE $\gamma$-rays (LAT; $0.3-300$\,GeV)  & 0.84 & 0.44 & -0.08 & 0.50 &  0.72 (9.45$\times$10$^{-8}$\,ph\,cm$^{-2}$s$^{-1}$) & 16.7 & 8.0 & 2.1 {(8.1$\times$10$^{-2}$)}\\
X-ray (BAT; $15-50$\,keV)             & 0.62 & 0.54 & -0.22 & 0.62 &  0.54 (0.27$\times$10$^{-2}$counts\,cm$^{-2}$s$^{-1}$)& 33.1  & 2.5 & 13.4 { (1.6$\times$10$^{-1}$)}\\
X-ray ($2-10$\,keV)               & 0.68 & 0.55 & -0.10 & 0.77  &  0.68 (3.67$\times$10$^{-10}$erg\,cm$^{-2}$\,s$^{-1}$)& 88.3 & 111.0 & 0.80 { (2.4$\times$10$^{-2}$)}\\
X-ray ($0.3-2$\,keV)              & 0.90 & 0.50 &  0.02 & 0.54 &  0.90 (6.82$\times$10$^{-10}$\,erg\,cm$^{-2}$\,s$^{-1}$)& 218.6 & 240.0 & 0.91 { (1.7$\times$10$^{-2}$)}\\
%UV-w1                            & 0.89 & 0.55 & -0.02 & -0.23 &  0.38 & 0.09 & -0.02  & -0.23 & 0.93 & 25.6   LN: 1.0 GLN: 1.1   \\ \hline
Optical (R-band)                  & 0.82 & 0.38 &  -0.12 & 0.41 & 0.75 (24.37\,mJy) & 103.5  & 68.0 & 1.52 {  (1.0$\times$10$^{-3}$)}\\
Radio (Mets{\"a}hovi; 37\,GHz)     & 0.96 & 0.32 & -0.01 & 0.34  & 0.90 (0.50\,Jy) & 4.7 & 21.3 & 0.22 { ($<$1.0$\times$10$^{-4}$)}\\
%Radio (OVRO; 15\,GHz)              & 0.88 & 0.28 & -0.08 & 0.23  & 0.87 (0.53\,Jy)& 234.3 & 212.2 & 1.21 { (N.A.)}\\ 
\hline %(N.A.)
\end{tabular}
\caption{The model parameters for the flux profiles in different energy bands fitted with Gaussian (G) and LogNormal (LN) distributions. The most probable states (MPS) are 
\textcolor{black}{retrieved from the function preferred by the flux profile, and} 
presented as fractions of the mean flux (given in the parentheses). The parameter R$^{G}_\mathrm{LN}$ is the ratio of \texttt{redchi} for LN to the \texttt{redchi} for G (see main text) and is used to estimate the goodness of fit. R$^{G}_\mathrm{LN}$ $>$ 1 means the profile is likely to be fitted better with a LN. The chance probability (p), given in the parentheses in the last column indicates the probability of wrongly reconstructing a LogNormal (Gaussian) distribution as a Gaussian (LogNormal). See Section  \ref{sec:typicalstate} and  
{the supplementary online material
(Appendix \ref{sec:A1})} for details.
The results for the 15\,GHz band have not been included here as the distribution is bimodal. See the main text and
{the supplementary online material}
(Appendix~\ref{sec:appendixB}) for details. 
}
\label{table:SS}
\end{table*}

\section{Determination of the MWL flux distributions using the flux profile method}
\label{sec:typicalstate}
The emission mechanisms in accreting sources like 
active 
\textcolor{black}{Galactic}
nuclei
and X-ray binaries have been found to be 
consistent with stochastic processes
\citep{2006Natur.444..730M, 2012MNRAS.419.2657C, 2013ApJ...773..177N,2014ApJ...786..143S}. 
For a linear stochastic process, one expects a Gaussian distribution \textcolor{black}{of fluxes}. However, a LogNormal distribution was found to be preferred (over a Gaussian one) in the long-term X-ray light curve of the blazar BL Lac where the average amplitude variability was found to be proportional to the flux \citep{2009A&A...503..797G}. The 
\textcolor{black}{Galactic}
X-ray binary Cygnus X-1 also showed such features in X-rays \citep{2001MNRAS.323L..26U}. Since then, LogNormal behaviour has been observed in several blazars primarily in optical/near IR, X-ray and $\gamma$-ray wavelengths \citep{2016A&A...591A..83S, 2017ApJ...836...83S, 2018Galax...6..135R, 2020ApJ...891..170V}. The presence of LogNormality indicates an underlying multiplicative process in blazars contrary to the additive physical process. It has been suggested that such multiplicative processes originate in the accretion disk \citep{1997MNRAS.292..679L, 2005MNRAS.359..345U, 2010LNP...794..203M}, however, \citet{2012MNRAS.420..604N} strongly argue the variability to originate within the jet. In case of Mrk 421, using data from 1991 to 2008, mostly from the old generation of VHE ground-based $\gamma$-ray instruments, the flux distribution above 1\,TeV was found to be consistent with a combination of a Gaussian and a LogNormal distribution  \citep{2010A&A...524A..48T}. The  improvement  of the sensitivity of the present day telescopes over last few years now 
\textcolor{black}{provides}
us with the opportunity to study the flux states with a much better accuracy, and a minimum energy as low as 0.2\,TeV, where the minimum energy is always above the analysis energy threshold.

\newpage
Here, we report on a detailed study of the flux distributions observed in different wave-bands, from radio to VHE $\gamma$-rays, using the data from the 2015--2016 campaigns, together with  previously published   \textcolor{black}{MWL} data from the 2007, 2008, 2009, 2010 and 2013 campaigns \citep{2012A&A...542A.100A, 2016A&A...593A..91A, 2015A&A...578A..22A, 2016ApJ...819..156B}, published multi-year optical R-band data \citep{2017MNRAS.472.3789C}, and unpublished data at radio (OVRO, Metasahovi), hard X-ray (\textit{Swift}-BAT) and GeV $\gamma$-rays (\textit{Fermi}-LAT). The multi-year light curves used for this study are reported in 
{the supplementary online material
(Appendix~\ref{sec:LongtermLC})}. 
\textcolor{black}{The two large VHE $\gamma$-ray flaring episodes of Mrk\,421 in 2010 February \citep{2020ApJ...890...97A} and 2013 April \citep{2020ApJS..248...29A} have been excluded to avoid large biases in the distributions. During these two time intervals of about 1 week, Mrk\,421 showed a VHE activity larger than 20 times its typical flux and, because of the exceptional activity, the number of X-ray and VHE observations were also increased by more than one order of magnitude with respect to the typical temporal coverage during the regular MWL campaigns. The inclusion of these two periods would create a large structure in the X-ray and VHE $\gamma$-ray distributions at fluxes of about ten times the typical ones, and would hamper any fit with a smooth function, like Gaussian or LogNormal.} 
The data used here relate to time intervals when Mrk\,421 showed typical or low activity (e.g. during years 2007, 2009, 2015, 2016) or somewhat enhanced activity, as it happened during year 2008  and 2 weeks in 2010 March. 
\textcolor{black}{Because of the high activity in 2008, some of the X-ray and VHE observations came from dedicated ToOs, which increased somewhat the number of observations that would not have been performed in the absence of high activity. The accurate identification of the "extra observations" is complicated because the dynamic scheduling that was being used at the time, and the fact that these observations occurred 12 years ago. We note that the inclusion of the 2008 data introduces a bias towards high fluxes in the X-ray and VHE flux distributions (because of the additional observations during a period of high activity). However, in 
{the supplementary online material
(Appendix~\ref{sec:LongtermLC})}, we show that the results about the shape of the distribution do not change in a substantial way, even when removing completely the data related to the entire year 2008.}

\textcolor{black}{In order to study the general shape of the flux-distribution and estimate the most-probable flux state, we developed a method largely inspired by the kernel density estimation (KDE), dubbed "flux profile construction".}
We treat each flux measurement, in a given energy band, as a Gaussian with the flux values as the mean and the flux uncertainty as the standard deviation. The amplitude is inversely proportional to the standard deviation, so that the area under each individual Gaussian is unity. A ``flux profile'' for a certain energy band is constructed by adding all individual flux measurements in that band.
In order to determine the preferred shape of a flux profile, we fit 
\textcolor{black}{the flux profile staring from the minimum flux}
with the following functions:\newline
1) Gaussian: G(x; $\mu_G,\sigma_G$) = $\frac{N_{G}}{\sigma_G\sqrt{2\pi}}  e^{-\frac{(x-\mu_G)^2}{2 \sigma_G^{2}}}$, and \newline
2) LogNormal: LN(x; $\mu_\mathrm{LN},\sigma_\mathrm{LN}$) = $\frac{N_\mathrm{LN}}{x \sigma_\mathrm{LN}\sqrt{2\pi}}  e^{-\frac{(log(x)-\mu_\mathrm{LN})^2}{2 \sigma_\mathrm{LN}^{2}}}$, \newline
\newline
where N$_{G}$ and N$_\mathrm{LN}$ are the normalization constants for the Gaussian and LogNormal profiles, respectively, and $\mu_{i}$ and $\sigma_{i}$ are the mean and standard deviation of the fitted profiles (i=G and LN for Gaussian and LogNormal, respectively). 
We used the \textsc{lmfit}\footnote{\url{https://lmfit.github.io/lmfit-py/fitting.html}} method to estimate the best fit and the goodness of fit. Here, 
the goodness of fit is given by the parameter \texttt{redchi}, which
is calculated from the ratio of the sum of the residuals to the degrees of freedom. A better fit is chosen based on the ratio of the corresponding \texttt{redchi} parameters named R$^{G}_\mathrm{LN}$. A LogNormal profile for the flux distribution is preferred if R$^{G}_\mathrm{LN}$ $>$ 1.
\textcolor{black}{The chance probability (p), based on toy Monte Carlo, indicates the probability of wrongly reconstructing a LogNormal (Gaussian) distribution as a Gaussian (LogNormal).}
The details and justification of this method can be found in
{the supplementary online material (Appendix \ref{sec:A1})}.

The flux profiles from radio to VHE $\gamma$-rays, along with the fits with the Gaussian and LogNormal functions, are shown in Figure~\ref{fig:SS1}. The fluxes were scaled with the average flux in the respective energy bands. The fit parameters are presented in Table \ref{table:SS}. 
\textcolor{black}{The flux profiles for X-ray observations in the  $0.3-2$\,keV and $2-10$\,keV energy bands show spikes. This is due to the very high SNR (average SNR above 60), which makes the available number of flux measurements insufficient to produce a smooth  convolved distribution. Despite this caveat, our simulations show that the number of measurements is sufficient to characterize the shape of the distribution, as well as to marginally distinguish between a Gaussian and LogNormal function.}
\textcolor{black}{Our findings suggest that the LogNormal is preferred over Gaussian for emissions in the VHE and HE $\gamma$-rays, hard X-rays in the $15-50$\,keV and optical band. The hard X-rays in the $15-50$\,keV shows a preference for a LogNormal profile, but with a chance probability (p) of only 0.16 (due to the large flux uncertainties), these results are not conclusive. 
The 37\,GHz radio band shows a clear preference for the Gaussian, while the flux profile for the X-rays in the $0.3-2$\,keV and $2-10$\,keV show a marginal preference for the Gaussian.
%%%%%%
%%%%%%
\textcolor{black}{The peak-position of the function (Gaussian/LogNormal) with which a flux profile is better fitted (depending on the value of the R$^{G}_\mathrm{LN}$) is considered as the most probable state (MPS).} The MPS for the energy bands above the synchrotron and IC peaks  (such as X-rays, $2-10$\,keV and $15-50$\,keV, and VHE $\gamma$-rays) are found to be in the range of $0.4-0.7$ times the average flux. On the other hand, the energy bands below the synchrotron and IC peaks (such as HE $\gamma$-rays, soft X-rays $0.3-2$\,keV, UV, optical, and \textcolor{black}{radio} emissions) lie in the range of $0.7-1.0$ \textcolor{black}{times} the average flux.}
The radio observations with OVRO at 15\,GHz show the emergence of an additional component at the high-flux end. A similar distribution has been reported in \citet{2016A&A...591A..83S} $\&$ \citet{2017MNRAS.467.4565L}. In our data set, the second peak in the high flux in the flux distribution with OVRO is due to the high flux state of the source during 2012--2013. 
Since the distribution is bimodal, we do not consider Gaussian and LogNormal distributions suitable for describing the flux distribution in this band. Therefore, the flux profile for OVRO data was not constructed.
{This is shown in the supplementary online material
(Fig.~\ref{fig:appenixB} in Appendix \ref{sec:appendixB})}. 
\textcolor{black}{The predictions of the flux profile method in different energy bands are backed by two additional methods: the (binned)  Chi-square fit and the (unbinned) log-likelihood fit. While the results of the Chi-square fit depend on the histogram binning, and do not take into account the flux measurement errors, the latter method does not depend on how the data are binned, and it considers the uncertainties of the fluxes. The detailed description of the methods and the results derived with them are reported in
{the supplementary online material (Appendix~\ref{sec:histandML})}. Similarly to the log-likelihood fit, the flux profile method is also unbinned, and considers the flux uncertainties; but it has the advantage over that it is easier to apply, and it leads to the shape of the distribution, regardless of any a-priori knowledge of the underlying shape (which is required for the log-likelihood fit).}
\textcolor{black}{The Table~\ref{table:SScomparisonbetweenMethod} reports the function preferred by the three methods (Gaussian or LogNormal) for all the bands probed.  Despite the different characteristics (and caveats) from these three methods, there is a very good agreement in the preferred shape for the flux distributions, with the LogNormal function being the most suitable shape for most of the energy bands probed. }

\section{Discussion and conclusions}
\label{sec:con}

This paper reports a detailed study of the broadband emission of Mrk\,421 during two observing campaigns, 2014 November to 2015 June, and 2015 
December to 2016 June. For simplicity, we dubbed them as  the 2015 and 2016 observing campaigns. The MWL data set used for this study was collected with 15 instruments, covering the emission of Mrk\,421 from radio (with OVRO, Mets{\"a}hovi, Medicina, and VLBA) to VHE $\gamma$-rays (with FACT and MAGIC), and including various instruments covering the optical and UV bands (KVA, ROVOR, West Mountain Observatory, iTelescopes network, and \textit{Swift}-UVOT), X-ray bands (\textit{Swift}-XRT and \textit{Swift}-BAT) and GeV $\gamma$-rays (with \textit{Fermi}-LAT). 
The sensitivity of the instruments used, and the large number of observations performed, 
enabled
the detailed characterisation of the 
  \textcolor{black}{MWL} variability and correlations during this period. 
A distinctive characteristic of this multi-year campaign is the large degree of simultaneity in the X-ray and VHE $\gamma$-ray observations, 
 which are two energy ranges where the variability is typically 
the
highest and can occur on the shortest timescales. We consider that the X-ray and VHE observations are simultaneous if taken within 0.3 days (i.e., the same night), although most of the observations were performed within 2 hours.  The large degree of simultaneity in the observations ensure reliability in the results reported, in contrast to other published works that use multiwavelengh data that are contemporaneous (taken within \textcolor{black}{one} or a few days), but not simultaneous. 
This simultaneity is particularly important for the X-ray and VHE $\gamma$-ray observations which, as we report in    \ref{sec:var} and   \textcolor{black}{Section}  \ref{sec:corr}  of this paper, show large variability and a large degree of correlated behaviour on timescales shorter than a day.

\subsection{Multi-band flux variability and correlations}
During the 2015 and 2016 observing campaigns, Mrk\,421 
showed a very low activity in the X-ray and VHE $\gamma$-rays \textcolor{black}{(see   \textcolor{black}{Section}  \ref{sec:Summary1516} and Fig.~\ref{fig:MWLC})}, which are the energy bands where the emitted power is the largest.
The spectral shape, quantified here with the 
HR$_\mathrm{keV}$ and HR$_\mathrm{TeV}$, also showed periods of extreme softness (very low   \textcolor{black}{HR$_{\mathrm{keV}}$ and HR$_{\mathrm{TeV}}$} values), like the one during the time interval of about MJD 57422 to MJD 57474, 
where the  \textcolor{black}{HR$_{\mathrm{TeV}}$} 
is $\leq$0.03, and the   \textcolor{black}{HR$_{\mathrm{keV}}$} 
is $\leq$0.25 (see Fig.~\ref{fig:HRXT} and Fig.~\ref{fig:HRXlong1}).
We found the typical {\em harder-when-brighter} trend in the X-ray and VHE $\gamma$-ray emission; although we also found a { deviation in the $HR$ vs. flux trend} for the largest X-ray and VHE $\gamma$-ray activity. The flattening \textcolor{black}{in the $HR$ vs. flux trend} 
for high (and low) X-ray fluxes had already been reported in  \citet{2016ApJ...819..156B}, but 
here we report, for the first time,  a similar behaviour in the VHE $\gamma$-ray 
band.

The fractional variability showed the typical double-bump structure reported in previous studies of the broadband (radio to VHE) emission of Mrk\,421 during low (non-flaring) activity \citep[e.g.][]{2015A&A...576A.126A,2016ApJ...819..156B}, and high (flaring) activity \citep[e.g.][]{2015A&A...578A..22A,2020ApJ...890...97A,2020ApJS..248...29A}. The highest variability is always observed in the highest X-ray  and VHE $\gamma$-ray energies at a similar level  (see Fig.~\ref{fig:Fvar}). 

We also searched for correlated behaviour among the emission from the various energy bands probed with these observations. We quantified these correlations (using Pearson and DCF) and 
evaluated the significance with Monte Carlo simulations. We detected a significant correlation between the emissions in the X-ray and VHE $\gamma$-ray bands. 
The positive correlation 
between 
these bands 
\textcolor{black}{has}
been reported with a high confidence level whenever the source 
showed
a
flaring activity \citep[e.g.][]{2015A&A...578A..22A,2020ApJS..248...29A}, but it is more elusive
during typical or low  \textcolor{black}{flux} \citep[e.g.][]{2015A&A...576A.126A,2016ApJ...819..156B}.
 Despite the strength of the correlation being similar for the various combinations of X-ray and VHE $\gamma$-ray energies probed, we report that the slope in the \textcolor{black}{VHE vs. X-ray flux plots} 
 changes with the specific energy band being used. In all cases, we found a slope lower than 1, with the 
 \textcolor{black}{largest}
 slope obtained for the highest 
 \textcolor{black}{energies (VHE $\gamma$-ray band; $>$1\,TeV)} versus the lowest X-ray band (0.3-2\,keV). These results 
 (see Fig.~\ref{fig:FF} and Table~\ref{table:DCF})
 are somewhat similar to those reported in \citet{2020ApJS..248...29A} during an extreme high activity in 2013 April. The results reported in this 
 paper further support that the X-ray and VHE emissions are closely related without any time delay, for all the energy bands probed, during high and during low activity. This indicates the presence of somewhat similar processes governing the emission of the source during a large range of activity, but showing also complexity in these processes, as is deduced from the diversity in the VHE vs. X-ray flux slopes when moving across nearby energy bands. \par

The strongly correlated zero-lag behaviour between the VHE and X-ray emissions, persistent during the 2015--2016 observing campaigns, indicates that the X-ray and VHE $\gamma$-ray emissions are dominated by leptonic scenarios (presumably SSC), where the same population of high-energy electrons radiate simultaneously at X-ray and VHE. The higher variability for the highest energies and the harder-when-brighter behaviour may be
interpreted as an indication of injection of high-energy particles dominating the flux variations over a large range of activity. But above a given flux, the spectral 
\textcolor{black}{shape no longer changes substantially with}
the flux, 
which suggests that the flux variations may be dominated by a different process 
yielding a variability that does not have a strong dependence with energy. One possibility could be a small change in the viewing angle, that would increase the 
\textcolor{black}{Doppler}
factor and, in first order approximation, produce a flux change that is similar in all energies\footnote{
The flux change would depend approximately on $\delta^{3.5}$ while the dependence in energy would relate to the energy shift that is proportional to $\delta$.}.
In order to produce flux changes of about a factor of two, one would need to change $\delta$ by about 20\%, which, for $\Gamma$=10 and a viewing angle of 5$^{\circ}$, could be achieved by a change in the viewing angle of about 1$^{\circ}$. Such change in $\delta$ would also produce 
\textcolor{black}{an}
energy-dependent flux change through the displacement of the broadband SED, but it would be a relatively small effect (e.g., 2.0\,keV would become 2.4\,keV).

We did not find a correlated behaviour
 between the optical and the X-ray bands, and did not find a
 correlation between the $\gamma$-ray emission below 2\,GeV and the one
 above 200\,GeV, hence indicating that the rising and falling segments
 of the two SED bumps may actually be produced by different particle
 populations, and even located at different regions.
 However, as reported in   \textcolor{black}{Section}  ~\ref{sec:vhe-vs-he}, we did observe, for the first time, a significant
 ($>3\,\sigma$) correlated behaviour, between the $>$2GeV emission measured with \textit{Fermi}-LAT and the VHE fluxes measured with FACT (E$_{th}$ $\sim$0.7\,TeV). The correlation, quantified in time steps of $\pm$3 days, occurs only for 
 \textcolor{black}{$\tau$=0,}
 indicating that the emission in these two energy bands is simultaneous within the resolution of the study.
 This observation suggests that the multi-GeV emission is produced (at least partially) by the same particle population that dominates the VHE emission, but such relation does not exist for the sub-GeV emission. A correlation between GeV and TeV energies for Mrk~421 had also been
 claimed by \citet{2016ApJS..222....6B}, using \textit{Fermi}-LAT and ARGO-YBJ. Apart from technical details
 in the quantification of the correlation, and the
 somewhat different energy bands considered in that study (median energy of 1.1\,TeV for ARGO-YBJ and energies above 0.3\,GeV for \textit{Fermi} LAT), the main practical difference is the temporal scale involved in these
 two studies, with \citet{2016ApJS..222....6B} reporting a positive
 correlation with 
 \textcolor{black}{$\tau$=0}
 within $\pm$30\,days, while we can ensure
 simultaneous emission within $\pm$3\,days.

 Owing to the substantially lower fractional variability and the longer variability timescales observed for the emission from the rising segments of the two SED bumps (namely radio, optical and GeV emission), 
 the 2015--2016 data set was complemented with data from years 2007--2014 (see 
 {the supplementary online material in}
 Appendix~\ref{sec:LongtermLC}) to enlarge the data set and better evaluate the correlations among these bands (see Section~\ref{sec:corr}). This correlation study, performed in the same fashion as done for the simultaneous X-ray/VHE fluxes, but with time-bins of 15 days instead of 1 day, yielded a number of interesting results, as reported in   \textcolor{black}{Section}  ~\ref{sec:he-vs-optical},    ~\ref{sec:he-vs-radio}, and    ~\ref{sec:optical-vs-radio}.

We found  a
positive correlation between the $>$0.3\,GeV emission (from \textit{Fermi}-LAT) and optical (R-band) emission for a range of about 60 days centered at 
\textcolor{black}{$\tau$=0}
(see   \textcolor{black}{Section}  \ref{sec:he-vs-optical}), which confirms and further strengthens the claim made by \citet{2017MNRAS.472.3789C}. Overall, this observation indicates that these two bands, belonging to the rising segments of the two SED bumps (and located somewhat close to the peak of the bumps) may indeed be produced (at least partially) by the same particle population and in the same region (or regions). The wide time interval with positive correlation may be due to a large size $R$ of the region dominating the optical and $\gamma$-ray emission. For instance, a variability timescale of about 30 days can be used to set an upper limit to the size $R$ of about 8$\times$10$^{17}$\,cm for a Doppler factor of 10. And, if the optical/GeV emitting region could be related to the radio emitting region, whose Doppler factor has often been estimated to be lower than 2 \citep[see e.g.][]{2010ApJ...723.1150P}, the upper limit to the size $R$ would be 2$\times$10$^{17}$\,cm.

Additionally, we found a positive correlation between the $>$0.3\,GeV emission (from \textit{Fermi}-LAT) and the radio emission at 15\,GHz and 37\,GHz (from OVRO and Mets{\"a}hovi) for a range of 
about 60 days centered at 
\textcolor{black}{$\tau \sim$ 45 days}
(see   \textcolor{black}{Section}  \ref{sec:he-vs-radio}), meaning that the radio emission occurs about 45 days after the GeV emission.
The same correlation with the same time lag occurs also for the optical and the 
radio emissions (see   \textcolor{black}{Section}  \ref{sec:optical-vs-radio}), which is expected given the correlation between $\gamma$-rays and optical emission mentioned above. Combining the time lags for the correlations among the GeV, R-band and the 15\,GHz fluxes, one obtains an overall time lag between optical/GeV and radio of
$43^{+9} _{-6}$ days. If instead one uses the 37\,GHz from Mets{\"a}hovi, where the DCF plots have less pronounced peak, the overall time lag between optical/GeV and radio is $37^{+15} _{-12}$ days (see 
{the supplementary online material in}
Appendix~\ref{sec:DCFLag} for details).

A positive correlation between the \textit{Fermi}-LAT and OVRO fluxes for a time lag of about 40 days had been first claimed by 
\citet{2014MNRAS.445..428M}.
The claim was only at 2.6\,$\sigma$ ($p$-value of 0.0104), and strongly affected by the large $\gamma$-ray and radio flares from July and September 2012, respectively.
In this paper, we report a correlation with a significance at the level of 11\,$\sigma$ when considering the entire data set, and, if we exclude the large flares, the significance is 9\,$\sigma$.
Therefore, we can confirm and further strengthen the correlation reported in  \citet{2014MNRAS.445..428M}, stating with reliability that this is an intrinsic characteristic in the multi-year emission of Mrk\,421, and not a particularity of a rare flaring activity.

Within the scenario of the emission being produced by plasma moving along the jet of Mrk\,421, the delay of the radio emission with respect to the $\gamma$-ray emission can be considered as an indication that the plasma (or jet disturbance) first crosses the surface of unit $\gamma$-ray opacity making the $\gamma$-ray emission visible, and then, about 0.2 pc down the jet \citep[assuming a common $\delta$ of 4 and $\Gamma$ of 2, see][for details of the calculation]{2014MNRAS.445..428M}, the radio emission is produced when the plasma (or disturbance) crosses the surface of unit radio opacity. \par

There are three distinct natures of correlation emerging from this study, a) correlation between X-ray and VHE $\gamma$-ray LCs at 
\textcolor{black}{$\tau$=0,}
b) correlation between optical and HE $\gamma$-rays at 
\textcolor{black}{$\tau$=0,}
and c) correlation between radio and HE (and optical) LCs at 
\textcolor{black}{$\tau \sim$ 45 days.}
The correlation in cases (b) and (c) have broader peaks compared to the case (a). The broader peaks for the radio, optical and GeV emission may be due to the lower variability and longer variability timescales related to the energy bands in consideration (because the emission involves lower energy particles), or it may related to the existence of two (or more) different radiation zones responsible for the production of the corresponding radiation components \citep[see][for description of the broadband SED variability of Mrk\,421 with these two theoretical scenarios]{2015A&A...578A..22A}.\par

\subsection{Multi-band flux distributions}
Using the historical   \textcolor{black}{MWL} data (from 2007 to 2016), we also quantified the flux variations with a methodology that allows us to estimate the flux distributions even for flux measurements with relatively large errors (see
{the supplementary online material in}
Appendix \ref{sec:A1} for details). Using this methodology, we determined the most  probable flux values and the dispersion in the flux values for all the bands probed (see Fig. \ref{fig:SS1} and Table \ref{table:SS}).
Among other things, we found that the most probable flux is close to the average flux for the energy bands below the synchrotron and inverse-Compton SED peaks (i.e. radio, optical and soft X-rays), while it substantially differs from it for the energy bands above the two SED peaks (i.e., hard X-rays and VHE $\gamma$-rays).
The flux distributions in radio and soft X-rays are better described with a Gaussian function, while the flux distributions in the optical, hard X-rays, HE and VHE $\gamma$-rays are preferably described with a LogNormal function. A LogNormal distribution of flux implies that the emission is being powered by a multiplicative process rather than an additive one. Suggestions have been put forward by several authors that LogNormality is a result of fluctuations in the accretion disk \citep{2005MNRAS.359..345U, 2010LNP...794..203M}. 
If the same behaviour is found in blazars, this may lead to the conclusion that the source of variations in blazars lie outside the jet, i.e., in the accretion disks which then modulate the jet emission.

\subsection{Radio flare at 37 GHz}
On 2015 September 11, the Mets{\"a}hovi telescope observed an increase by a factor of two in the 37\,GHz radio flux, from about 0.5\,Jy to about 1.1\,Jy (see Fig.~\ref{fig:MeFlare} and   \textcolor{black}{Section}  \ref{sub:MetFlare}). It is the first time that such a large flux change, with a temporal timescale shorter than 3 weeks, is observed in the 37\,GHz radio emission of Mrk\,421. But the quasi-simultaneous flux density measurements at 5\,GHz and 24\,GHz from the Medicina radio telescope, 
performed also on September 11, show an enhanced flux density at 5\,GHz only, while the 24\,GHz flux density is in line with the usual values for the source. As the data are not strictly simultaneous, it is possible that some very short term  fluctuation affected the measurement during only some of the observations. This would argue for externally induced short-term variability (scintillation, or instrumental) rather than an episode of flaring from the source, although the current data remain insufficient to make any strong claims on this episode. 
The VLBA observations show an increase in the polarization fraction on  September 22, while it returns to normal values on December 5 (see Fig.~\ref{fig:MeFlare}). The flare could then be explained via a kink instability that momentarily disrupts the ordering of the field and accelerates particles to cause an increase in flux and a decrease in polarization. The disturbance propagates down the jet, causing first a high-energy flare, followed by a millimeter-wave flare, as observed. After the flare, the polarization returns to its normal radial pattern.
Other simultaneous observations are those from {\it Fermi}-LAT, and {\it Swift}-BAT where there is no substantial enhancement in the $\gamma$-ray or X-ray flux activity around the time of the radio flare. 
There is, however, some structure in the GeV and keV light curves about 40 days before the radio flare, which is similar to the time lags reported in Fig.~\ref{fig:DCF_main1} between multi-year GeV and radio emission.

\subsection{Hard X-ray component}
During the 7-day time interval MJD 57422--57429 (2016 February 4--11) where Mrk\,421 showed
a very low X-ray  \textcolor{black}{flux} and low 
HR$_\mathrm{keV}$ (i.e. soft X-ray spectra), we 
noted a  $15-50$\,keV flux (from \textit{Swift}-BAT) that is well above the emission that one would expect if the optical to X-ray emission (from 1 eV to 10\,keV), characterized with a log-parabola function, is extrapolated to the hard X-ray range above 15\,keV 
(see Fig.~\ref{fig:BATexcess}).
This is the first time that BAT measures a flux significantly above the one expected from the simple extrapolation of the XRT spectral data. But an excess in the hard X-ray with respect to the expected flux from the synchrotron component has already been reported by \citet{2016ApJ...827...55K} for Mrk\,421, using {\it NuSTAR} data during a period of very low X-ray and VHE activity in 2013, and considered to be the onset of the SSC component. Such hard X-ray excesses, considered to be the beginning of the SSC component, have also been observed in another blazar, PKS~2155--304, also using {\it NuSTAR} observations during a period of very low X-ray  \textcolor{black}{flux}  \citep{2016ApJ...831..142M,2019arXiv191207273H}. On the other hand, the hard X-ray {\it NuSTAR} excess in the Mrk\,421 data from 2013 was also interpreted within the scenario of the spine/layer jet structure, and considered to be an indication of inverse-Compton emission produced by high-energy electrons from the spine region up-scattering the synchrotron photons from the layer, as was proposed by \citet{2017ApJ...842..129C}.  Another possible origin of the hard X-ray excess could be a Bethe-Heitler cascade, which is expected to occur in many of the hadronic scenarios, such as the ones that were used to explain the broadband emission of TXS~0506+056 contemporaneous to a high-energy astrophysical neutrino detected by IceCube in 2017 September \citep[e.g.][]{2018ApJ...863L..10A}. Moreover, the BAT excess reported here may also be related to the presence of an additional (and narrow) spectral component that appears occasionally, as has been recently reported for Mrk\,501 at multi-TeV energies \citep{2020A&A...637A..86M}, and interpreted as a indication for pile-up in the electron energy distribution, or an indication for electrons accelerated in the vacuum gaps close to the super-massive black hole that is powering the source. 
The probability of occurrence of hard-X-ray excesses, the theoretical interpretation of the broadband SED of Mrk\,421 for the time period of very low X-ray and VHE fluxes, as well as for the time intervals before and after this time period, will be discussed in a forthcoming paper.

\subsection{Outlook}
Overall, the data set presented in this 
paper, which focuses on the two observing campaigns in 2015--2016, when Mrk\,421 showed very low  \textcolor{black}{flux} at keV and TeV energies, and without any prominent flare, allowed us to derive a good number of new observational results. The continuation of these multi-instrument observations in the upcoming years, with at least the same depth in temporal and energy coverage, will be important to determine whether these novel features that we report in this 
paper are rare, or whether they repeat over time.
\newline

\section*{Acknowledgements}
%MAGIC 
%
% MAGIC STANDARD ACKNOWLEDGEMENTS
% LAST UPDATE: 9 Feb 2020
% Author: M. Doro
%
The MAGIC collaboration would like to thank the Instituto de
Astrof\'{\i}sica de Canarias for the excellent working conditions 
at the Observatorio del Roque de los Muchachos in La Palma. The 
financial support of the German BMBF and MPG; the Italian INFN 
and INAF; the Swiss National Fund SNF; the ERDF under the Spanish 
MINECO (FPA2017-87859-P, FPA2017-85668-P, FPA2017-82729-C6-2-R, 
FPA2017-82729-C6-6-R, FPA2017-82729-C6-5-R, AYA2015-71042-P, 
AYA2016-76012-C3-1-P, ESP2017-87055-C2-2-P, FPA2017-90566-REDC); 
the Indian Department of Atomic Energy; the Japanese ICRR, the 
University of Tokyo, JSPS, and MEXT;  the Bulgarian Ministry of 
Education and Science, National RI Roadmap Project 
DO1-268/16.12.2019 and the Academy of Finland grant nr. 320045 is 
gratefully acknowledged. This work was also supported by the 
Spanish Centro de Excelencia ``Severo Ochoa'' SEV-2016-0588 and 
SEV-2015-0548, the Unidad de Excelencia ``Mar\'{\i}a de Maeztu'' 
MDM-2014-0369 and the "la Caixa" Foundation (fellowship 
LCF/BQ/PI18/11630012), by the Croatian Science Foundation (HrZZ) 
Project IP-2016-06-9782 and the University of Rijeka Project 
13.12.1.3.02, by the DFG Collaborative Research Centers SFB823/C4 
and SFB876/C3, the Polish National Research Centre grant 
UMO-2016/22/M/ST9/00382 and by the Brazilian MCTIC, CNPq and 
FAPERJ.
% Fermi

The \textit{Fermi} LAT Collaboration acknowledges generous ongoing support from a number of agencies and institutes that have supported both the development and the operation of the LAT as well as scientific data analysis. These include the National Aeronautics and Space Administration and the Department of Energy in the United States, the Commissariat \`a l'Energie Atomique and the Centre National de la Recherche Scientifique / Institut National de Physique Nucl\'eaire et de Physique des Particules in France, the Agenzia Spaziale Italiana and the Istituto Nazionale di Fisica Nucleare in Italy, the Ministry of Education, Culture, Sports, Science and Technology (MEXT), High Energy Accelerator Research Organization (KEK) and Japan Aerospace Exploration Agency (JAXA) in Japan, and the K.~A.~Wallenberg Foundation, the Swedish Research Council and the Swedish National Space Board in Sweden.

Additional support for science analysis during the operations phase is gratefully  acknowledged from the Istituto Nazionale di Astrofisica in Italy and the Centre  National d'\'Etudes Spatiales in France. This work performed in part under DOE Contract DE-AC02-76SF00515.

The authors are very grateful to {Matthew} Kerr for his guidance and various discussions about the unbinnned log-likelihood fits using Gaussian and LogNormal PDFs.

The authors acknowledge the use of public data from the Swift data archive, and are particularly thankful to the Swift operations team for maximizing the simultaneity of the VHE and X-ray observations. This research has made use of the XRT Data Analysis Software (XRTDAS) developed under the responsibility of the ASI Science Data Center (ASDC), Italy.

%Other groups
% From boston group 
Some of the results reported are based on data taken and assembled by the WEBT collaboration and stored in the WEBT archive at the Osservatorio Astrofisico di Torino - INAF \url{(http://www.oato.inaf.it/blazars/webt/)}. The authors are particularly grateful to M.I.~Carnerero for discussions about the WEBT data. 

The research at Boston University was supported in part by  NASA Fermi Guest Investigator grant 80NSSC17K0649. The authors thank H.\ Zhang for enlightening correspondence.

This publication makes use of data obtained at the Mets\"ahovi Radio Observatory, operated by Aalto University in Finland.

The Medicina radio telescope is funded by the Department of University and Research (MIUR) and is operated as National Facility by the National Institute for Astrophysics (INAF).

The important contributions from ETH Zurich grants ETH-10.08-2 and ETH-27.12-1 as well as the funding by the Swiss SNF and the German BMBF (Verbundforschung Astro- und Astroteilchenphysik) and HAP (Helmholtz Alliance for Astroparticle Physics) are gratefully acknowledged. Part of this work is supported by Deutsche Forschungsgemeinschaft (DFG) within the Collaborative Research Center SFB 876 "Providing Information by Resource-Constrained Analysis", project C3. We are thankful for the very valuable contributions from E. Lorenz, D. Renker and G. Viertel during the early phase of the project. We thank the Instituto de Astrofísica de Canarias for allowing us to operate the telescope at the Observatorio del Roque de los Muchachos in La Palma, the Max-Planck-Institut für Physik for providing us with the mount of the former HEGRA CT3 telescope, and the MAGIC collaboration for their support.

This research has made use of data from the OVRO 40-m monitoring program (Richards, J. L. et al. 2011, ApJS, 194, 29) which is supported in part by NASA grants NNX08AW31G, NNX11A043G, and NNX14AQ89G and NSF grants AST-0808050 and AST-1109911.

\section*{DATA AVAILABILITY}
The data underlying this article will be shared on reasonable request to the corresponding author.

\bibliographystyle{mnras.bst} % style aa.bst
\bibliography{ref.bib} % your references Yourfile.bib

%%%%%%%%%%%%%%%%%%%%%%%%%%%%%%
%%%%%% Affiliation:%%%%%%%%%%%
%%%%%%%%%%%%%%%%%%%%%%%%%%%%%%
\vspace{2cm}
$^{1}$ {Inst. de Astrof\'isica de Canarias, E-38200 La Laguna, and Universidad de La Laguna, Dpto. Astrof\'isica, E-38206 La Laguna, Tenerife, Spain} \\
$^{2}$ {Universit\`a di Udine, and INFN Trieste, I-33100 Udine, Italy} \\
$^{3}$ {National Institute for Astrophysics (INAF), I-00136 Rome, Italy} \\
%$^{4}$ {ETH Zurich, CH-8093 Zurich, Switzerland} \\
$^{5}$ {Japanese MAGIC Consortium: ICRR, The University of Tokyo, 277-8582 Chiba, Japan; Department of Physics, Kyoto University, 606-8502 Kyoto, Japan; Tokai University, 259-1292 Kanagawa, Japan; Physics Program, Graduate School of Advanced Science and Engineering, Hiroshima University, 739-8526 Hiroshima, Japan; Konan University, 658-8501 Hyogo, Japan; RIKEN, 351-0198 Saitama, Japan} \\
$^{6}$ {Technische Universit\"at Dortmund, D-44221 Dortmund, Germany} \\
$^{7}$ {Croatian Consortium: University of Rijeka, Department of Physics, 51000 Rijeka; University of Split - FESB, 21000 Split; University of Zagreb - FER, 10000 Zagreb; University of Osijek, 31000 Osijek; Rudjer Boskovic Institute, 10000 Zagreb, Croatia} \\
$^{8}$ {IPARCOS Institute and EMFTEL Department, Universidad Complutense de Madrid, E-28040 Madrid, Spain} \\
$^{9}$ {Centro Brasileiro de Pesquisas F\'isicas (CBPF), 22290-180 URCA, Rio de Janeiro (RJ), Brasil} \\
$^{10}$ {University of Lodz, Faculty of Physics and Applied Informatics, Department of Astrophysics, 90-236 Lodz, Poland} \\
$^{11}$ {Universit\`a  di Siena and INFN Pisa, I-53100 Siena, Italy} \\
$^{12}$ {Deutsches Elektronen-Synchrotron (DESY), D-15738 Zeuthen, Germany} \\
$^{13}$ {Centro de Investigaciones Energ\'eticas, Medioambientales y Tecnol\'ogicas, E-28040 Madrid, Spain} \\
$^{14}$ {Istituto Nazionale Fisica Nucleare (INFN), 00044 Frascati (Roma) Italy} \\
$^{15}$ {Max-Planck-Institut f\"ur Physik, D-80805 M\"unchen, Germany} \\
$^{16}$ {Institut de F\'isica d'Altes Energies (IFAE), The Barcelona Institute of Science and Technology (BIST), E-08193 Bellaterra (Barcelona), Spain} \\
$^{17}$ {Universit\`a di Padova and INFN, I-35131 Padova, Italy} \\
$^{18}$ {Universit\`a di Pisa, and INFN Pisa, I-56126 Pisa, Italy} \\
$^{19}$ {Universitat de Barcelona, ICCUB, IEEC-UB, E-08028 Barcelona, Spain} \\
$^{20}$ {The Armenian Consortium: ICRANet-Armenia at NAS RA, A. Alikhanyan National Laboratory} \\
$^{21}$ {Universit\"at W\"urzburg, D-97074 W\"urzburg, Germany} \\
$^{22}$ {Finnish MAGIC Consortium: Finnish Centre of Astronomy with ESO (FINCA), University of Turku, FI-20014 Turku, Finland; Astronomy Research Unit, University of Oulu, FI-90014 Oulu, Finland} \\
$^{23}$ {Departament de F\'isica, and CERES-IEEC, Universitat Aut\`onoma de Barcelona, E-08193 Bellaterra, Spain} \\
$^{24}$ {Saha Institute of Nuclear Physics, HBNI, 1/AF Bidhannagar, Salt Lake, Sector-1, Kolkata 700064, India} \\
$^{25}$ {Inst. for Nucl. Research and Nucl. Energy, Bulgarian Academy of Sciences, BG-1784 Sofia, Bulgaria} \\
$^{26}$ {now at University of Innsbruck} \\
$^{27}$ {also at Port d'Informaci\'o Cient\'ifica (PIC) E-08193 Bellaterra (Barcelona) Spain} \\
$^{28}$ {also at Dipartimento di Fisica, Universit\`a di Trieste, I-34127 Trieste, Italy} \\
$^{29}$ {also at INAF-Trieste and Dept. of Physics \& Astronomy, University of Bologna} \\
$^{30}$ {Universit\`a di Padova and INFN, I-35131 Padova, Italy} \\
%MAGIC (above)
$^{31}$ ETH Zurich, Institute for Particle Physics and Astrophysics, Otto-Stern-Weg 5, 8093 Zurich, Switzerland \\
$^{32}$ TU Dortmund, Experimental Physics 5, Otto-Hahn-Str. 4a, 44227 Dortmund, Germany \\
$^{33}$ University of Geneva, Department of Astronomy, Chemin d\'Ecogia 16,  1290 Versoix, Switzerland \\
$^{34}$ University of W{\"u}rzburg, Institute for Theoretical Physics and Astrophysics, Emil-Fischer-Str. 31, 97074 W{\"u}rzburg, Germany \\
$^a$ also at RWTH Aachen University \\
$^b$ also in MAGIC \\
%FACT (above)
$^{35}$Laboratoire Leprince-Ringuet, École Polytechnique, CNRS/IN2P3, F-91128 Palaiseau, France, \\ 
$^{37}$INAF – Istituto di Radioastronomia, Bologna, Italy, \\
%Fermi (above)
$^{38}$Space Science Data Center (SSDC) - ASI, via del Politecnico, s.n.c., I-00133, Roma, Italy\\
$^{39}$INAF—Osservatorio Astronomico di Roma, via di Frascati 33, I-00040 Monteporzio, Italy\\
$^{40}$Italian Space Agency, ASI, via del Politecnico snc, 00133 Roma, Italy\\
%other MWL collab-1 (aboveA.C. Sadun$^{41}$)
$^{41}$Department of Physics, University of Colorado Denver, Denver, Colorado, CO 80217-3364, USA\\
$^{42}$Department of Physics and Astronomy, Brigham Young University, Provo, UT 84602, USA\\
$^{43}$Institute for Astrophysical Research, Boston University, 725 Commonwealth Avenue, Boston, MA 02215, USA\\
$^{44}$Astronomical Institute, St.Petersburg State University, Universitetskij Pr. 28, Petrodvorets, 198504 St.Petersburg, Russia\\
$^{45}$Aalto University Mets\"ahovi Radio Observatory, Mets\"ahovintie 114, FIN-02540 Kylm\"al\"a, Finland\\
$^{46}$Aalto University Department of Electronics and Nanoengineering, P.O. Box 15500, FIN-00076 Aalto, Finland\\
$^{47}$Astronomy Department, Universidad de Concepción, Casilla 160-C, Concepción, Chile\\
$^{48}$European Space Agency, European Space Astronomy Centre, C/ Bajo el Castillo s/n, 28692 Villanueva de la Cañada, Madrid, Spain\\
$^{49}$INAF Istituto di Radioastronomia, Stazione di Medicina, via Fiorentina 3513, I-40059 Villafontana (BO), Italy\\
$^{50}$ Astrophysics Science Division, NASA Goddard Space Flight Center, 8800 Greenbelt Road, Greenbelt, MD 20771, USA\\
$^{51}$ Department of Physics, University of Maryland, Baltimore County, 1000 Hilltop Circle, Baltimore, MD 21250, USA\\
%%%%%%%%%%%%%%%%%%%%%%%%%%%%%%%%%%%%

\newpage
\appendix

\section{Intra-night variability at VHE}
\label{sec:INV}

This section reports the single-night LCs at VHE $\gamma$-rays that show intra-night variability (INV),
\textcolor{black}{ considered to occur when the fit with a constant value to the available intra-night flux measurements (time bins of 20 minute for FACT and 15 min for MAGIC) yield a $p_{value}$ below 0.003 (i.e. more than 3$\sigma$ significance). In case of FACT, the 20-minute binned light curves of all nights with a minimum observation time of 1 hour (196 nights) were checked for INV.}
From all the observations performed, INV was observed on only two nights, 2015 January 27 (MJD~57049) and 2015 March 12 (MJD~57093). In the first night, there were observations with both MAGIC (above 0.2\,TeV) and FACT (E$_{\mathrm{th}}\sim$0.7\,TeV).
The INV is statistically significant only in the LC from MAGIC. In the case of FACT, the flux variations are not significant (less than 2\,$\sigma$) because of the larger flux uncertainties and the \textcolor{black}{different} temporal coverage. 
It seems that the flux of Mrk\,421 dropped by 50\% sometime between MJD~57049.20 and  MJD~57049.25. In the second night, there
are only FACT observations. Mrk\,421 shows a decrease in the VHE flux by about a factor of 3 in the 3.5 hours that the observation spans.

\begin{figure}
    \includegraphics[height=6cm, width=\linewidth]{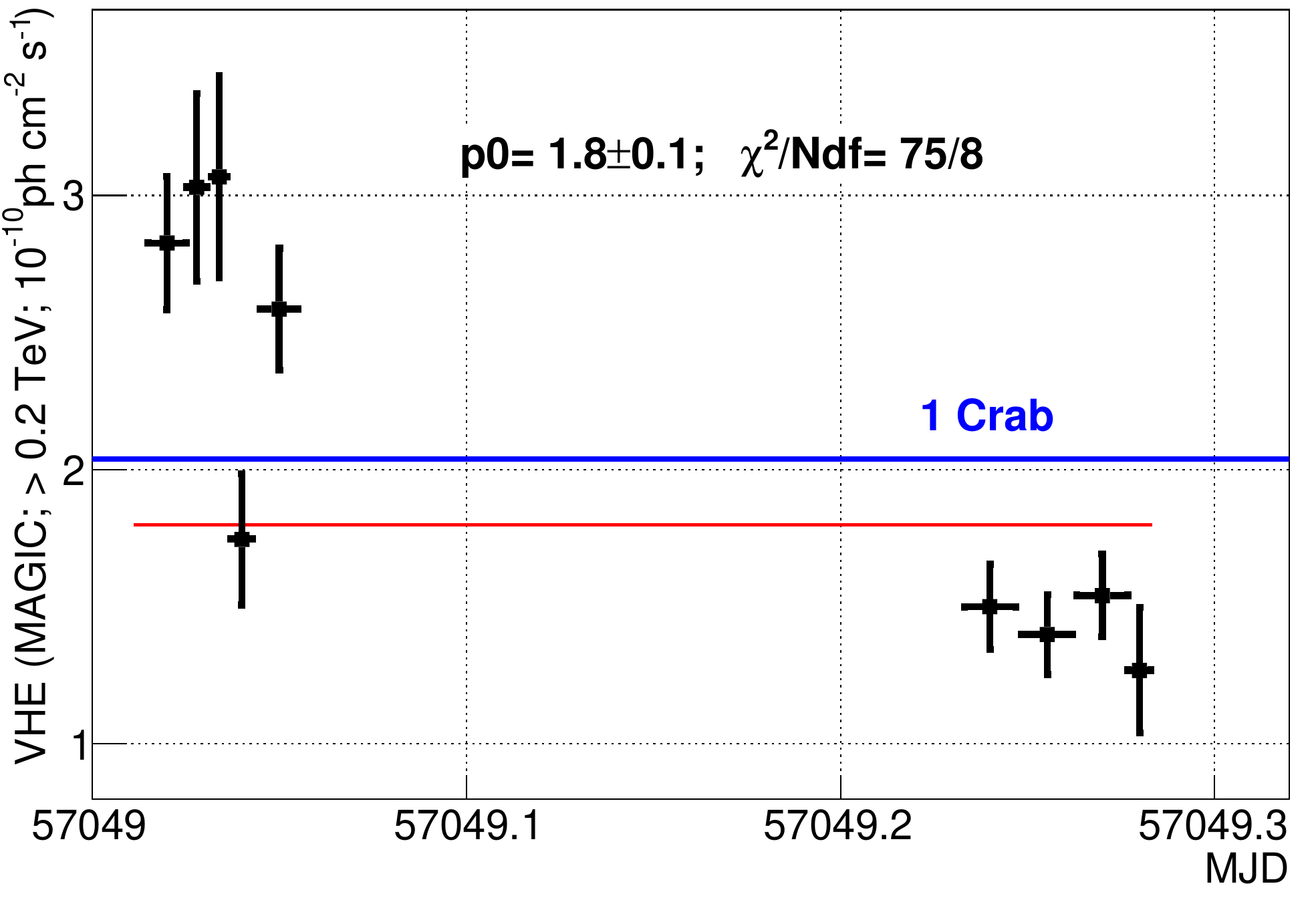}
    \includegraphics[height=6cm, width=\linewidth]{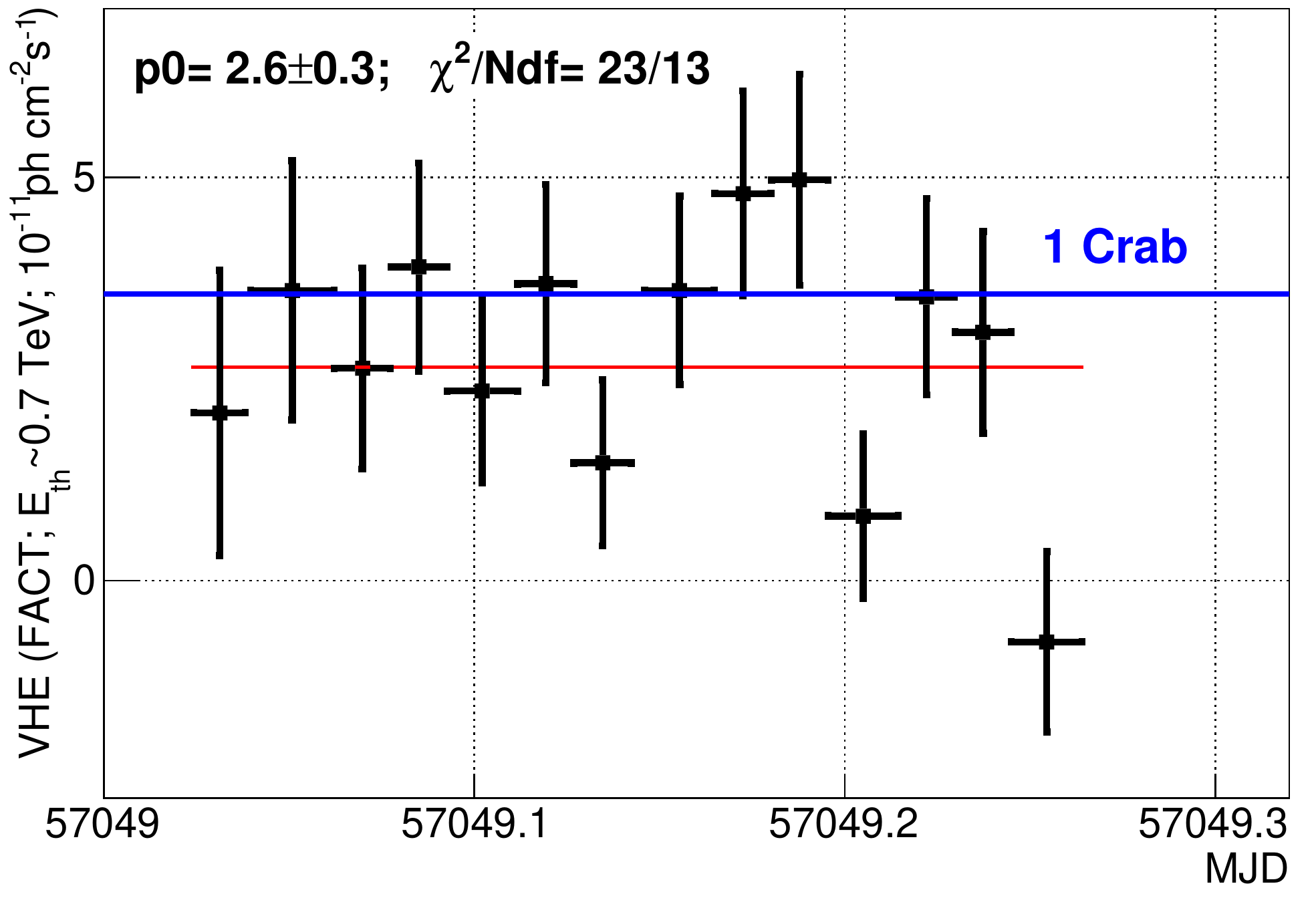}
    \includegraphics[height=6cm, width=\linewidth]{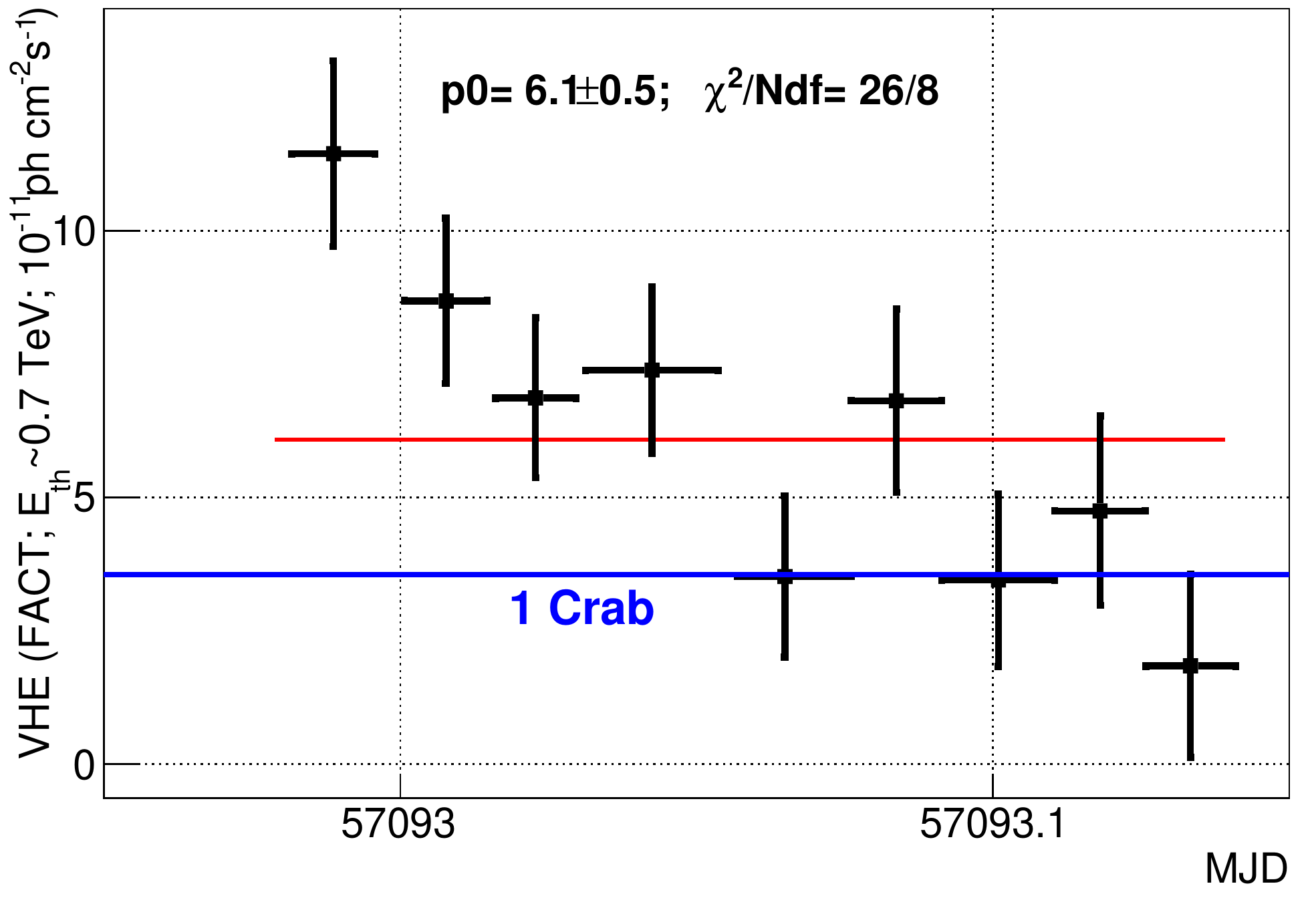}
    \caption{Single night VHE $\gamma$-ray LCs that show statistically significant intra-night variability. The first two panels show the MAGIC (above 0.2\,TeV) and the FACT (E$_{\mathrm{th}}\sim$0.7\,TeV) LCs for 
    \textcolor{black}{2015 January 27}
    (MJD~57049). The lower panel shows the FACT (E$_{\mathrm{th}}\sim$0.7\,TeV) LC for 2015 March 12 (MJD~57093). The blue horizontal lines depict the Crab \textcolor{black}{N}ebula flux in the respective energy band, and the red horizontal line represents a constant fit to the VHE $\gamma$-ray \textcolor{black}{flux}, with the resulting fit parameters and goodness of the fit reported in the panels.}
    \label{fig:INV}
\end{figure}

\clearpage

\section{Multi-year light curves}
\label{sec:LongtermLC}

The studies reported in this paper are derived mostly with the extensive MWL data set collected during the campaigns in the years 2015 and 2016, when Mrk\,421 showed very low  \textcolor{black}{flux} at X-ray and VHE $\gamma$-rays. This 2-year data set is described in Section~\ref{sec:data_analysis}. However, for the correlation studies reported in Sections~\ref{sec:vhe-vs-he}, \ref{sec:he-vs-optical}, \ref{sec:he-vs-radio}, and \ref{sec:optical-vs-radio}, and the characterization of the flux distributions reported in   \textcolor{black}{Section}~\ref{sec:typicalstate}, the 2015–2016 data set is complemented with data from the years 2007–2014. This appendix provides a description of this additional (complementary) 2007–2014 data set.

\begin{landscape}
\begin{figure}
\centering
\includegraphics[width=\linewidth, height=16cm]{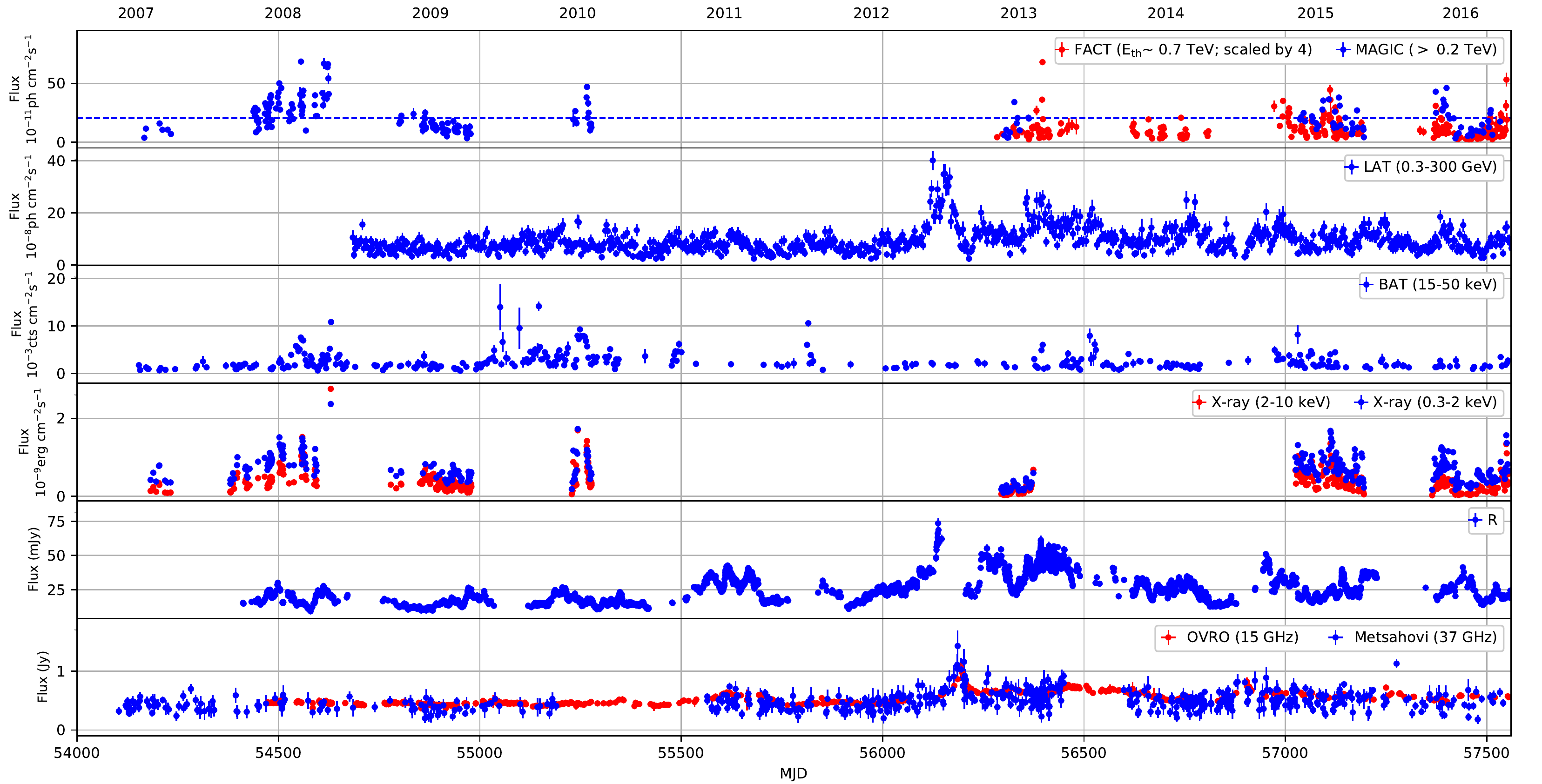}
\caption{Multi-year data set {(with flux measurements with SNR$>$2)} used in the study reported in Sections~\ref{sec:he-vs-optical}, \ref{sec:he-vs-radio}, \ref{sec:optical-vs-radio}, and \ref{sec:typicalstate}. In the top panel, the horizontal blue line represents 1 Crab flux in above 0.2\,TeV.}
\label{fig:LongtermMWLC}
\end{figure}
\end{landscape}

The 2007--2016 data set, used for the above-mentioned correlation and flux-profile studies, is depicted in Fig.~\ref{fig:LongtermMWLC}.
The MAGIC VHE $\gamma$-ray and the \textit{Swift}-XRT X-ray LCs are retrieved from various published works \citep{2012A&A...542A.100A, 2015A&A...578A..22A, 2016A&A...593A..91A, 2016ApJ...819..156B}. {The FACT fluxes from 2012 December to 2016 June were produced with the analysis described in Section~\ref{FACTDescription}.}  The \textit{Fermi}-LAT fluxes in the band $0.3-300$\,GeV were analyzed as described in   \textcolor{black}{Section}~\ref{sec:FermiLAT}. The \textit{Swift}-BAT fluxes were retrieved from the BAT website\footnote{\url{http://heasarc.nasa.gov/docs/swift/results/transients/}}, and treated as explained in   \textcolor{black}{Section}~\ref{sec:BAT}. The optical data in the R-band were retrieved from \citet{2017MNRAS.472.3789C}. The 37\,GHz radio fluxes from Mets{\"a}hovi were provided by the instrument team, and the 15\,GHz radio fluxes from OVRO were
retrieved from the website of the instrument team\footnote{\url{http://www.astro.caltech.edu/ovroblazars/index.php?page=home}}. 
As done in   \textcolor{black}{Section} \ref{sec:Fvar}, we only consider fluxes with the relative errors \mbox{(flux-error/flux)} smaller than 0.5 (i.e. SNR$>$2). In this way, we ensure the usage of reliable flux measurements, and minimize unwanted effects related to unaccounted (systematic) errors. \par
There are seven MAGIC VHE fluxes from the year 2007,  from the time interval MJD~54166--54438, and five VHE fluxes from the year 2009, from the time interval MJD~54800--54835, that relate to energies above 0.4\,TeV \citep[published in][]{2016A&A...593A..91A}, and all the MAGIC VHE fluxes from the 4.5-months long MWL campaign in year 2009, from the time interval MJD~54851--54977, relate to energies above 0.3\,TeV. \citep[published in][]{2015A&A...578A..22A}. The reason for the higher minimum energy in these two publications with respect to other publications that relate to observations performed after year 2010 (where the light curves are produced with energies above 0.2\,TeV) is the operation of MAGIC in mono mode (with a \textcolor{black}{single-telescope}). The MAGIC observations of Mrk\,421 in  stereo mode, which started in the MWL campaign from year 2010, provide additional sensitivity and a lower analysis energy threshold, which allows one to reliably produce light curves with a minimum energy of 0.2\,TeV. During the year 2008, Mrk\,421 showed high VHE \textcolor{black}{flux} and, despite MAGIC operating with a \textcolor{black}{single-telescope}, the large VHE $\gamma$-ray fluxes and the longer exposures, permitted the reliable reconstruction of the VHE fluxes above 0.2\,TeV, as reported in  \citet{2012A&A...542A.100A}.  In order to properly compare the published VHE fluxes from the years 2007 and 2009 with 
those from 2008 and from 2010 onwards, we scaled VHE fluxes above 0.4\,TeV and 0.3\,TeV (and their related errors) 
by a factor of 2.83 and 1.84, respectively. 
These scaling factors were calculated by considering that the VHE spectral shape of Mrk~421 around the energy of 0.3\,TeV can be well described with a power-law function with index 2.5, when Mrk~421 is in its typical (non-flaring) state \citep{2011ApJ...736..131A}. They can then be used to convert the VHE fluxes above 0.4\,TeV and 0.3\,TeV to that above 0.2\,TeV.
The spectral shape of the VHE emission of Mrk\,421 does vary over time, and it is known to be related to the flux (e.g. {\em harder-when-brighter} behaviour). However, owing to the relatively small energy range over which one needs to extrapolate, and the relatively low VHE flux and low variability from years 2007 and 2009, including these spectral variations would vary the reported VHE fluxes by less than $\pm$10\% 
in most cases. These additional flux variations are typically
smaller than the statistical uncertainties of the flux measurements
during these low-\textcolor{black}{flux} periods, and hence they do not affect the reported study in any significant manner.

\begin{figure*}
\centering
\includegraphics[width=\linewidth, height=10cm]{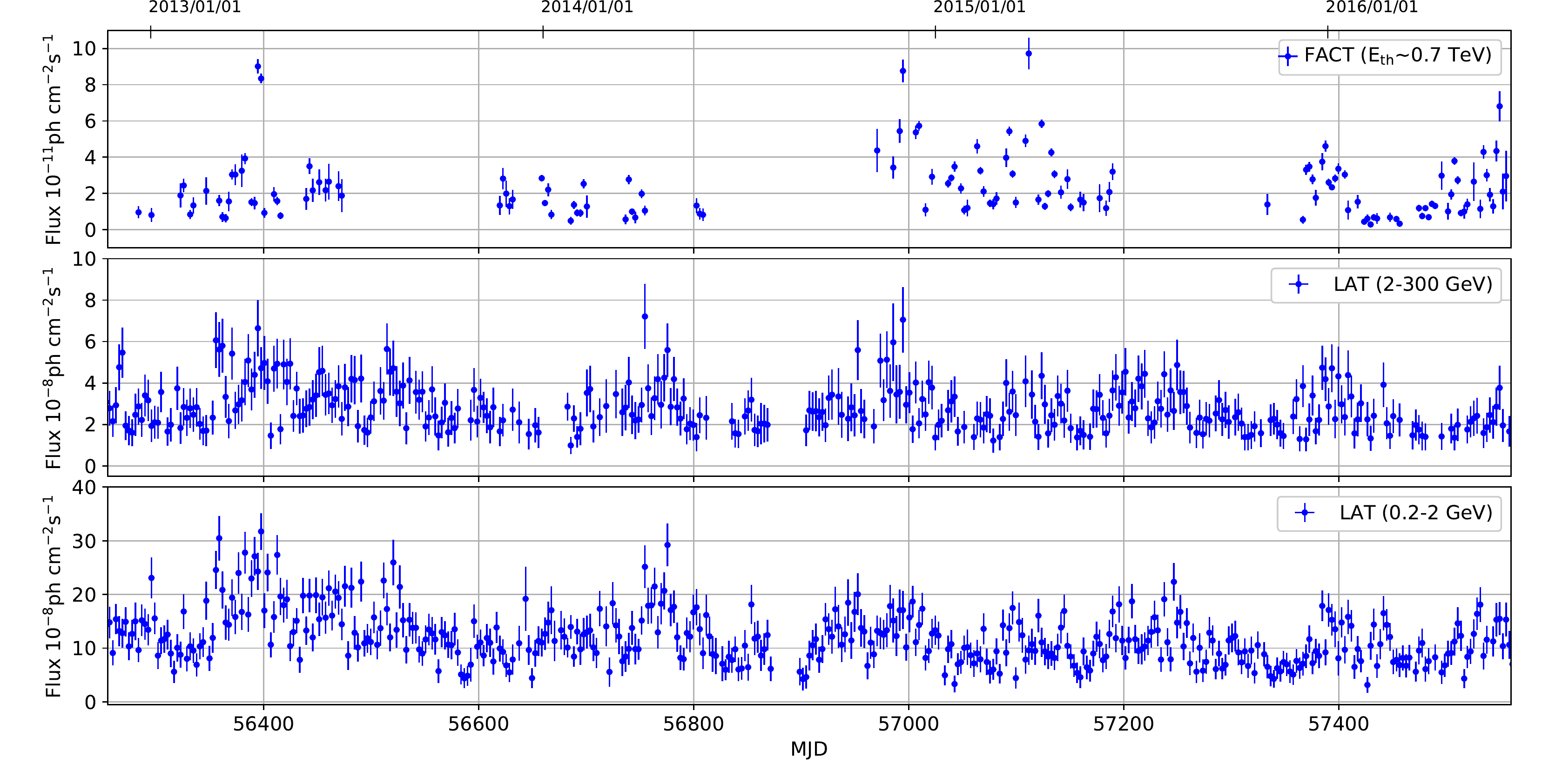}
\caption{{The 3-day binned LCs (with flux measurements with SNR$>$2) measured with FACT and \textit{Fermi-}LAT in the energy bands E$_\mathrm{th}\sim$0.7\,TeV (top panel), $2-300$\,GeV (middle panel), and $0.2-2$\,GeV (bottom panel), that were used in the study reported in Section \ref{sec:vhe-vs-he}.}}
\label{fig:LongtermMWLC_FACT-LAT}
\end{figure*}

{Figure \ref{fig:LongtermMWLC_FACT-LAT} shows the 3-day binned light curves during 2012 December to 2016 June used in the correlation studies presented in Section~\ref{sec:vhe-vs-he}. 
The VHE fluxes from FACT were derived with the analysis described in Section~\ref{FACTDescription}, but this time in 3-day time intervals. The data from \textit{Fermi-}LAT were analyzed as described in Section~\ref{sec:FermiLAT}.  }

\clearpage

\section{Multi-band flux-flux relations}
\label{sec:FluxFlux}

\begin{figure*}
\centering
\includegraphics[width=0.49\linewidth,height=5.5cm]{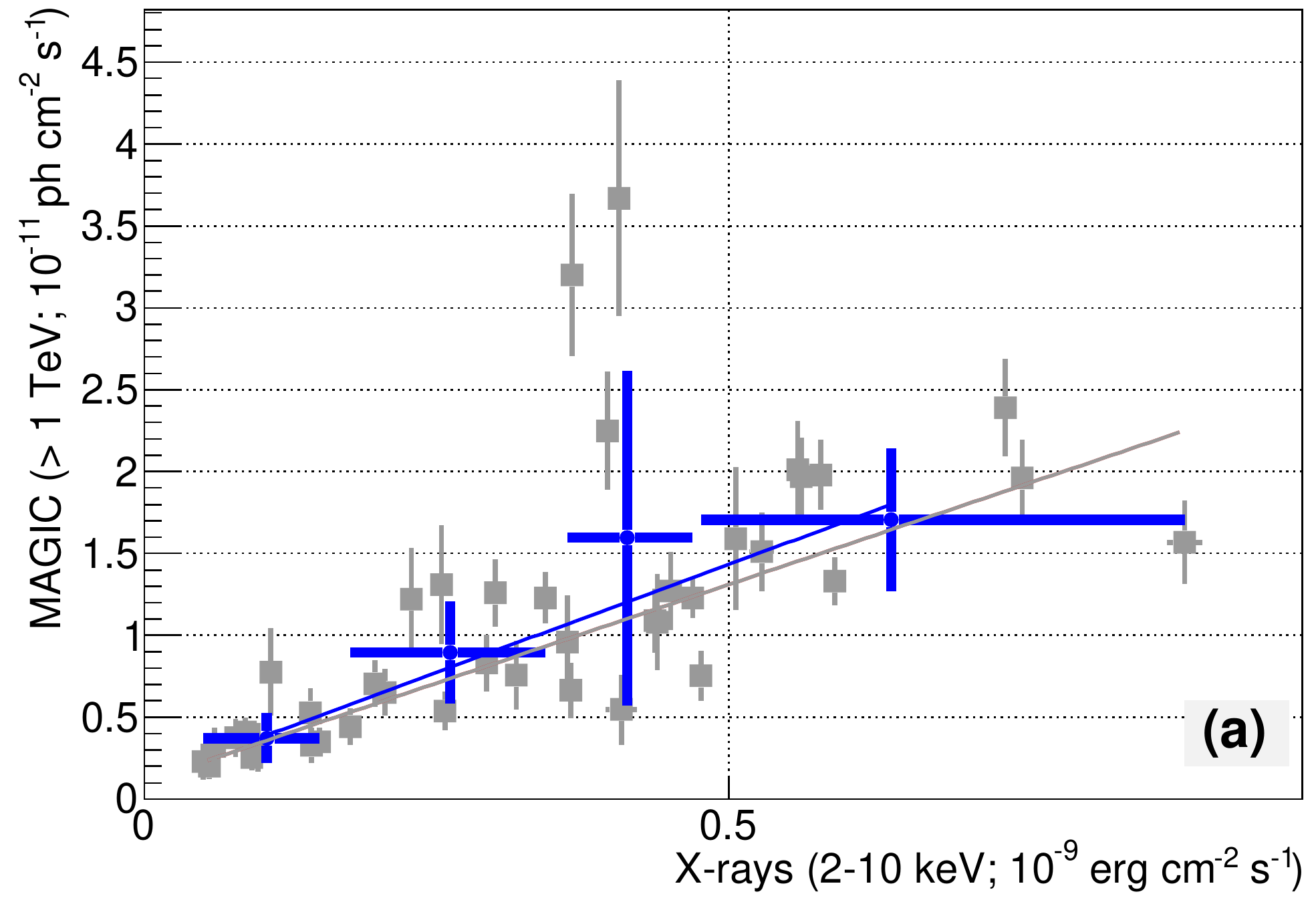}
\includegraphics[width=0.49\linewidth,height=5.5cm]{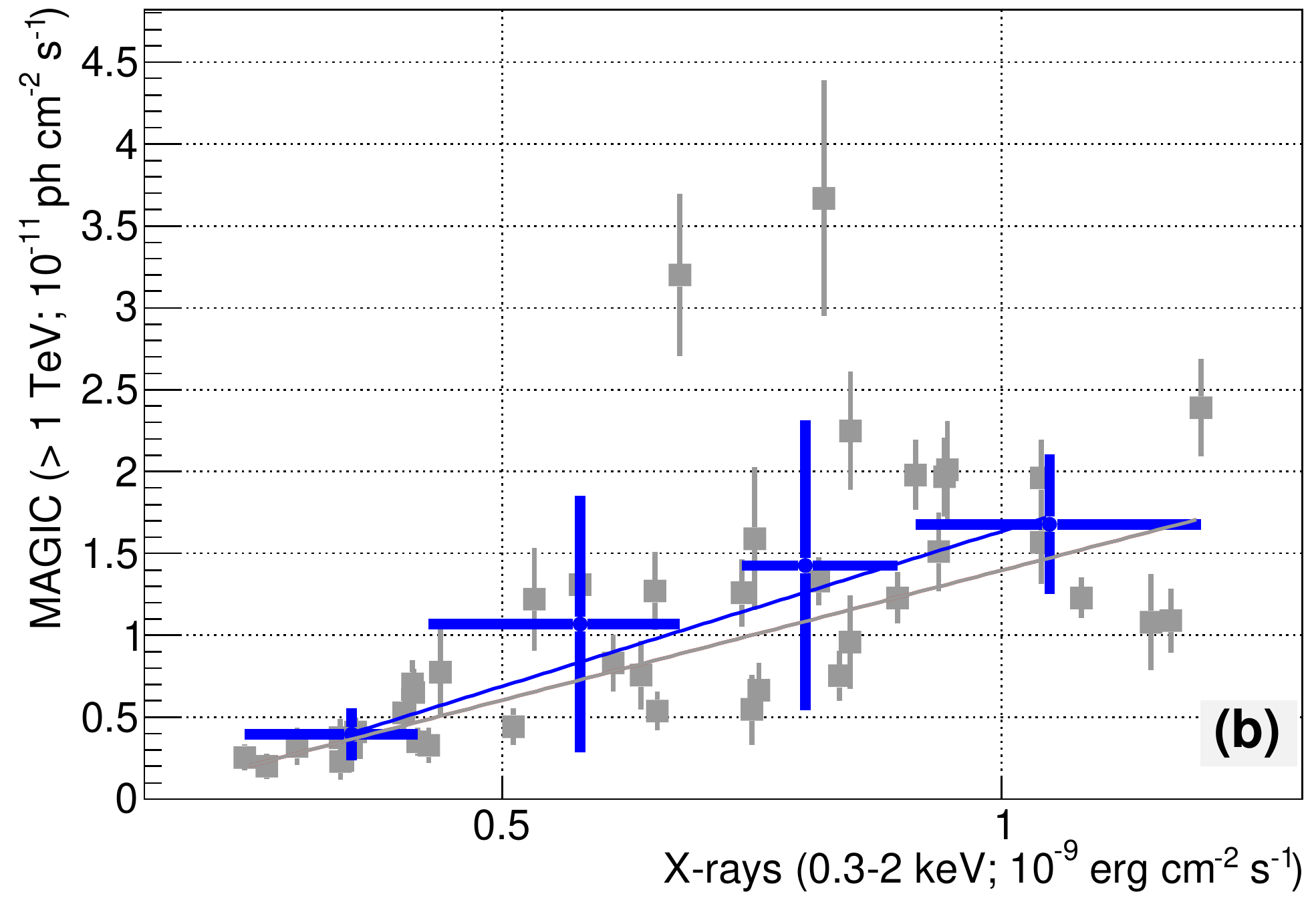}
\includegraphics[width=0.49\linewidth,height=5.5cm]{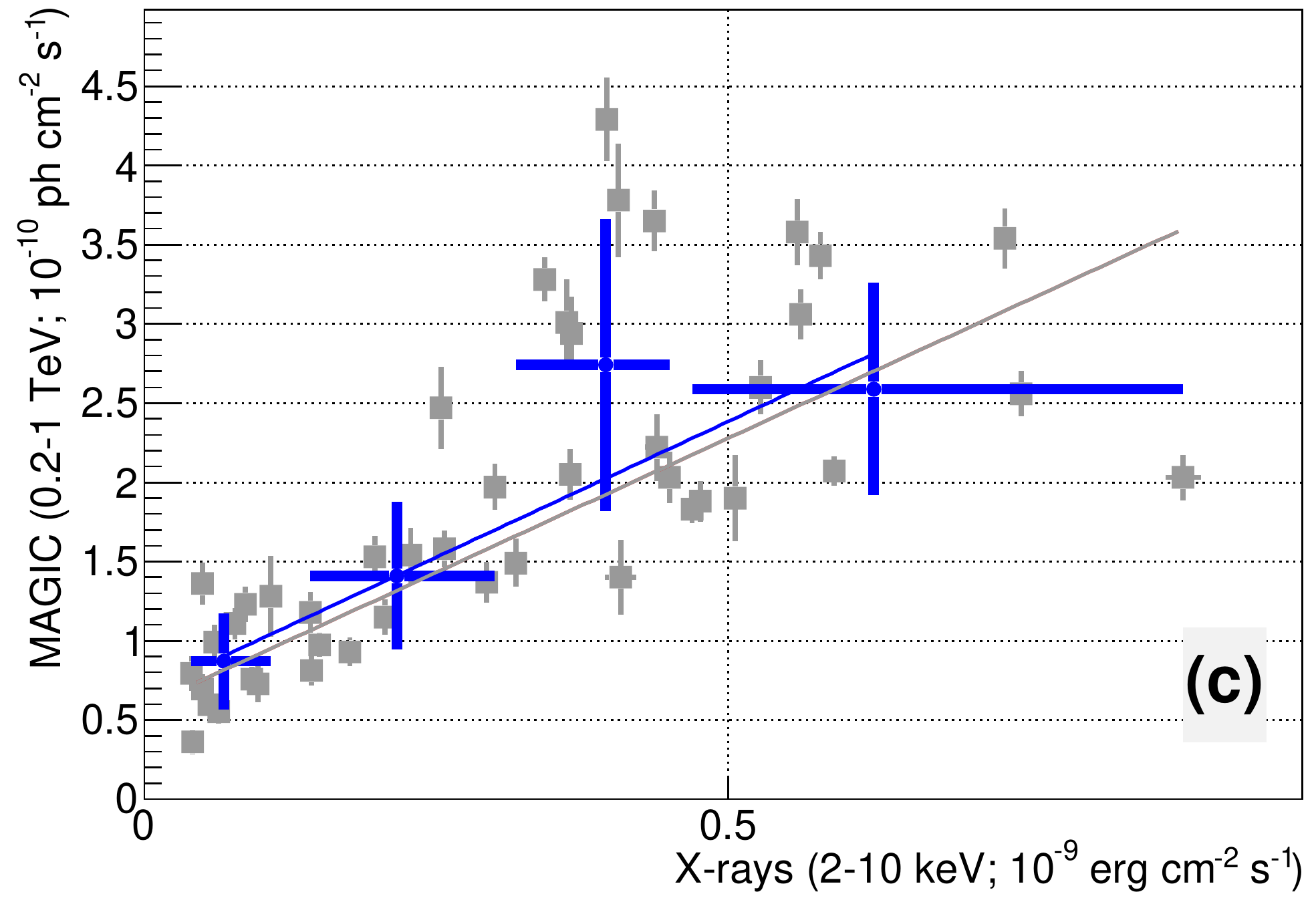}
\includegraphics[width=0.49\linewidth,height=5.5cm]{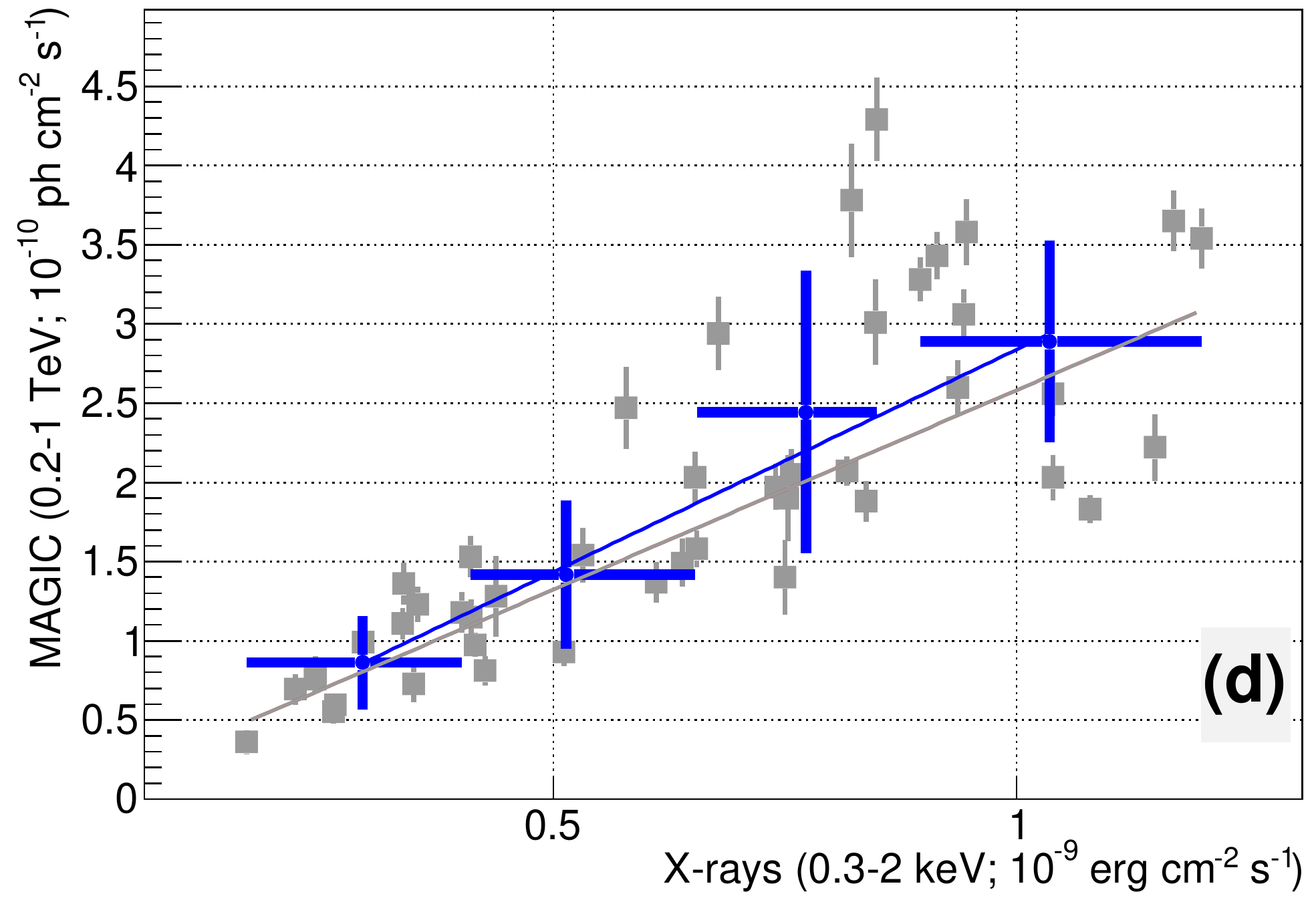}
\includegraphics[width=0.49\linewidth,height=5.5cm]{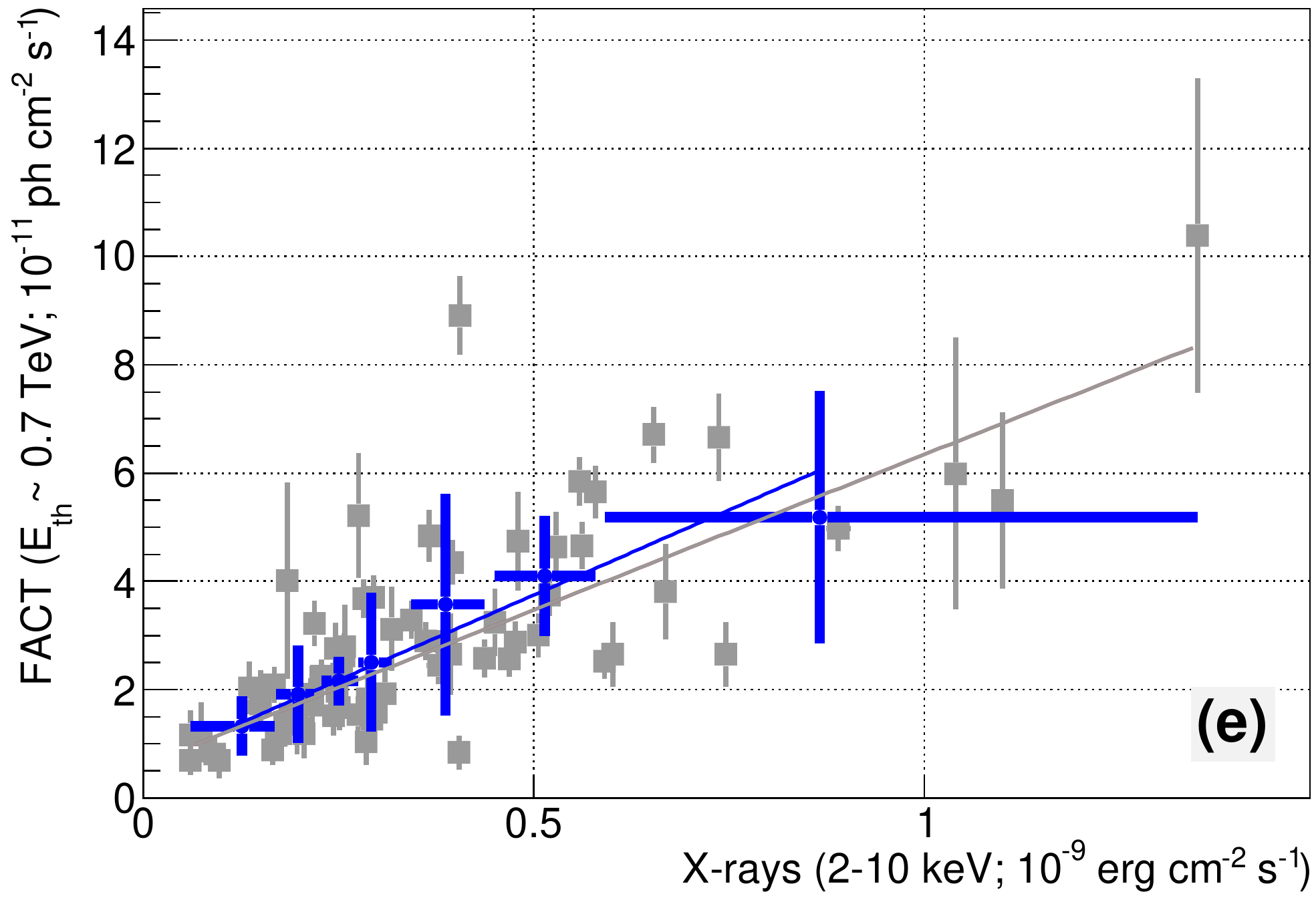}
\includegraphics[width=0.49\linewidth,height=5.5cm]{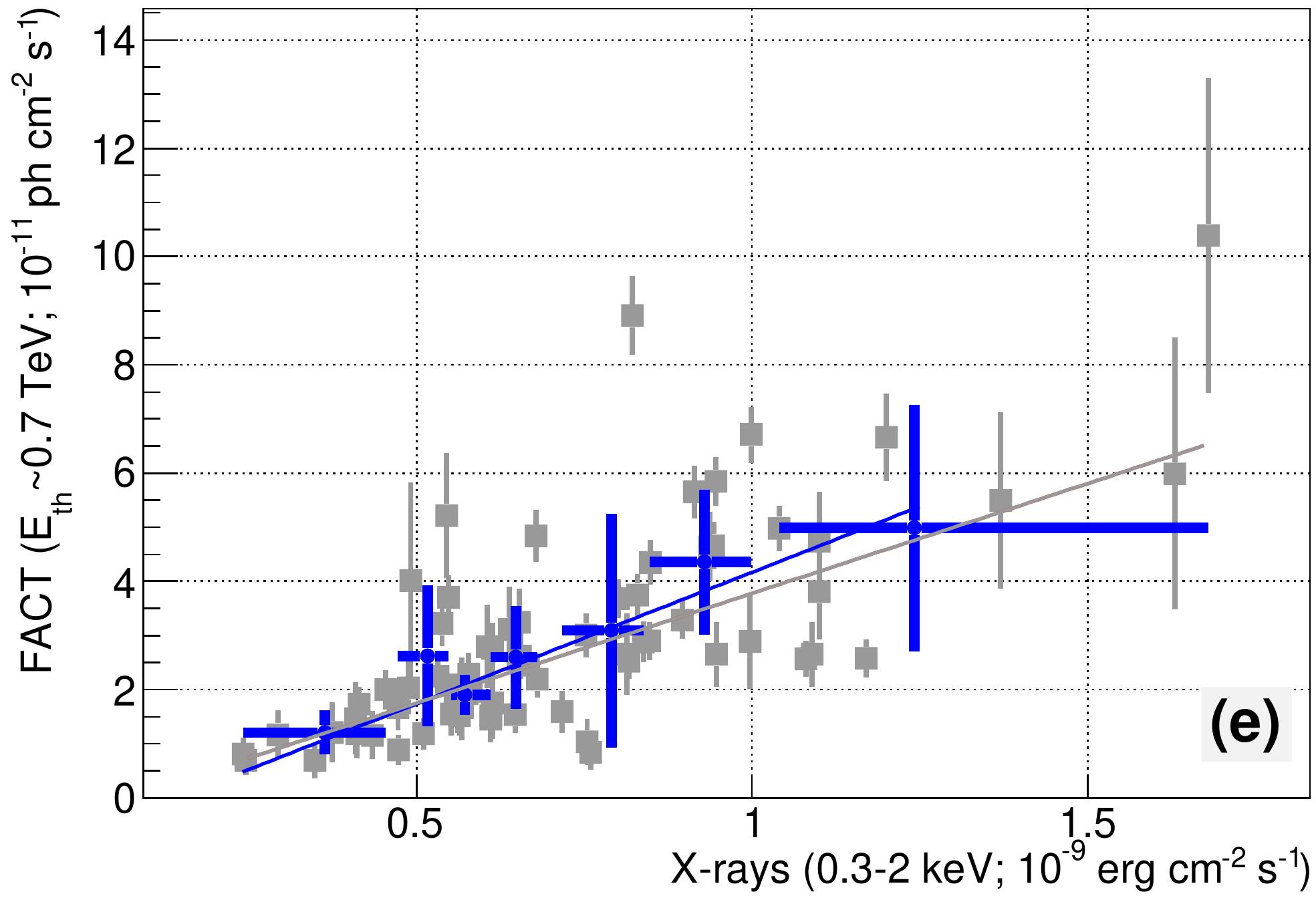}
\caption{
VHE vs. X-ray flux correlation plots during 2015--2016 campaign.
The grey markers denote the individual flux measurements and related errors (unbinned data), while the blue
markers show the average and the standard deviation computed with 
data subsets of 10 observations, binned according to their flux (binned data). The 
grey and blue lines depict the best linear fit to the unbinned and binned data, with the slopes reported in Table~\ref{table:DCF}.
Only simultaneous VHE-X-ray data (taken within 0.3 days) 
were used. 
See   \textcolor{black}{Section} \ref{sec:vhe-vs-xrays} for details.}
\label{fig:FF}
\end{figure*}

This section reports the multi-band flux-flux plots related to the correlations discussed in   \textcolor{black}{Section}~\ref{sec:corr}.

Panels (a)--(d) 
in Fig.~\ref{fig:FF} show the integral VHE $\gamma$-ray flux from the two energy bands measured with MAGIC (reported in Fig~\ref{fig:MWLC}, namely $0.2-1$\,TeV and above 1\,TeV), plotted against the X-ray flux in the two energy bands from \textit{Swift}-XRT (reported in  Fig~\ref{fig:MWLC}). The panels (e)-(f) 
of Fig.~\ref{fig:FF} show the VHE vs. X-ray flux relations when using the VHE fluxes with E$_{\mathrm{th}}$ $\sim$0.7\,TeV measured with FACT. Only simultaneous observations are used in these figures. Besides the display of all the flux measurements 
(roughly equivalent to
{\em unbinned} data), the panels also show the average and the standard deviation computed with data subsets of 10 observations, binned according to their flux ({\em binned} data). The binned data allow 
\textcolor{black}{us}
to better visualize the main trend, as well as the dispersion in the single-day flux measurements. Both the unbinned and binned data are fitted with a linear function to quantify the slope in the VHE vs. X-ray flux relation.
These slopes are
reported in Table~\ref{table:DCF}. Despite the large dispersion in the VHE vs. X-ray flux values, there is a
roughly
linear trend for all the bands, with the slope of the trend increasing for increasing VHE energy band, or for decreasing X-ray energy band.

\begin{figure*}
\centering
\includegraphics[width=0.49\linewidth,height=5.5cm]{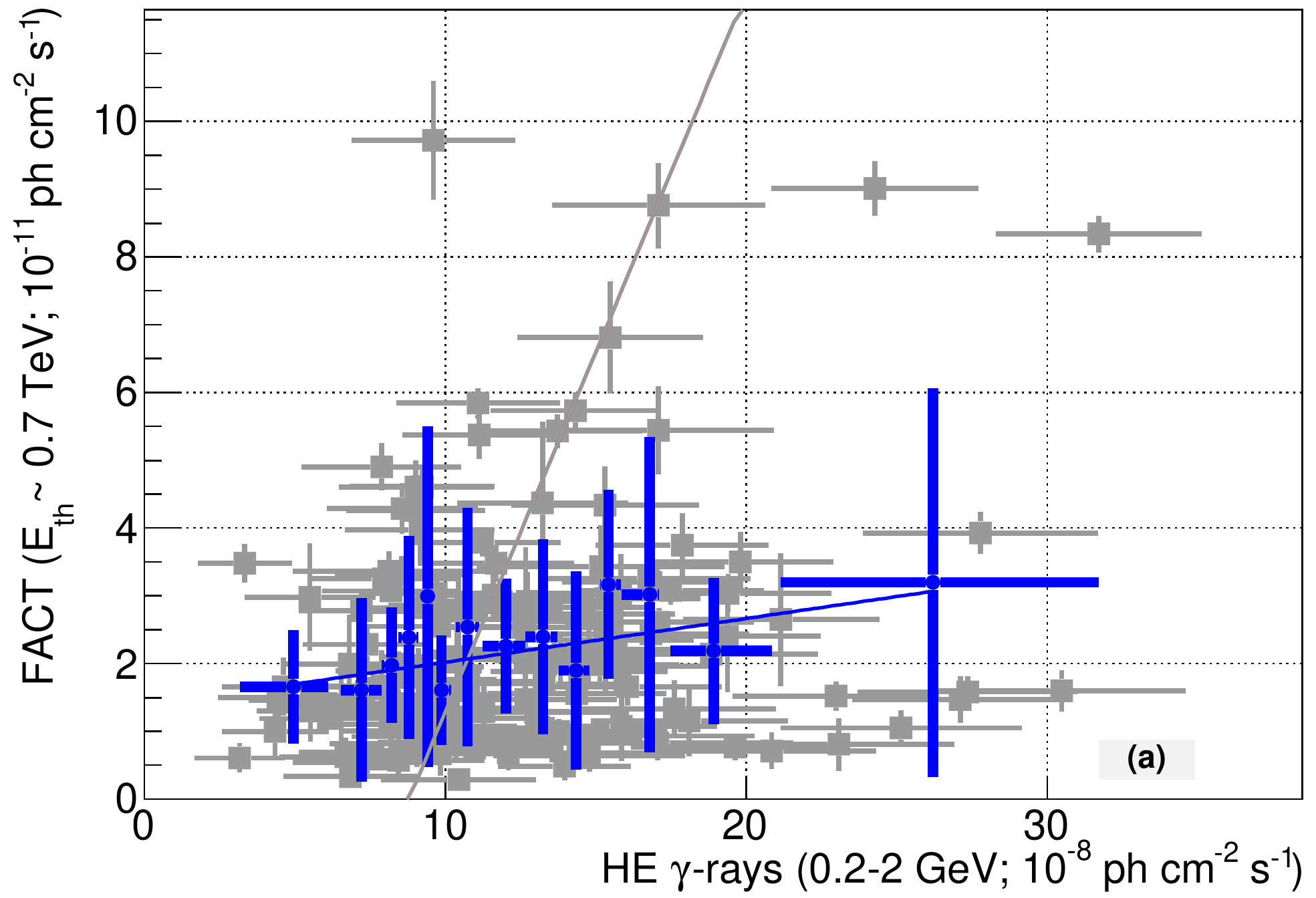}
\includegraphics[width=0.49\linewidth,height=5.5cm]{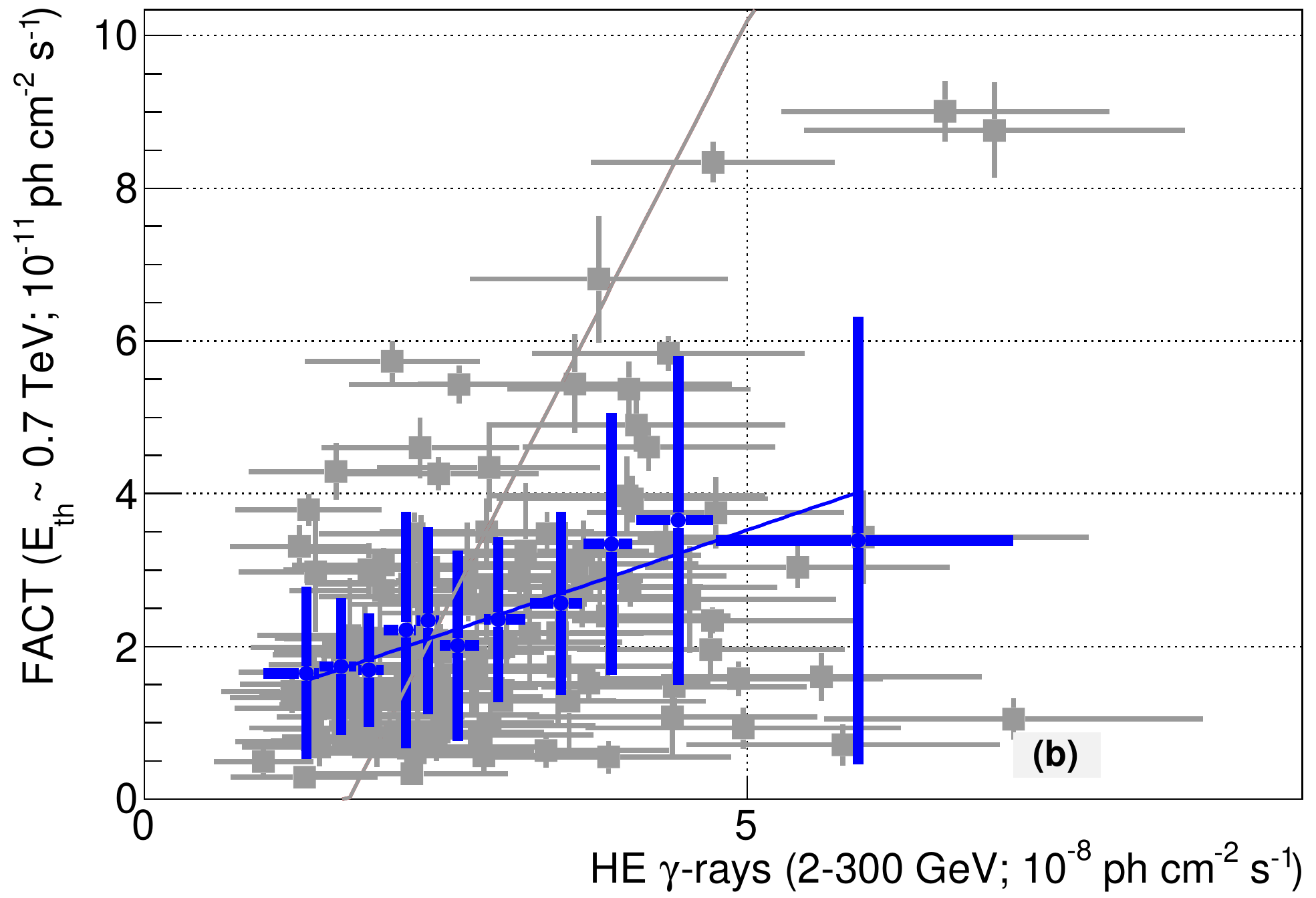}
\caption{VHE vs. HE flux correlation plots during the 
{period from 2012 December to 2016 June. 
For the description of the grey and blue markers, see
the caption of Fig.\ref{fig:FF}.}
The grey and blue lines depict the best linear fit to the unbinned (grey) and binned (blue) data, with the slopes reported in Table~\ref{table:DCF_he_vhe}. The flux values relate to 3-day time intervals. See \textcolor{black}{Section} \ref{sec:vhe-vs-he} for details.}
\label{fig:FF-VHE-HE}
\end{figure*}

The panels in Fig.~\ref{fig:FF-VHE-HE} show the VHE $\gamma$-ray flux from FACT (E$_{\mathrm{th}}$ $\sim$0.7\,TeV) {during the period {from} 2012 December to 2016 June (see Fig.~\ref{fig:LongtermMWLC_FACT-LAT})}, plotted against the HE flux
from \textit{Fermi}-LAT in two energy bands, {$0.2-2$\,GeV and $2-300$\,GeV (see Fig.~\ref{fig:LongtermMWLC_FACT-LAT}).}
%The energy bands used are those employed in Fig~\ref{fig:LongtermMWLC_FACT-LAT}, and the flux values relate to 3-day time intervals
%(see \textcolor{black}{Section} \ref{sec:vhe-vs-he}). 
As with the panels in Fig.~\ref{fig:FF}, besides showing {all of} the 3-day flux measurements ({\em unbinned} data), the panels also show the average and the standard deviation computed with data subsets of 10 observations, binned according to their flux ({\em binned} data). Both the unbinned and binned data are fitted with a linear function to quantify the slope in the VHE vs. HE flux relation. These slopes are
reported in Table~\ref{table:DCF_he_vhe}.
In contrast to what happens in the panels 
of Fig.~\ref{fig:FF}, there is a large difference between the slopes in the linear functions fitted to the unbinned and binned data. The difference is ascribed to VHE vs. HE flux pairs which are well outside the main trend (outliers), which have a large impact 
on the fit to the unbinned data, but not to the binned data. The difference is also partly due to the weak (if not absent) correlation between these energy bands (see   \textcolor{black}{Section} \ref{sec:vhe-vs-he} for further details).

\begin{figure*}
\centering
\includegraphics[width=0.51\linewidth,height=5.5cm]{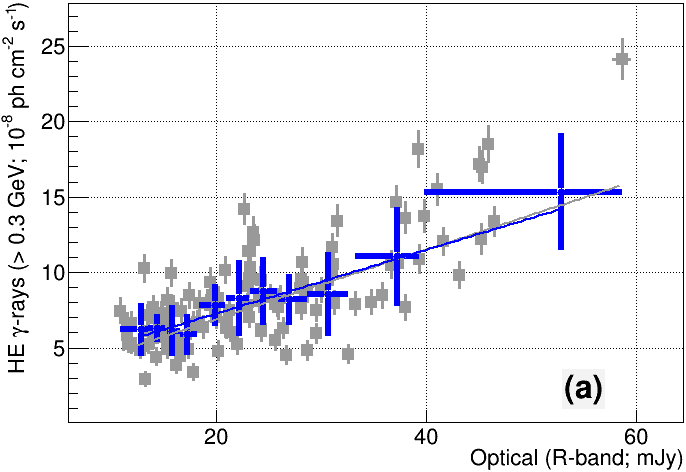}
\includegraphics[width=0.49\linewidth,height=5.5cm]{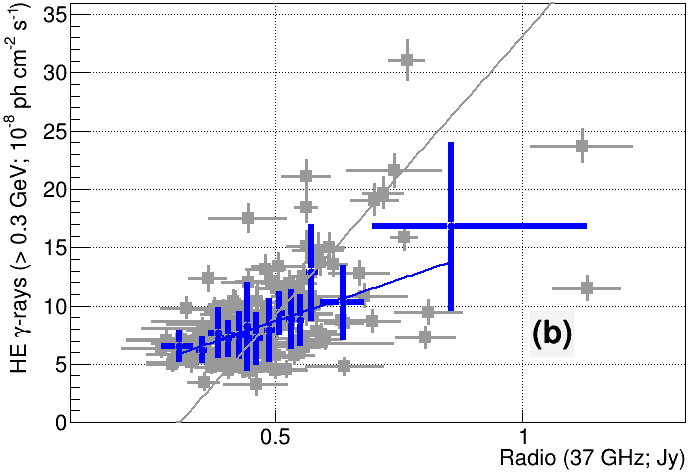}
\includegraphics[width=0.49\linewidth,height=5.5cm]{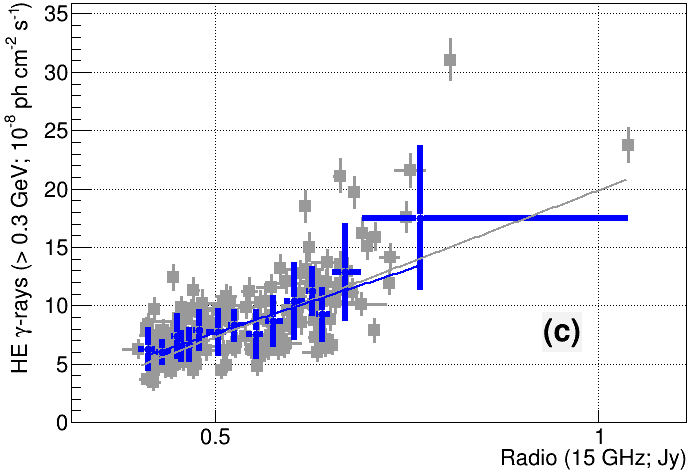}
\includegraphics[width=0.49\linewidth,height=5.5cm]{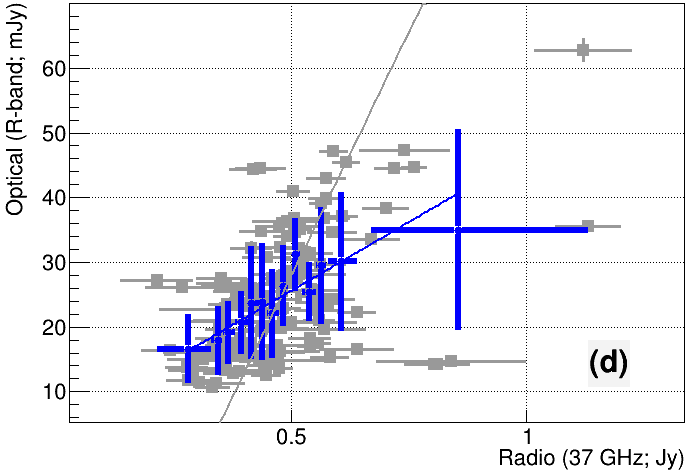}
\includegraphics[width=0.49\linewidth,height=5.5cm]{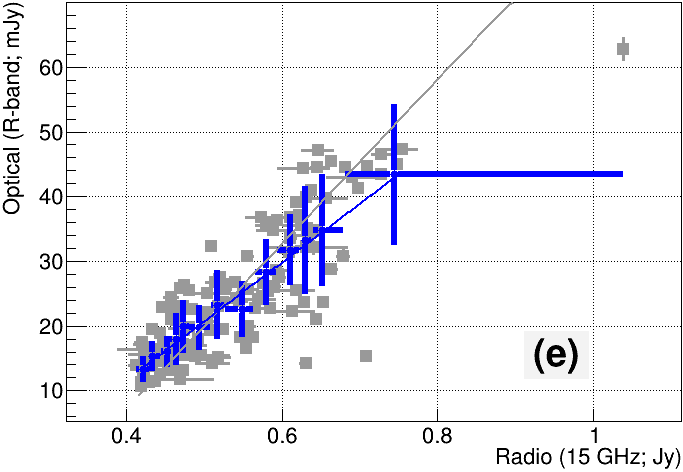}
\caption{
Flux-flux plots for several energy bands. 
The grey and blue markers as in Fig.\ref{fig:FF}. The grey and blue lines depict the best linear fit to the unbinned (grey) and binned (blue) data, with the slopes reported in Table~\ref{table:DCF1}.
Panel (a) shows the flux-flux cross-correlation between the HE $\gamma$-rays (LAT; $>$0.3\,GeV) and optical (R-band) fluxes, computed for 15-day time intervals at a zero timelag.
The panels (b)-(e) report fluxes computed for 15-day time intervals, where the radio (15\,GHz and 37\,GHz) have been shifted 45 days earlier in order to match the time lag observed in the correlation
plots from Fig.~\ref{fig:DCF_main1}. 
See   \textcolor{black}{Section} \ref{sec:he-vs-optical}, \ref{sec:he-vs-radio}, and \ref{sec:optical-vs-radio}, for details.}
\label{fig:FF1}
\end{figure*}

Figure~\ref{fig:FF1} shows the HE vs. optical flux correlation plots for the HE $\gamma$-rays vs. optical for 
\textcolor{black}{$\tau$=0,}
the HE $\gamma$-rays vs. radio for a time shift of 45 days, and the optical vs. radio for a time shift of 45 days. The time shift of 45 days is the time for which the correlation between these two bands is the highest (see   \textcolor{black}{Section} \ref{sec:he-vs-radio} and \ref{sec:optical-vs-radio}). The panels (b) and (c) of Fig.~\ref{fig:FF1} show that, for a time shift of 45 days, the relation between the GeV and the radio fluxes can be approximated by a linear function.  As with Fig.~\ref{fig:FF}, the panels also show the average and the standard deviation computed with data subsets of 10 observations, binned according to their flux ({\em binned} data). In the case of LAT vs. Mets{\"a}hovi, there is a large difference between the slopes from the linear functions fitted to the unbinned and binned data. This is produced by a few HE vs. optical flux pairs which are well outside the main trend; they have a substantial impact 
on the fit to the unbinned data, while they do not affect the binned data.

\clearpage

\section{Estimation of the most representative time lag}
\label{sec:DCFLag}

\begin{figure}
\centering
\includegraphics[width=\linewidth,height=6cm]{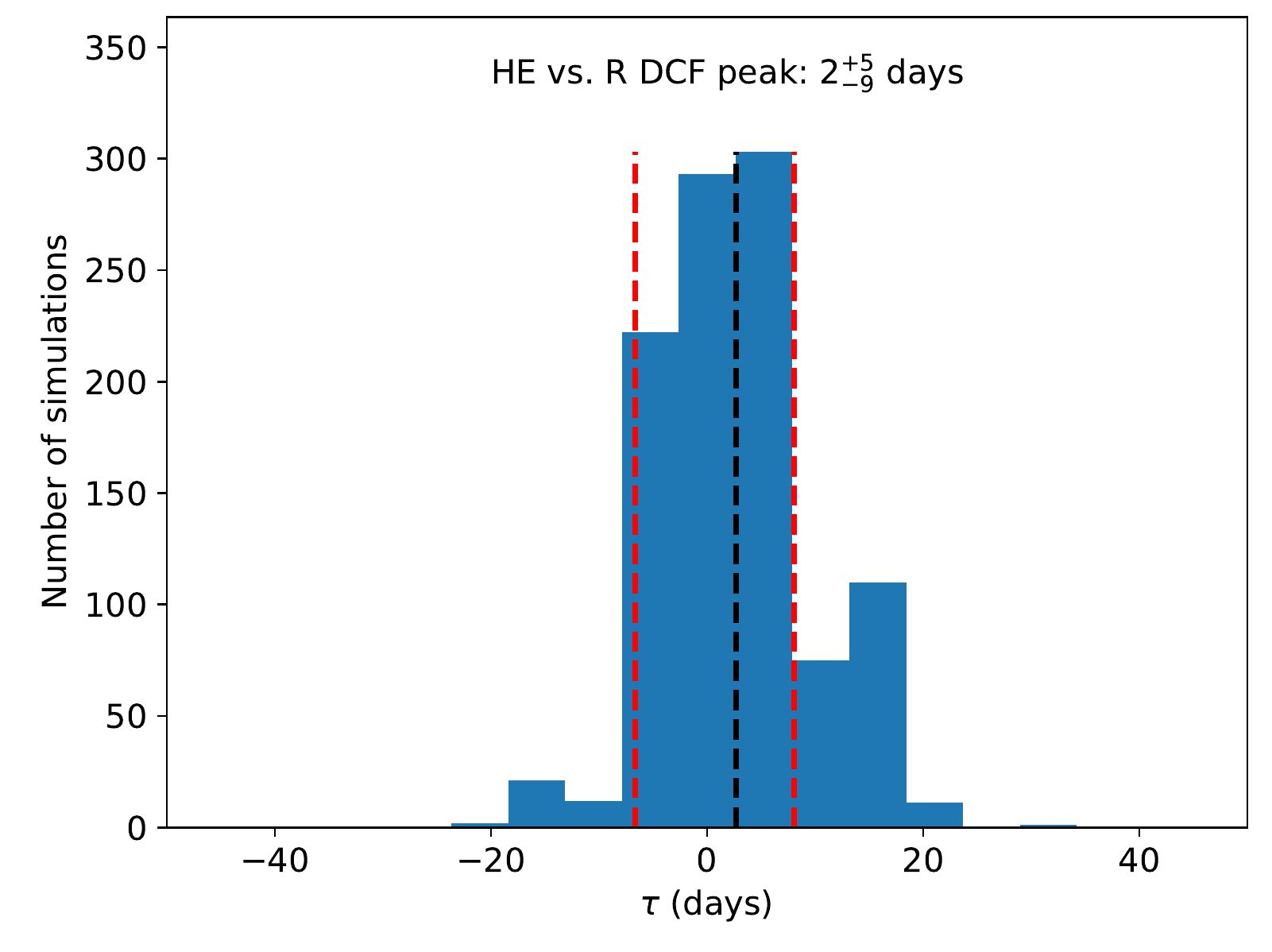}
\caption{
Distribution of DCF$_{cen}$ derived with 1000 Monte Carlo FR/RSS simulations to estimate the time lag between the HE $\gamma$-ray and optical R-band LCs that were used to compute the DCF reported in panel \textit{a} 
of Fig.~\ref{fig:DCF_main1}. The average and the 68\% containment are depicted with the black and red lines, respectively, and are used as the estimate of the time lag between these two bands. See text in Appendix~\ref{sec:DCFLag} for further details.
}
\label{fig:DCFlagCalc1}
\end{figure}

\begin{figure*}
\centering
\includegraphics[width=0.49\linewidth,height=6cm]{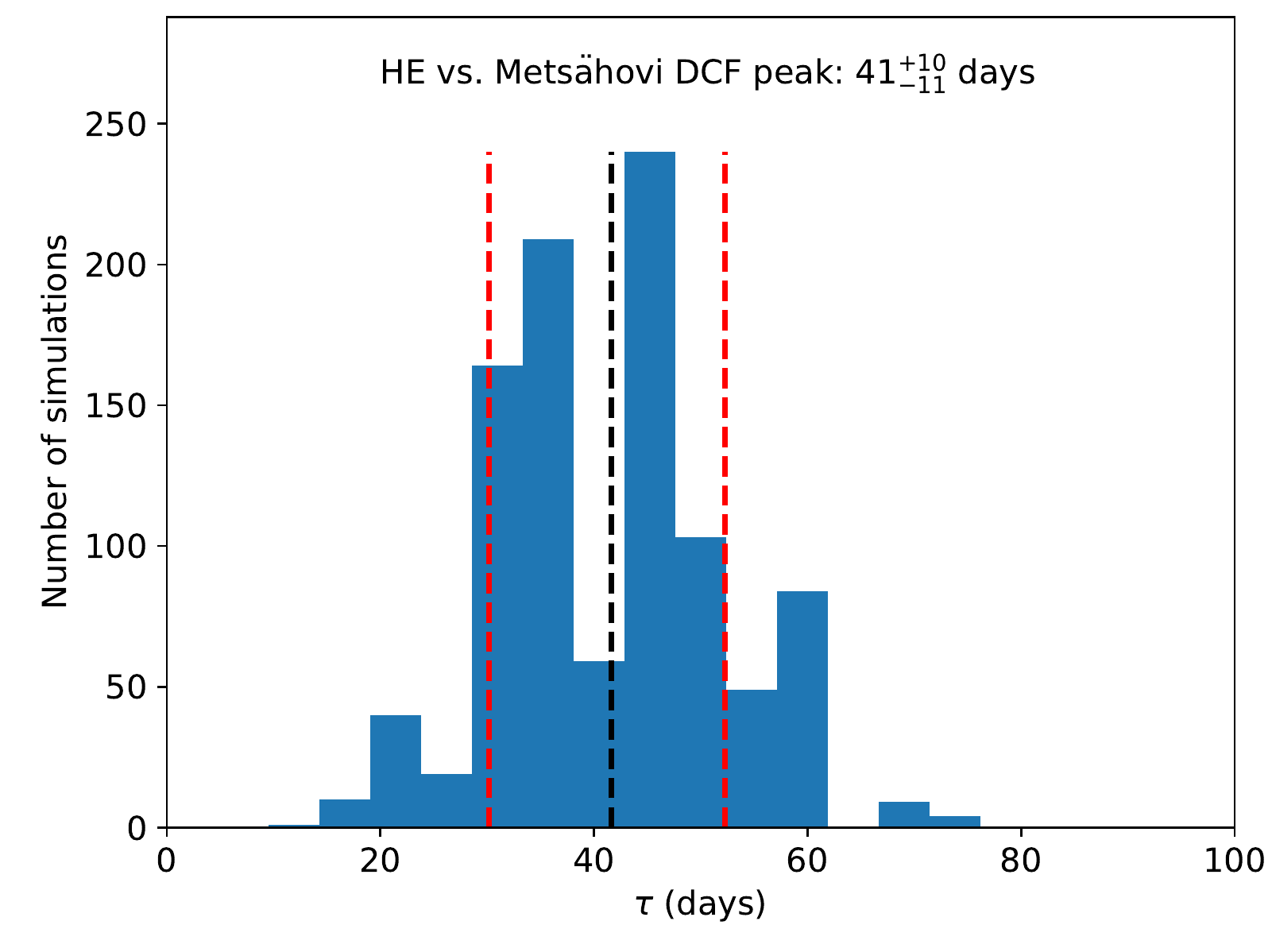}
\includegraphics[width=0.49\linewidth,height=6cm]{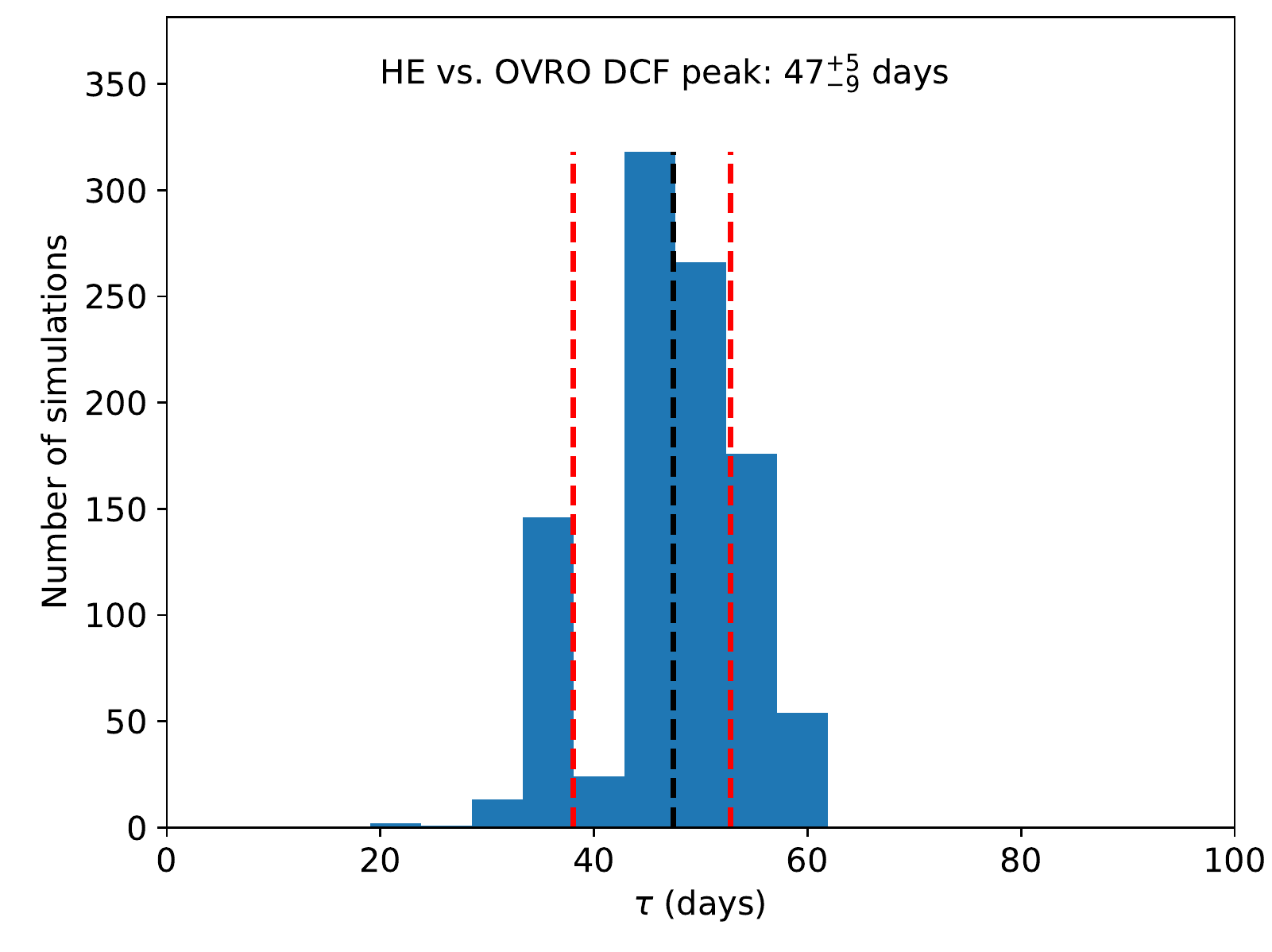}
\includegraphics[width=0.49\linewidth,height=6cm]{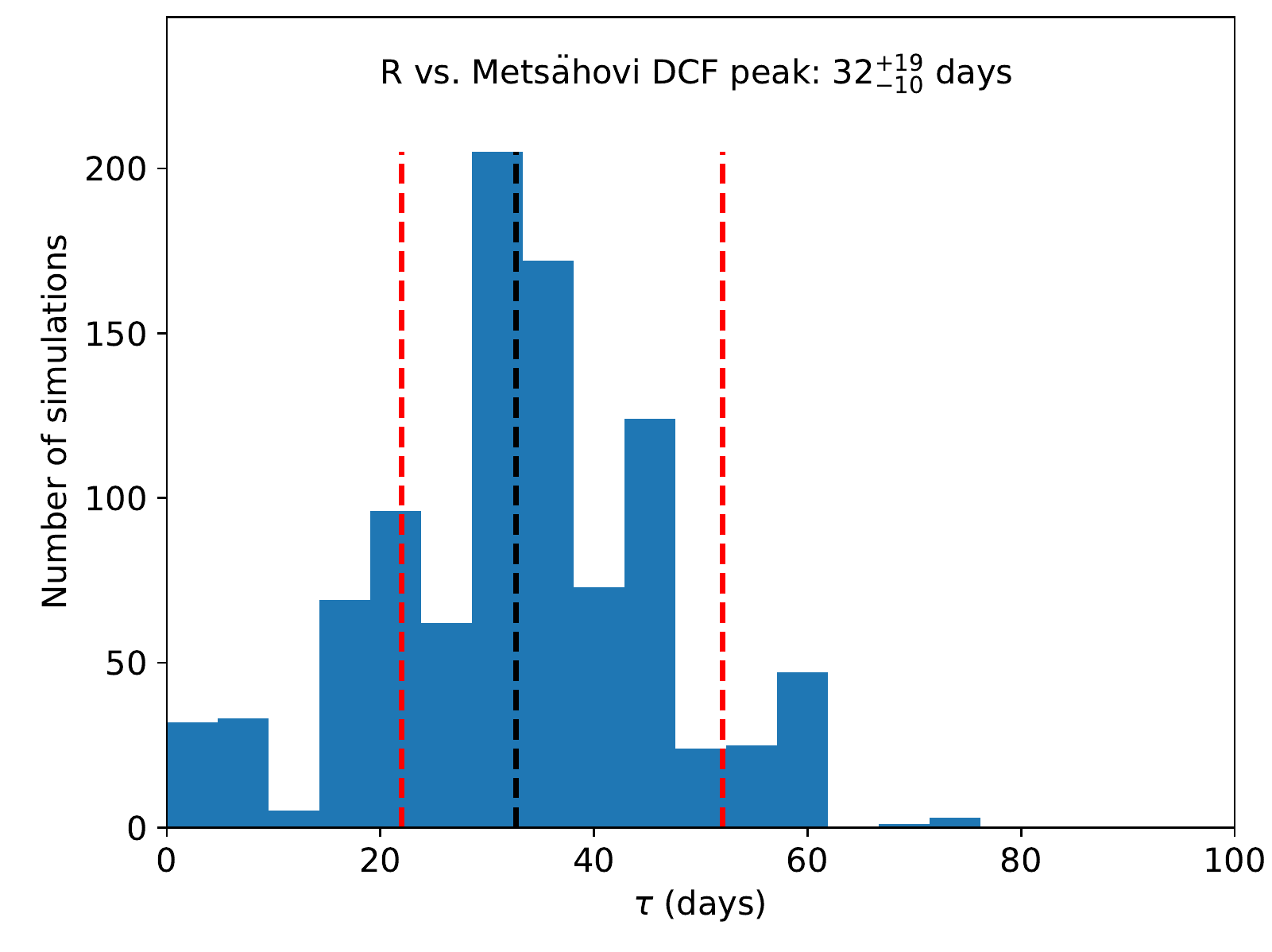}
\includegraphics[width=0.49\linewidth,height=6cm]{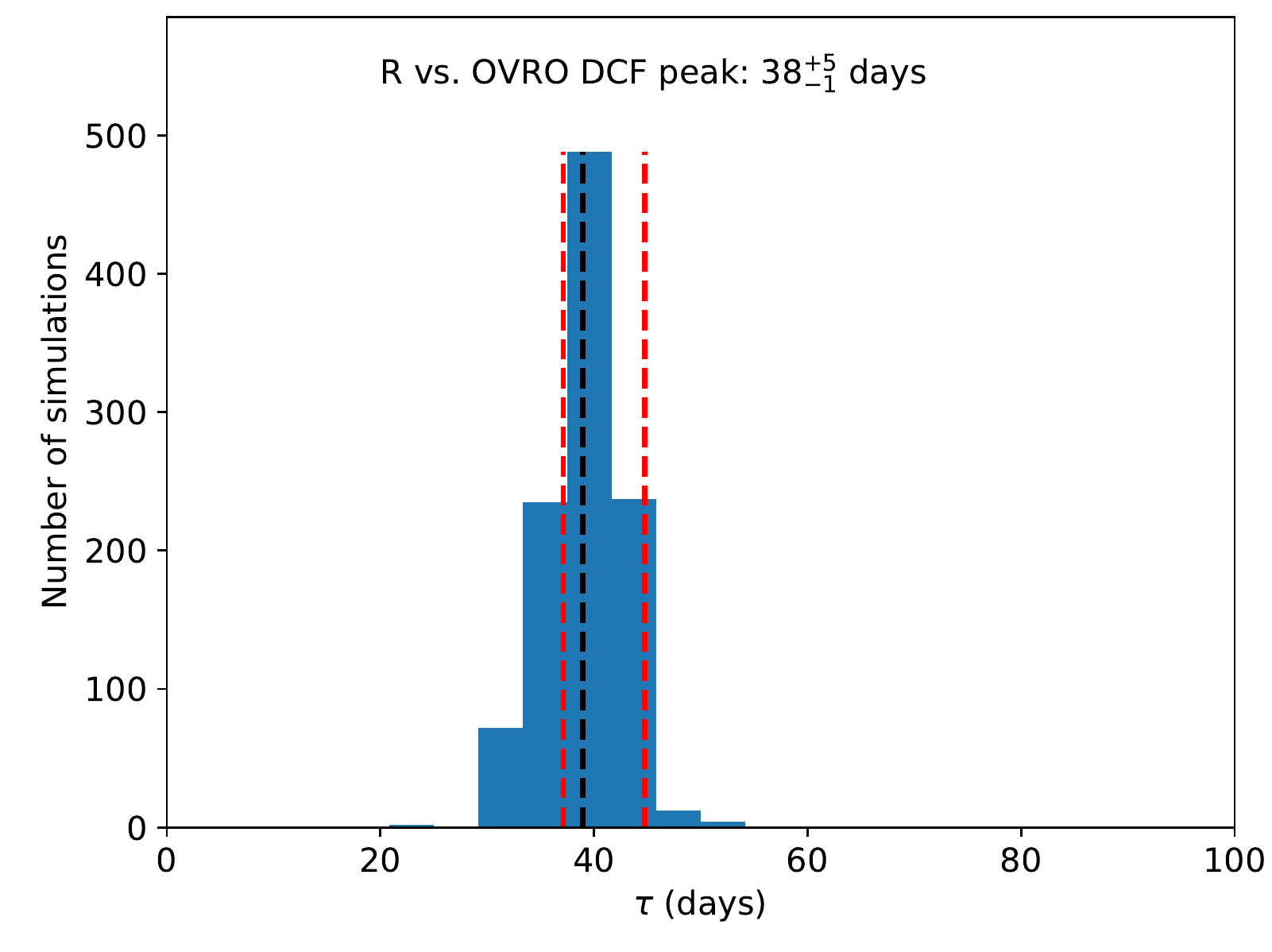}
\caption{
Distributions of DCF$_{cen}$ derived with 1000 Monte Carlo FR/RSS simulations to estimate the time lag between the various multi-band LCs that were used to compute the DCFs reported in panels \textit{b, c, d,} and \textit{e} 
of Fig.~\ref{fig:DCF_main1}: a) HE vs. Mets\"{a}hovi (top-left), b) HE vs. OVRO (top-right), c) R vs. Mets\"{a}hovi (bottom-left) and d) R vs. OVRO (bottom-right).  The average and the 68\% containment are depicted with the black and red lines, respectively, and are used as the estimate of the time lag between the bands. See text in Appendix~\ref{sec:DCFLag} for further details.}
\label{fig:DCFlagCalc2}
\end{figure*}

In this section, we 
report an estimate of the most representative time lag and its related uncertainty for the multi-band fluxes used in the correlation studies reported in   \textcolor{black}{Sections}~\ref{sec:he-vs-optical}, ~\ref{sec:he-vs-radio}, and  ~\ref{sec:optical-vs-radio}. We use the 
model-independent Monte Carlo flux randomization (FR) and random subset selection (RSS) method described in \citet{1998PASP..110..660P} and \citet{2004ApJ...613..682P}, which is the methodology used by \citet{2014MNRAS.445..428M} to estimate the time lag of 40$\pm$9 days between the \textit{Fermi}-LAT and OVRO fluxes. Briefly, the method 
employed in this study is as follows: we 
perform
RSS of the first LC and 
select 
 the simultaneous observations between the first and second LC. Then, we perform
FR according to the flux uncertainties of
 both LCs. 
In this way, through this process of RSS and FR, we generate a set of 1000 Monte Carlo simulated LC pairs. Then we perform the DCF study for these 1000 simulated pairs. 
As in \citet{1998PASP..110..660P}, a cross-correlation is considered successful if the maximum correlation coefficient is large enough such that the correlation between the LC pairs is significant above 95\% confidence level. 
Instead of using the peak of the DCF (DCF$_{max}$), following the prescriptions from \citet{2004ApJ...613..682P}, we used the centroid of the DCF (DCF$_{cen}$), computed with the DCF values above 0.8$\times$DCF$_{max}$, which is expected to provide better results when the DCF has a broad peak. 
 The distributions of 
DCF$_{cen}$ 
are then obtained. The most representative value of the time lag is estimated by considering the mean of the distribution, and the uncertainties are computed using the 68\% containment, that would correspond to 1\,$\sigma$ error for a normal distribution.

Figure~\ref{fig:DCFlagCalc1} shows  
the distribution of DCF$_{cen}$ 
for the 
1000 simulated LCs for the HE and optical R-band. 
The average and the 68\% containment (depicted with the black and red lines in Fig.~\ref{fig:DCFlagCalc1}) is $2\substack{+5\\-9}$, which can be considered as good estimate of the time lag and related uncertainty between the fluxes for these two energy bands. This is perfectly consistent with no time lag, and hence simultaneous emission in these two energy bands.

\begin{figure*}
\centering
\includegraphics[width=0.48\linewidth,height=6cm]{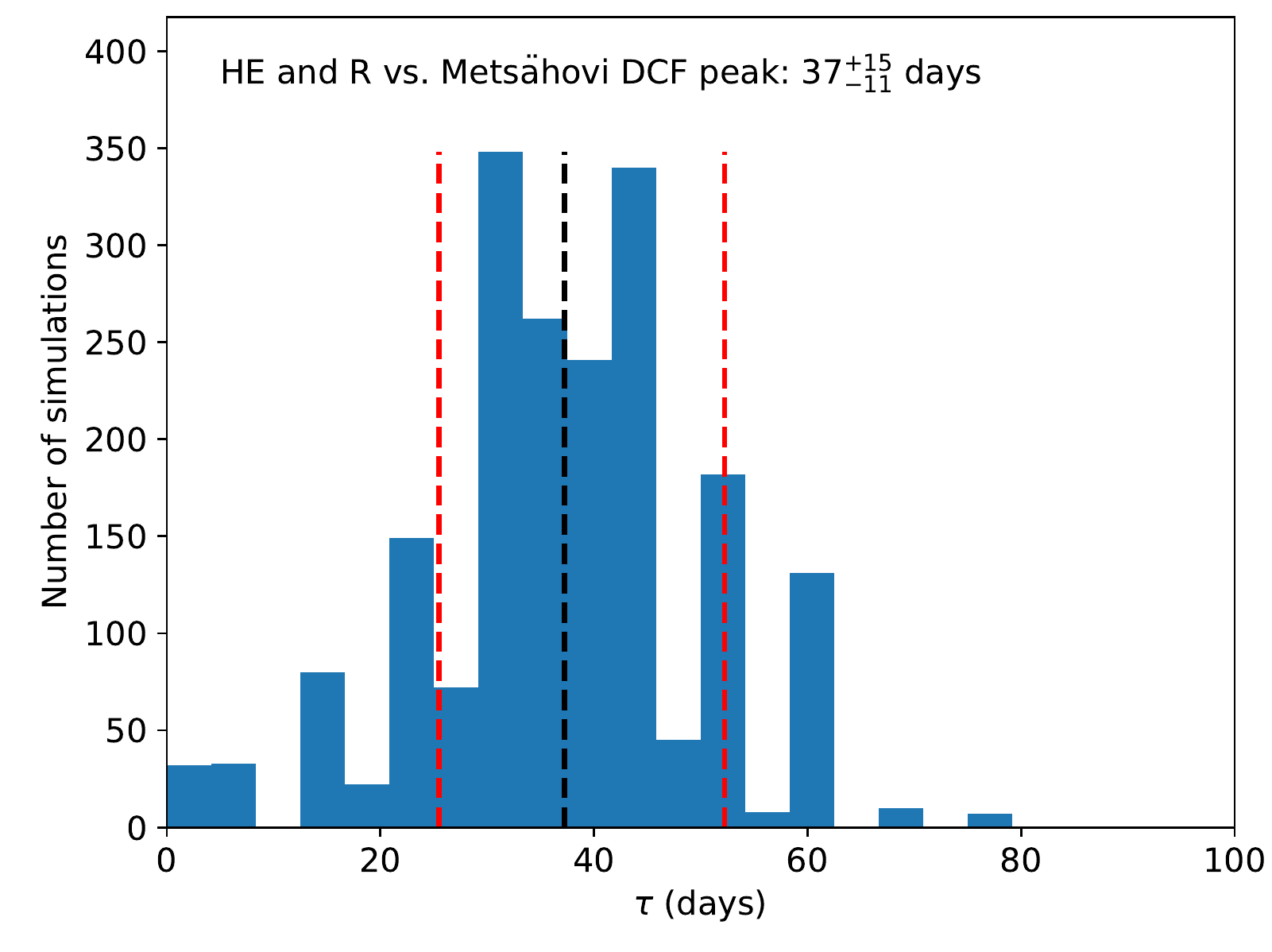}
\includegraphics[width=0.48\linewidth,height=6cm]{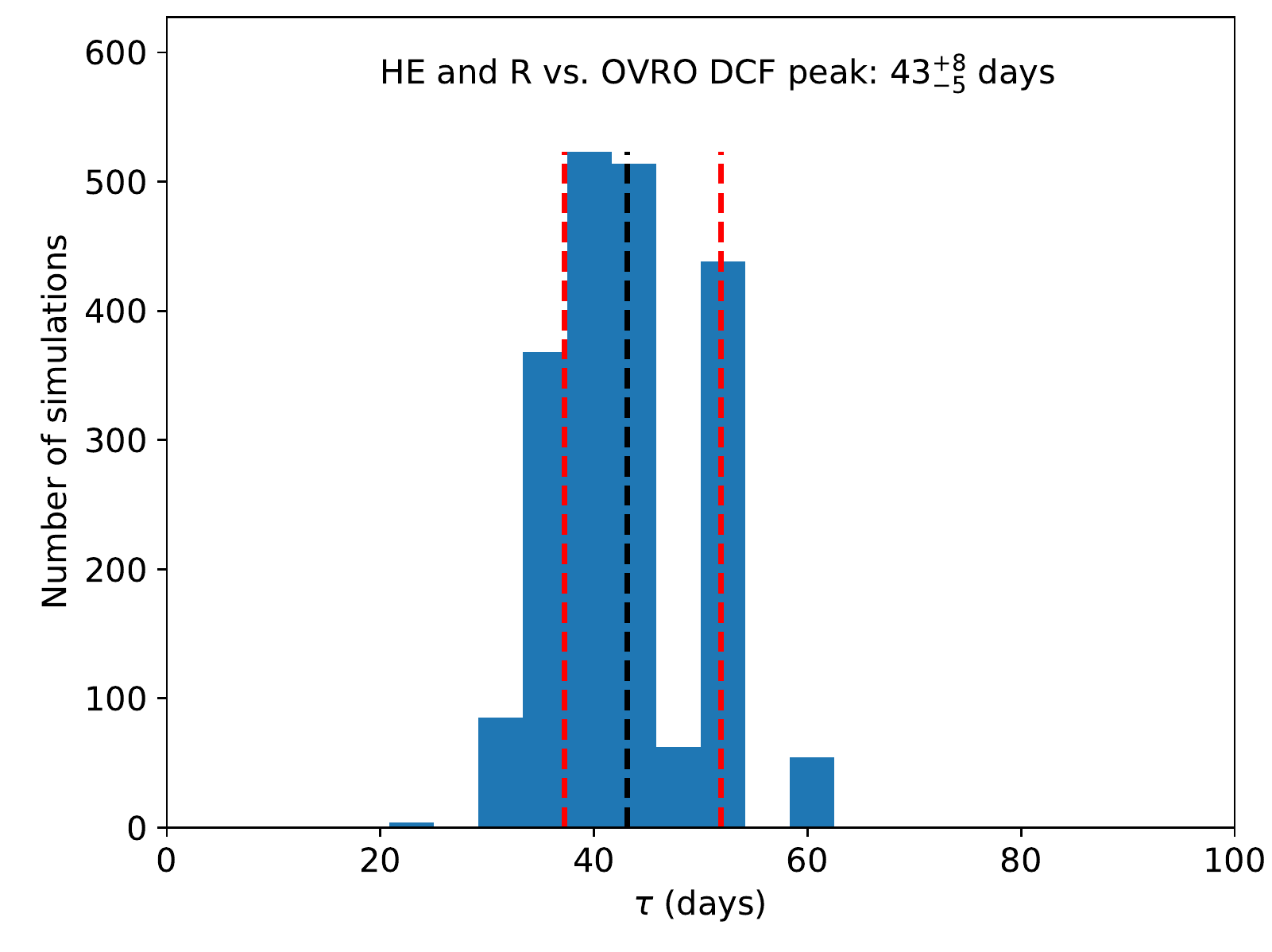}
\caption{
Distribution of DCF$_{cen}$ from the combinations of the two panels with 37\,GHz Mets\"{a}hovi data (left) and the two panels with 15\,GHz OVRO data (right) from Fig.~\ref{fig:DCFlagCalc2}. See text in Appendix~\ref{sec:DCFLag} for further details. }
\label{fig:DCFlagCalc3}
\end{figure*}

The panels in Fig.~\ref{fig:DCFlagCalc2} show the 
distributions of DCF$_{cen}$ 
for the 1000 simulated LCs for the HE and R band vs. the two radio bands observed with Mets\"{a}hovi and OVRO. Since the time lags shown in  Fig~\ref{fig:DCFlagCalc2} are statistically compatible, we decided to combine the GeV and R-band with the 37\,GHz and with the 15\,GHz cases, in order to estimate combined time lags for the GeV/optical and 37\,GHz, and the GeV/optical and 15\,GHz. 
The combined 
distributions of DCF$_{cen}$, 
derived with the 2000 simulated LCs, are shown in Fig.~\ref{fig:DCFlagCalc3}, 
leading to the estimation of combined time lags of
$37\substack{+15\\-11}$ days and 
$43\substack{+8\\-5}$ days, respectively.

\clearpage

\section{Flux profile}
\label{sec:A1}
\begin{figure}
\centering
\includegraphics[width=\linewidth,height=6cm]{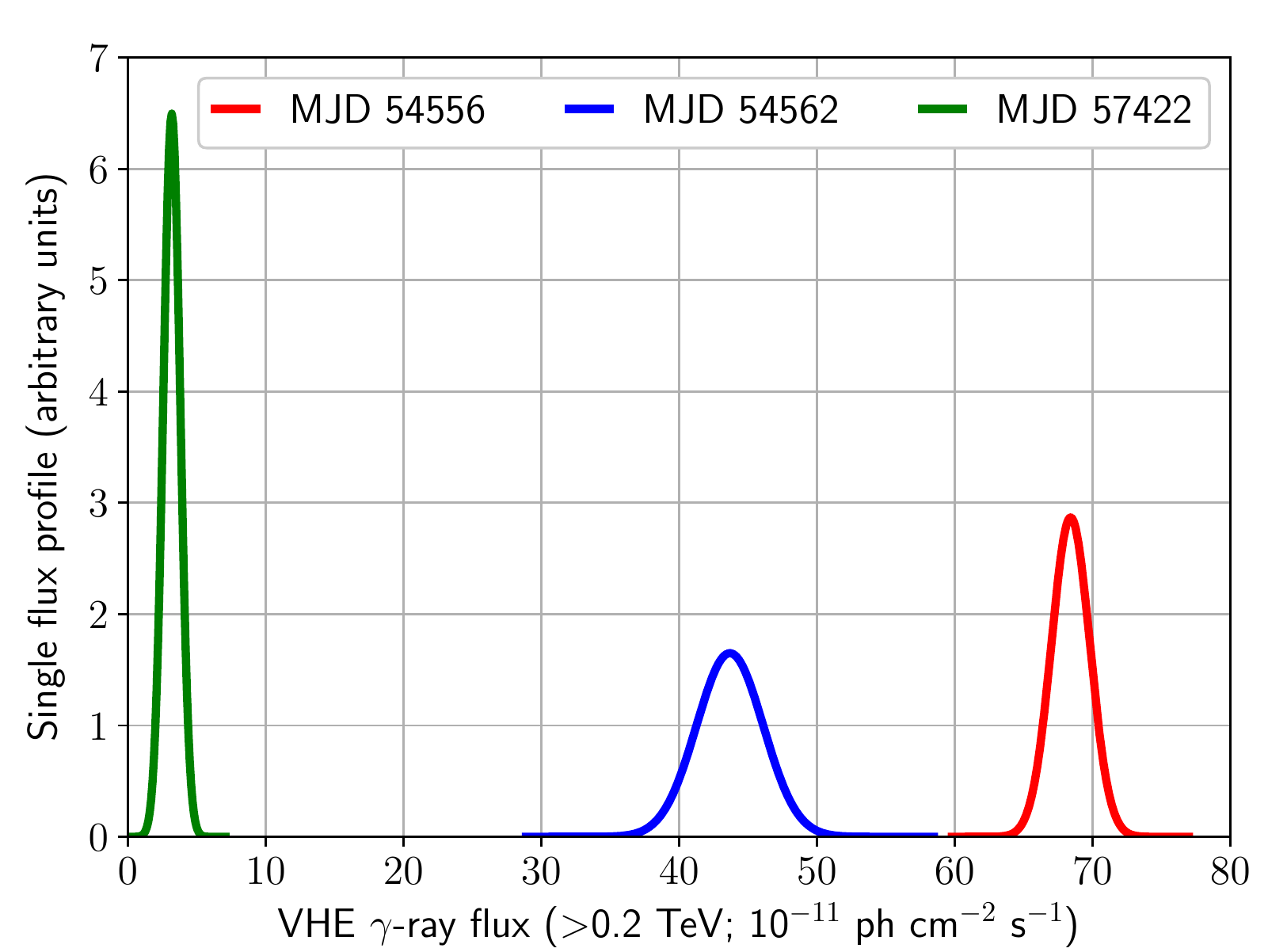}
\caption{Examples of contributions to the VHE $\gamma$-ray flux profiles from three selected flux measurements with the MAGIC telescopes.}
\label{fig:FPA20}
\end{figure}

\begin{figure*}
\centering
\includegraphics[width=0.49\linewidth,height=6cm]{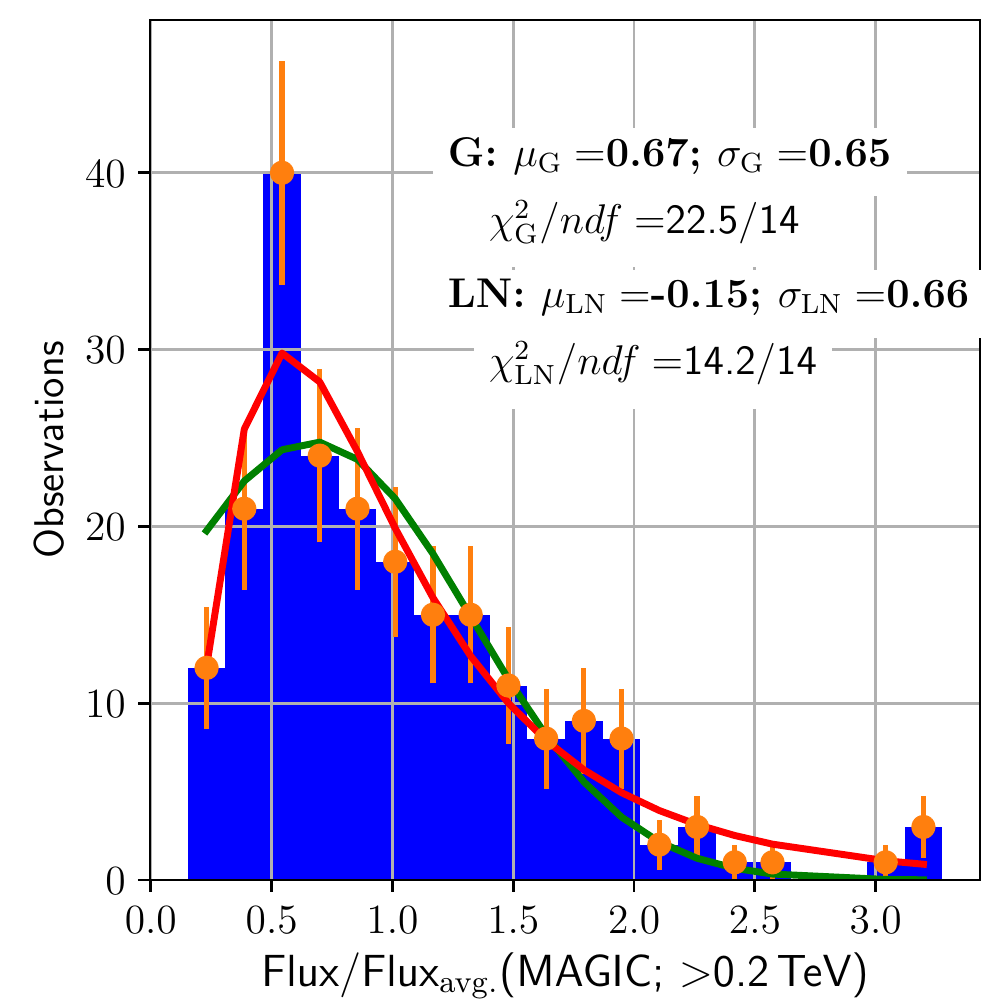}
\includegraphics[width=0.49\linewidth,height=6cm]{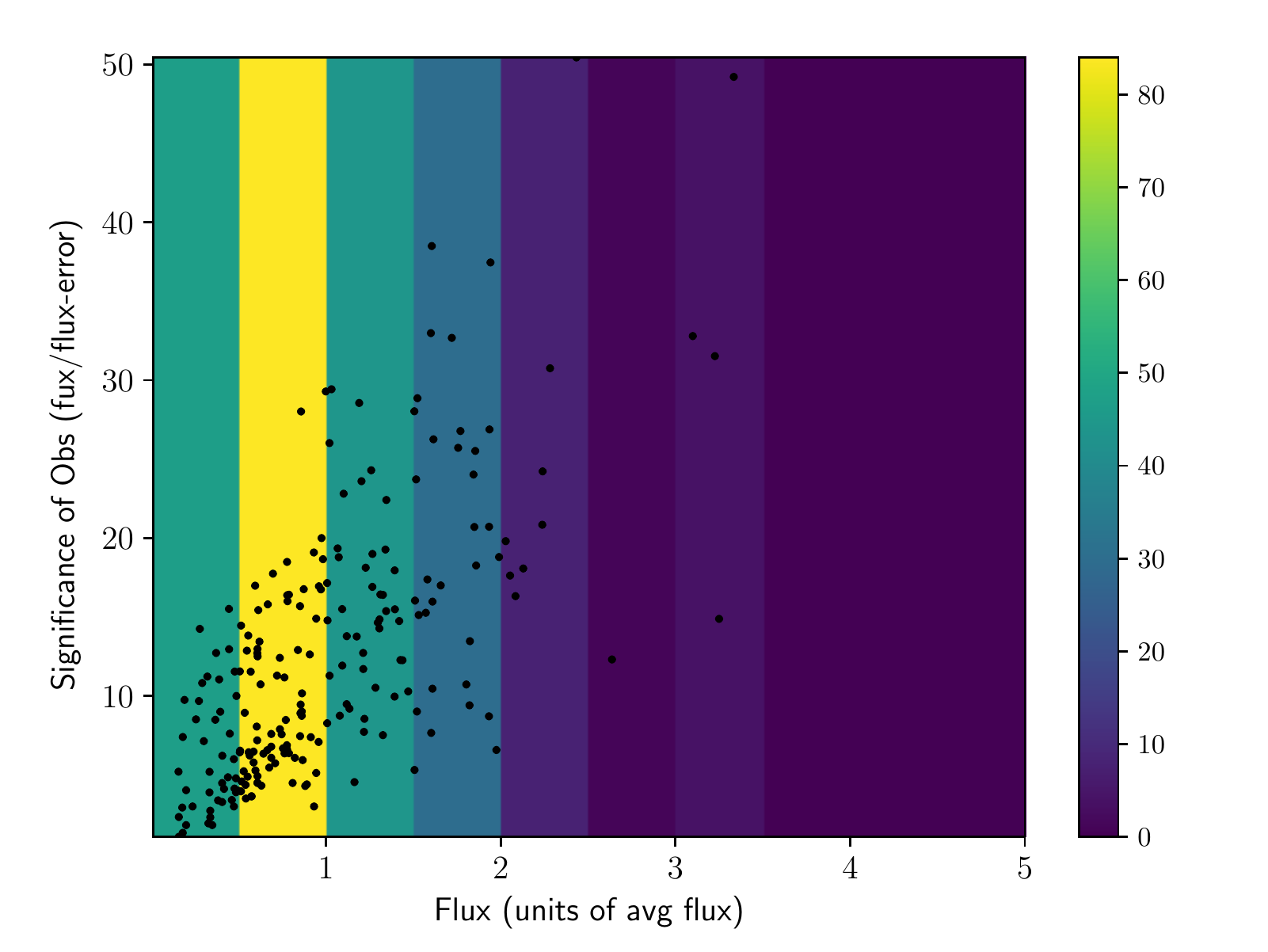}
\includegraphics[width=0.49\linewidth,height=6cm]{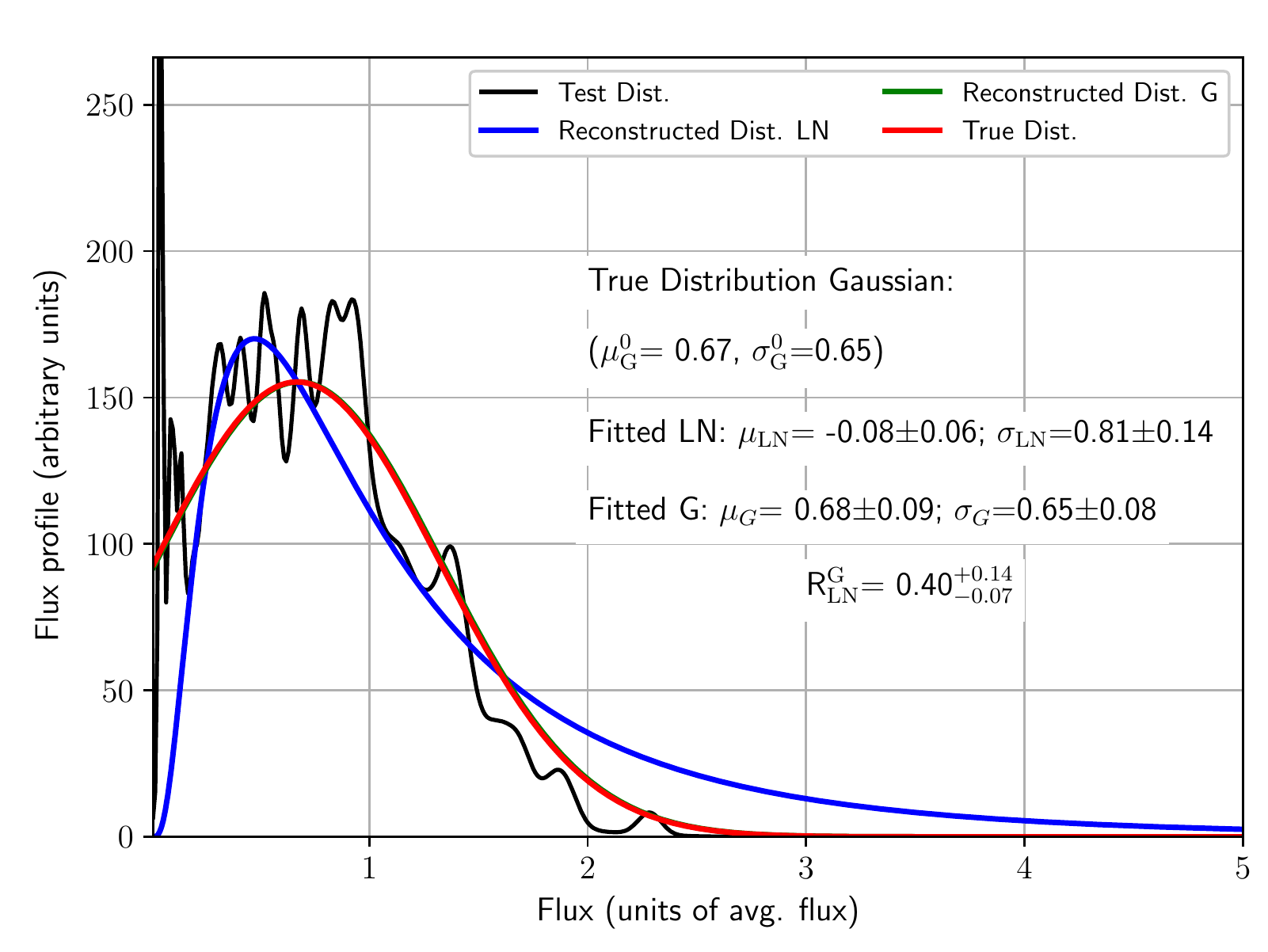}
\includegraphics[width=0.49\linewidth,height=6cm]{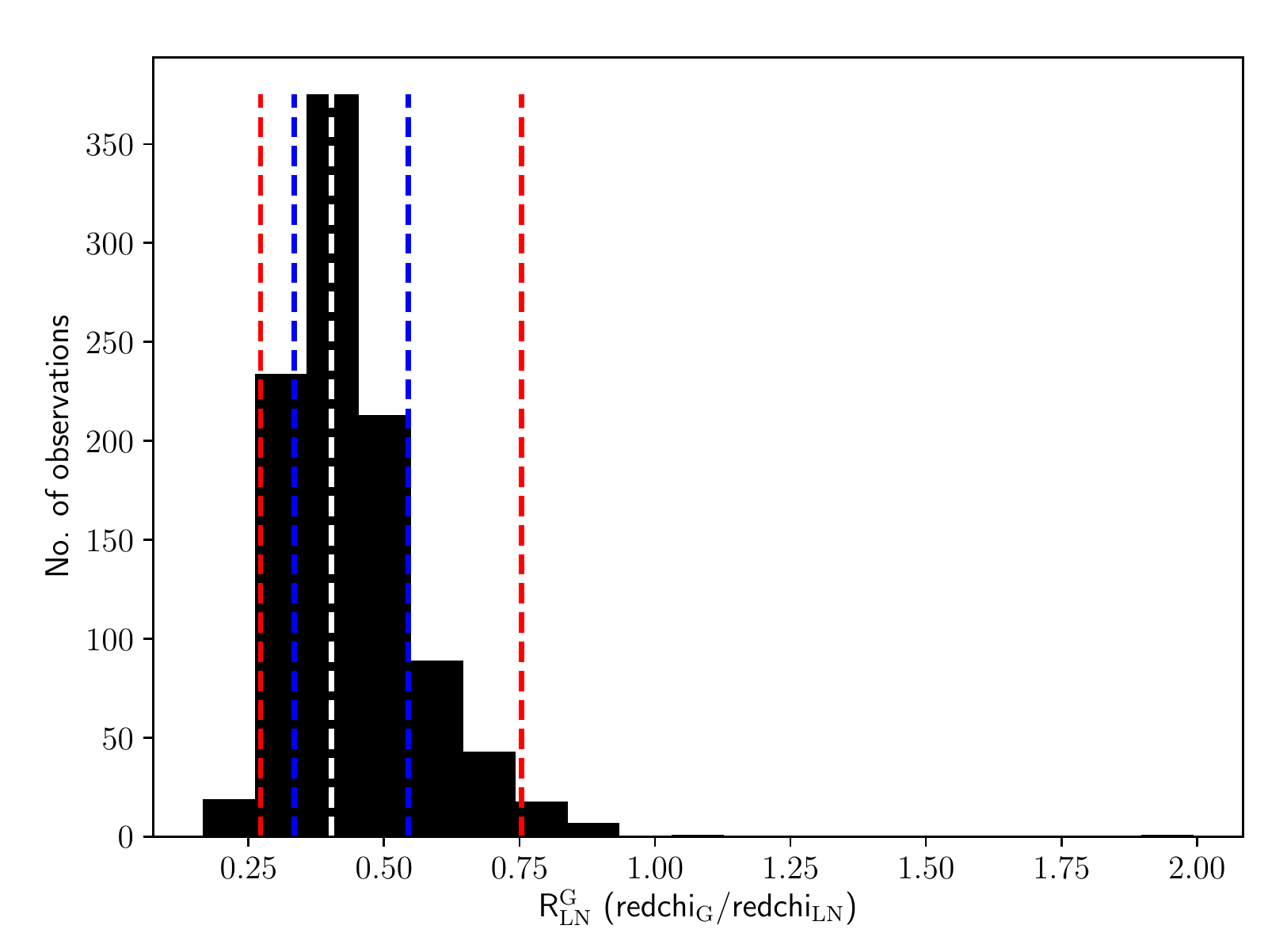}
\includegraphics[width=0.49\linewidth,height=6cm]{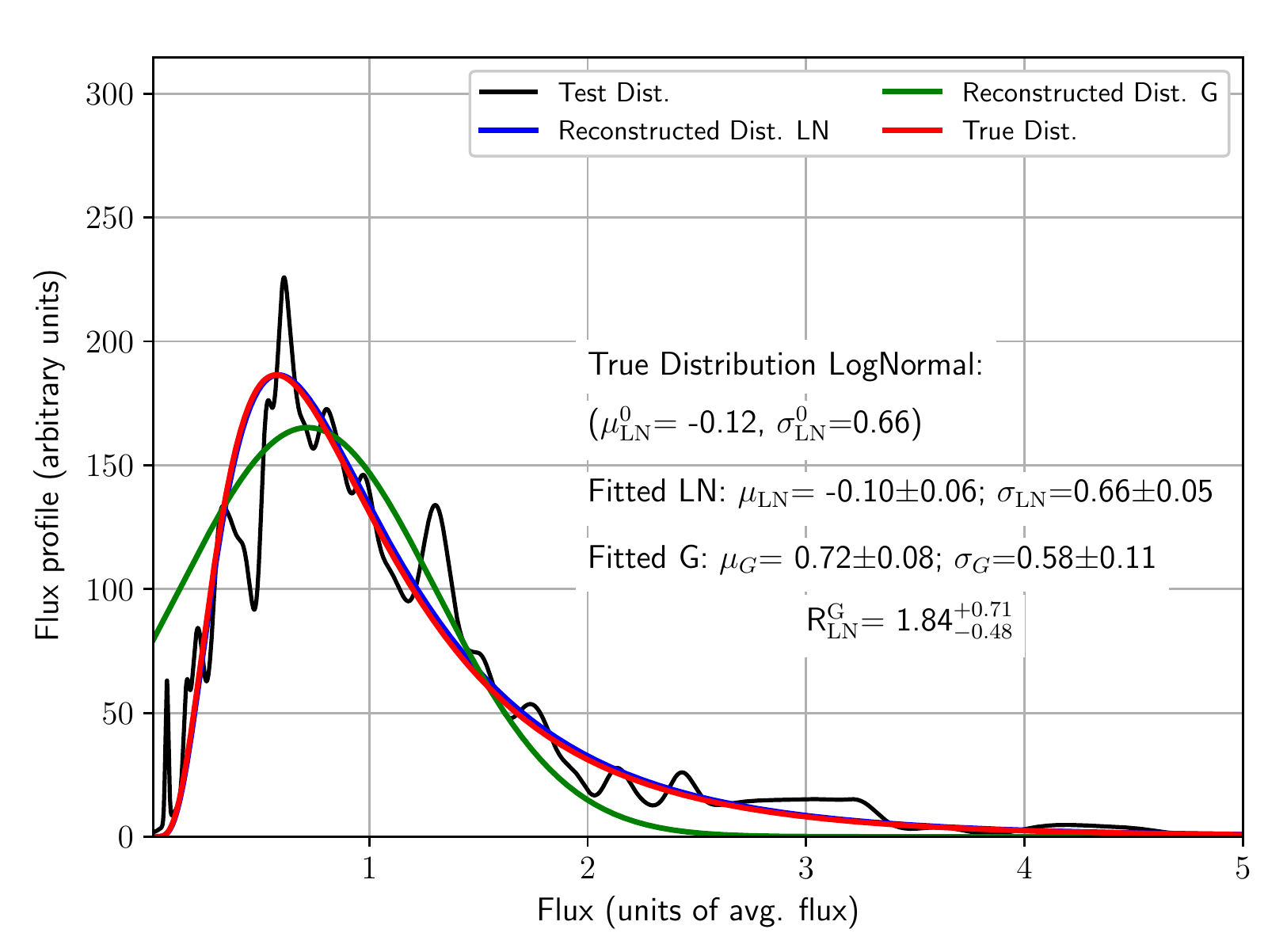}
\includegraphics[width=0.49\linewidth,height=6cm]{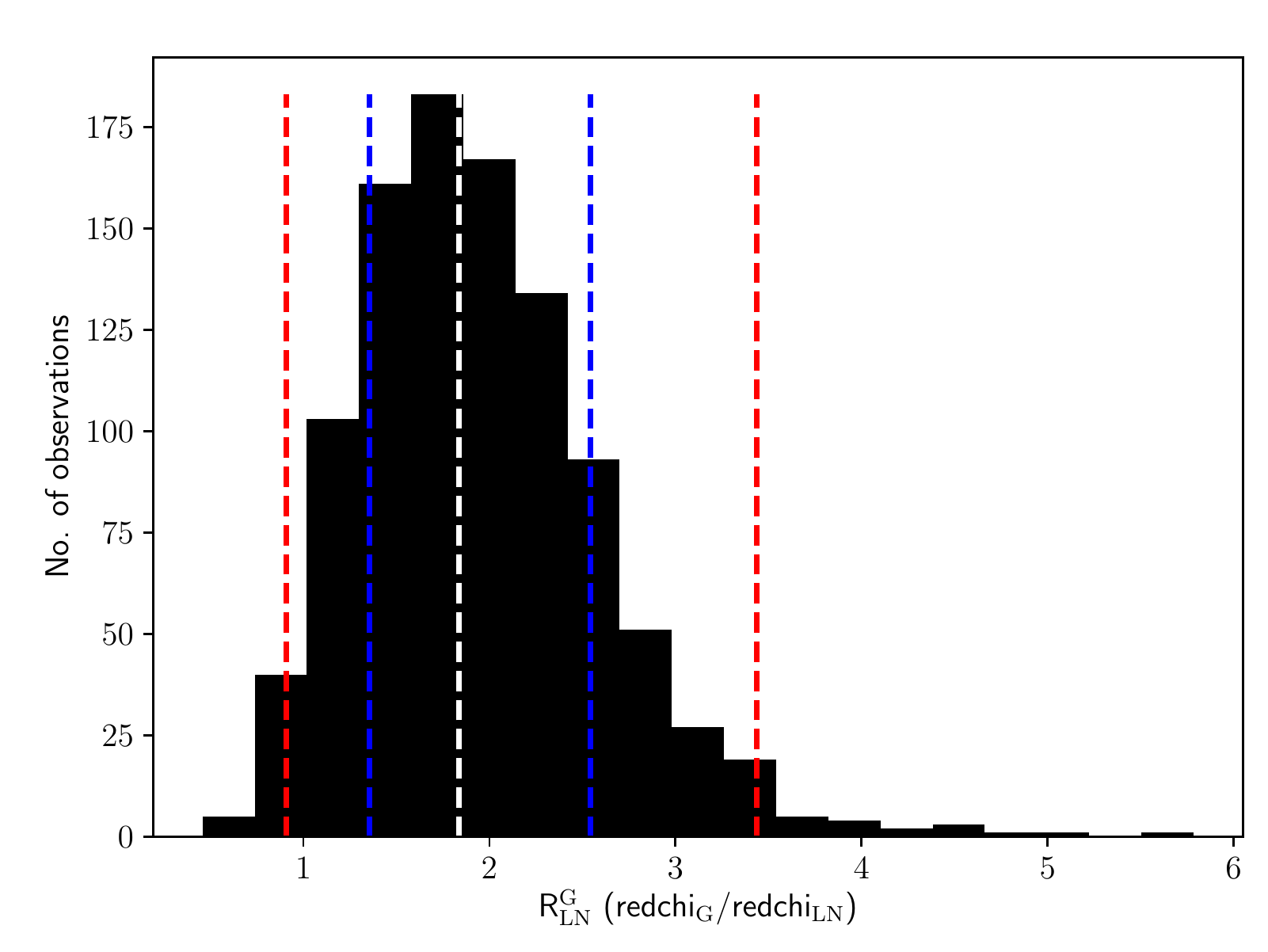}
\caption{Validation of flux profile-method using long-term VHE observed flux. The histogram of real data (fluxes) along with fits with the Gaussian (red line) and LogNormal (green line) functions is shown in the top left plot. These functions are shown in red in the middle left (Gaussian) and bottom left (LogNormal) plots. The top right plot shows the distribution of the flux/flux-error ratio (SNR) vs. flux. The colour scale indicates the number of flux measurements in each flux bin. In the middle left and bottom left plots, the red lines represent fits to the true flux distribution with a Gaussian (middle) and LogNormal (bottom), while the green and blue lines show fits to simulated flux profiles with Gaussian and LogNormal, respectively. In each of these two plots one example of the 1000 simulated Gaussian (LogNormal) flux profiles is presented with black line. The middle right and bottom right plots the distributions of the parameter R$^\mathrm{G}_\mathrm{LN}$ for Gaussian and LogNormal distributions, respectively. The white, blue, and red vertical dashed lines represent the 
%centriod of the histogram
\textcolor{black}{weighted average of the histograms bins}
, the 1\,$\sigma$ and 2\,$\sigma$ confidence intervals.}
\label{fig:FPA21}
\end{figure*}

\begin{figure*}
\centering
\includegraphics[width=0.49\linewidth,height=6cm]{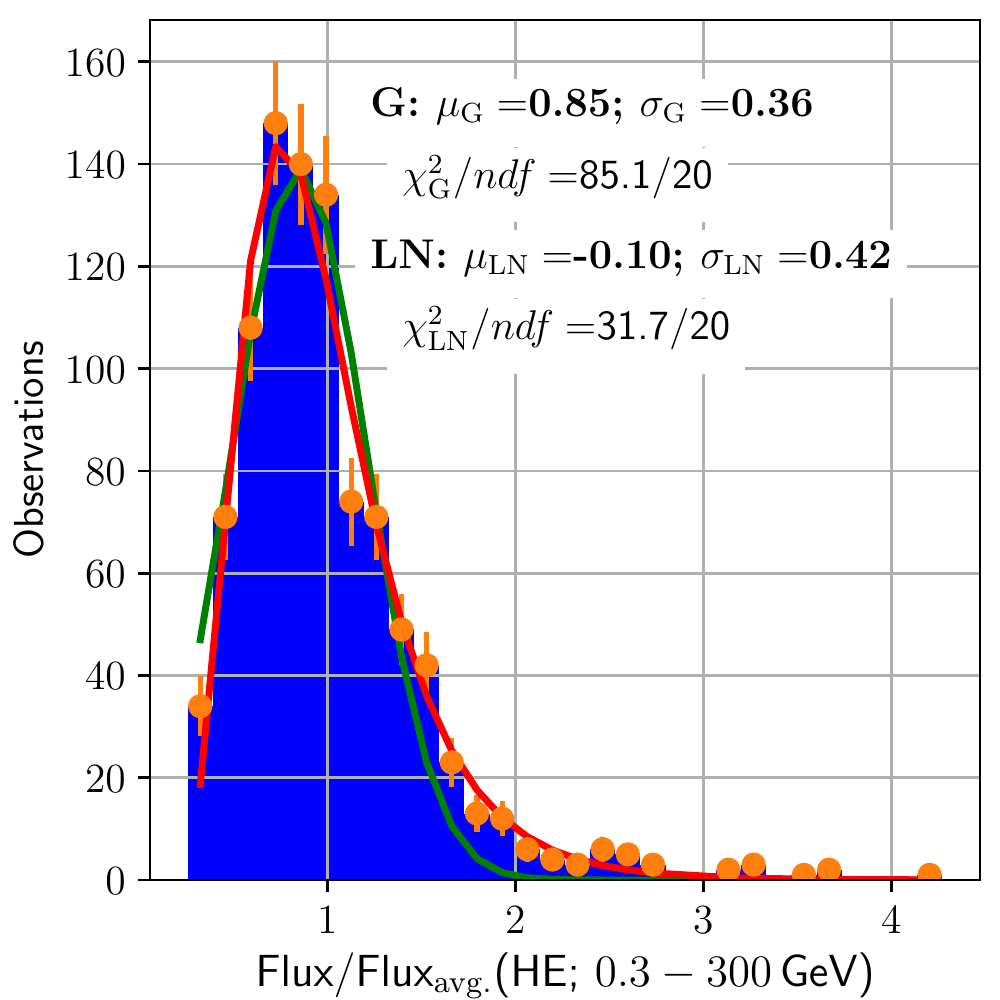}
\includegraphics[width=0.49\linewidth,height=6cm]{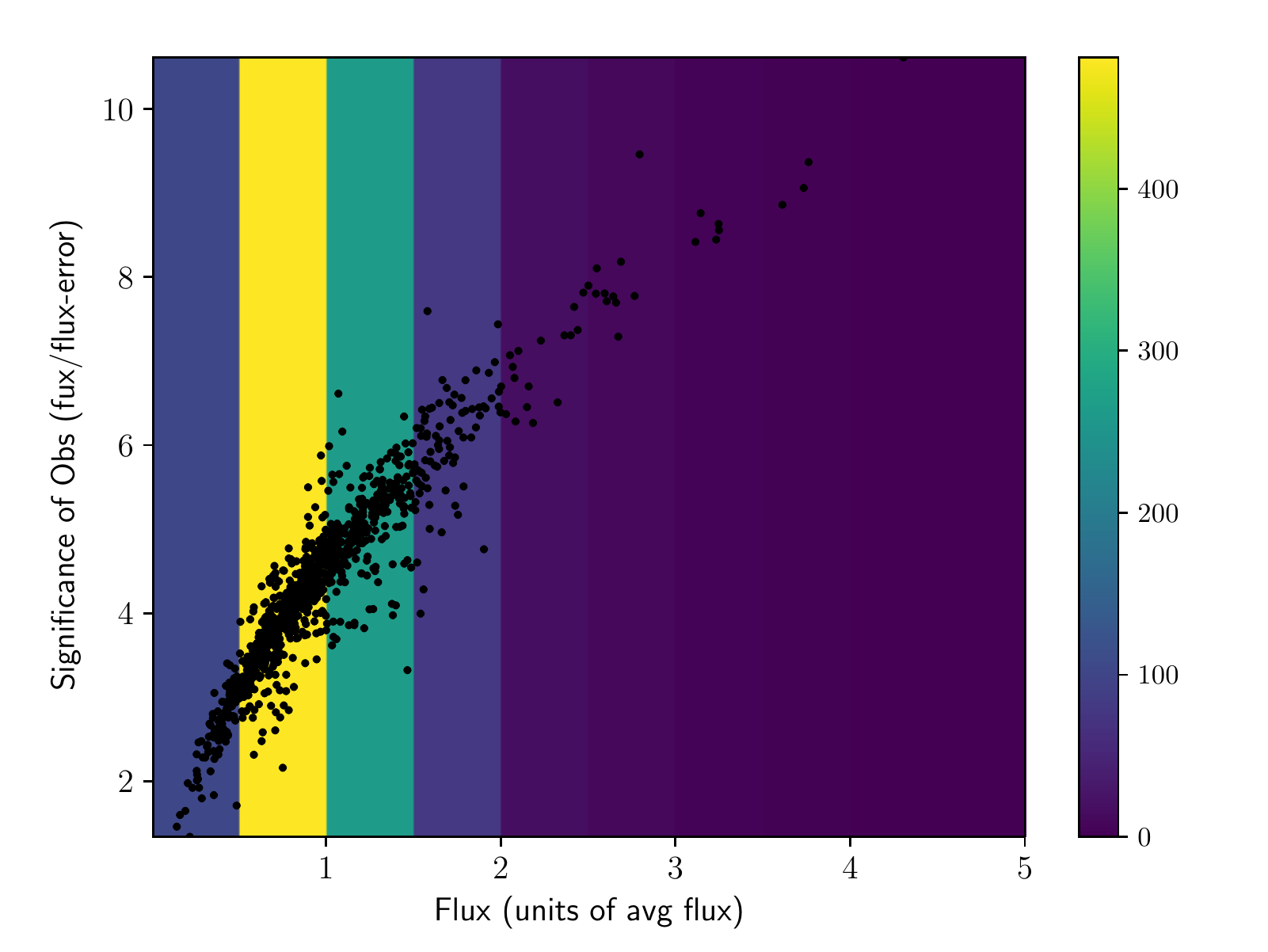}
\includegraphics[width=0.49\linewidth,height=6cm]{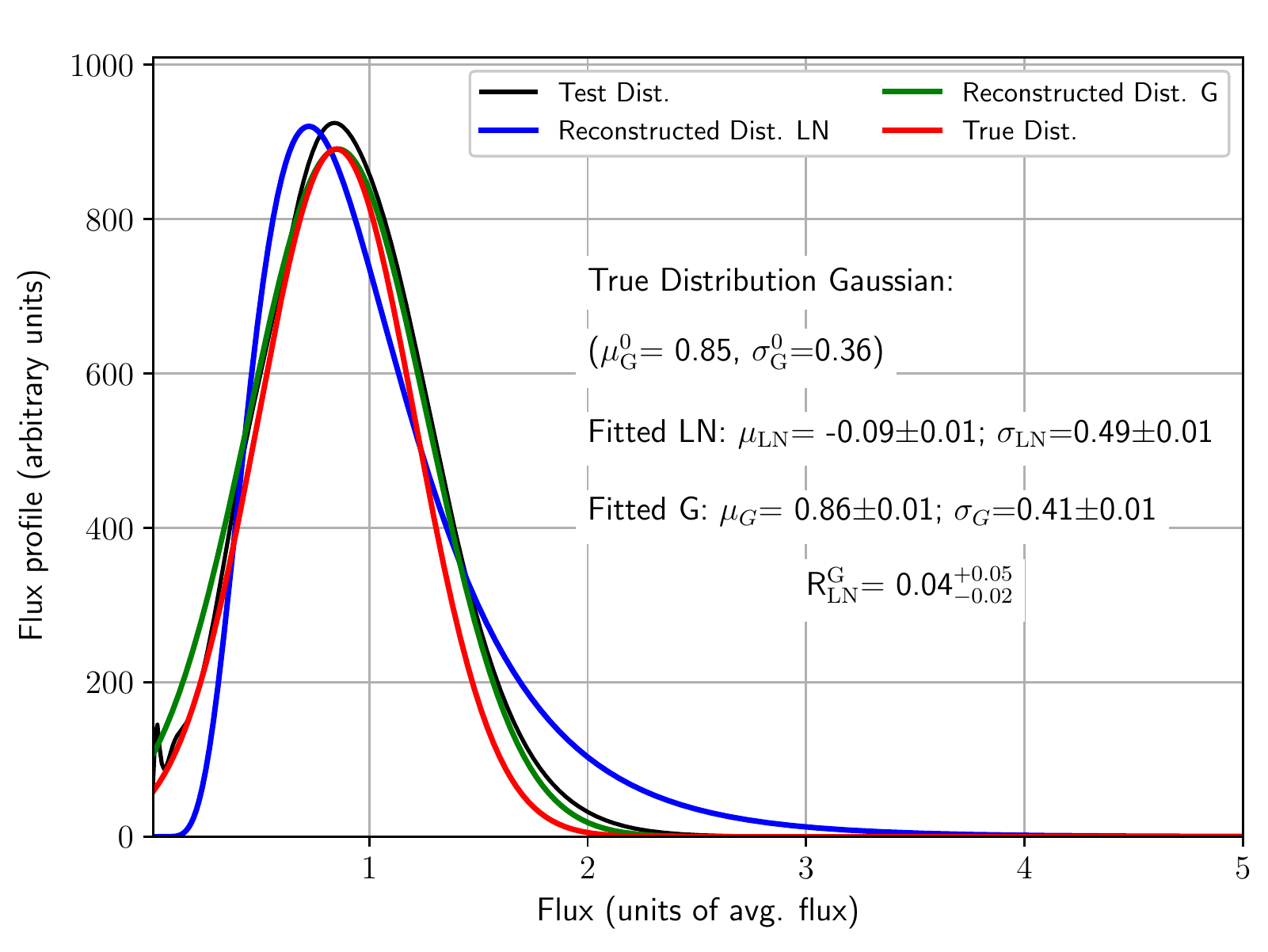}
\includegraphics[width=0.49\linewidth,height=6cm]{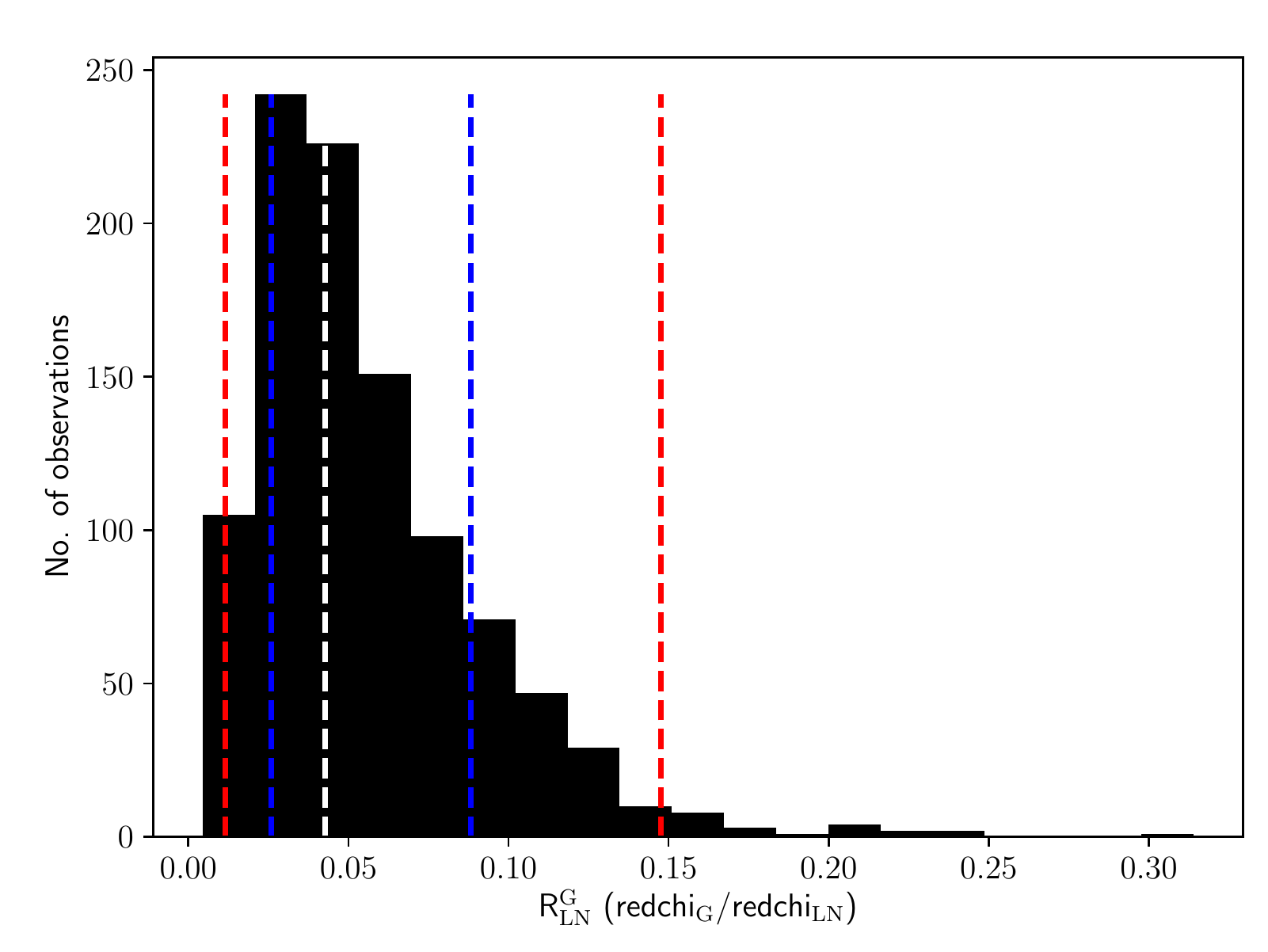}
\includegraphics[width=0.49\linewidth,height=6cm]{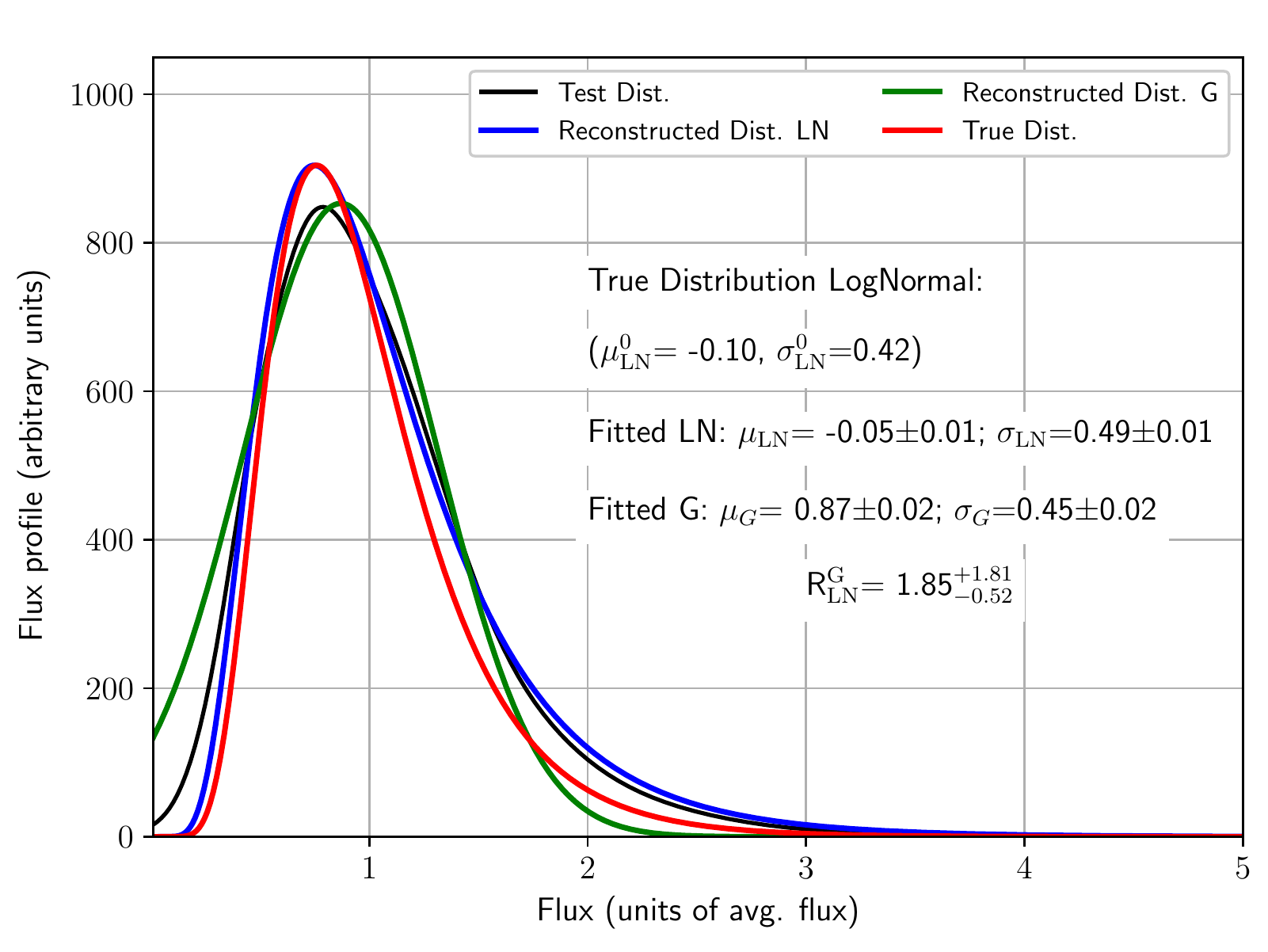}
\includegraphics[width=0.49\linewidth,height=6cm]{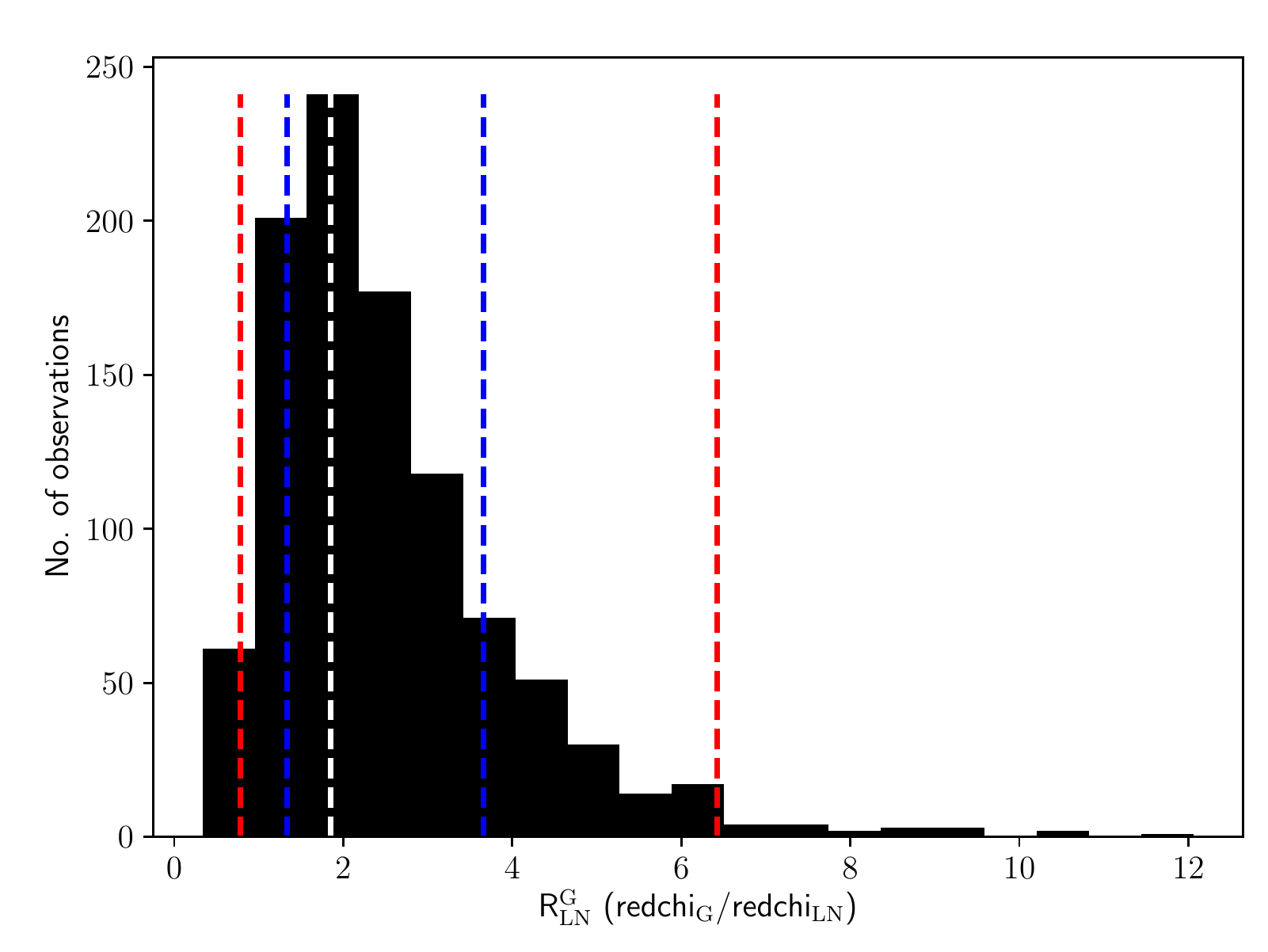}
\caption{Validation of flux profile-method using long-term HE $\gamma$-ray (0.3-300\,GeV) data}\label{fig:FPA23}
\end{figure*}

\begin{figure*}
\centering
\includegraphics[width=0.49\linewidth,height=6cm]{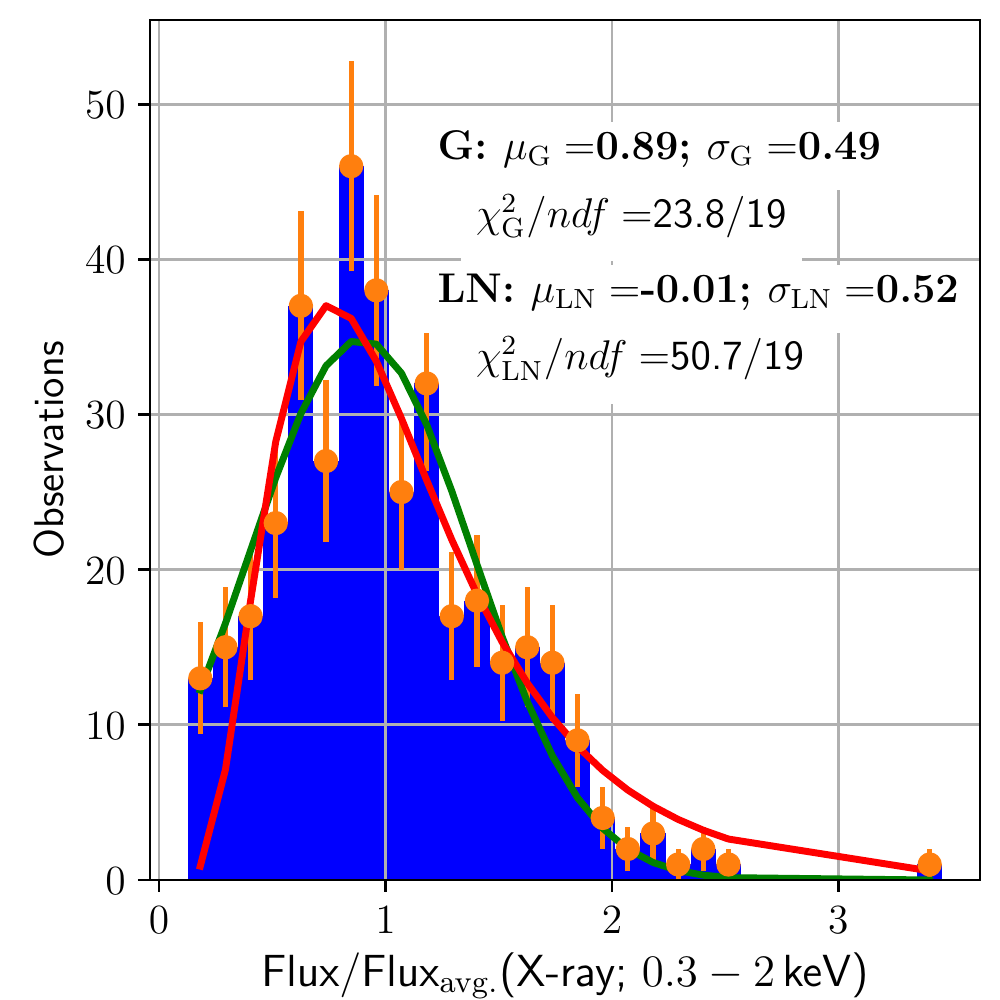}
\includegraphics[width=0.49\linewidth,height=6cm]{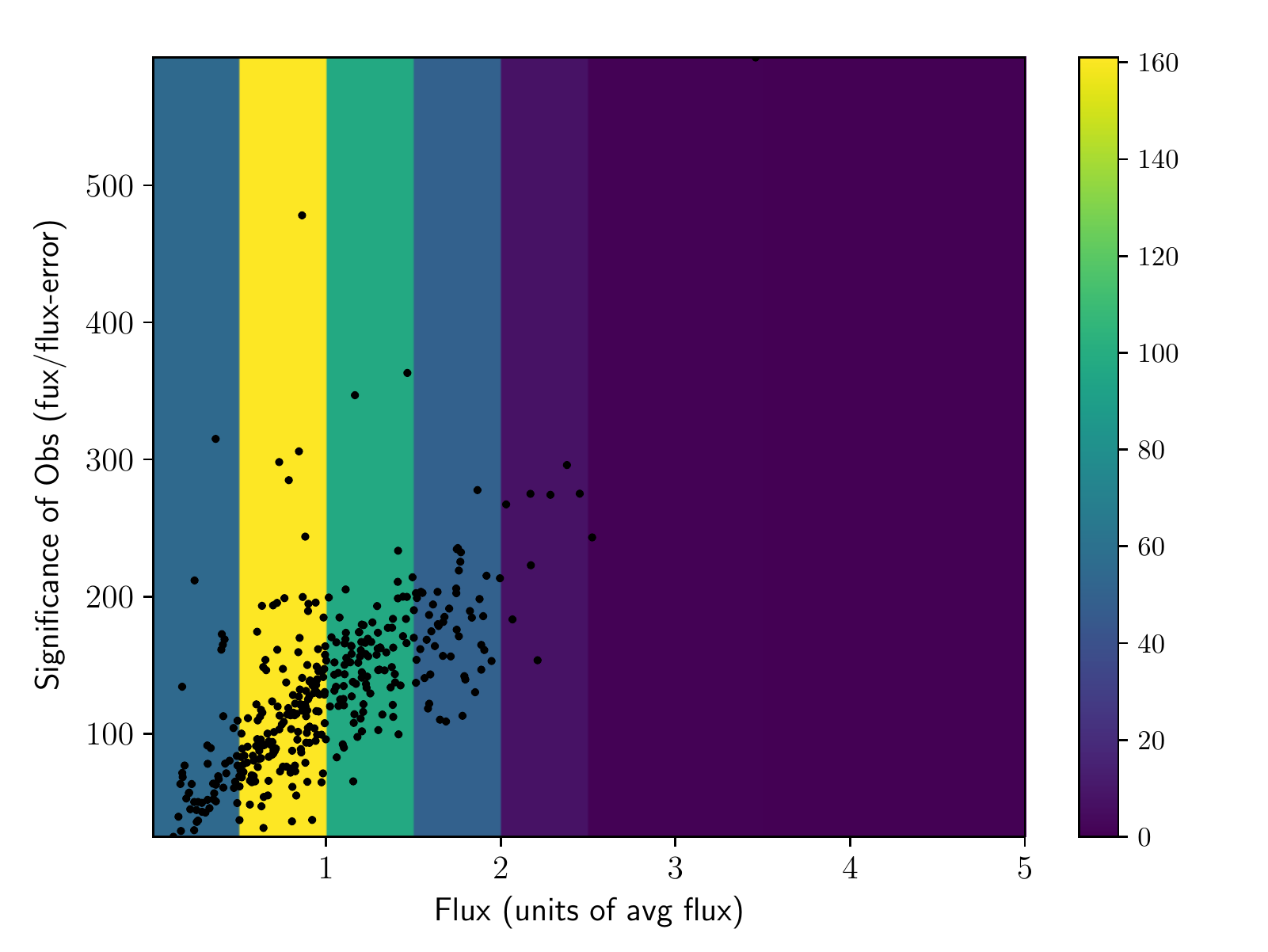}
\includegraphics[width=0.49\linewidth,height=6cm]{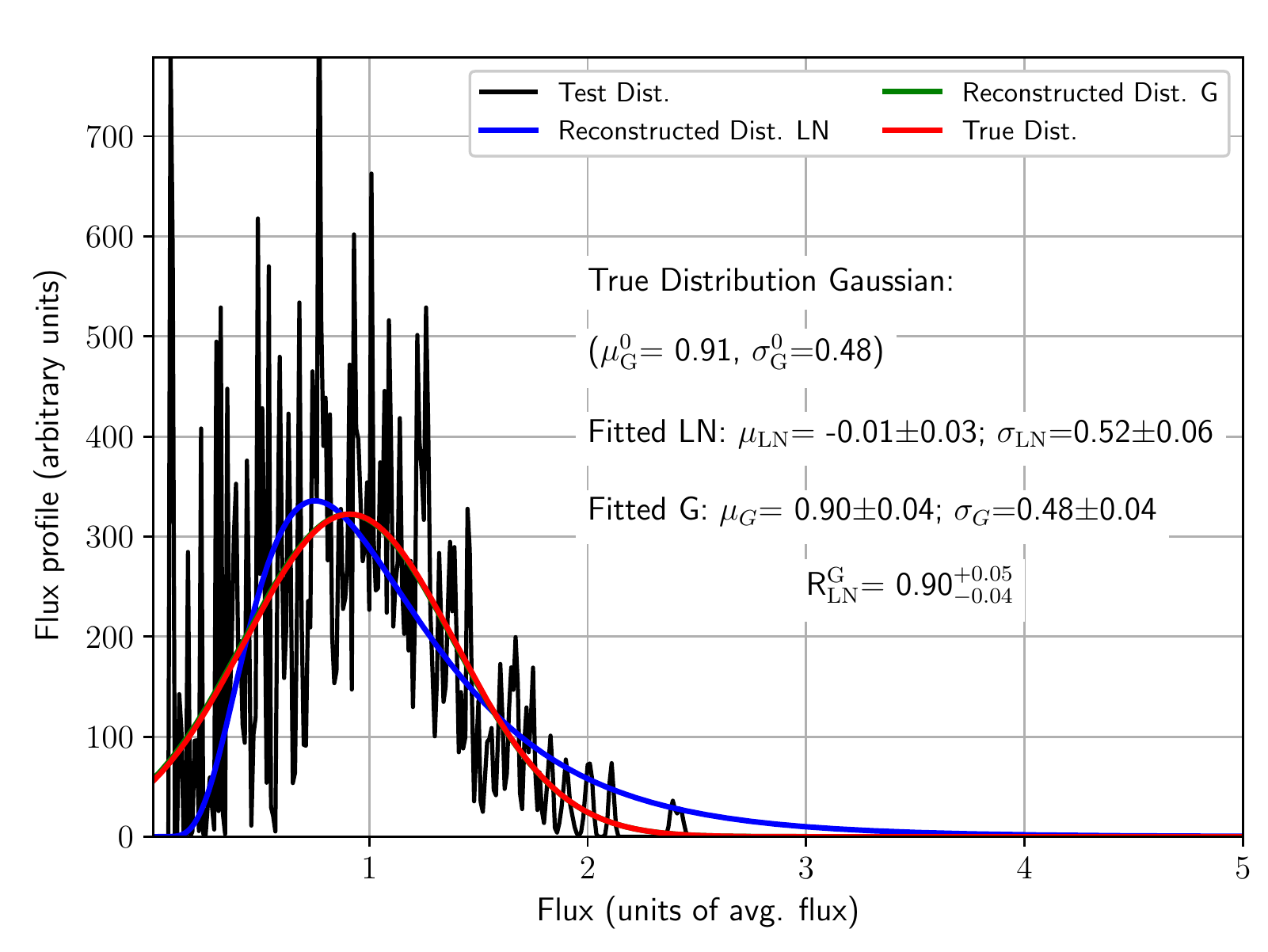}
\includegraphics[width=0.49\linewidth,height=6cm]{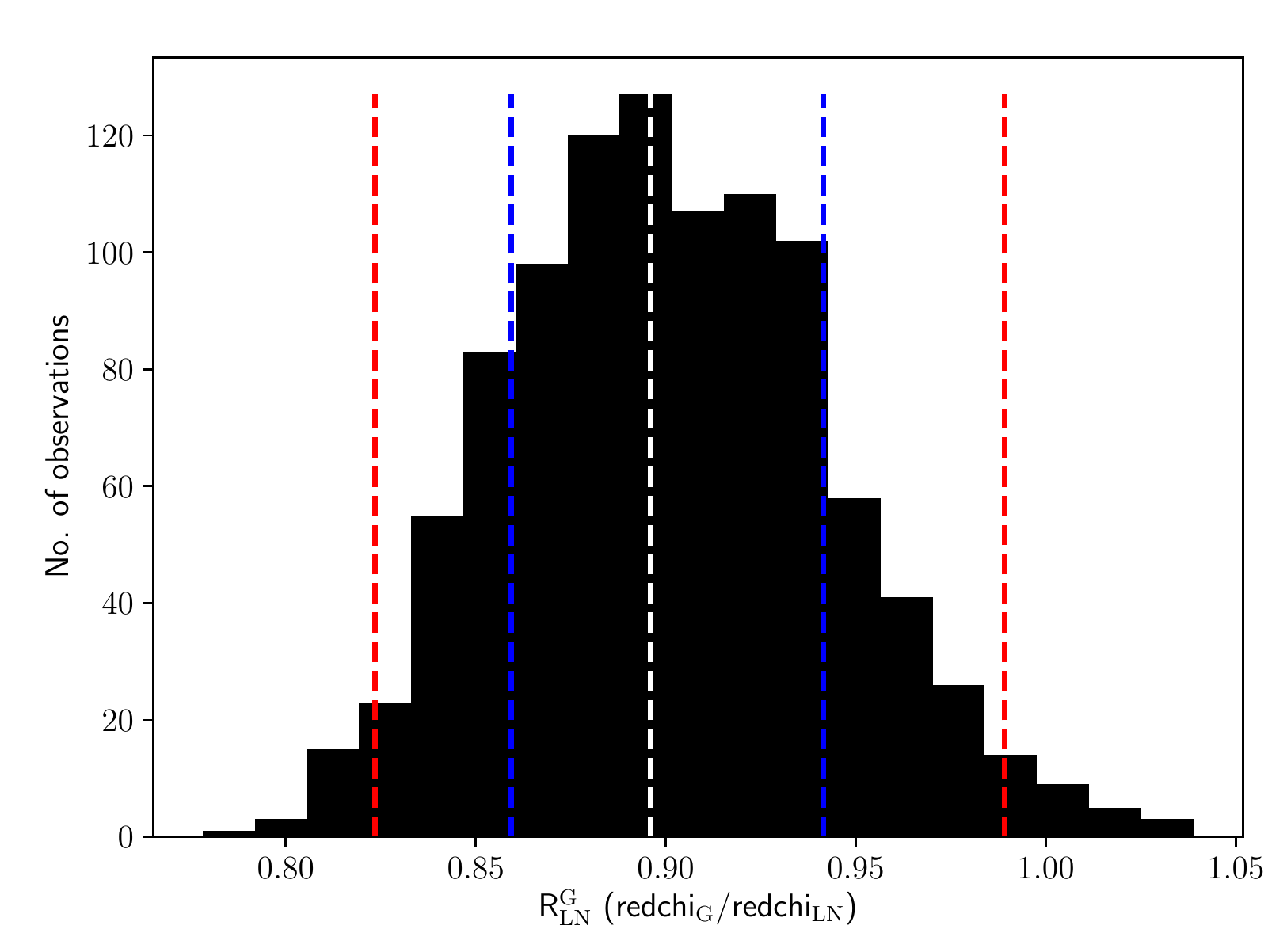}
\includegraphics[width=0.49\linewidth,height=6cm]{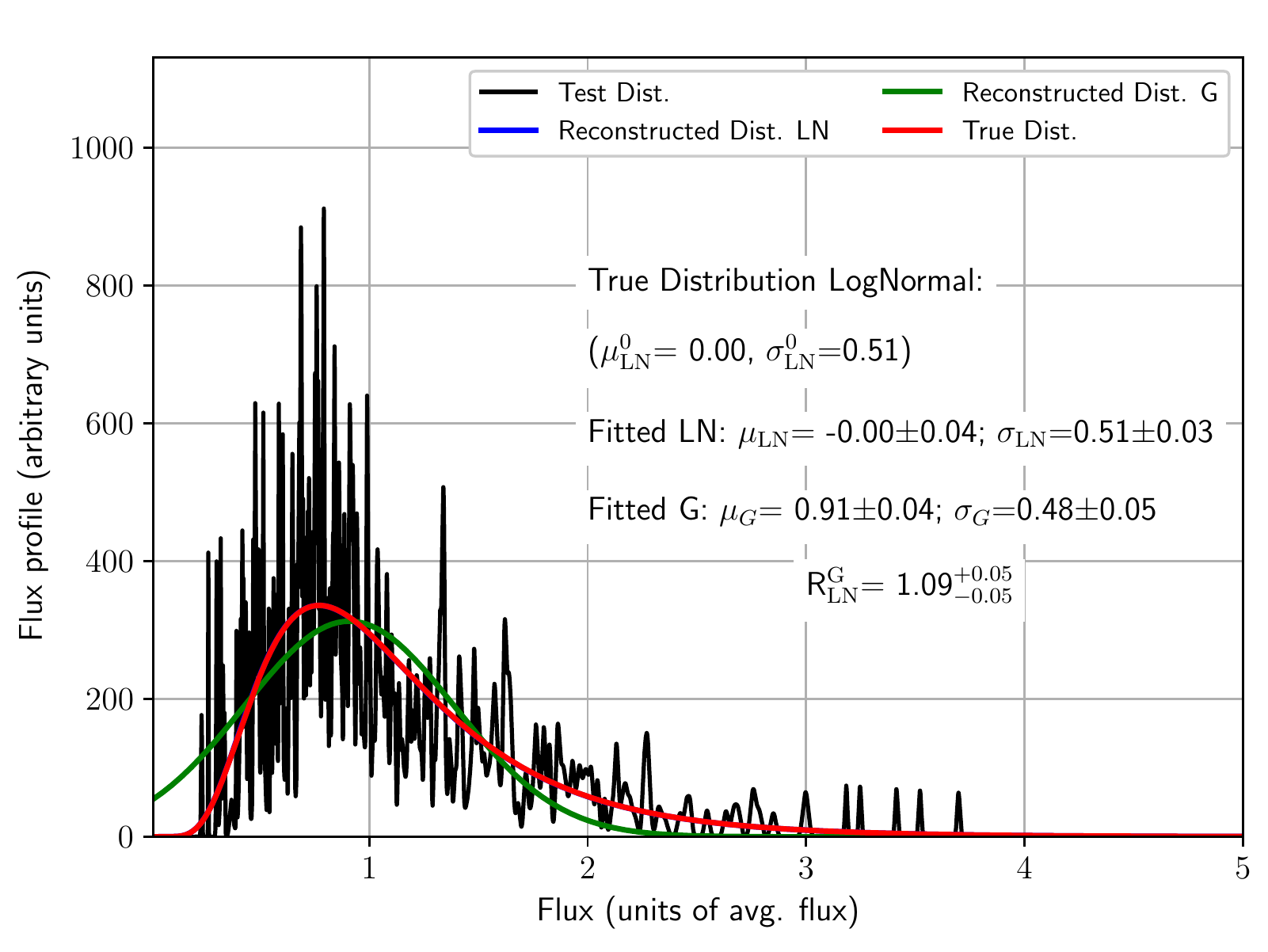}
\includegraphics[width=0.49\linewidth,height=6cm]{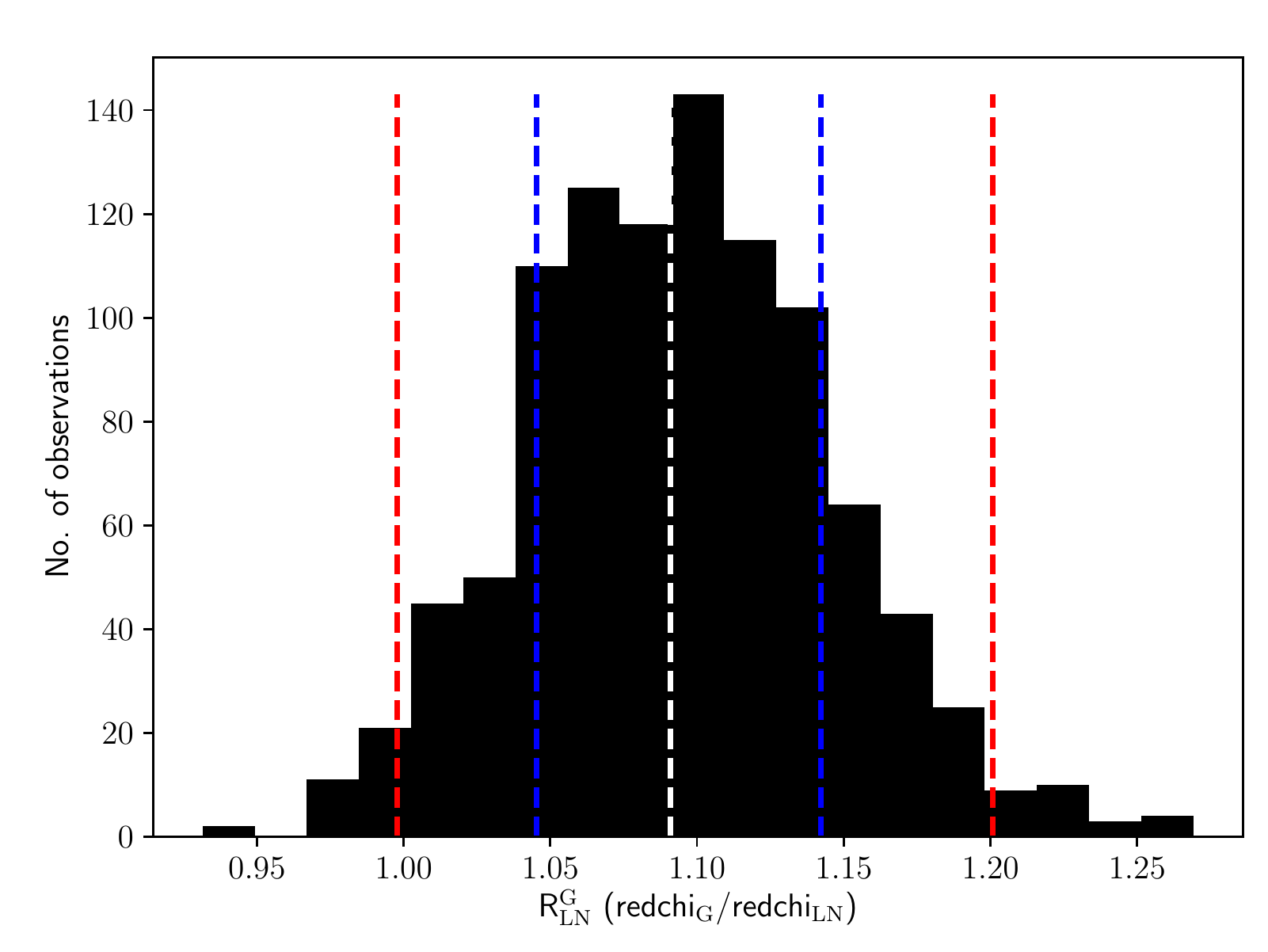}
\caption{Validation of flux profile-method using long-term X-ray ($0.3-2$\,keV) data}\label{fig:FPA22}
\end{figure*}
The shape of flux distribution of a source is a useful tool to study the nature of the underlying variability processes in the source. Almost all the studies done so far in this respect involve construction of 
\textcolor{black}{Chi-square fit to the} flux histograms \citep{2010A&A...524A..48T,2017ApJ...841..100A,2016A&A...591A..83S,galaxies7020057}. However, generating flux distributions from histograms has certain inaccuracies and biases  related to the selection of the bin-width and the flux measurement errors, which are not  considered when making a simple flux distribution. In order to address this issue, we have developed a new method, in which, we construct "flux profiles" instead of histograms.

\subsection{Flux profile from a light curve:}\label{sec:FPcon}

We create the flux profile by adding contributions from individual flux measurements. We assume that for individual observations, flux errors are normally distributed around the mean. 
At VHE $\gamma$-rays ($>$ 0.2\,TeV) during 2015--2016, the lowest number of excess events was found to be around 40, supporting this assumption.
Therefore, for each individual measurement we create a Gaussian profile \mbox{$G (x: \mu,\sigma)$}, where $\mu$ and $\sigma$ are the flux and flux error, respectively. The amplitude of the profile is normalised to \mbox{$1/(\sigma \sqrt{2\pi})$}, so that the area under each individual flux profile is unity. Therefore, a high uncertainty measurement will result in a smaller amplitude, but will contribute to a wider range of flux values. Finally, the overall flux profile for the whole observation period is obtained by adding contributions from individual flux profiles. 
\textcolor{black}{A few examples}
of such individual flux profiles are presented in Fig.~\ref{fig:FPA20}.
\par
In order to create the flux profile in the VHE band for the 2007--2016 period, we have selected only flux points for which the detection significance (flux-error/flux) is less than 0.5. The highest flux in this data set, (86.1$\pm$3.2)$\times$10$^{-11}$\,ph\,cm$^{-2}$\,s$^{-1}$, was observed on MJD 54555.9, while the lowest flux state of (3.2$\pm$0.6)$\times$10$^{-11}$\,ph\,cm$^{-2}$\,s$^{-1}$ was observed on MJD 57422. The corresponding flux profiles are presented in Fig.~\ref{fig:FPA20}. \textcolor{black}{Also,} the flux profile for MJD 54562 is also shown which has a rather large flux uncertainty (78.4 $\pm$8.1)$\times$10$^{-11}$\,ph\,cm$^{-2}$\,s$^{-1}$. 
In addition,
we construct the flux profiles using the individual fluxes and flux errors scaled with the average flux of the entire observation period reported in Fig.~\ref{fig:LongtermMWLC} (e.g. at VHE the fluxes and errors are scaled with  2.09$\times$10$^{-10}$\,ph\,cm$^{-2}$\,s$^{-1}$ which is the long-term average flux). From the overall flux profile, we \textcolor{black}{determine} the most probable state to be around 60 per cent of the average flux.

\subsection{Validation of the flux profile method using VHE $\gamma$-ray data:}\label{sec:method} 

In this section, we present the validation of the flux profile method by assuming the flux distribution of the source as i) Gaussian and ii) LogNormal. We explain the procedure for this exercise for the Gaussian case and for the VHE $\gamma$-ray data set, but the same procedure also applies to LogNormal case, as well as for all the energy bands. The steps are as follows:

Step 1: We create a histogram of the fluxes in the VHE band using the long-term (2007--2016) data set, as shown in Fig.~\ref{fig:FPA21} (top left panel), and fit it with a Gaussian \textcolor{black}{using Chi-square minimization}.

Step 2: We assume that the fluxes from our source are distributed according to the fitted distribution from Step 1. 
We simulate 226 flux values as present in the real VHE LC.

Step 3: We then use real measurements to create a 2-D histogram of the flux vs. SNR, with 10 bins in flux and 5 bins in SNR (top right panel of Fig.~\ref{fig:FPA21}). The SNR bins are not the same for each flux bin, rather, in each flux bin, we take the range between minimum and maximum values of the SNR and divide it in 5 bins. Finally, we take the number of points in each SNR bin and divide it with the total number of points in the whole flux bin to estimate the distribution of SNR in each flux bin.

Step 4: Using fractions of SNR in each flux bin (obtained in Step 3), we generate flux errors for each of the 226 fluxes generated in Step 2.
Some high flux bins in the real data histogram are empty (see top left panel of Fig.~\ref{fig:FPA21}). In such cases, we take the SNR to be the average SNR of the first lower flux bin.

Step 5: The 226 generated flux and flux-error
pairs are now used to create a simulated flux profile.

Step 6: Steps 2--5 are repeated 1000 times in order to create 1000 generated flux profiles. 

Step 7: Every generated flux profile is fitted with both the Gaussian and LogNormal functions. The fit parameters are $\mu^{i}_G$ and $\sigma^{i}_G$ ($\mu^{i}_\mathrm{LN}$ and $\sigma^{i}_\mathrm{LN}$) for fitting with Gaussian (LogNormal), where $i$ is the flux profile index. In addition, a parameter \texttt{redchi} is calculated (see   \textcolor{black}{Section} \ref{sec:typicalstate} for details) for each flux profile and both the functions, as well as a ratio of \texttt{redchi} parameters  R$^{G}_\mathrm{LN}$ (\texttt{redchi}(G)/ \texttt{redchi}(LN)).

Step 8: Using the fit parameters for individual flux profiles, we calculate the average values of the fit parameters ($\mu_G$ and $\sigma_G$) and their standard deviations ($\Delta\mu_G$ and $\Delta\sigma_G$). A Gaussian function with $\mu_G$ and $\sigma_G$ as mean and standard deviation is plotted in Fig.~\ref{fig:FPA21} as the reconstructed Gaussian distribution (green lines in middle left and bottom left panels). The errors on mean and standard deviation of the Gaussian are quoted as $\Delta\mu_G$ and $\Delta\sigma_G$ in the same panels. The same procedure is followed for LogNormal distribution (shown as blue lines in Fig.~\ref{fig:FPA21}).

Step 9: We make a distribution of the R$^\mathrm{G}_\mathrm{LN}$ which quantifies the goodness of fit.

First, we perform the described analysis by fitting the real data fluxes with a Gaussian function (Step 1). The resulting fit parameters are 
($\mu_G$= 0.67, $\sigma_G$= 0.65),
and the corresponding function is shown in the top left plot of Fig.~\ref{fig:FPA21} with the red line. These parameters are used to generate 
\textcolor{black}{213}
flux values (Step 2),
and later to generate 1000 flux profiles (Step 6). Fitting each flux profile with Gaussian and LogNormal (Step 7) and averaging over all flux profiles (Step 8) results in the average 
\textcolor{black}{fit parameters ($\mu_G$= 0.68 $\pm$ 0.09, $\sigma_G$= $0.65 \pm 0.08$) and ($\mu_\mathrm{LN}$= $-0.08 \pm 0.06$, $\sigma_\mathrm{LN}$= $0.81 \pm 0.14$),}
for Gaussian and LogNormal distributions, respectively.
The results are shown in the middle left plot of Fig.~\ref{fig:FPA21}. The red line indicates the fit of the real data set with the Gaussian distribution, while the green and blue lines indicate the Gaussian and LogNormal functions, respectively. The average fit parameters for the Gaussian are consistent with the fit parameters of the initial real data distribution (the red and green lines overlapping). In addition, the ratio of the parameters \texttt{redchi} for Gaussian to the LogNormal for this case is  
\textcolor{black}{
R$^\mathrm{G}_\mathrm{LN}$ = $0.40^{+0.14}_{-0.07}$},
indicating that the Gaussian distribution is the preferred one, and thus 
proving that we correctly recovered the initial distribution. 
\textcolor{black}{The chance probability (p), based on toy Monte Carlo, indicates the probability of wrongly reconstructing a LogNormal (Gaussian) distribution as a Gaussian (LogNormal). We calculate this by the distribution of the parameter R$^\mathrm{G}_\mathrm{LN}$.
For an initial true LogNormal (Gaussian) distribution, we calculate the survival function ({\em sf} \footnote{\url{https://docs.scipy.org/doc/scipy/reference/generated/scipy.stats.skewnorm.html}}) of R$^\mathrm{G}_\mathrm{LN}$ below (above) 1 assuming the distribution to be a {\it skew-normal$^{11}$}. This survival fraction indicates the chance probability of obtaining a Gaussian (LogNormal) flux distribution from a true LogNormal (Gaussian) distribution.
The chance probability for the flux distribution in VHE $\gamma$-rays is 1.1$\times$10$^{-4}$.}
The distribution of the R$^\mathrm{G}_\mathrm{LN}$ for individual simulated flux profiles is shown in the middle right plot.
Next, we repeat the analysis, this time fitting the real data fluxes with the LogNormal function (Step 1), resulting in parameters
($\mu_\mathrm{LN}$= -0.12, $\sigma_\mathrm{LN}$= 0.66). The corresponding function is shown in the top left plot of Fig.~\ref{fig:FPA21} with the green line. The final results of the analysis are shown in the bottom plots of Fig.~\ref{fig:FPA21}. In the left plot, the red line indicates the fit of the real data set with the LogNormal distribution, while the green and blue lines again indicate the Gaussian and LogNormal functions, respectively. This time, the simulated flux profiles were generated using parameters of the initial LogNormal distribution. 
\textcolor{black}{The average fit parameters in this case are ($\mu_G$= $0.72 \pm 0.08$, $\sigma_G$= $0.58 \pm 0.11$) and ($\mu_\mathrm{LN}$= $-0.10 \pm 0.06$, $\sigma_\mathrm{LN}$= $0.66 \pm 0.05$), while R$^\mathrm{G}_\mathrm{LN}$ = $1.84^{+0.71}_{-0.48}$ (chance probability of having a Gaussian distribution is 4.4$\times$10$^{-2}$).}
We can see that the average fit parameters for the LogNormal are consistent with the fit parameters of the initial real data distribution  (the red and blue lines overlapping), and that the LogNormal distribution is the preferred one. Therefore, we again correctly recovered the initial distribution. The distribution of the R$^\mathrm{G}_\mathrm{LN}$ for individual simulated flux profiles is shown in the bottom right plot.

We inspected our method on flux profiles in HE and X-ray bands. HE was chosen as an example of 
a band with larger relative flux uncertainties and lower variability, while the X-ray band is an example of the opposite (smaller relative flux uncertainties and higher variability). 
The procedure used in HE (X-ray) is exactly the same as that of the VHE band, with the exception of using 955 (374) flux points in Step 2 and 15 (10) flux bins in the 2-D histogram in Step 3.
The results are shown in Fig.~\ref{fig:FPA23} and \ref{fig:FPA22} for the HE and X-ray bands, respectively. 
In the HE band, the parameters of the initial flux distributions are not recovered. This is mainly because of the relatively large flux uncertainties. 
\textcolor{black}{However, the chance probability of having LogNormal (Gaussian) from a true Gaussian (LogNormal) is 0.0 (8.1$\times$10$^{-2}$)
which indicates that we do correctly reconstruct and distinguish between Gaussian and LogNormal shapes of the initial distribution. 
In the X-ray band, the chance probability of having LogNormal (Gaussian) from a true Gaussian (LogNormal) is 1.7$\times$10$^{-2}$ (3.3$\times$10$^{-2}$).}
Therefore, in the VHE and X-ray bands, we were able to recover the initial flux distributions (including the parameters), thus validating our method for measurements with higher sensitivity.

We recognize the following types of biases that can affect our results:\\
{\bf a) A cut on the relative error:} In this study, we only use flux measurements with a SNR $>$ 2.0. This will bias towards slightly higher values of flux for some of the distributions (those with the largest errors), e.g., FACT, {\it Fermi} and BAT.
It affects only the rising part of the flux distribution. For FACT, there will be some distortion in the distribution because we remove 25\% of the data. 
In any case, it is the high fluxes what dominates the distinction between G and LN, and those remained unaffected. For the flux distribution of data from the \textit{Fermi}-LAT, the impact is negligible (only 3\% of data removed).\\
{\bf b) The bias for including the observations during alert (ToO) for the high flux states:} The MAGIC and \textit{Swift}-XRT observations triggered by the target of opportunity (ToO) programs during the high  \textcolor{black}{flux} of the source may bias the flux distribution. Ideally, the unbiased observations should only be considered. The data set under consideration includes the following campaigns, 2008 \citep{2012A&A...542A.100A} and 2010 \citep{2015A&A...578A..22A, 2020ApJ...890...97A} where the source showed high flux states. While the 2008 flaring episode had many ToOs involved, the 2010 March flaring activity observed consisted on observations that had been coordinated with {\it Swift} and {\it RXTE} several weeks in advanced. We have performed a study by removing all the high flux states observed during 2008 \citep{2012A&A...542A.100A} to check for LogNormality. We have found that even with this extreme 
condition, a LogNormal is preferred over a Gaussian flux profile.
This proves that the LogNormal distribution of the flux at VHE is a feature of the source and does not depend on the ToOs. We also note that this bias is negligible for MAGIC and \textit{Swift}-XRT and has no effect on the observations with the FACT, \textit{Fermi}-LAT and \textit{Swift}-BAT.

\clearpage

\section{Year-wise variation of flux profiles}
\label{sec:appendixB}

\begin{figure}%
    \centering
   \includegraphics[width=\linewidth,height=6.5cm]{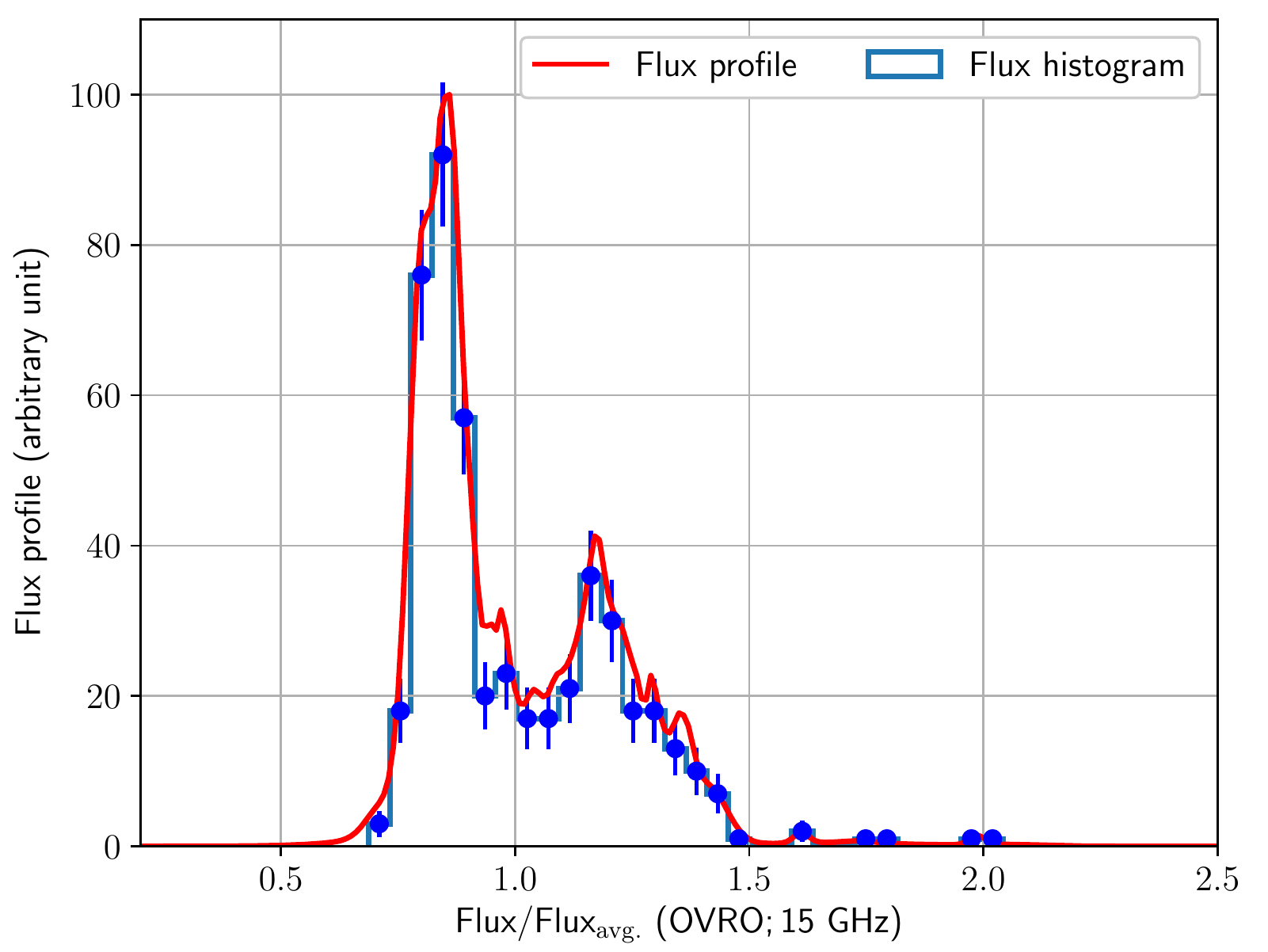}
    \caption{The flux profile and flux histogram of Mrk\,421 in radio (15\,GHz) band. See \textcolor{black}{text} for details.}
    \label{fig:OVRO_fluxdistribution}%
\end{figure}

\begin{figure}%
    \centering
   \includegraphics[width=\linewidth,height=6.5cm]{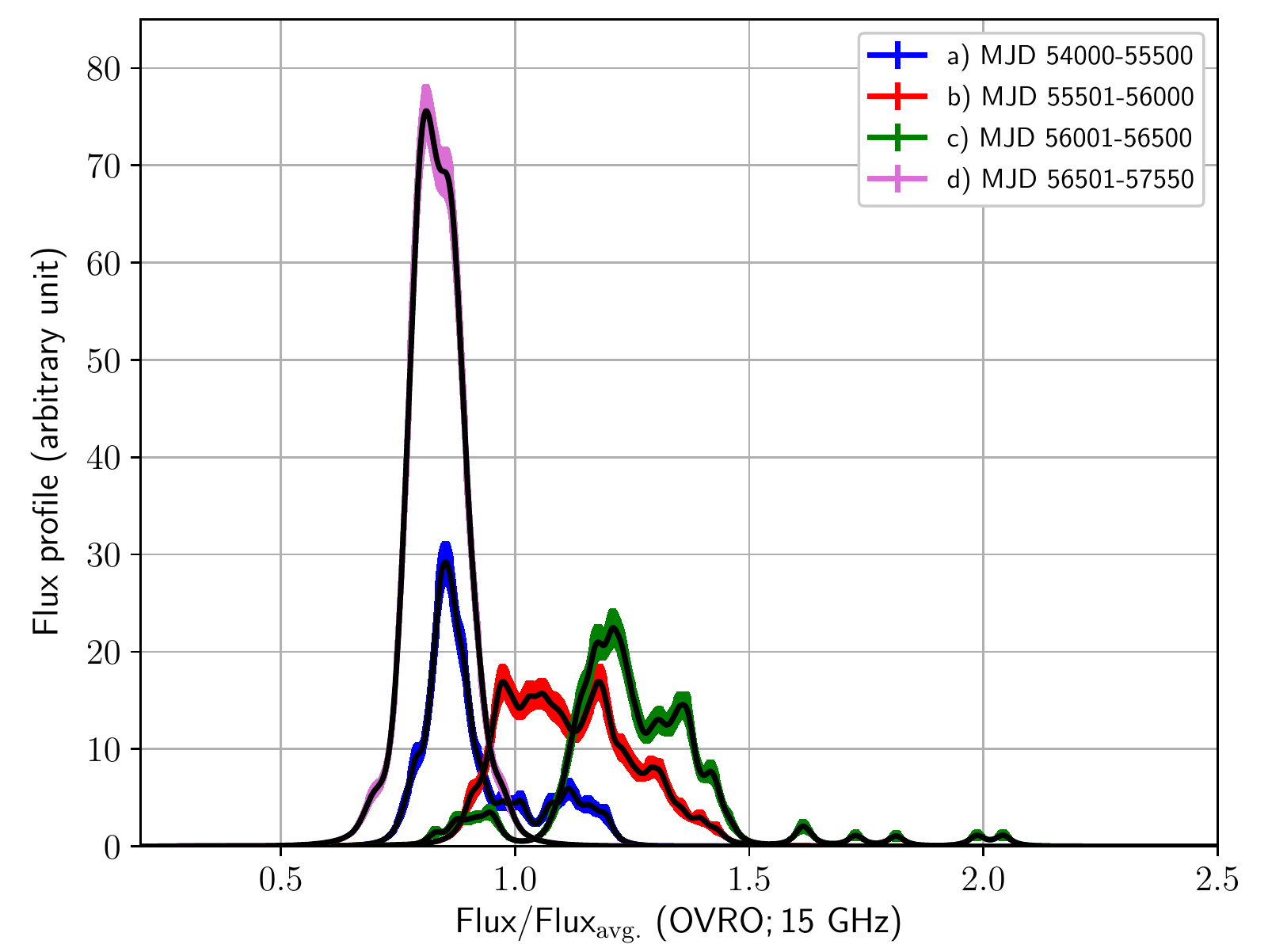}
    \caption{Year-wise variation of flux profiles of Mrk\,421 in radio (15\,GHz) band. The bands for different colors indicate 
    1\,$\sigma$ confidence intervals for different years. We have divided the multiyear radio LC into four different periods of observations. See   \textcolor{black}{Appendix}  \ref{sec:appendixB} for details.}%
    \label{fig:appenixB}%
\end{figure}

The flux distribution of the 15\,GHz radio band is shown in Fig.~\ref{fig:OVRO_fluxdistribution}, where the flux histogram and the flux profile are presented. This suggests that the flux distribution for this band is a bimodal distribution, hence, the Gaussian and LogNormal functions are not suitable. The  year-wise variation of the flux profiles for the 15 \,GHz radio band, observed by OVRO, has been reported in Fig.~\ref{fig:appenixB}.
We have divided the multiyear radio LC into four different periods: \\
a) 2006 September 22 to 2010 October 31 (MJD 54000--55500),\\
b) 2010 November 01 to 2012 March 14 (MJD 55501--56000),\\
c) 2012 March 15 to  2013 July 27 (MJD 56001--56500), and \\
d) 2013 July 28 to 2016 June 11 (MJD 56501--57550). \\
For each of the periods stated above, we shuffled the uncertainties on the flux and added to the flux in order to construct a simulated flux profile. We repeated this exercise for
1000 times for a single period in order to estimate the standard deviation on the flux profiles. The bands in Fig. \ref{fig:appenixB} represent the standard deviation on the flux profile (68\% confidence limit) estimated from the simulations mentioned above.
This study indicates 
that the most probable states of the source in different years are not unique. The flux profile also changes according to the flux states in different years. For example, the flux profile for the period (c) shows an isolated peak at higher flux. 
This is 
due to the huge radio flaring event in 2012.  The variation in flux profiles in Fig. \ref{fig:appenixB} indicates a shift form the low-flux state in period (a) to a high flux state (c) via an intermediate state (b). During period (c) the low/ typical state can also be identified.

\clearpage
\section{Characterization of the MWL flux distributions using a (binned) Chi-square fit and a (unbinned) log-likelihood fit}\label{sec:histandML}

This section reports the characterization of the
flux distributions using a binned Chi-square fit and an unbinned
log-likelihood fit. They are conventional ways of quantifying the shape of a distribution, and complement the results obtained with the flux profile method reported in Section  \ref{sec:typicalstate} and  Appendix \ref{sec:A1}. In both exercises, we use fluxes and their errors scaled by the average flux for each of the energy bands, and present them as F and $\Delta$F.

\begin{figure*}
\centering
\includegraphics[width=0.33\linewidth,height=5cm]{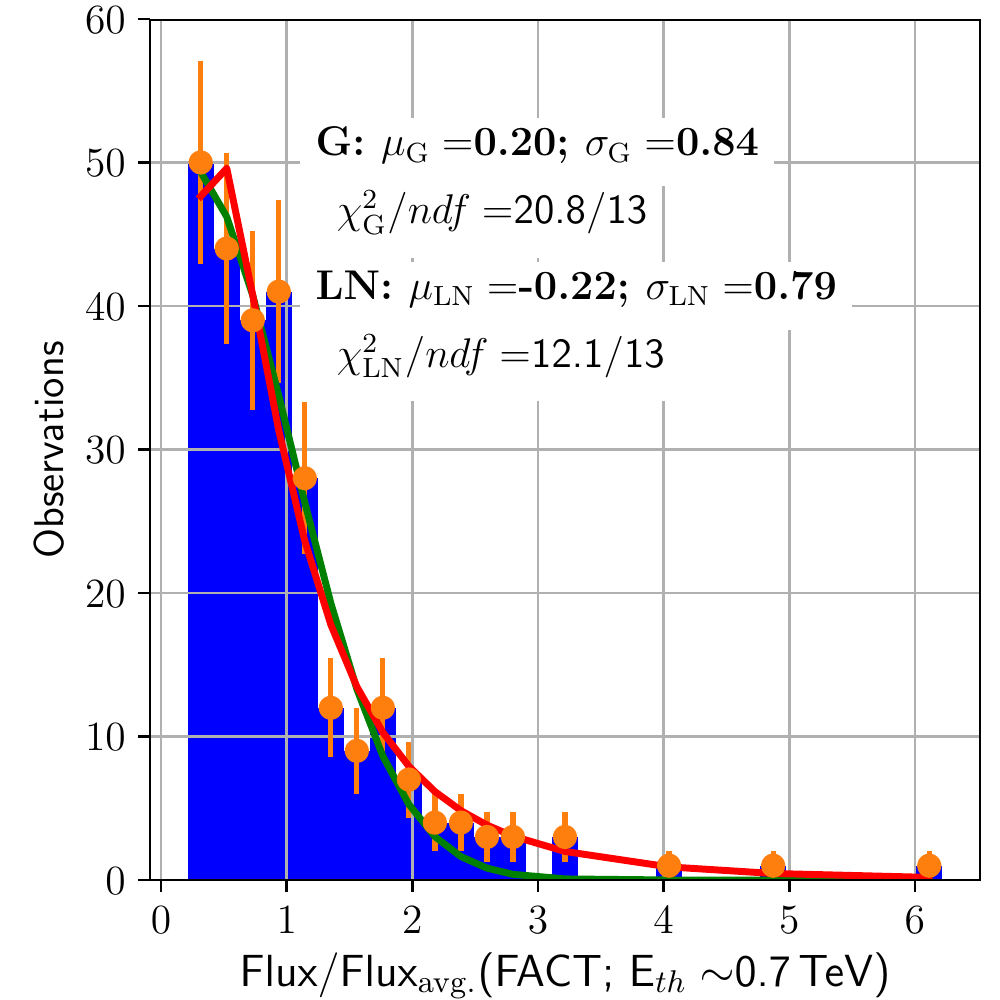}
\includegraphics[width=0.33\linewidth,height=5cm]{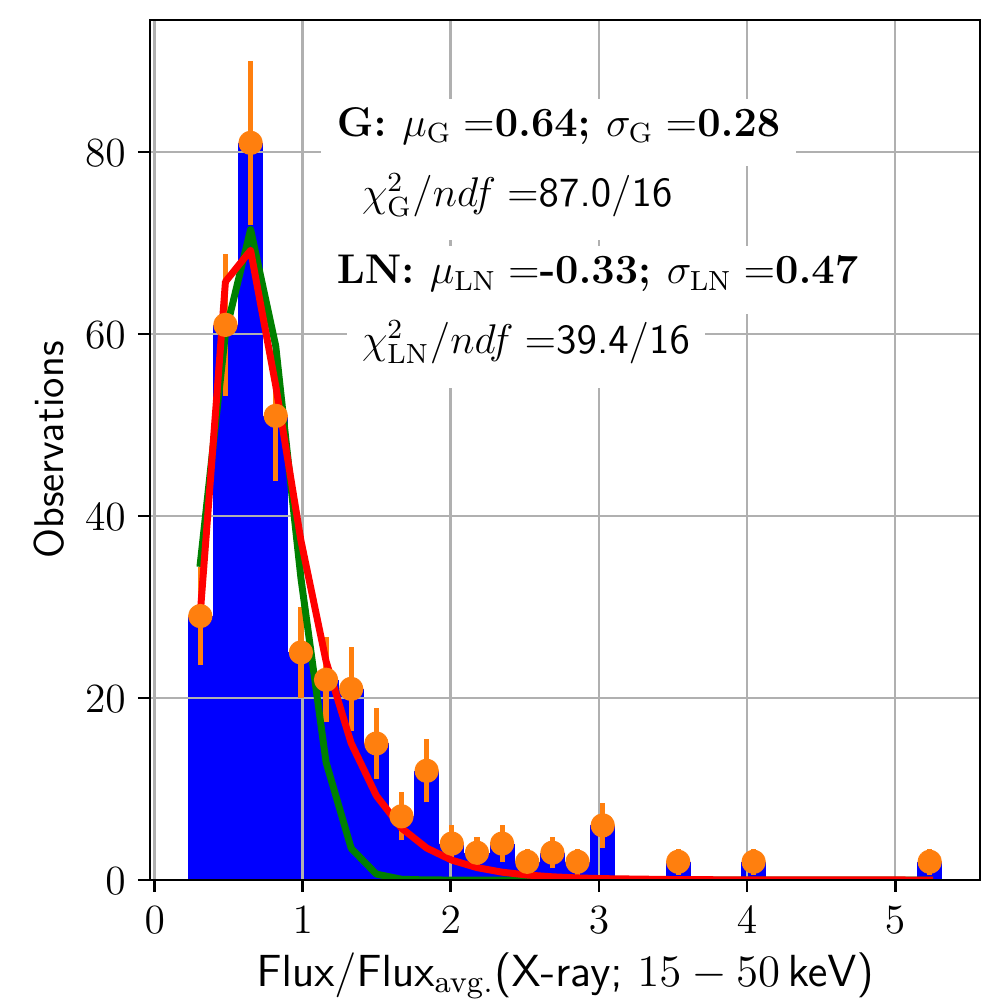}
\includegraphics[width=0.33\linewidth,height=5.0cm]{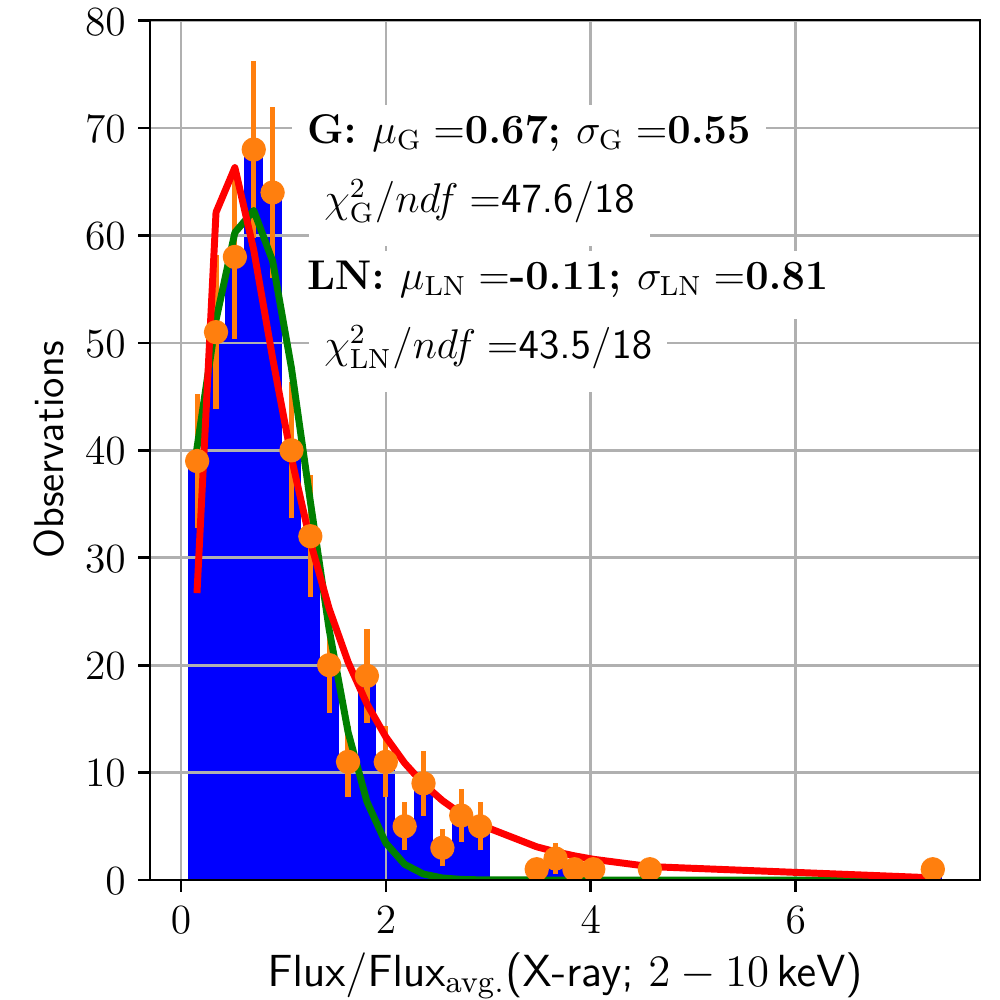}
\includegraphics[width=0.33\linewidth,height=5.0cm]{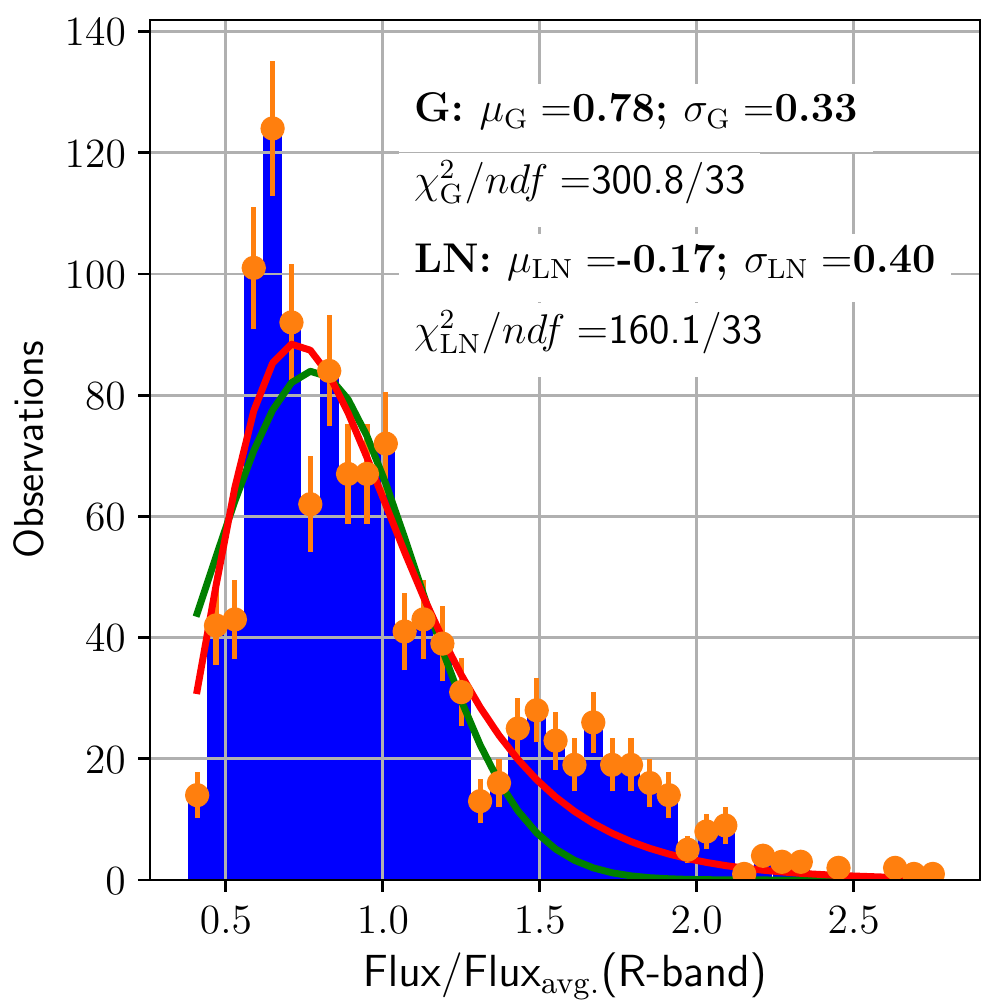}
\includegraphics[width=0.33\linewidth,height=5.0cm]{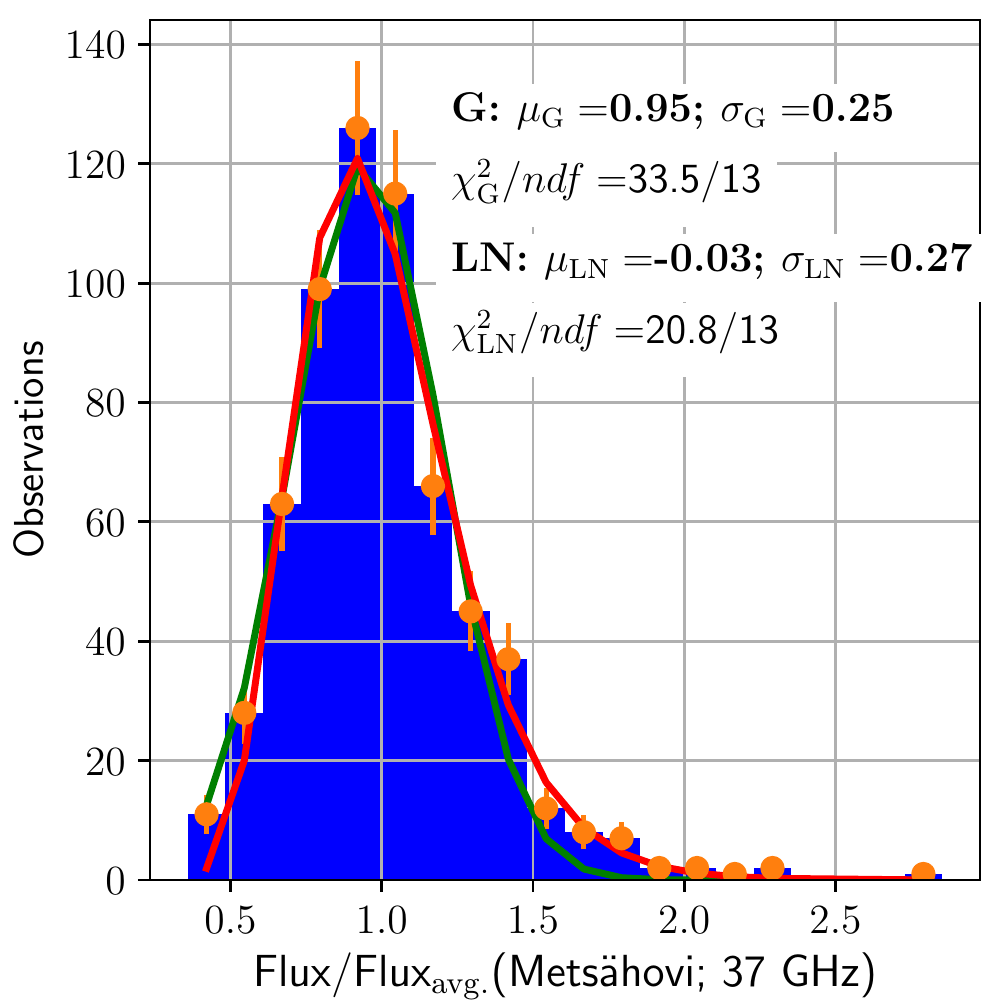}
\caption{Characterization of the MWL flux distributions with a Chi-square fit. The X-axis shows the scaled flux and the Y-axis presents the number of observations. The green and red lines represent the best fit with the Gaussian and LogNormal functions for flux histograms (presented in blue). See text in Appendix  \ref{sec:histandML} for details.}
\label{fig:Fhistogram}
\end{figure*}

In order to perform the Chi-square fit, we first
bin the scaled flux F. For each of the energy bands, the number of
histogram bins employed permits to show the overall shape of the
distribution, while keeping sufficient statistics (more than 10
entries) in most of the bins. Afterwards, we
performed a regular fit with a Gaussian and LogNormal functions, starting from the minimum flux F$_{min}$, and obtaining the function parameters mean ($\mu$) and standard deviation
($\sigma$) for which the Chi-square is minimum. It must be
noted that the outcome of the Chi-square fit can depend on the
histogram binning, and does not consider the flux uncertainties.
Figure~\ref{fig:Fhistogram} shows the results of the Chi-square fit for all the bands, except for the VHE $\gamma$-ray with MAGIC, HE $\gamma$-ray with {\it Fermi}-LAT and X-ray ($0.3-2$\,keV) with {\it Swift}-XRT, which are presented in Fig.~\ref{fig:FPA21}, \ref{fig:FPA22}, and \ref{fig:FPA23}, respectively.

\begin{figure*}
\centering
\includegraphics[width=0.33\linewidth,height=5.0cm]{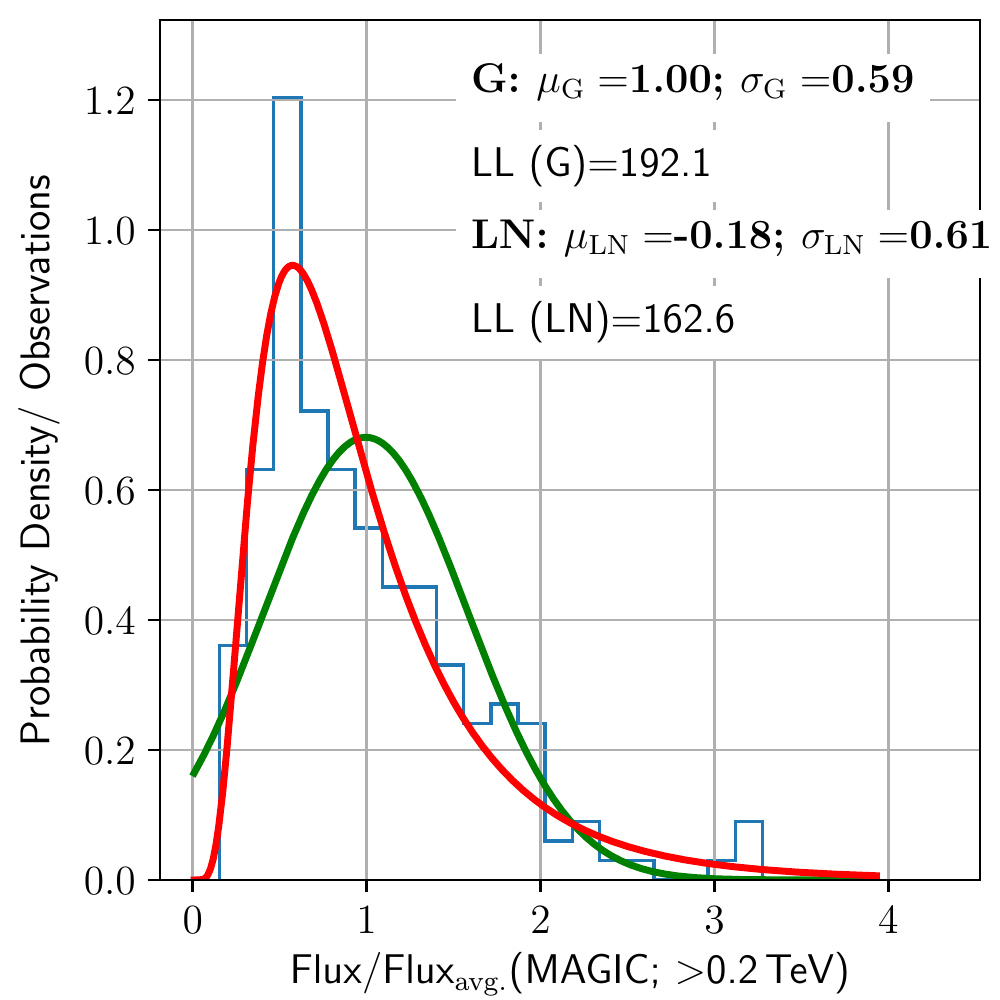}
\includegraphics[width=0.33\linewidth,height=5.0cm]{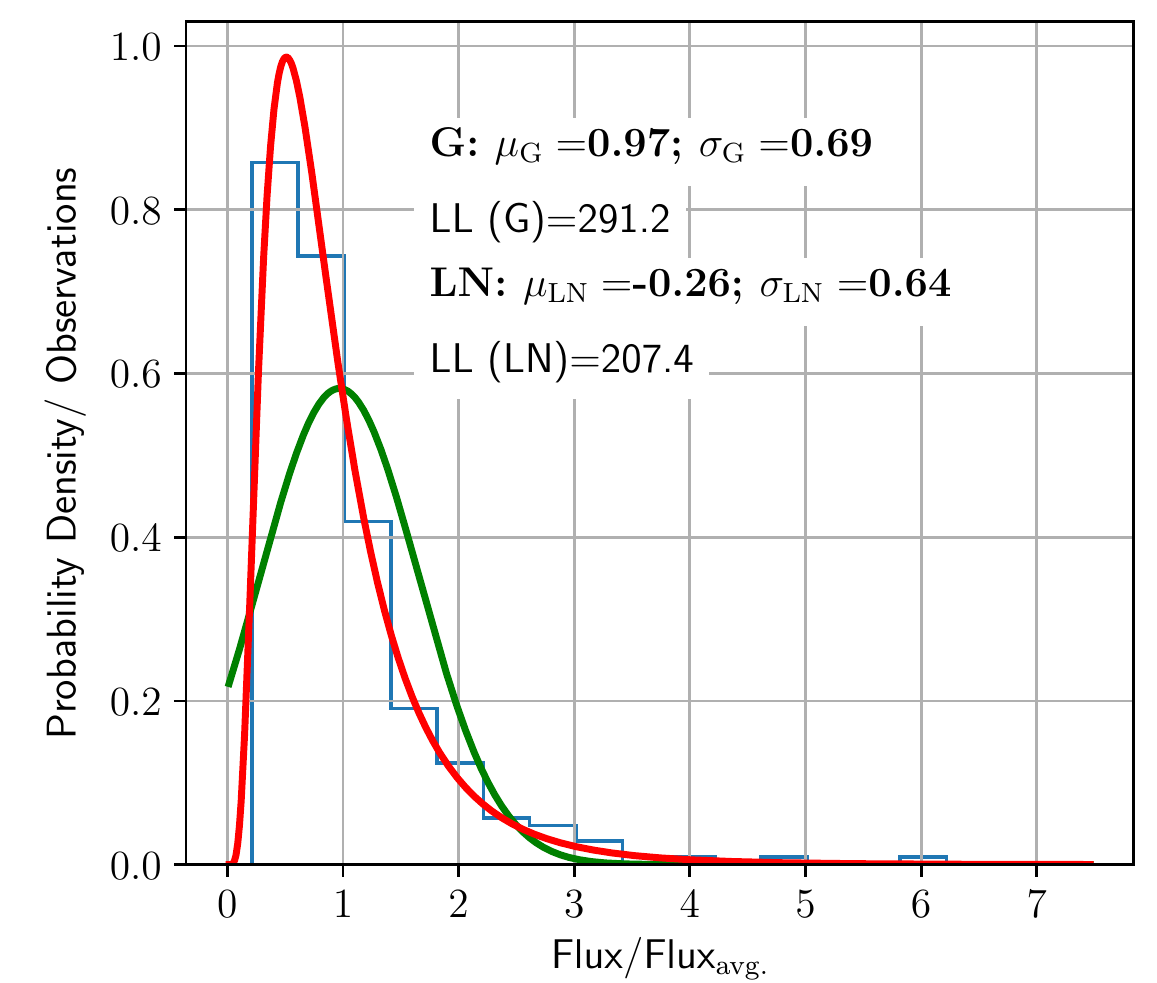}
\includegraphics[width=0.33\linewidth,height=5.0cm]{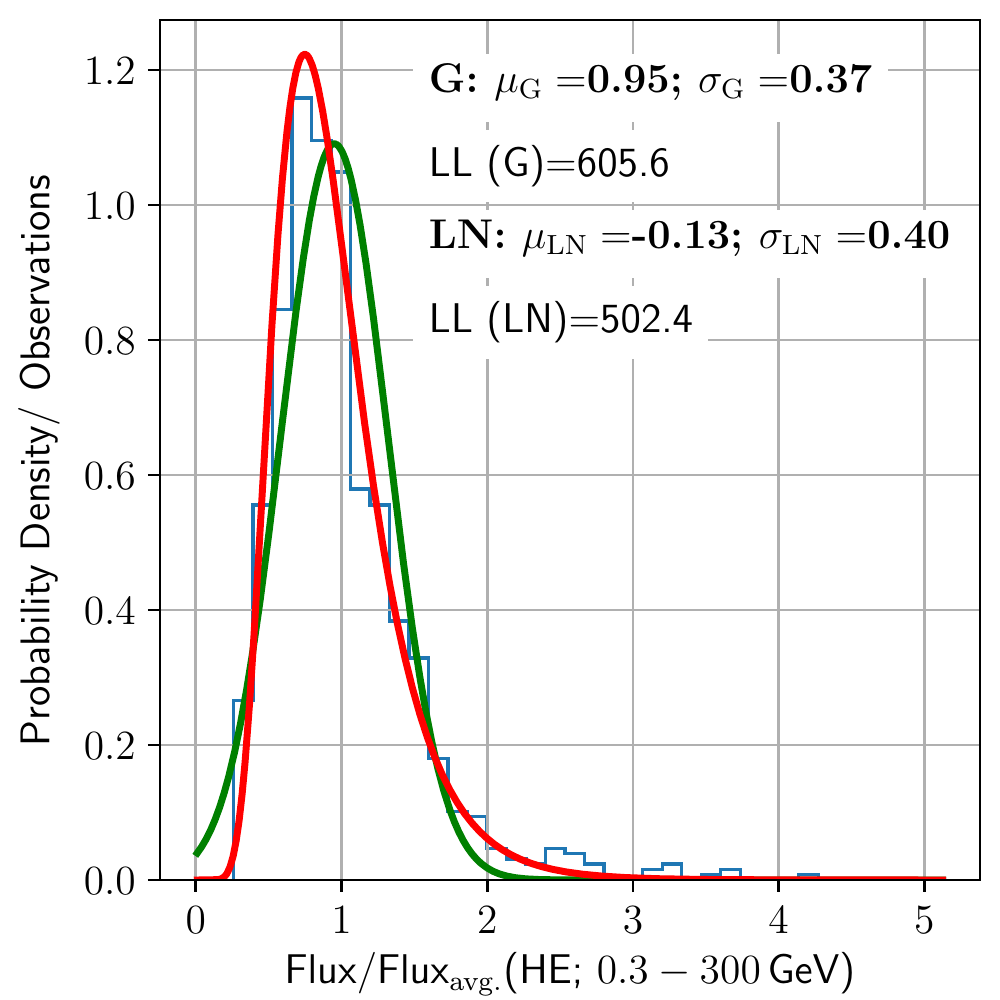}
\includegraphics[width=0.33\linewidth,height=5.0cm]{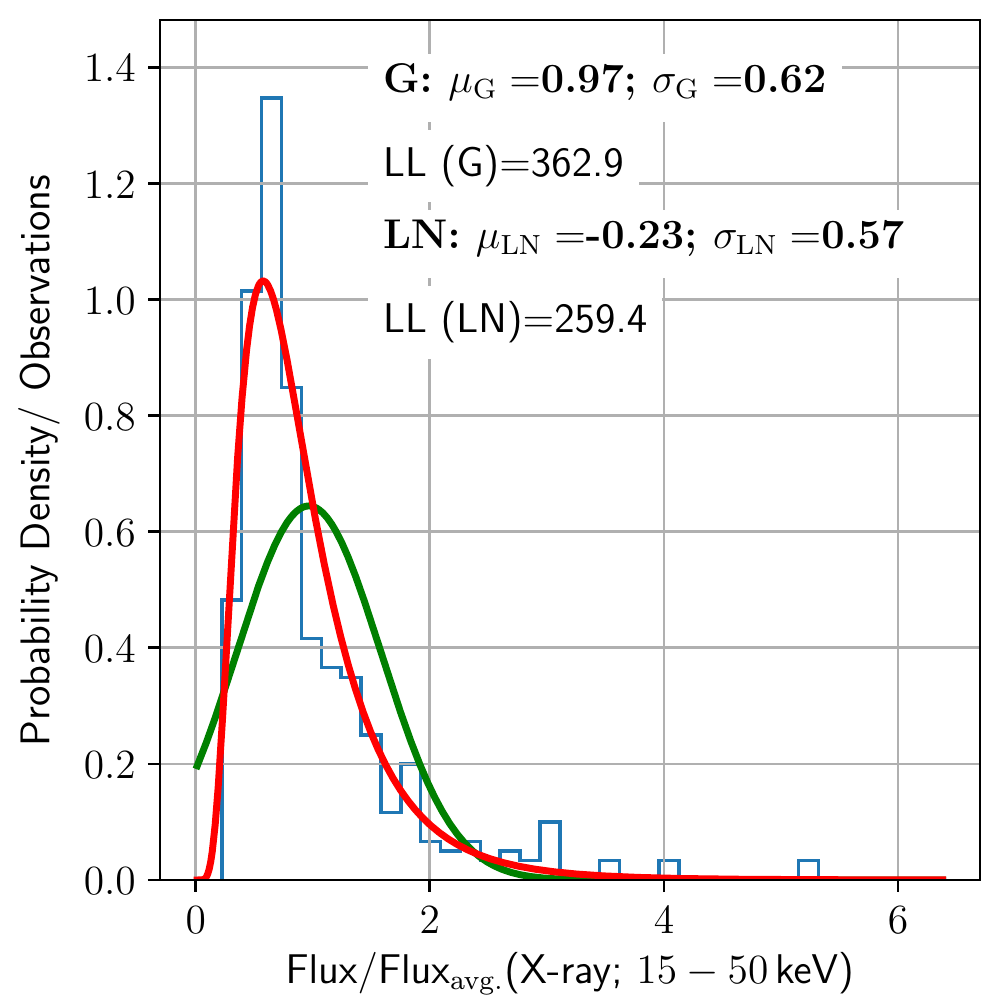}
\includegraphics[width=0.33\linewidth,height=5.0cm]{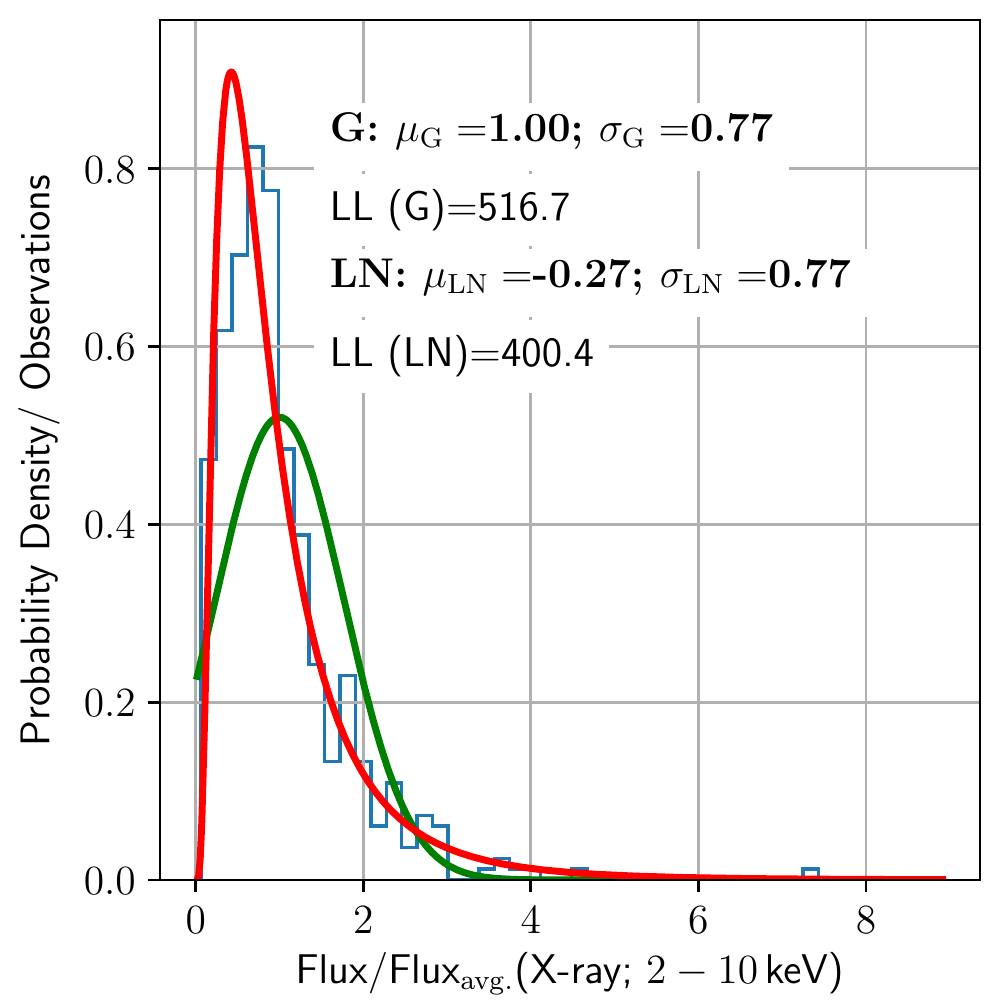}
\includegraphics[width=0.33\linewidth,height=5.0cm]{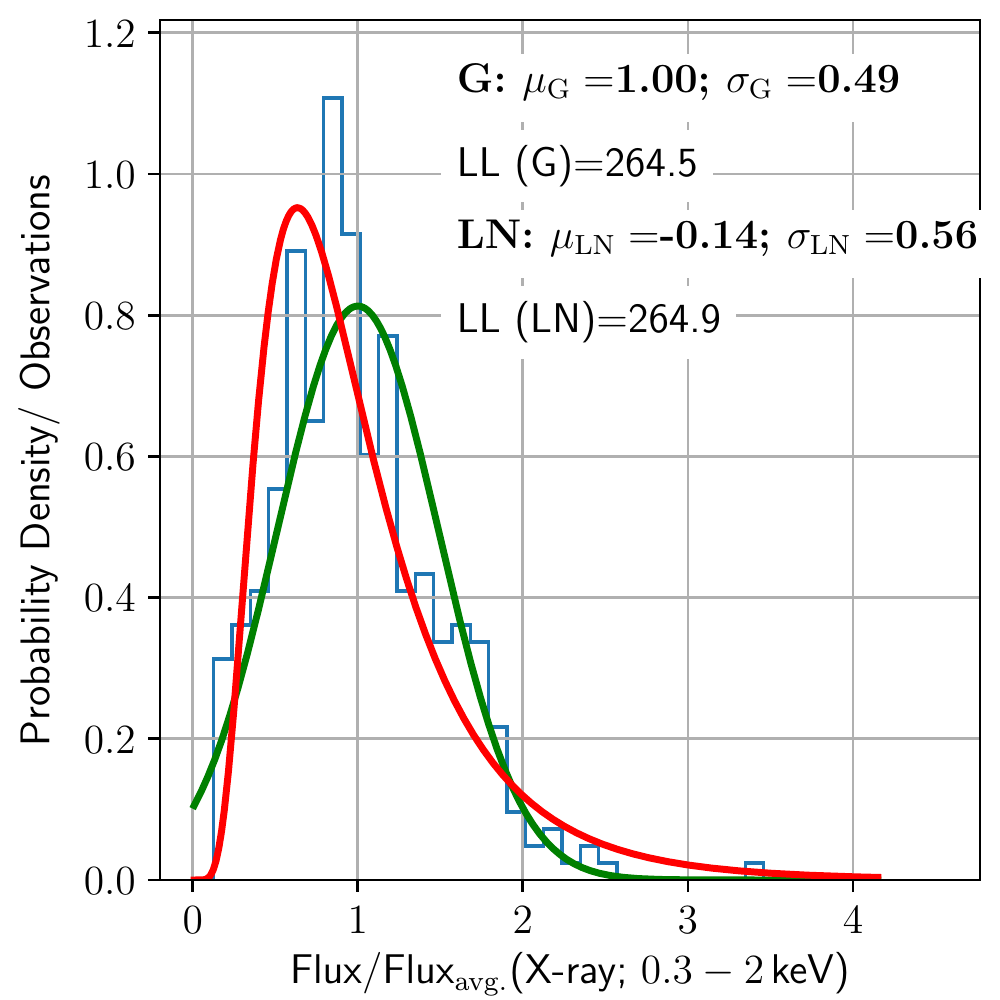}
\includegraphics[width=0.33\linewidth,height=5.0cm]{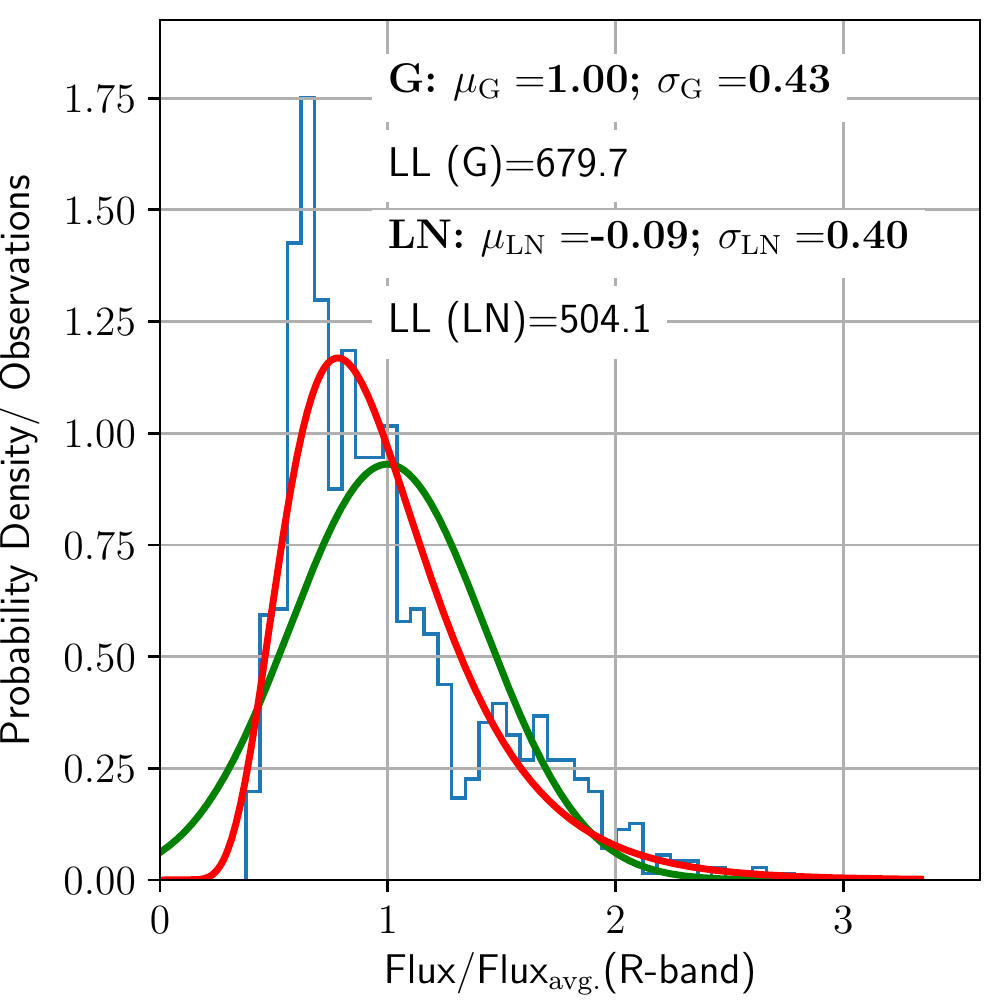}
\includegraphics[width=0.33\linewidth,height=5.0cm]{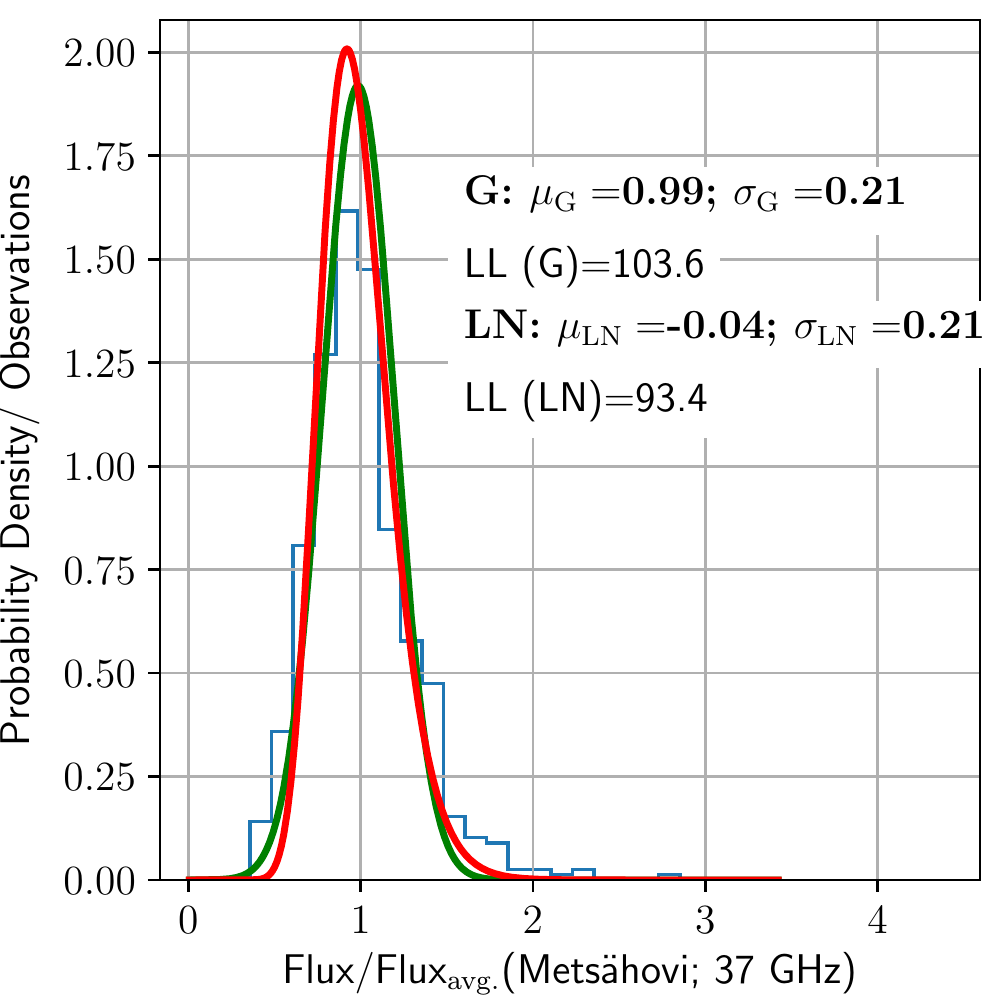}

\caption{Characterization of the MWL flux distributions with a log-likelihood fit. The X-axis shows the scaled flux and the Y-axis the probability density. The green and red lines represent the Gaussian and LogNormal functions for which the log-likelihood is minimum. For completeness, the flux histograms used in Fig. \ref{fig:Fhistogram} are also shown in blue. See text in Appendix  \ref{sec:histandML} for details.} 
\label{fig:FML}
\end{figure*}

In the log-likelihood fit, the log-likelihood function used for the Gaussian PDF, as a function of the parameters $\mu$ and $\sigma$, is given as:
%%%%%%%%%%%%%%%%%%%%%%%%%%%%%%%
\begin{equation}
L_g(f(F_{i},\Delta F_{i})|\sigma, \mu )
= -\frac{1}{2}\sum \left[ log(2\pi (\Delta F^2_{i}+\sigma^2)) + \frac{(F_i -\mu)^2}{\Delta F^2_i +\sigma^2} \right]
\end{equation}\label{eq:ML1}
In order to calculate the log-likelihood of the LogNormal distribution, we consider a grid with 3000 points (x$_{i}$), using a dynamic grid resolution, ranging from log(F$_\mathrm{min}$)-5 to log(F$_\mathrm{max}$)+5, where F$_\mathrm{min}$ and F$_\mathrm{max}$ are the minimum and maximum scaled fluxes in the corresponding energy bands. The exponential of the grid points (e$^{x_i}$) are then used for the defining the LogNormal PDF as a function of the parameters $\mu$ and $\sigma$, in the form given below:
%%%%%%%%%%%%%%%%%%%%%%%%%%%%%%%%
\begin{equation}
L^i_\mathrm{LN}(f(x_{i}|\sigma, \mu )
= -\frac{1}{\sqrt{2 \pi}} \frac{1}{e^{x_i} \sigma} exp\left[- \frac{(x_i-\mu)^2}{2\sigma^2} \right] 
\end{equation}\label{eq:ML2}
Next, we calculate the Gaussian probability $G^j_{i}(F_j, \Delta F_j)$ for each of the flux measurements with measured flux ($F_j$) and flux-error  ($\Delta F_j$) using the following equation (Eq. \ref{eq:ML3}3) at different grid-points (x$_{i}$).
\begin{equation}
G^j_{i}(F_j, \Delta F_j)
= -\frac{1}{\sqrt{2 \pi}} \frac{1}{\Delta F_j} exp\left[- \frac{(e^{x_i}-F_j)^2}{2\Delta F^2_j} \right]
\end{equation}\label{eq:ML3}
We obtain the log-likelihood by the convolution of these two terms and integrating over the grid-range. Finally, we minimize the log-likelihood and obtain the optimal parameters for $\mu$ and $\sigma$. The results are presented in Fig. \ref{fig:FML} where, for completeness, the flux histograms used in Fig. \ref{fig:Fhistogram} are also shown. By construction, the log-likelihood fit considers the flux uncertainties, and does not require to bin the data, both representing big advantages over the Chi-square fit. However, we note that log-likelihood fit applied here is very simple and generic because we are using the same PDF functions for all the energy bands. One could exploit the full potential of the log-likelihood method by using dedicated PDFs for each energy band, which would allow one to address the problem in a more efficient manner.  However, that would require introducing instrument response functions and physical models of emission into the PDFs, which is out of the scope of this work.

Table \ref{table:SScomparisonbetweenMethod} lists the preferred flux-distributions (Gaussian or LogNormal) from the Chi-square fit, log-likelihood fit and the flux profile methods. The flux distribution for the OVRO (15\,GHz) is not included in the table and in the figures because it has a bimodal shape due to the strong flare in 2012 (see Appendix \ref{sec:appendixB}). The entries marked with "*" denote cases where the preference is not clear, either because both options are roughly equally probable, or because the methods suffer from some caveats.
In the case of the Chi-square fit, this happens for FACT (E$_{th}\sim$0.7\,TeV), where the resulting Chi-square values show equally probable fits. In the case of the log-likelihood fit, this occurs for FACT (E$_{th}\sim$0.7\,TeV), \textit{Swift}-BAT ($15-50$\,keV) and X-ray in the $0.3-2$\,keV band. In the first two cases, the applicability of the PDF (Gaussian or LogNormal) suffers from the truncation of these two distributions at low flux values (given the limited sensitivity to measure low fluxes)\footnote{The Chi-square fit and the flux profile method are less sensitive to this effect because the fits are performed above the minimum flux F$_{min}$, and hence do not need to apply the entire distribution shape to the available data.}, and in the latter case, the resulting log-likelihood values are equal (within one unit) for both the functions.
In the case of the flux profile method, the preference is not clear for \textit{Swift}-BAT ($15-50$\,keV) because the chance probability (p; see Section \ref{sec:typicalstate} and Appendix \ref{sec:appendixB} for details) for a Gaussian distribution when the true distribution is a LogNormal is only 0.16. The table shows preference for the LogNormal distribution shape in all of the energy bands, apart from the X-rays in the $0.3-2$\,keV, $2-10$\,keV and the 37\,GHz radio band. The three methods prefer the Gaussian shape for the $0.3-2$\,keV (although the preference is not clear in the case of the log-likelihood method), while for the other two bands,
the Chi-square and log-likelihood fits prefer a LogNormal shape, while the flux profile method prefers a Gaussian shape.

\begin{table*}
\centering
\begin{tabular}{ c  c c c }  \hline
\hline
Energy-bands & Chi-square fit & Log-likelihood fit& Flux profile \\ \hline
VHE $\gamma$-rays ($> 0.2$\,TeV)      & LogNormal & LogNormal & LogNormal\\
VHE $\gamma$-rays (FACT; E$_\mathrm{th}\sim0.7$\,TeV) 
                                      & LogNormal$^{*}$ & LogNormal$^{*}$  & {LogNormal$^{*}$}\\
HE $\gamma$-rays (LAT; $> 0.3$\,GeV)  & LogNormal & LogNormal & LogNormal\\
X-ray (BAT; $15-50$\,keV)             & LogNormal & LogNormal$^{*}$ & LogNormal$^{*}$ \\
X-ray ($2-10$\,keV)                   & LogNormal & LogNormal & Gaussian\\
X-ray ($0.3-2$\,keV)                  & Gaussian & Gaussian$^{*}$  & Gaussian\\
Optical (R-band)                      & LogNormal & LogNormal & LogNormal\\
Radio (Mets{\"a}hovi; 37\,GHz)        & LogNormal & LogNormal & Gaussian \\
\hline %(N.A.)
\end{tabular}
\caption{The preferred flux-distributions based on the three methods namely Chi-square fit, log-likelihood fit and the flux profile method. Entries in the table that are marked with "$^{*}$" do not have a clear preference for Gaussian or LogNormal and are discussed in Appendix \ref{sec:histandML} . The flux distribution for OVRO (15\,GHz) has a bimodal shape, hence, it is not included in this comparison table. See Section \ref{sec:typicalstate} and text in Appendix \ref{sec:histandML} for details.}
\label{table:SScomparisonbetweenMethod}
\end{table*}

% Don't change these lines
\bsp	% typesetting comment
\label{lastpage}
\end{document}